\documentclass[a4paper,oneside,12pt,openright]{memoir}

\usepackage{amsmath}
\usepackage{amsfonts}
\usepackage{amssymb}
\usepackage{setspace}
\usepackage{bm}
\usepackage{indentfirst}
\usepackage[numbers]{natbib}
\usepackage{graphicx}
\usepackage{palatino}
\usepackage{oxfordthesis}
\usepackage{cancel}
\thetitle{The Zero-Turbulence Manifold in Fusion Plasmas}
\theauthor{Edmund Highcock}
\degreedate{April 2012}
\degree{Doctor of Philosophy}
\college{Merton College}

\usepackage{memhfixc}

\newcommand{\vth}[1][s]{\ensuremath{v_{\mathrm{th}_#1}}}

\newcommand{\pd}[3][]{\ensuremath{ \frac{\partial^{#1} #2} {\partial #3} } }
\newcommand{\gyror}[1]{\ensuremath{ {\left< #1 \right>}_{\bm{r}}}}

\newcommand{\gyroR}[2][s]{\ensuremath{{\left< #2 \right>}_{\bm{R}_{#1}}}}

\newcommand{\secref}[1]{Sec.\ \ref{#1}}

\newcommand{\uv}[1]{\bm{\hat{#1}}} % uv = unit vector
\newcommand{\Oi}{\Omega_{i}}

\newcommand{\fsr}{\gamma_{E}}
\newcommand{\tprim}{\kappa}

\newcommand{\gstwolnorm}{R_0} % Length normalisation
\newcommand{\gstworefspec}{i} % Reference species letter
\newcommand{\gstwofsrnorm}{} % Flow shear normalisation
\newcommand{\gstwoQnormflatbd}{n_\gstworefspec T_\gstworefspec v_{th\gstworefspec} \rho_{\gstworefspec}^2 / 2 \sqrt{2} \gstwolnorm^2 } % Heat flux normalisation for graphs
\newcommand{\pa}{_{\parallel}}
\newcommand{\pp}{_{\perp}}

\newcommand{\oov}[1]{\frac{1}{#1}}
\newcommand{\doubleunderline}[1]{\underline{\underline{#1}}}
\newcommand{\figref}[1]{Fig.~\ref{#1}}
\newcommand{\figrefs}[1]{Figs.~\ref{#1}}

\newcommand{\vct}[1]{\bm{#1}}

\newcommand{\pdt}[1]{\pd{#1}{t}} % partial wrt t
\newcommand{\dbd}[2]{\frac{d#1}{d#2}} % derivative
\newcommand{\ddt}[1]{\dbd{#1}{t}} % derivative wrt t

\newcommand{\lp}{\left(}
\newcommand{\rp}{\right)}

\newcommand{\lsq}{\left[}
\newcommand{\rsq}{\right]}

\newcommand{\md}[1]{\left| #1 \right|}

\newcommand{\btor}{B_{\phi}}
\newcommand{\bpol}{B_{\theta}}

\newcommand{\myfig}[4][5.5in]{
\begin{figure}[htpd]
	\centering
\includegraphics[width=#1]{#2}%
\caption[#3]{#4\label{#2}}%
\end{figure}
}
\newcommand{\myfiga}[5][4.5in]{
\begin{figure}[htpd]
	\centering
\includegraphics[width=#1]{#2}%
\caption[#4]{#5\label{#3}}%
\end{figure}
}

\newcommand{\prt}{\mathrm{Pr}_t}

\newcommand{\mypart}[1]{\part{\hspace{2pt}#1}}
\newcommand{\field}{\varphi}

\newcommand{\webonly}[1]{#1}

\newcommand{\eref}[1]{\eqref{#1}}

\newcommand{\Eqref}[1]{Equation \eref{#1}}

\newcommand{\figsref}[1]{figures \ref{#1}}
\newcommand{\Figref}[1]{Figure \ref{#1}}

\newcommand{\fl}{}

\newcommand{\dpanel}[2]{
\centerline{\includegraphics[width=3.0in]{#1}
\hskip0.1in
\includegraphics[width=3.0in]{#2}}
}

\newcommand{\bea}{\begin{eqnarray}}
\newcommand{\eea}{\end{eqnarray}}
\newcommand{\beq}{\begin{equation}}
\newcommand{\eeq}{\end{equation}}
\newcommand{\lt}{\left}
\newcommand{\rt}{\right}
\newcommand{\la}{\langle}
\newcommand{\ra}{\rangle}

\newcommand{\dd}{\partial}

\newcommand{\ephi}{\varphi}

\newcommand{\ee}{\epsilon}
\renewcommand{\Re}{\mathrm{Re}}
\renewcommand{\Im}{\mathrm{Im}}
\newcommand{\Zfn}{\mathcal{Z}}

\newcommand{\bi}[1]{\mathbf{#1}}
\newcommand{\vr}{\bi{r}}
\newcommand{\vR}{\bi{R}}
\newcommand{\vk}{\bi{k}}

\newcommand{\vE}{\bi{E}}

\newcommand{\tphi}{\tilde\ephi}
\newcommand{\vB}{\bi{B}}

\newcommand{\avgR}[1]{\la#1\ra_{\vR}}
\newcommand{\avgr}[1]{\la#1\ra_{\vr}}

\newcommand{\kpar}{k_\parallel}
\newcommand{\kperp}{k_\perp}

\newcommand{\LB}{L_s}

\newcommand{\LT}{L_T}
\newcommand{\Ln}{L_n}
\newcommand{\omn}{\omega_{*}}

\newcommand{\zz}{\bar{\omega}}
\newcommand{\gbar}{\bar{\gamma}}
\newcommand{\zn}{\zz_*}

\newcommand{\zS}{\zz_S}
\newcommand{\eS}{\eta_S}
\newcommand{\tz}{\tilde{\omega}}
\newcommand{\tg}{\tilde{\gamma}}

\newcommand{\nuii}{\nu_{ii}}

\newcommand{\vthi}{v_{\mathrm{th}i}}
\newcommand{\rhoi}{\rho_i}

\newcommand{\vw}{\bi{v}}

\newcommand{\wperp}{v_\perp}
\newcommand{\wpar}{v_\parallel}

\newcommand{\tbar}{\Delta t}

\newcommand{\tpvg}{t_0^\mathrm{(PVG)}}

\newcommand{\Nmax}{N_\mathrm{max}}

\usepackage[
pdftitle={\oxfthetitle},
pdfauthor={\oxftheauthor},
pdfsubject={Thesis for the Degree of \oxfdegree, \oxfdegreedate},
pdfborder=0,
bookmarks=true,
bookmarksnumbered=true,
bookmarksopen=true,
bookmarksopenlevel=1,
plainpages=false,
pdfpagelabels=true
]{hyperref}

\begin{document}

\titlepage

\frontmatter

\begin{dedication}
To my family
\end{dedication}

\begin{abstract}
The transport of heat that results from turbulence is a major factor limiting
%the ion temperature gradient, and hence the core temperature and
the temperature gradient, and thus
the performance, of fusion devices.
We use nonlinear simulations to
%calculate the effect of a toroidal equilibrium scale sheared flow on turbulence within a toroidal fusion device.
%We demonstrate that such a flow
show that a toroidal equilibrium scale sheared flow
can completely suppress the turbulence across a wide range of flow gradient and temperature gradient values.
We demonstrate the existence of a bifurcation across this range whereby the plasma may transition from a low flow gradient and temperature gradient state to a higher flow gradient and temperature gradient state.
We show further that the maximum temperature gradient that can be reached by such a transition is limited by the existence, at high flow gradient, of subcritical turbulence driven by the parallel velocity gradient (PVG).
 We use linear simulations and analytic calculations to examine the properties of the transiently growing modes which give rise to this subcritical turbulence,
and conclude that there may be a critical value of the ratio of the PVG to the suppressing perpendicular gradient of the velocity 
(in a tokamak this ratio is equal to $q/\epsilon $ where $q $ is the magnetic safety factor and $\epsilon$ the inverse aspect ratio) below which the PVG is unable to drive subcritical turbulence.
In light of this, we use nonlinear simulations to calculate, as a function of three parameters (the perpendicular flow shear, $q/\epsilon $ and the temperature gradient),
the surface within that parameter space which divides the regions where turbulence can and cannot be sustained: the zero-turbulence manifold.
We are unable to conclude that there is in fact a critical value of $q/\epsilon $ below which PVG-driven turbulence is eliminated.
Nevertheless, we demonstrate that at low values of $q/\epsilon $, the maximum critical temperature gradient that can be reached without generating turbulence (and thus, we infer, the maximum temperature gradient that could be reached in the transport bifurcation) is dramatically increased.
Thus, we anticipate that a fusion device for which, across a significant portion of the minor radius, the magnetic shear is low, the ratio $q/\epsilon $ is low and the toroidal flow shear is strong, will achieve high levels of energy confinement and thus high performance.
\end{abstract}

\tableofcontents

\listoffigures

\begin{acknowledgements}
Reaching the end of nearly twenty-two years of formal education is a strange moment, producing a feeling akin to disbelief, but also of immense gratitude for all that I have been given in that time.

Beginning, like David Copperfield, at the beginning, I start with my parents. From the very first they have provided their children with an atmosphere of the deepest love and trust, but one also pervaded with a love of knowledge: literature, music, history and science, that has remained with me to this day and shaped everything that I have done.
To my sisters, my oldest comrades, I extend my sincere thanks for their love, and for their never-ending imagination and sense of adventure, which made my childhood the perfect place to start from.

From my school I would like in particular to thank my teachers David Rawkins, Stephen Oliver, and Fintan O'Reilly, who drove their pupils forward with such relentless energy and enthusiasm, and my schoolmates Hisashi Arawaka and Mark Cassin for their rivalry as well as their friendship.

There are too many inspirational and brilliant lecturers and teachers from my time at Cambridge to count, but I would like to thank my friends at Pembroke who made it so unforgettable, and with whom I hope to be friends to my dying day.
From my course I would like in particular to thank Jonty Lovat and Kath Toney, who weathered the dry hard early years of an undergraduate physics course with a great deal of humour and optimism.

To begin a Ph.D. is to realise that despite what you thought to the contrary, you know nothing. Fortunately, there were many great people who made starting again completely bearable. From my group at Imperial, there were Nuno Loureiro and Tarek Yousef, who provided a great deal of humour and enthusiasm; from further afield Colin Roach my co-supervisor, who has been unfailingly helpful and supportive, Greg Hammett and Bill Dorland, who fly in and out with bucketfuls of ideas, and Steve Cowley, who drops in from time to time to knock us into shape.
From Imperial I would also like to thank Vasa Curcin and the other wardens of Piccadilly Court who made me at home there.

From Oxford I would like to thank all those at Merton who have been such good friends and companions and all the members of my group, Greg Colyer, Young-chul Kim, Alfred Mallet, Joseph Parker and Alessandro Zocco, who have been supportive colleagues and friends. Special mention must be made of Felix Parra (and not just for those memorable margaritas in Los Angeles) and of Michael Barnes, who has been a good friend and extraordinarily patient with the myriad of questions I have asked him about GS2. Last, but not least, I would like to thank Ian Abel, who from the beginning has suffered the slings and arrows of outrageous Ph.D. fortune with me, who has been my office mate and my friend and an excellent political sparring partner to boot.

With regard to my Ph.D., I own my greatest debt to one person, my supervisor Dr. Alexander Schekochihin. 
All those who know him will recognise the description that follows.
As a supervisor he has provided the perfect mix of enthusiasm for physical enquiry, acerbic and forceful criticism, patience and motivation which makes the experience of a Ph.D. what it should be. As leader of our group in Oxford, and as chief organiser of our extended group in exile, he has been in large part responsible for the excellence, the drive and the energy which has surrounded his students. As a friend, barring his ruinous taste in restaurants, he has been responsible for many good and memorable evenings in many parts of the world.

And now it remains to thank one person, Amber, whose love, humour, spirit, support and companionship have made a profound difference to my life, particularly in the Sisyphean final months of my Ph.D., and to whom I owe much more than can be expressed here. I hope that in the joyful years ahead I will be able to repay her in full.
\end{acknowledgements}

\clearpage

\mainmatter

\mypart{Introduction}
\chapter{Fusion, Turbulence and Sheared Flows}
\section{Fusion Devices}

If one could persuade a nucleus of deuterium
%(heavy hydrogen)
to collide with a nucleus of tritium at sufficient velocity, they would
fuse together to produce helium, a neutron and 17.6 MeV of energy. Of this
energy 3.5 MeV would be given to the helium nucleus, and 14.1 MeV would be given
to the neutron \cite{rebhan2002thermonuclear}.

If one could convince around $10^{21}$ such pairs of nuclei to collide per second
(that is, a total weight of just 10 milligrams of fuel per second),
one could generate a gigawatt of power.\footnote{
Assuming a reactor efficiency of 30\% \cite{maisonnier2007power}.
}
A mere 100 tonnes per year of fuel would power the United Kingdom,\footnote{Based on gross UK energy consumption in 2010 of around $220$ million tonnes of oil equivalent (Mtoe), with 1 Mtoe equivalent to approximately 42 GJ  \cite{uknationalstatistics}.}
and 3500 tonnes per year the whole world.\footnote{
Based on world energy consumption of 8400 Mtoe in 2008 \cite{internationalenergyagency}.
}

The only product of this reaction is helium, a harmless and useful gas.
The reaction produces neither carbon dioxide 
nor any of the radioactive byproducts of a nuclear fission power station.
Although a small proportion of the vessel surrounding the reaction itself would become radioactive (at a very low level compared to the products of a fission reaction)
over the life of the power station,
a careful choice of design would ensure that
the radioactivity of all but a negligible portion of that material would decline to safe levels
within 50 years of the power station being closed \cite{cook2001safety}.
Of the two fuels required, deuterium is abundant in seawater,
and tritium can be bred from lithium (making use of the neutron)
which is also abundant in seawater \cite{ongena2010prospects}.

This, then, is the case for fusion:
a method of generating power which would require
a mere 3.5 kilotonnes of fuel to power the entire world for a year,
which produces no carbon dioxide and no radioactive waste.
The fact that this source of energy still remains,
after more than sixty years of research,
beyond the reach of humankind,
is owing to the problem of confinement: the problem of containing the fuel. 

%Sufficient velocity, as described above, means something of the order of 1000 kilometres per second.
In order to achieve the rate of $10^{21}$ reactions per second needed to generate a gigawatt of power,
16 milligrams of fuel would need to be heated up (in a reaction volume of 10 cubic metres) to 7 million $ ^{\circ} $C.\footnote{
Assuming a reaction cross-section of $10^{-21}\mathrm{m}^{3}\mathrm{s}^{-1}$ \cite{rebhan2002thermonuclear} and a core density of $10^{21}\mathrm{m}^{-3}$
\cite{maisonnier2007power}.}
At such temperatures, no material known would be sufficient
to confine the fuel directly.
Therefore, another way of confining the fuel
--- which at such temperatures becomes a plasma ---
must be found.

The most promising method to date is the use of a magnetic field.
The fuel particles in a fusion plasma are fully ionised;
as a result of their charge the electric and magnetic fields exert a force on them according to the Lorentz force law:\footnote{
Except when stated otherwise, Gaussian units are used throughout this work.} 

\begin{equation}
	\vct{F} \equiv Ze\left( \vct{E} + \frac{1}{c}\vct{v}\times\vct{B} \right),
	\label{eqn:lorentzforcelaw}
\end{equation}
where $Ze$ is the charge of the particle, $\vct{E}$ is the electric field, $c$ is the speed of light, $\vct{v}$ is the velocity of the particle and $\vct{B}$ is the magnetic field. 
A single particle in a strong magnetic field with no electric field is constrained by this law to move helically along a magnetic field line (see \figref{helical}(a)).
It cannot move across it.
This means that if one could design a set of magnetic field lines that formed a closed surface, the field lines would act like the bars of the cage 
--- the particle would never be able to cross that surface and escape.
A well-known theorem \cite{alexandroff1935topologie} states that the only
practicable closed surface on which a set of magnetic field lines can lie (without ever crossing the surface) is a torus.
Thus, a magnetically confined fusion device principally comprises 
a strong equilibrium magnetic field
whose field lines lie on a set of nested toroidal surfaces called flux surfaces,
each of which forms a cage barring the way of a hot particle attempting to escape confinement (\figref{toroidalcage}).
There are two principle types of such devices: tokamaks and stellarators.
Tokamaks are devices which are roughly axisymmetric around the centre of the torus, in which the toroidal magnetic field
(the component of the field around the major circle of the torus)
is generated by external coils,
and the poloidal field (the component of the field around the minor circle of the torus) is generated by a current which flows toroidally about the plasma \cite{todd1993build, wesson2004tokamaks}.
Stellarators are non-axisymmetric devices whose equilibrium field is generated 
almost entirely by external coils \cite{hirsch2008major, motojima2003recent}.
%Their geometry is in general significantly more complicated.
This work focuses exclusively upon tokamaks.

\myfig[2.5in]{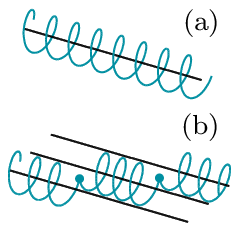}{Helical Motion}{(a) The helical path taken by a charged particle along a straight magnetic field line. (b) The path of a particle moving across magnetic field lines via collisions with other particles at the marked points (an illustration only since head-on collisions as shown would be vanishingly rare; see Footnote \ref{note:collisiondef}).}
\myfig[3.5in]{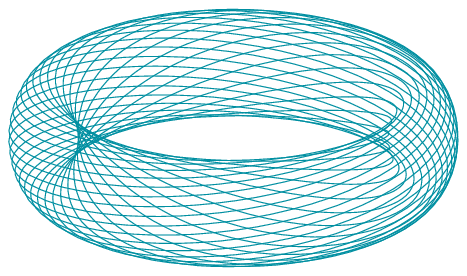}{The Toroidal Cage}{The toroidal cage of magnetic field lines which confines the plasma.}

The absence of a working fusion device based upon the principle of a magnetic cage is clear evidence that this cage is far from perfect.
Two effects which cause a lessening of the confinement provided by the magnetic field are the effect of collisions and the effect of drifts.
In a typical fusion device there is not one particle in the vessel, but of the order of $10^{22}$.
These particles collide with each other frequently: about $4\times10^{13}$ such collisions\footnote{\label{note:collisiondef}A collision here means a sequence of interactions which will change the momentum of the particle by order unity.} occur a second per cubic centimetre in the scenario described above (
%a number which depends strongly on the density and temperature, 
see equation \eqref{nuiidef}). 
A collision between two particles can cause them
to cross the magnetic field lines,
and so in this way, through multiple collisions,
a particle can eventually escape (see \figref{helical}(b)).
This mode of transport of particles
(and with them heat and energy) across magnetic field lines
is known as classical transport.

Classical transport itself is a minor problem.
%If classical collisional transport was the only type of transport,
%a fusion device only 15 cm in minor radius could achieve ignition (that is, support a self-sustaining reaction with net power output).\footnote{
%This simple estimate assumes that fusion reactions happen in the central 10\% by radius of the plasma, that the temperature and density gradients are flat within this region and that the energy of the helium products is deposited there.
%We further assume a magnetic field strength of 1 Tesla and the density and temperature  in the central region given above.}
The effect of collisions is made far worse by the existence of drifts.
Drifts, which can result from the electric field (the $\vct{E}\times\vct{B}$ drift),
and the shape of the magnetic field (the gradient and curvature drifts),
cause a particle to spiral slowly across magnetic field lines \cite{hastie1993plasma}.
Drifts by themselves are not fatal.
A fusion device can be designed so that on average a particle which drifts out somewhat will eventually drift back \cite{blank2010guiding}.
It is the combination of drifts and collisions that is deleterious:
a particle can drift some way out,
collide with another particle and enter a new orbit
which allows that particle to drift further.
Eventually the particle escapes.
The transport which results from the combination of drifts and collisions is known as neoclassical transport. Neoclassical transport has been extensively studied \cite{helander2002ctm}.
It is possible to directly calculate for many simpler fusion devices a good estimate of how large
neoclassical transport is \cite{hintonwong1985nit};
for the case of tokamaks, it is found that, although much larger than classical transport, neoclassical transport is not fatal: even including its effects, a fusion device only approximately 1 m in minor radius would be viable \cite{galambos1995commercial}. 

%\footnote{To make this estimate we use the same assumptions as previously, but take the neoclassical estimate of the thermal diffusivity given by \eqref{chindef}.}

The reason for the failure to produce a working fusion device lies in an altogether different phenomenon: turbulence.
\section{Turbulence}

%According to Schekochihin et. al. \cite{schekcrete},
%\begin{quote}
%\emph{turbulence is multiscale disorder: we tend to say that we are dealing with a turbulent 
%system if we have detected (measured, observed, simulated, intuited) chaotic fluctuations 
%of some field(s) over a broad range of scales}.
%\end{quote}
Turbulence, in essence, is formed of fluctuations: fluctuations in the density and temperature; fluctuations in the electric and magnetic fields.
Fluctuations in the magnetic field, by bending the bars of the magnetic cage, 
can cause very large degradations in the confinement of the plasma; 
in the core of many fusion devices, however, they are fortunately kept to a minimum.\footnote{At least currently, see e.g. Ref.~\cite{colas1998internal}, for the simple reason that the particle pressure is in general low compared to the magnetic pressure: this may change as this ratio is increased ever higher in future devices.}
Fluctuations in electric field can lead to eddies which lie across the magnetic field lines  and 
which can convect heat, momentum and particles out of the plasma.
The rate at which heat can escape via the turbulence is an
order of magnitude greater than that due to neoclassical transport.
It is the heat loss due to turbulence that has necessitated
the building of ever larger fusion devices,
in order to have a greater distance from the core to the edge of the plasma
(and thus a greater distance for the heat to travel to escape).
The large sizes made necessary by turbulence create a raft of new challenges,
particularly that of finding materials that can withstand the enormous heat loads, neutron fluxes and mechanical stresses that larger devices entail.
It seems clear, therefore, that a vital task is to find ways of reducing or eliminating this turbulence.

The turbulence in a fusion device is driven by the steep gradient in pressure between the core and the edge.
Fluctuations in electric field are driven unstable by the pressure gradient
and the resulting eddies grow to large enough amplitude to interact
with each other, leading to turbulence. 

The pressure gradient can be split into the gradients of
temperature and density of
the ions and electrons.
The temperature gradients of
both the ions and the electrons
and the density gradient of the electrons
can all drive turbulence \cite{horton1980fluid,waltz1988three,fonck1989ion,wootton1990fluctuations,cowley1991considerations,kotschenreuther1995quantitative,carreras1997progress,dimits:969,dorlandETG,jenko:1904,dannert2005gyrokinetic}	 
but the temperature gradient of the ions in particular
is a strong driver of turbulence and 
eliminating the turbulent transport that results from
the ion temperature gradient is the main focus of this work.
The turbulent transport be eliminated both by preventing the growth of fluctuations, and by shearing apart the large eddies that result from the turbulence.
Fortunately, there is a phenomenon which can achieve both:
a sheared flow perpendicular to the magnetic field lines.

\section{Sheared Flows}
\label{sec:shearedflows}

A significant body of experimental work has shown that an equilibrium scale flow that
runs perpendicular to the magnetic field lines
which has a steep enough gradient across the magnetic flux surfaces can reduce,
or even eliminate, turbulence \cite{burrell1997effects}.
Such flows are often referred to as $\vct{E}\times\vct{B}$ flows as they can be thought of as a drift which results from a strong equilibrium radial electric field.
They can be found both in the edge
(where they are believed \cite{wagner1984development,putterich2009evidence,synakowski1999comparative,oost2003turbulent} to add to the formation of
the edge transport barriers responsible for the high confinement mode \cite{wagner1984regime})
and in the core, where they can lead to a general reduction in turbulence \cite{akers2003transport}
or to the formation of internal transport barriers,
regions in the core with strongly sheared flows and excellent confinement properties \cite{connor2004itbreview, wolf2003internal, synakowski1997local}.
In this work we consider sheared flows within the core of a fusion device.

Numerical work has confirmed the effectiveness of perpendicular sheared flows in reducing both the levels of turbulence and the associated heat loss \cite{waltz1994toroidal,waltz1997gyro,dimits2001parameter, kinsey2005flowshear, camenen2009impact, roach2009gss, barnes2011turbulent, highcock2010transport}.
However, if there is a component of the flow parallel to the magnetic field, its gradient can drive another instability, in addition to the ion temperature gradient instability, known as the parallel velocity gradient (PVG) instability \cite{catto1973parallel}.
As shown in Refs.~\cite{dimits2001parameter, kinsey2005flowshear, roach2009gss}, the PVG instability can start to drive turbulence at higher flow gradients.
This matters because in general, the equilibrium flow within a fusion device is toroidal,
and the angle between the flow and the magnetic field, and hence the ratio of the destabilising parallel flow shear and the stabilising perpendicular flow shear is fixed by the magnetic geometry,
in fact by the ratio $\btor/\bpol $ where $\btor $ is the toroidal magnetic field and $\bpol $ is the poloidal magnetic field. This ratio is typically much greater than one.
%In Refs.~\cite{kinsey2005flowshear} and \cite{roach2009gss}, it was demonstrated that the PVG could prevent the flow shear from quenching the turbulence.
Nonetheless, Barnes et. al. \cite{barnes2011turbulent} showed that even at $\btor/\bpol=7.8 $, i.e., with the parallel velocity gradient nearly 8 times bigger than the perpendicular velocity gradient,
it was still possible to quench turbulence completely,
provided that the flow shear was large enough 
(and the ion temperature gradient was moderate).
\myfig{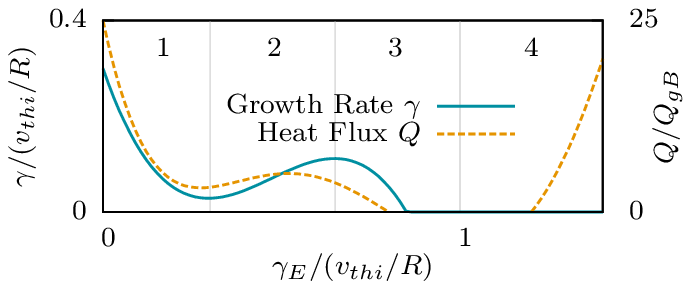}{A Cartoon of the Effects of Perpendicular Flow Shear}{A cartoon of the effects of perpendicular flow shear, based on data from Ref.~\cite{barnes2011turbulent}, showing the growth rate $\gamma$ and the heat flux $Q$ vs the perpendicular flow shear $\gamma_E$. The behaviour may be loosely divided into four regions (see text). (The normalising quantities are the ion thermal velocity $v_{thi}$, the tokamak major radius $R$ and the gyro-Bohm estimate of the heat flux $Q_{gB}$ (see \eqref{gpdefs}).}

The situation can be thought of as a three-way competition between the ion temperature gradient which drives turbulence, the perpendicular flow shear which suppresses turbulence, and the parallel flow shear which drives turbulence once again.
Considering Fig. 2 of Ref.~\cite{barnes2011turbulent}, of which we provide a cartoon in \figref{cartoon}, we may divide the interaction of the these effects, as a function of the gradient of the flow, into four regions.
At zero flow gradient, an ion temperature gradient above critical drives strong turbulence, leading to the large heat fluxes at the left of the figure. 
In the first region, as the flow shear increases, the perpendicular flow shear reduces the levels of turbulence dramatically;
in the second region the parallel velocity gradient causes a slight resurgence of the turbulence;
in the third region the perpendicular flow gradient once again dominates, and can, at low temperature gradients, completely suppress the turbulence;
at still higher flow gradients, the fourth region, the PVG is dominant and the turbulence is rekindled.

At lower flow shears, the strength of the turbulence is strongly affected by the growth rate of the underlying instability,
which in turn is strongly affected by the flow shear (\figref{cartoon}, after Fig. 1 of Ref.~\cite{barnes2011turbulent}, see also Ref.~\cite{artun1993integral}).
At zero flow shear, the perturbations grow strongly for temperature gradients above the linear critical temperature gradient for instability.
As the flow shear increases from zero, the growth rate drops precipitously owing to shear suppression in the first region, and then (at low temperature gradients) rises owing to the PVG in the second.
In the third region the behaviour of the growth rate diverges from that of the turbulence; we find that for all temperature gradients  the growth rate drops completely to zero. It remains zero even in the fourth region, where the turbulence starts growing again.

There are two points of great interest in the behaviour of the linear growth rate. The first is that it drops to zero at all temperature gradients in the third region. This behaviour is explained by Newton et. al. \cite{newton2010understanding} as follows. At low flow shears, where there is a finite magnetic shear or twist in the magnetic field lines, it is possible for the magnetic shear to exactly cancel out the flow shear, provided the perturbations move along the magnetic field lines at a speed proportional to the flow shear divided by the magnetic shear. As the flow shear increases, this speed becomes higher than the sound speed in the plasma and it is no longer possible for the perturbations to move quickly enough; they are sheared apart by the flow and die away.

The second point of interest is that it is possible to have strong turbulence even when the growth rate of the instability is zero. This is possible because even though the perturbations eventually die away, they can grow transiently for long enough for their energy to be scattered by nonlinear action into other perturbations which are still growing (as illustrated in a simple matrix model by  Baggett \cite{baggett1995mostly}). Turbulence which exists where there are no linear instabilities is known as \emph{subcritical turbulence} and will be discussed extensively in this work.

\section{The Zero Magnetic Shear Limit}

In this work we consider the effects of flow shear when the magnetic shear is zero.
When the magnetic shear is zero the speed with which a perturbation would need to move in order to grow without limit is infinite for all non-zero values of the flow shear.
Thus, all perturbations grow only transiently before decaying.
We find that the first and second regions as described above disappear and that all turbulence is subcritical.
We find also that in the absence of growing modes the turbulent transport levels for any given flow or temperature gradient are always lower than at the finite value of magnetic shear ($\hat{s}=0.8$) considered in Ref.~\cite{barnes2011turbulent}.
We find, in addition, that at zero magnetic shear conditions are favourable for a bifurcation from low to high flow gradients, and from low to high temperature gradients, of a sort that may lead to the formation of a transport barrier.
This agrees with the experimental observation that internal transport barriers tend to form in regions of low magnetic shear \cite{sips1998operation,vries2009internal}.

 The rest of this work is organised as follows.
 In Part \ref{sec:modellingturbulence}, we discuss the way that turbulence and the effects of flow shear are modelled.
 In Part \ref{sec:bifurcationtoareducedtransportstate} we calculate the effect of the flow shear on the turbulent transport using nonlinear gyrokinetic simulations.
We demonstrate the existence firstly
of a large range of values of the flow shear
and the ion temperature gradient
where the turbulence is completely suppressed, and secondly of a transport bifurcation from low to high flow gradients
and from low to high temperature gradients.
We present a parametric model of the turbulent transport and use it to investigate further properties of the high-gradient, reduced-transport state reached by the bifurcation,
as well as the values of the heat and momentum fluxes (which, assuming steady state, are the control parameters of a real fusion device) for which we expect such bifurcations to occur.

In Part \ref{sec:subcriticalturbulence} we investigate the phenomenon of subcritical turbulence in more detail.
 We start by using a linear model in simplified geometry to calculate the way the strength of the transiently growing modes is affected by the parameters of the system. We  identify three key parameters, the ion temperature gradient, the flow shear and the ratio $q/\epsilon $, which strongly affect the strength of these transient modes.
 We then use nonlinear simulations to map out, as a function of these three parameters, the \emph{zero-turbulence manifold}, the dividing surface between the area of that parameter space where the turbulence can exist and the area where it cannot.
If we cast the problem of improving confinement as that of increasing the ion temperature gradient (and hence the core fuel temperature) as much as possible without generating turbulence, we find that the optimum strategy is firstly to minimise the ratio $q/\epsilon $ (as might be expected), and then to increase the toroidal flow shear to an optimum value which we calculate.

In the final section of this work, Part \ref{sec:summaryanddiscussion}, we present our conclusions and our suggestions for further work.

\mypart{Modelling the Turbulence: A Concise Description of \texttt{GS2}}
\label{sec:modellingturbulence}
%\chapter{Equations and Numerics}
\chapter{Modelling the Turbulence}
\section{Introduction}

The aim of modelling plasma turbulence in a fusion device is to determine, for a given set of plasma properties, for example a given temperature gradient, how much heat, momentum or how many particles escape through the plasma as a result of that turbulence.
The task of modelling the turbulence can be divided into two parts: deriving an equation which describes the way the turbulence evolves in time, and solving that equation. In general, the equations that govern plasma turbulence cannot be solved analytically and must be solved numerically.

In this work we use the gyrokinetic equation to describe the evolution of the turbulence.  The gyrokinetic equation is derived from the Vlasov-Landau equation, which describes the evolution of a general plasma.
The gyrokinetic equation is specifically designed for modelling the particular phenomena of microturbulence, for example the turbulence that is generated by the ion temperature gradient in a tokamak. It uses various properties of the microturbulence to simplify the Vlasov-Landau equation and make it easier to solve, in particular reducing the number of dimensions in the problem from 6 to 5. These properties are, in summary, that it evolves on timescales that are much faster than the evolution of the plasma equilibrium, but much slower than the plasma frequency or the gyration of the particles about the magnetic field lines;
that the turbulent structures are much smaller than the scale on which the equilibrium changes,
and that the turbulent structures are strongly elongated along the magnetic field lines.
The derivation of the gyrokinetic equation based on these and other assumptions has been extensively covered elsewhere \cite{flowtome1}, and is only covered in summary here in Section \ref{sec:thegyrokineticequation}.

The tool which will be used to solve the gyrokinetic equation and calculate the behaviour of the turbulence is the code \texttt{GS2}.
We will use \texttt{GS2} to calculate the behaviour of the turbulence within a small region of the fusion device called a flux tube.
To do this we make a set of further simplifications known as the local approximation.
In particular, we assume that the equilibrium density, temperature and flow gradients are constant across the flux tube.

\section{The Gyrokinetic Equation}
\label{sec:thegyrokineticequation}

In order to model the turbulence, it is necessary to have an equation which describes how the turbulence evolves in time.
In this work we will use the gyrokinetic equation \cite{frieman1982nge} in the high-flow low-Mach limit.
The derivation of this equation, including all the assumptions and approximations that go into it has been extensively covered, in particular by Abel et. al. \cite{flowtome1} and Sugama and Horton \cite{sugama1998neg}.
Here, we only provide a brief summary of the derivation, and refer our readers to Ref.~\cite{flowtome1} and references therein for a comprehensive description.

%We use the gyrokinetic approximation \cite{frieman1982nge}
%to calculate the way in which the equilibrium gradients of the temperature and flow 
%affect the transport of heat and momentum by turbulent fluctuations.
%For a detailed description of high-flow gyrokinetics, 
%the reader is referred to \cite{flowtome1,sugama1998neg,artun1992gai} and references therein. 
%Here a brief summary is presented.

We consider a toroidal plasma with an axisymmetric equilibrium magnetic field $\vct{B}$. 
The field lines lie on closed flux surfaces,
which can be labelled by the poloidal magnetic flux $\psi$
contained within each surface
(see Section \ref{sec:geometry} for a much more detailed discussion of 
the equilibrium magnetic field). 
The magnetic field can be written as
\begin{equation}
%$\vct{B} = \nabla \phi \times \nabla \psi + I \nabla \phi$,
\vct{B} = \nabla \phi \times \nabla \psi + I \nabla \phi,
\label{eqn:Bdef}
\end{equation}
where \(\phi\) is the toroidal angle,  
$R$ is the radial coordinate measured from the central axis of the torus (the major radius),
and $I/R \equiv \btor$ is the toroidal magnetic field. 
Owing to the fast motion of the particles along the magnetic field lines,
equilibrium quantities such as the density
and temperature are constant on each flux surface.\footnote{
For the density, this is only true if the toroidal flow is subsonic,
which is what we assume below, see Eq. \ref{flowmagordering}.} 
%In the current study, which considers the Cyclone Base Case regime \cite{dimits:969}, 
%the magnetic flux surfaces are concentric and circular in cross-section, 
%so that $\nabla \psi = \bpol R \nabla r$, 
%where $r$ is the minor radius of the torus and $\bpol$ the poloidal magnetic field. 
%Thus, $r$ may be used as the flux label, and the magnetic field may be written as $\vct{B} = \bpol R \nabla \phi \times \nabla r + \btor R \nabla \phi$.

We allow an equilibrium plasma flow $\vct{u}$, 
of the same order as the ion thermal velocity 
\begin{equation}
	v_{thi} = \sqrt{\frac{2T_i}{m_i}}, 
	\label{vthidef}
\end{equation}
where $T_i$ is the temperature of the ions and $m_i$ is their mass.\footnote{
The definition of the thermal velocity 
given in \eqref{vthidef} is chosen in accordance with the analytical works
\cite{flowtome1,hintonwong1985nit} cited in this thesis;
maintaining correspondence with \cite{barnes2011turbulent,highcock2010transport},
where the ion thermal velocity was defined as $\sqrt{T_i/m_i}$,
results in various factors of $\sqrt{2}$ in the normalisations 
of the output from simulations.}
It can be shown that such a flow is toroidal, 
as any poloidal component will be quickly damped \cite{catto1987ion,hintonwong1985nit}. 
The flow can then be expressed as \(\vct{u}= \omega R^2 \nabla \phi\), 
where \(\omega\) is the angular velocity of the flow (which must be constant on a flux surface).

The state of each species $s$ of charged particles can be described by its distribution function, $f_s$, which is, very loosely, the probability that there is a particle at a given location $\vct{x}$ and travelling at a given speed $\vct{v}$. Thus, $f_s$ is a function with seven parameters; three spatial coordinates, three velocity coordinates, and time. 
The evolution of $f_s$ is given by the Vlasov-Landau equation:

\begin{equation}
	\pdt{f_s} + \vct{v}\cdot\nabla f_s + \ddt{\vct{v}}\cdot\pd{f_s}{\vct{v}} = C \lsq f_s \rsq, 
	\label{<++>}
\end{equation}
where  $C$, the collision operator, comprises the effects of collisions between particles. 
The Vlasov-Landau equation could, in theory, be solved directly.
However, since it is a six dimensional equation, involving scales which range from the tiny orbit size of particles gyrating around the magnetic field lines to the size of the device, its direct solution for a fusion device is impossible. 
In order to proceed, we make certain assumptions which lead to an equation which is in fact soluble: the gyrokinetic equation.

Firstly, the Vlasov-Landau equation can be expanded by splitting $f_s$ into an equilibrium part $F_s$ and a perturbation $\delta f_s$: 
\begin{equation}
f_s = F_s + \delta f_s.	
\end{equation}
The latter is further split into averaged and fluctuating parts: $\delta f_s = \left< \delta f_s \right> + \delta \tilde{f}_s$, so that

\begin{equation}
	f_s = F_s + \left< \delta f_s \right> + \delta \tilde{f_s} 
	\label{deltafsdef}
\end{equation}
where the angle brackets denote a spatial and temporal average over fluctuations \cite{flowtome1}.
Secondly, we assume that:
\begin{itemize}
	\item the perturbations to the distribution function (and the background electromagnetic fields) are small compared to the equilibrium values of those quantities;
	\item owing to the fact that particles can travel quickly along field lines but only slowly across them via collisions or drifts, the perturbations vary slowly along the field lines but quickly across them;
	\item the perturbations vary very quickly compared to the rate at which the equilibrium changes but slowly compared to the speed with which the particles gyrate around the magnetic field lines;
	\item the spatial scale of the perturbations is small compared to the scale of variation of equilibrium quantities.
\end{itemize}
These assumptions give rise to the gyrokinetic ordering \cite{frieman1982nge,flowtome1} which may be written: 

\begin{eqnarray}
	\frac{\delta f_s}{F_s} \sim \frac{k\pa}{k\pp} \sim \frac{\gamma}{\Oi} \sim  \frac{\rho_i}{L} \equiv \epsilon_{\mathrm{gk}} \ll 1, 
	\label{orderings}
\end{eqnarray}
where
\(k\pa\) and \(k\pp\) are the typical parallel and perpendicular wavenumbers of the fluctuations,
\(\gamma\) is the growth rate of the fluctuations,
\(\Oi\) and $\rho_i$ are the gyrofrequency and Larmor radius of the ions, respectively,
\(L\) is scale length of the variation of $F_s$, 
and $\epsilon_{\mathrm{gk}}$, the gyrokinetic expansion parameter, is small.
Using the gyrokinetic ordering, it is then possible to show\footnote{
With the additional assumption
that the like-particle collision frequency $\nu_{ss}$
satisfies $\epsilon_{\mathrm{gk}}^2 \gamma \ll \nu_{ss} \lesssim \gamma$.}
that to lowest order,
$F_s$ is a local Maxwellian in the frame of the equilibrium flow $\vct{u}$:
\begin{equation}
  F_s (\psi, \varepsilon_s) =	 n_s (\psi) \left( \frac{m_s}{2\pi T_s (\psi)} \right)^{3/2} e^{-\varepsilon_s/T_s},
	\label{eqn:Fsdef}
\end{equation}
where
$\varepsilon_s$ is the energy
of the particle,
$m_s$ is the mass of species $s$, and $n_s$ and $T_s$ are the equilibrium density and temperature of species $s$, respectively.
It may be further shown that to first order in $\epsilon_{\mathrm{gk}}$,

\begin{equation}
	\delta \tilde{f_s}  =  - \frac{Z_s e\varphi}{T_s} F_s + h_s \lp t, \vct{R}_s, \mu_s, \varepsilon_s\rp,
	\label{hsdef}
\end{equation}
where \(Z_s e\) is the charge of species $s$ and 
\(\varphi\) is the perturbed electrostatic potential. 
The non-Boltzmann part of the distribution function, $h_s$,
the gyrocentre distribution,
is independent of the gyrophase, that is,
the exact position of a particle 
within its orbit around a magnetic field line. 
Thus $h_s$ is only a function 
of the centre of that orbit $\vct{R}_s$,
the perpendicular velocity $v_{\perp}$ of the particle 
around that orbit (for which we use the coordinate 
$\mu_s = m_s v_{\perp}^2 / 2 B$), 
%the energy $\varepsilon_s=m_2 ( v_{\parallel}^2 + v_{\perp}^2)$ 
the energy $\varepsilon_s$
of the particle,
and the time, $t$ \cite{flowtome1}.
%the magnetic of the 
%and a function of time $t$, 
%the gyrocentre position $\vct{R}_s$,
%the magnetic moment $\mu_s$ 
%and the energy $\varepsilon_s$ of the particle \cite{flowtome1}. 
The gyrocentre position $\vct{R}_s=\vct{r} - \uv{b}\times \vct{v} / \Omega_s $, where 
$\vct{r}$ is the position coordinate,
$\Omega_s$ is the gyrofrequency of species $s$,
the unit vector $\uv{b} = \vct{B}/B$ is the direction of the equilibrium magnetic field,
and $B$ is its magnitude. 
 Thus, the problem has been reduced from a six-dimensional to a five-dimensional one by only considering timescales longer than the fast orbit of the particles around the magnetic field.

The turbulence can be affected both by the gradient of the flow and by its magnitude, 
the latter through the Coriolis and centrifugal drifts. 
Here we neglect the effects of the magnitude of the flow
(these effects are described in detail in Refs.~\cite{peeters2007toroidal,casson2010gyrokinetic}) 
but allow the gradient of the flow to be of order the growth rate of the fluctuations,
$R d\omega / dr \sim \gamma$,
as is necessary for the flow gradient to affect the turbulence non-negligibly \cite{waltz1994toroidal}.
This can be systematised
by assuming
that the magnitude of the flow is small but the gradient of the flow is large 
in a Mach-number ordering subsidiary to \eqref{orderings}, namely 
\begin{equation}
	\frac{R\omega}{v_{thi}} \sim M \ll 1,\; 
	\frac{R}{\omega} \dbd{\omega}{r} \sim \oov{M} \gg 1
	\label{flowmagordering}
\end{equation}
where $\epsilon_{\mathrm{gk}} \ll M \ll 1$.

Under these combined assumptions, it can be shown
that $h_s$ evolves according to the following gyrokinetic equation:

\begin{multline}
	\left( \frac{\partial }{\partial t} + \vct{u}\cdot \nabla \right)h_s + \lp v\pa \uv{b} + \vct{V}_{Ds} + \frac{c}{B}\uv{b}\times \nabla\gyroR{\varphi} \rp \cdot \nabla h_s - \gyroR{C\lsq h_s \rsq}\\
	%\frac{Z_s e}{T_s} \lsq  	\left( \frac{\partial }{\partial t} + \vct{u}\cdot \nabla \right) \gyroR{\varphi}  - \frac{\btor v\pa R}{B \Omega_s}\dbd{\omega}{r} \lp \uv{b} \times \nabla 
%\gyroR{\varphi}\rp \cdot \nabla r \rsq F_{s} \\
%- \frac{cF_s}{B}  \lp \uv{b} \times \nabla \gyroR{\varphi} \rp \cdot \nabla r \lsq \oov{n_s} \dbd{n_s}{r} + \oov{T_s}\dbd{T_s}{r} \lp  \frac{\varepsilon_s}{T_s} - \frac{3}{2} \rp \rsq \\  + \gyroR{C\lsq h_s \rsq}, 
= \frac{Z_se F_s}{Ts}\left( \frac{\partial }{\partial t} + \vct{u}\cdot \nabla \right)\gyroR{\varphi} - \left[ \frac{\partial F_s}{\partial \psi} + \frac{m_sIv_{\parallel}F_s}{T_s B}\frac{d \omega}{d \psi} \right] \left( \uv{b} \times \nabla \gyroR{\varphi} \right)\cdot \nabla\psi,
	\label{gyrokineticeq1}
\end{multline}
where
$\gyroR{}$ is a gyroaverage 
(an average over a particle orbit)
at constant $\vct{R}_s$,
%$d/dt = \partial / \partial t + \vct{u}\cdot \nabla$, 
\begin{equation}
	\vct{V}_{Ds} = \frac{\uv{b}}{\Omega_s}\times\left[ v_{\parallel}^2 \uv{b}\cdot\nabla\uv{b} + \frac{1}{2}\nabla\ln B \right],
	\label{eqn:VDsdef}
\end{equation}
is the magnetic drift velocity, 
$v_{\pa} = \sigma \sqrt{\lp \varepsilon_s - \mu_s B \rp/m_s}$ is the parallel  velocity, 
$\sigma$ is the sign of $v_{\pa}$, 
and $c$ is the speed of light.
We further simplify the problem by assuming purely electrostatic perturbations ($\delta \vct{B}=0$),
so the system is closed by the quasineutrality condition:
\begin{equation}
	\sum_s \frac{  Z_s^2 e \varphi}{T_s}n_s = \sum_s Z_s \int d^3  \vct{v} \gyror{h_s},
	\label{quasin}
\end{equation}
where $\gyror{}$ is a gyroaverage at constant $\vct{r}$.

In general, the gyrokinetic equation must be solved separately for each species $s $,
and the resulting perturbed distribution function for each species $h_s $
entered into the quasineutrality condition
to compute how each species affects the electric field.
However, in this work, we do not solve the gyrokinetic equation for electrons,
but instead treat them using a modified Boltzmann response \cite{hammett1993developments}:

\begin{equation}
	\int d^3 \vct{v} \gyror{h_e} = \frac{e \lp \varphi - \overline{\varphi}\rp}{T_e} n_e,
	\label{boltzmanne}
\end{equation}
where the overbar denotes averaging over the flux surface.
Thus, Eq. \eqref{gyrokineticeq1} is only solved for ions, $s=i$, and together with \eqref{quasin} comprises a closed system of equations for the evolution of $h_i$.

Knowing \(h_i\), 
we can calculate the turbulent heat and angular momentum fluxes
associated with a given minor radius and given values of the gradients,

\begin{equation}
	Q_t =  \overline{\left< \int d^3 \vct{v} \oov{2} m_i v^2 \frac{c}{B}  \gyror{h_i}\lp \uv{b}\times \nabla \varphi \rp \cdot \nabla r \right>},
	\label{heatfluxdef}
\end{equation}

\begin{equation}
	\Pi_t = \overline{\left< \int d^3 \vct{v}  m_i \gyror{h_i\vct{v}} \cdot \nabla{\phi}  R^2\frac{c  }{B} \lp  \uv{b}\times \nabla \varphi \rp \cdot \nabla r \right>}.
	\label{momfluxdef}
\end{equation}
All that remains is to solve the equations \eqref{gyrokineticeq1} and \eqref{quasin} and find $h_i $.

\section{An Overview of \texttt{GS2}}

In this work, the gyrokinetic equation derived above is solved numerically using
the freely available simulation code \texttt{GS2} \cite{gyrokineticswebsite}.
\texttt{GS2} is an initial value code which solves the nonlinear $\delta f$ continuum gyrokinetic equation in a small region of the plasma known as a flux tube, in which it is assumed that the gradients of equilibrium quantities are constant. It is one of the earliest in a class of codes which solve the $\delta f $ gyrokinetic equation in similar ways; other notable members of this class include the codes
\texttt{GENE} \cite{jenko:1904,jenko2001nonlinear,gorler2011global},
\texttt{GYRO} \cite{candy2003eulerian,fahey2004gyro} and \texttt{GKW} \cite{gkwRef}.
%which, like \texttt{GS2}, also solve the $\delta f$ gyrokinetic equation, but which possess in addition the ability to solve it in a global, rather than a local, manner, considering a (notionally) larger region of the plasma and allowing equilibrium gradients to vary across that region and the code \texttt{GKW}, which, as mentioned earlier, includes the effect of the magnitude of an equilibrium flow as well as its gradients.
\begin{table}[h]
	\caption{A summary of which parts of \texttt{GS2} are documented in each source. The letters are defined as \texttt{F} --- full: covered in full detail; \texttt{C} --- concise: the whole functionality of the code is covered, but briefly; \texttt{P} --- partial: only a (possibly large) part of the functionality of the code is documented; \texttt{O} --- old: the description does not correspond to the code in its current form.
} 
\begin{center}
	\begin{tabular}{ r | c | c | c | c | c | c | c | c | c | c}
		    %\hline
		    %\hline
				Source & 
				\rotatebox{90}{Linear Algorithm} &
				\rotatebox{90}{Nonlinear Terms} &
				\rotatebox{90}{Collisions} &
				\rotatebox{90}{Electromagnetic Terms} &
				\rotatebox{90}{Coordinates} &
				\rotatebox{90}{Geometry} & 
				\rotatebox{90}{Boundary Conditions}&
				\rotatebox{90}{Velocity Space}&
				\rotatebox{90}{Flow Shear}&
				\rotatebox{90}{Normalisations}
				\\
			 \hline
			 \hline
			 Here & \texttt F & \texttt F & \texttt C & \texttt -& \texttt F & \texttt P & \texttt F & \texttt C & \texttt F & \texttt P\\
			 Beer et. al. \cite{beer1995field} & \texttt - & \texttt - &\texttt  - & \texttt -&\texttt  F &\texttt  P &\texttt  P &\texttt  - &\texttt  - &\texttt  - \\ 
			 Kotschenreuther et. al. \cite{gs2ref} & \texttt O & \texttt - &\texttt  O & \texttt -&\texttt  - &\texttt  P &\texttt  P &\texttt  O &\texttt  - &\texttt  - \\ 
			 Dorland \& Kotschenreuther  \cite{dorland-microinstabilities} & \texttt - & \texttt - &\texttt  - & \texttt - &\texttt  P &\texttt  F &\texttt  P &\texttt  - &\texttt  - &\texttt  P \\ 
			 Belli  \cite{belli2006studies} & \texttt F & \texttt - &\texttt  - & \texttt F&\texttt  - &\texttt  - &\texttt  - &\texttt  - &\texttt  - &\texttt  - \\ 
			 Barnes \& Abel  \cite{abel2008linearized, numerics} & \texttt - & \texttt - &\texttt  F & \texttt -&\texttt  - &\texttt  - &\texttt  - &\texttt  - &\texttt  - &\texttt  P \\ 
			 Numata et. al. \cite{numata2010astrogk} & \texttt F & \texttt C &\texttt  C & \texttt F&\texttt  - &\texttt  - &\texttt  - &\texttt  C &\texttt  - &\texttt  P \\ 
			 Barnes  \cite{michaelthesis} & \texttt - & \texttt - &\texttt  F & \texttt -&\texttt  - &\texttt  F &\texttt  - &\texttt  F &\texttt  - &\texttt  P \\ 
%MAST & 0.6e+03& 3.9e+19& 8.0e-01& 5.3e-01& 5.6e-01\\
			 \hline
			 \hline
	\end{tabular}
\end{center}
\label{documentationstatus}
\end{table}

There is no one existing source which fully describes the exact equations and algorithms of \texttt{GS2}; as the majority of of the results contained within this thesis are generated using it,
Chapters \ref{sec:geometry}-\ref{flowshearapp} are devoted to describing how \texttt{GS2} works.\footnote{
In the absence of electromagnetic perturbations.}\footnote{ It should be noted that a simulation code like \texttt{GS2}  is continuously evolving. The description given in this work corresponds to revision 1555 of the trunk in the Subversion repository on the Sourceforge Gyrokinetics project page \cite{gyrokineticswebsite}.}
Here we give a brief overview of \texttt{GS2} and the extensive, if fragmented, existing documentation.
%We aim to give a concise summary of \texttt{GS2} and directions to further information. 
%\subsection{Existing Documents about \texttt{GS2}}
The main algorithms of \texttt{GS2} may be roughly summarised as:
\begin{itemize}
	\item the implicit algorithm which solves the linearised set of equations \eqref{gyrokineticeq1} and \eqref{quasin}\footnote{With the addition of Amp\`ere's Law if magnetic perturbations are included.},
	\item the addition of the nonlinear term (the term proportional to $\uv{b}\times \nabla\gyroR{\varphi} \cdot \nabla h_i$ in \eqref{gyrokineticeq1}) as a source in the linear equation,
	\item the addition of collisions.
\end{itemize}
The original exposition of the 
implicit algorithm that solves the linear equation in the absence
of collisions is presented by Kotschenreuther et. al. \cite{gs2ref}, but, as is documented by Belli \cite{belli2006studies},
the actual implementation of the algorithm is different.
The clearest and most compact description of the linear algorithm
as it currently stands is given in the paper documenting the \texttt{AstroGK} code \cite{numata2010astrogk},
but as \texttt{AstroGK} is a straight-field-line
simplification of \texttt{GS2},
it does not cover some of the complications
that follow from the geometry,
such as the parallel boundary conditions,
which affect \texttt{GS2}.
The inclusion of the nonlinear terms is noted in Ref.~\cite{jenko:1904} and described more fully in Ref.~\cite{numata2010astrogk}, 
while the collision operator is derived by Abel et. al. \cite{abel2008linearized}
and implemented according to Barnes et. al. \cite{numerics}.

In addition to the main algorithms, there are additions and subtleties such as:
%\needspace{16\baselineskip}
\begin{itemize}
	\item the definition of the coordinates,
	\item the treatment of the geometry of the magnetic field,
	\item the treatment of boundary conditions,
	\item the layout of the velocity space grids,
	\item the implementation of perpendicular flow shear,
	\item the normalisation of input parameters and calculated quantities.
\end{itemize}
Again, the documentation of these is fragmented; the treatment of boundary conditions is discussed in Ref.~\cite{gs2ref}, following the work of Beer et. al. \cite{beer1995field};
the layout of the velocity space grids and the calculation of velocity integrals
is described by Barnes \cite{michaelthesis}
and the implementation of flow shear is briefly described in Ref.~\cite{gs2flowshear}. 
Descriptions of coordinates, geometry and normalisations are given in an unpublished paper \cite{dorland-microinstabilities} which was included as an appendix to Ref.~\cite{michaelthesis}.
A summary of the current documentation of \texttt{GS2}, as described below, is given in Table \ref{documentationstatus}.

%\subsection{Documentation within the Current Work}

The rest of Part \ref{sec:modellingturbulence} is organised as follows.
In Chapter \ref{sec:geometry}, we describe how
the geometry of the equilibrium magnetic field is dealt with,
the choice of coordinates and the variables used within \texttt{GS2}
and the way they are normalised.
In Chapter \ref{sec:gs2algorithm}, we describe the algorithms that are used to solve the gyrokinetic equation in the absence of collisions between particles. In Chapter \ref{sec:velocityspace}, 
we describe the way that the velocity integrals of 
the distribution function 
(used to calculate, for example, 
the perturbed field generated by the particles) are treated, and how the effect of collisions between particles is included. 
In Chapter \ref{flowshearapp}, we describe how the effect of a sheared equilibrium flow is incorporated within \texttt{GS2}.

\label{modelsec}

%\section{High-Flow Gyrokinetics}
%\section{High-Flow Low-Mach Gyrokinetics}

%\chapter{A Concise Description of \texttt{GS2} }

\chapter{Geometry}
\label{sec:geometry} 
\section{Introduction}

Many of the terms in the gyrokinetic equation \eqref{gyrokineticeq1}
depend on the shape or strength of the magnetic field; they are given in Table \ref{tab:fieldoperators}. In a fusion device, the magnetic field can have a very complicated shape indeed, and the task of calculating terms that depend on it becomes equally complicated.
Fortunately, in a well-behaved equilibrium in a fusion device,
that is, in the absence of magnetic islands and stochasticity,
the magnetic field lines have properties, namely that they are
continuous and do not cross, that make them an excellent basis
for an  curvilinear coordinate system, with one unit
vector along the magnetic field line, and two perpendicular to it.
In any such coordinate system, many operators such as $\uv{b}\cdot\nabla$ can be expressed simply.
 Such a coordinate system also allows a simulation code to take advantage of the fact that structures in the turbulence tend to be elongated along a background magnetic field: this allows grid points along the magnetic field to be much more widely spaced.
  This saving in grid size is indispensable, and it is principally for this reason that field line following coordinate systems are used.

\begin{table}
	\centering
	\begin{tabular}{c | l}
		Operator  & Description\\
		\hline
		\hline
		 $\vct{u}\cdot\nabla$ & Equilibrium Flow \\	
		 $\vct{b}\cdot\nabla$ & Parallel Streaming \\	
		 $\vct{b}\times\left( \vct{b}\cdot\nabla\vct{b} \right)$ & Curvature Drift \\	
		 $\vct{b}\times\left( \nabla\ln B \right)$ & Gradient Drift \\	
		 $\left( c/B \right)\left( \vct{b}\times\nabla \gyroR{\field} \right) \cdot \nabla h_s$ & Nonlinear Term \\	
		 $\left( c/B \right)\left( \vct{b}\times\nabla\gyroR{\field} \right)\cdot \nabla\psi$ & Perturbed Flow (radial component)\\	
		 $\partial F_s/\partial \psi $ & Background Gradients \\	
		 $k_{\perp}$ & Perpendicular Wavenumber  \\	
		 $B$ & The Magnetic Field Strength \\	
		\hline
	\end{tabular}
	\caption{The operators in the gyrokinetic equation which depend on the size or shape of the magnetic field, along with their physical significance or the physical significance of the terms in which they appear.  }
	\label{tab:fieldoperators}
\end{table}

The two tasks that must be faced, therefore, when dealing with
the geometry are the definition of a set of field line
following coordinates, and the calculation of those terms
in the gyrokinetic equation which depend on the shape or strength
of the magnetic field and which are \emph{not} rendered trivial to
calculate by such a coordinate system. The first task can be 
easily achieved (at least, once it has been done once: Ref.~\cite{beer1995field});
the second is Herculean.
The reason for this is that while the choice of a field line following coordinate system renders most of the gyrokinetic equation easier to solve, the calculation of the remaining terms such as the magnetic drifts requires one
to keep track, firstly, of which points in real space the grid points in this new coordinate system correspond to, and secondly, what the shape and strength of the magnetic field actually is at those points.

\section{Defining a Coordinate System}
\label{sec:definingacoordinatesystem}

When defining a coordinate system,
we assume that the magnetic field vector, $\vct{B}$, 
is known at every point. 
All coordinates will be defined with respect to this vector. 
%We start by defining our three orthogonal basis vectors. 
We will now define our three basis vectors and our three coordinates.
One of our basis vectors is, naturally, 
the magnetic field direction: $\uv{b} = \vct{B}/B$ 
(we choose the corresponding coordinate below).
The second follows from the fact that in any axisymmetric  toroidal system,
the magnetic field lines will all lie on nested surfaces,
and each of these nested surfaces encloses a well-defined \emph{poloidal flux},
that is, a magnetic flux going around the magnetic axis,
which is defined as the integral of the poloidal magnetic field $B_{\theta} \equiv \vct{B}\cdot\nabla\theta$ across a surface of constant poloidal angle $\theta$ (see \figref{coordinatesystem})
which extends from the magnetic axis to the flux surface itself:\footnote{
In fact, it is perfectly possible to have a 
\emph{non-axisymmetric} system where the magnetic field lines 
still lie on well-defined closed flux surfaces 
(for example, a stellarator). In this case, 
we can use a more general definition of $\psi$:
\begin{equation}
	\psi =  \left( \frac{1}{2\pi} \right)^2 \int_{0}^V dV \vct{B}\cdot\nabla\theta
	\label{eqn:psigeneraldef}
\end{equation}
where $V$ is the volume enclosed by the flux surface. This formula reduces to \eqref{eqn:psidef} for the case of an axisymmetric system.}
\begin{equation}
	\psi\left( r \right) = \int_{0}^{r} dr^{'}\;r^{'}R ( r^{'} )\vct{B}( r^{'} )\cdot\nabla\theta .
	\label{eqn:psidef}
\end{equation}
The magnetic axis is at the centre of the nested flux surfaces, the point where $B_{\theta}=0$ and the field is purely toroidal. The minor radius $r$ is the distance from the magnetic axis, and $R$ is the major radius at $r$ (see \figref{coordinatesystem}).
Since the quantity $\psi$ is constant on a flux surface, and is therefore constant on any magnetic field line, it will serve as our second coordinate, and its gradient will  serve as our second basis vector: $\uv{\psi} = \nabla \psi / \md{\nabla\psi}$.
%Our third basis vector is then simply the vector product of the first two: $\uv{\alpha} \equiv \uv{\psi}\times\uv{b}$.

%We must now choose the coordinates that correspond to each basis vector. For $\uv{\psi}$, it is natural to choose $\psi$.
%For $\uv{\alpha}$, 
For our third basis vector $\uv{\alpha}$, we require $\uv{b}\cdot\uv{\alpha}=0$ (so that coordinate system follows the field lines) and $\uv{b}\cdot(\uv{\alpha}\times\uv{\psi})\neq0$ (to prevent the coordinate system from diverging) at every point except at the magnetic axis.
A coordinate $\alpha$ (with basis vector $\uv{\alpha} = \nabla\alpha/\left| \nabla\alpha \right|$) defined such that $\vct{B} = \nabla \alpha \times \nabla \psi$ and $\nabla\alpha\cdot\nabla\psi\neq B$ will satisfy these conditions.\footnote{
In this case, $\alpha$ is a surface potential \cite{hazeltine2003plasma}.}
%that we may write the equilibrium field as $ \vct{B} = \nabla \alpha  \times \nabla \psi$.
Kruskal and Kulsrud \cite{kruskal1958equilibrium} showed that for a magnetic equilibrium composed of closed flux surfaces, we may write:
\begin{equation}
	\alpha = \phi + q\left( \psi \right)\theta+\nu\left( \phi, \theta, \psi \right),
	\label{eqn:alphadef}
\end{equation}
where $\phi$ is the toroidal angle (see \figref{coordinatesystem}), 
$\theta$ is the poloidal angle, 
$\nu$ is a function which depends on the geometry and which is periodic in $\phi$ and $\theta$,
and the magnetic safety factor 
\begin{equation}
	q = \pd{\psi_t}{\psi},
	\label{eqn:qdeffull}
\end{equation}
where $\psi_t$ is the toroidal magnetic flux:\footnote{
No 
equivalent of the \eqref{eqn:psidef} exists for the toroidal flux as in general 
the flux surfaces in a tokamak are not symmetric in $\theta$.}
\begin{equation}
	%\psi_t\left( r \right) = \int_{0}^{r} dr^{'}\;r^{'}R ( r^{'} )\vct{B}( r^{'} )\cdot\nabla\phi .
	\psi_t =  \left( \frac{1}{2\pi} \right)^2 \int_{0}^V dV \vct{B}\cdot\nabla\phi
	\label{eqn:psitdef}
\end{equation}
Lastly, for the coordinate along the magnetic field, \texttt{GS2} follows Ref.~\cite{beer1995field} and many others in choosing the poloidal angle $\theta$.
 This may seem counterintuitive: why not just choose a coordinate such as ``the distance along the field line''?
In fact, we note firstly that at fixed $\psi$ and $\alpha$, 
any coordinate which is not fixed will serve as the parallel coordinate, and secondly that, at fixed $\psi$, many equilibrium quantities (such as $B$) can be written (in some limits) as quite simple functions of $\theta$.
%Furthermore, choosing a physical coordinate with a well-defined origin (see \figref{coordinatesystem})
%allows one to appreciate more quickly whereabouts in the tokamak a particular value of ($\psi$, $\alpha$, $\theta$) corresponds to.
%(note that in an axisymmetric system, at fixed $\psi$ and $\theta$, no equilibrium quantity can depend on $\alpha$, so it is not particularly important.
\myfig[5.0in]{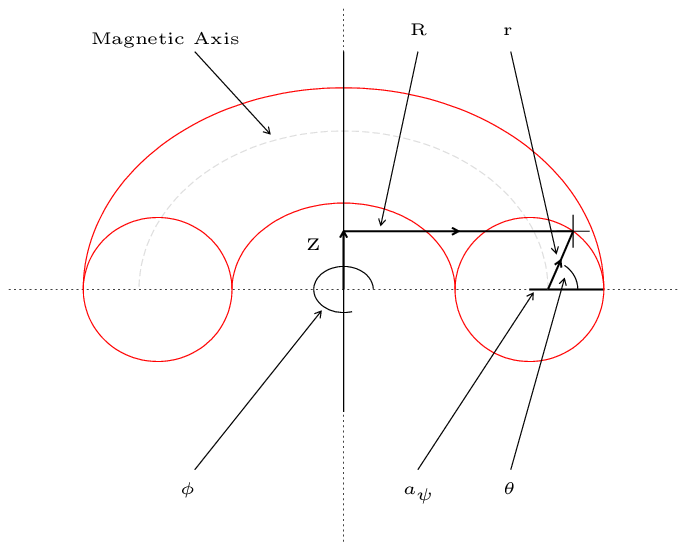}{Toroidal Coordinates}{Toroidal coordinates. Note that the signs of $\theta$ and $\phi$ are not fixed (see Section \ref{sec:noteonsigns}), though the origin of $\theta$ is fixed at the height of the magnetic axis on the outboard side. Note also that $a_{\psi}$, the half radius of the flux surface at the height of the magnetic axis, is not in general equal to $r$, the minor radius (except in the very special case of zero Shafranov shift and circular flux surfaces). }

\subsection{The Local Approximation}
In order to take full advantage of the separation between the scale of the turbulence and the scale of the equilibrium \texttt{GS2} solves the gyrokinetic equation in a small region of the plasma known as a flux tube.
A flux tube is designed to be large enough to include several turbulence decorrelation lengths in both the parallel and perpendicular directions (i.e. large enough that the turbulence in the centre cannot ``see'' the edge of the box) but small enough that high spatial resolution can still be achieved. Since the turbulent eddies themselves are elongated along the magnetic field line, a flux tube likewise follows the magnetic field lines and is elongated along them (see Fig. 1 of Ref.~\cite{beer1995field}).
Within the flux tube we define two local perpendicular coordinates, $x$ and $y$, which measure the distance from the magnetic field line at the centre of that flux tube, located at $(\psi_0,\alpha_0)$:

\begin{eqnarray}
	x & \equiv& \frac{1}{a B_a} \frac{q_0}{ \rho_0  }\left( \psi -\psi_{0} \right) = a \frac{q_0}{\rho_0 }\left( \psi_N -\psi_{0N} \right)\label{eqn:xdef}\\
	y &\equiv& \frac{1}{a B_a} \left. \frac{\partial \psi}{\partial \rho} \right|_{\rho_0}\left( \alpha-\alpha_0 \right)  = a \left. \frac{\partial \psi_N}{\partial \rho} \right|_{\rho_0}\left( \alpha-\alpha_0 \right)\label{eqn:ydef}
\end{eqnarray}
where $a$ is is the half radius of the last closed flux surface (LCFS) at the height of the magnetic axis, $B_a$ is used to normalise all magnetic quantities,\footnote{
$B_a$ is defined to be the magnitude of the magnetic
 field at the centre of the LCFS.
 %In the high aspect ratio, concentric circular flux surface limit, $B_a = \btor $, where $\btor \equiv \vct{B}\cdot\nabla\phi$  is the toroidal magnetic field.
 See the discussion of normalising quantities, Section \ref{sec:normalisingquantities}.
 } $\psi_N=\psi/a^2B_a$, $\rho_0 \equiv \rho(\psi_0) $, $q_0 \equiv q (\psi_0) $ and $\rho $ is the \texttt{GS2} flux surface label, which can have varying definitions, but which is always 0 at the magnetic axis, increasing to 1 at the last closed flux surface (LCFS).
In this work $\rho$ is defined to be $a_{\psi}/a$ (the default choice), where $a_{\psi}(\psi)$ is the half radius of the flux surface at the height of the magnetic axis (note that $a_{\psi}(\psi_{LCFS})=a$).

The solution of the gyrokinetic equation depends on (among other quantities) the equilibrium gradients,
$dn_s/d\psi$, $dT_s/d\psi$ and $d\omega/d\psi$.
%(the latter of which is the gradient the of parallel component of the background flow).
To simplify the treatment of these gradients,
we expand the equilibrium quantities
around the flux surface $\psi=\psi_0$.
%around a reference flux surface near the centre of the domain, labelled by  $\psi_0$. 
Thus we may write:
\begin{eqnarray}
	n_s \lp \psi \rp &=&  n_s \lp \psi_0 \rp + \lp \psi - \psi_0 \rp\left. \dbd{n_s}{\psi} \right|_{\psi_{0}}
	%\nonumber
	%\\ 
	%&\equiv& n_{s0} \lsq 1 - \lp \psi_N - \psi_{N0} \rp \left. \frac{\partial \rho}{\partial \psi_N} \right|_{\psi_{N0}} \frac{a}{L_{ns}}\rsq
	%\nonumber\\ 
	=  n_{s0} \lsq 1 - \frac{\rho_0}{q_0}\frac{x}{a} \left. \frac{\partial \rho}{\partial \psi_N} \right|_{\psi_{N0}} \frac{a}{L_{ns}}\rsq,
	\label{localngrad} \\
	T_s \lp \psi \rp &=&  T_s \lp \psi_0 \rp + \lp \psi - \psi_0 \rp \left. \dbd{T_s}{\psi} \right|_{\psi_0}
	%\nonumber\\
	%&\equiv& T_{s0} \lsq 1 - \lp \psi_N - \psi_{N0} \rp\left. \frac{\partial \rho}{\partial \psi_N} \right|_{\psi_{N0}}  \frac{a}{L_{Ts}}\rsq 
	%\nonumber\\
	=  T_{s0} \lsq 1 - \frac{\rho_0}{q_0}\frac{x}{a}\left. \frac{\partial \rho}{\partial \psi_N} \right|_{\psi_{N0}}  \frac{a}{L_{Ts}}\rsq,
	\label{localTgrad}\\
	\omega \lp \psi \rp &=&  \omega \lp \psi_0 \rp + \lp \psi - \psi_0 \rp \left. \dbd{\omega}{\psi} \right|_{\psi=\psi_0}
	%\nonumber \\
	%&\equiv& \omega_0  + \lp \psi_N - \psi_{N0} \rp \left| \frac{\partial \rho}{\partial \psi} \right|_{\psi_{N0}}\frac{q_0}{\rho_0}\fsr
	%\nonumber \\
	= \omega_0  + \frac{x}{a} \left| \frac{\partial \rho}{\partial \psi} \right|_{\psi_{N0}}\fsr,
	\label{localomegagrad}
\end{eqnarray}
where we have defined the density and temperature gradient scale lengths, 
%where in the last line we have introduced
\begin{equation}
	\frac{a}{L_{ns}} \equiv - \frac{1}{n_s}\left. \frac{\partial n_s}{\partial \rho} \right|_{\rho_0},
	\label{eqn:Lnsdef}
\end{equation}
and 
\begin{equation}
	\frac{a}{L_{Ts}} \equiv - \frac{1}{T_s}\left. \frac{\partial T_s}{\partial \rho} \right|_{\rho_0},
	\label{eqn:LTsdef}
\end{equation}
and the perpendicular flow shear 

\begin{equation}
	\gamma_E \equiv \sigma_{\psi}\frac{\rho_0}{q_0}\left. \frac{d\omega}{d\rho} \right|_{\rho_0},
	\label{eqn:gammadef}
\end{equation}
where $\sigma_\psi $, the sign of $\psi$,  is defined in Section \ref{sec:noteonsigns}.\footnote{ The
introduction of this sign maintains the relation $\vct{u}\cdot\nabla = x \gamma_E \partial / \partial y $ (see Section \ref{sec:equilibriumflowoperator}) regardless of the sign of $\psi $, which determines the sign of $y $.	}
 It will be immediately obvious to the reader that (owing to a historical quirk) $\gamma_E $ is not in fact the perpendicular flow shear, but is equal to the perpendicular flow shear multiplied by $B/\btor $. However, in general $B/\btor \sim 1 $, so the parameter $\gamma_E $ may reasonably be referred to as ``the perpendicular flow shear''.
%where $L_{ns}$ is the equilibrium density scale length, $L_{Ts}$ the equilibrium temperature scale length.
%In the presence of a strong magnetic field,
%most perturbations are extended along the field line.
%Therefore, \texttt{GS2} solves the gyrokinetic equation in a domain
%called a \emph{flux tube} (see Fig of \cite{beer1995field}),
%which has a small perpendicular extent
%(though it can extend up to several times the connection length, $qR$, in the parallel direction). 

\subsection{Spectral Coordinates}
\label{sec:spectralcoordinates}

Since the gyrokinetic equation (in the absence of flow shear) has no explicit dependence on either $x $ or $y $, it can be solved spectrally in both perpendicular directions.
The reasons for doing this are firstly that the linearised gyrokinetic equation can be solved separately for each Fourier mode
(see Section \ref{sec:linearimplicitalgorithm}), secondly that the pseudo-spectral calculation of the nonlinear term leads to faster convergence with perpendicular grid size (see Section \ref{sec:nonlinearterms}) and thirdly that the quadratic nonlinearity in the gyrokinetic equation is exactly energy conserving for a spectral system \cite{orszag1970transform,platzman1960spectral}.
A perturbed quantity $A $ is therefore written as 

\begin{equation}
	A = \sum_{k_x,k_y} \hat{A}_{k_x,k_y}(\theta)e^{i\left( k_x x + k_y y \right)} \equiv \mathcal{F}^{-1} \hat{A}\left( \theta \right)
	\label{eqn:kxkyeikonal}
\end{equation} 
where $k_x $ and $k_y $ are the actual perpendicular coordinates used by \texttt{GS2} and we have defined the inverse Fourier transform $\mathcal{F}^{-1}$. Historically, however, to make contact with the ballooning approximation,  \eqref{eqn:kxkyeikonal} was written \cite{dorland-microinstabilities}: 
\begin{equation}
	A = \sum_{n_0,\theta_0} \hat{A}_{n_0, \theta_0}(\theta)e^{i S},
	\label{eqn:not0eikonal}
\end{equation} where \begin{equation}
	S = n_0 \left( \alpha + q \theta_0\right).
	\label{eqn:Seikdef}
\end{equation} If we Taylor expand $q $, we can write \begin{equation}
	S = n_0 \left[ \left( \alpha-\alpha_0 \right)+ \left( \psi-\psi_0 \right)\left.\frac{\partial q}{\partial \psi}\right|_{\psi_0}\theta_0 \right]+ n_0\left( \alpha_0 + q_0\theta_0 \right),
\end{equation}
and if we absorb the term $n_0\left( \alpha_0 + q_0\theta_0 \right) $, which is constant over the simulation domain, into $\hat{A} $, we may write \begin{eqnarray}
	S& =& \frac{n_0}{a}\left. \frac{d \rho}{d \psi_N} \right|_{\psi_{0N}}
	\left[ a \left( \alpha-\alpha_0 \right) \left. \frac{d \psi_N}{d \rho} \right|_{\psi_0} 
	+ a \hat{s}\theta_0\left( \psi_N - \psi_{0N} \frac{q_0}{\rho_0} \right) \right]\nonumber\\
	&= &\left[ \frac{n_0}{a}\left. \frac{d \rho}{d \psi_N} \right|_{\psi_{0N}} \hat{s}\theta_0 \right] x + \left[ \frac{n_0}{a}\left. \frac{d \rho}{d \psi_N} \right|_{\psi_{0N}} \right] y \nonumber\\
	&=& k_x x + k_y y
\end{eqnarray} 
where $k_y =  \left( n_0/a \right)\left. d \rho / d \psi_N \right|_{\psi_{0N}} $, $k_x = k_y \hat{s}\theta_0 $ and $\hat{s}$ is the magnetic shear (see below). Thus, the expressions \eqref{eqn:kxkyeikonal} and \eqref{eqn:not0eikonal} are equivalent up to a constant factor.\footnote{My thanks to M. Barnes for his notes on this subject.}

\subsection{Magnetic Shear}
In the case of finite magnetic shear, that is, where $\hat{s}=(\rho/q)dq/d\rho>0$, the magnetic field lines are twisted between one flux surface and another, as is illustrated by \figref{kxshear}, which shows sets of magnetic field lines on two different flux surfaces (that is, at two different values of $x$).
At finite magnetic shear the field line following coordinate system ($\psi$,$\alpha$,$\theta$) becomes non-orthogonal as $q$, and hence $\alpha$, becomes a function of $\psi$.
Thus, the twisted shape of the magnetic field lines has been absorbed into the coordinate system. 
In the local coordinate system ($x$, $y$, $\theta$), $y$ becomes a function of $x\theta$, and as a result the full radial derivative of the eikonal $dS/dx$ becomes a function of $\theta$. This has important consequences for the parallel boundary conditions, as is discussed in Section \ref{sec:solvingthediscretizedgyrokineticequation}.
%The choice of $\alpha$ as a coordinate was motivated by the desire for a field line following coordinate system. , it results in a non-orthogonal coordinate system . 

\myfig{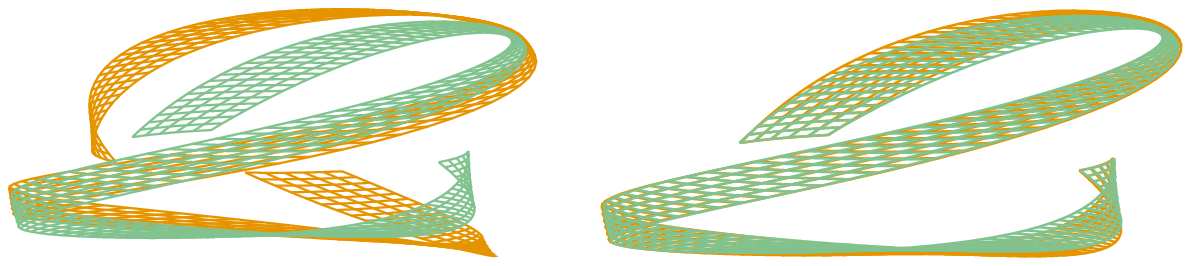}{Magnetic Shear}{The shape of magnetic field lines on two neighbouring flux surfaces with (left) and without (right) magnetic shear. Each picture shows two planes of constant $x$  within a flux tube, each of which encompasses a set of magnetic field lines. It can be seen that in the case of finite magnetic shear the gradient of any quantity taken between the two flux surfaces (that is, the effective radial gradient) can change dramatically along the field lines, even though the radial gradient in the local coordinates may be constant. }

%\webonly{
\section{A Note on Signs}
\label{sec:noteonsigns}

Although \texttt{GS2} uses the ($x$, $y$, $\theta$) coordinate system, at various points it is mathematically convenient to use any one of 8 other coordinates: $\psi $, $\alpha $, $\phi $, $\theta $, $r $, $\rho $, $R$ and $Z$, the vertical distance above the plane $\theta=0$ (the midplane).
 Although all of these coordinates have been defined above, the sign relation between all of them is not immediately clear: for example, in which directions, relative to the magnetic field, do the toroidal and poloidal angles increase?

 \begin{table}[h]
	 \centering
	 \begin{tabular}{c}
		 Signs \\
		 \hline 
		 \hline
		 $\sigma_B = \mathrm{sign}\left( \vct{I}_p \cdot \vct{B}_t \right)$\\
		 $\sigma_I = \mathrm{sign}\left( \vct{I}_p \cdot \nabla \phi \right)$\\
		 $\sigma_t = \mathrm{sign}\left( \vct{B}\cdot \nabla \phi \right)$\\
		 $\sigma_{\theta}=\mathrm{sign}\left( \vct{B}\cdot\nabla\theta \right)$\\
		 $\sigma_{\psi}=\mathrm{sign}\left( \nabla\psi\cdot\nabla r \right)$\\
		 $\sigma_{\rho}=\mathrm{sign}\left( \nabla\rho\cdot\nabla r \right)$\\
		 $\sigma_q = \mathrm{sign}\left( q \right)$\\
		 $\sigma_Z = \mathrm{sign}\left( (\nabla\phi\times \nabla R)\cdot \nabla Z \right)$\\
		 \hline
	 \end{tabular}
	 \caption{Signs which relate the various coordinates used by \texttt{GS2}.}
	 \label{tab:signs}
 \end{table}
 We start by noting that there are two fundamental vectors to which all other vectors can be related:
%$ R \vct{B}\cdot \nabla\phi$
$ \btor \uv{\phi}$
, the toroidal magnetic field, and $\vct{I}_p $, the plasma current which gives rise to the poloidal magnetic field.\footnote{
 In a stellarator, of course, there is no plasma current, and so $\sigma_I$ does not exist. The choice of signs is much less restricted because the device is not axisymmetric, and there is no equivalent of the relation \eqref{eqn:sIsp}.
 }
 We next  define a set of signs (Table \ref{tab:signs}). The sign $\sigma_B $ is determined by the equilibrium field and is thus a given. Clearly 
 $
 %\begin{equation}
	 \sigma_I = \sigma_t \sigma_B,
	 %\label{eqn:sistsB}
 %\end{equation}
 $
 and 
 $
 %\begin{equation}
	 \sigma_{\psi} = \sigma_\theta
	 %\label{eqn:spst}
 %\end{equation}
$
 by definition. Since also by definition 
 \begin{equation}
	 \sigma_q = \mathrm{sign}\left( \frac{\uv{b}\cdot\nabla\theta}{\uv{b}\cdot\nabla\phi} \right),
 \end{equation}
we find that 
$
%\begin{equation}
	\sigma_q=\sigma_t\sigma_\theta.
	%\label{eqn:sqstst}
%\end{equation}
$
For the sign of $Z $, we can write 
\begin{equation}
	\sigma_Z = \sigma_t \, \mathrm{sign}\left[  \left( \vct{B}_t \times \nabla R \right)\cdot \nabla Z\right] 
	\equiv \sigma_t \sigma_{\uparrow},
	\label{eqn:sZstbtgR}
\end{equation}
where $\sigma_{\uparrow} $ is determined by the magnetic field. In a system with up-down symmetry, $\sigma_Z $ is, of course, irrelevant.
Finally, although $\rho$ can be defined in various ways (usually $r/a$) its always defined to be 0 at the magnetic axis and 1 at the LCFS; thus, $\sigma_\rho=1$, and $\sigma_\psi = \mathrm{sign}\left( \nabla\psi \cdot \nabla \rho \right)$.

In a non-axisymmetric system, this is all that can be determined. In other words, given a magnetic field we can choose both $\sigma_t $ and $\sigma_\theta $, that is, we can choose the definitions of both $\phi $ and $\theta $.

In an axisymmetric system, we can go further. Using Amp\`ere's law, $\nabla\times\vct{B}_p = \vct{I}_p$ , we can write 
\begin{equation}
	\mathrm{sign}\left[ \left( \vct{B}\times\vct{I} \right)\cdot \nabla r \right] = 1,
\end{equation}
which means that 
\begin{equation}
	\mathrm{sign}\left[ \left( \vct{B}\times\nabla\phi \right)\cdot \nabla \psi \right] = \sigma_I\sigma_\psi.
\end{equation} 
Using the usual expression $\vct{B} = \nabla\alpha\times\nabla\psi $, we can write \begin{equation}
	\mathrm{sign}\left( \left| \nabla\psi \right|^2 \nabla\alpha\cdot\nabla\phi \right) = \sigma_I\sigma_\psi,
\end{equation}
which in an axisymmetric system means that 
\begin{equation}
	\sigma_I\sigma_\psi=1,
	\label{eqn:sIsp}
\end{equation}
which means that 
\begin{eqnarray}
	\sigma_q = \sigma_B,
	\label{eqn:sqsB}\\
	\sigma_I = \sigma_\theta = \sigma_\psi = \sigma_t\sigma_q.
	\label{eqn:signsallrelated}
\end{eqnarray}
In other words,  we only have the freedom to choose the direction of \emph{either} $\phi$ \emph{or} $\theta$.
%\footnote{ In many real tokamaks,  $\theta$ and $\phi$ are often chosen so that $\nabla\theta$ always points downwards on the outboard side, and $\nabla\phi$ is anticlockwise when viewed from above. In order for the \texttt{GS2} to adopt these definitions, we would have to redefine 
%\begin{equation}
	%\psi^{real} = Int sign = \mathrm{sign}() \psi \mathbf{!!!!!!!!!!!!!!!!!!!!!!!! BROKEN !!!!!}
	%\label{}
%\end{equation}
%Having this definition of $\psi$, however, would cause $\mathrm{sign}()$, which depends on the specific field being considered, to propagate all through the equations, which would be clearly undesirable}

This restriction comes from the fact that in an axisymmetric up-down symmetric system, all the equations must be invariant under a $180^{\circ}$ rotation of the magnetic field about a horizontal axis which passes through the centre of the  tokamak. Such a rotation would change $\sigma_I $ and $\sigma_t $: if we change the sign of $\phi $ to undo this, we must also change the sign of $\theta $ to keep the equations the same.

Lastly, we note that in \cite{flowtome1}, where $\vct{B} = \nabla\check{\psi} \times \nabla\alpha$ (where $\check{\psi} = -\psi$),
\begin{eqnarray}
	\check{\sigma}_q = -\sigma_B,
	\label{eqn:sqsBian}\\
	\check{\sigma}_I = -\check{\sigma}_\theta = -\check{\sigma}_\psi = -\check{\sigma}_t\check{\sigma}_q,
	\label{eqn:signsallrelatedian}
\end{eqnarray}
and so in Ref.~\cite{flowtome1}, the sign of $q$ is \emph{always} opposite to this work, and the sign of \emph{either} the toroidal field $\check{\btor} = R\vct{B}\cdot\nabla\check{\phi}$ \emph{or} the poloidal field $\check{\bpol} = r \vct{B}\cdot\nabla\check{\theta}$ must be opposite.

%} % end \webonly

\section{Geometric Operators in the \texttt{GS2}  Coordinate System}

Certain of the operators listed in Table \ref{tab:fieldoperators} can now be written explicitly in the ($x $, $y $, $\theta $) coordinate system, specifically $\vct{u}\cdot\nabla $,
$\vct{b}\cdot\nabla $ (in certain circumstances),
$\left( \vct{b}\times\nabla \gyroR{\field} \right) \cdot \nabla h_s$,
$\nabla\psi\cdot\left(\vct{b}\times\nabla \right) $
and
$\partial F_s / \partial\psi$.
\subsection{The Gradient Operator $\nabla$}
In this section, we make heavy use of the gradient operator

\begin{equation}
	\nabla = \uv{x}\left| \nabla x \right|\frac{\partial }{\partial x} + 
\uv{y}\left| \nabla y \right|\frac{\partial }{\partial y} +
\uv{b}k_p\frac{\partial }{\partial \theta} 
	\label{eqn:gradoperator}
\end{equation}
where $\uv{x} = \nabla x / \left| \nabla x \right|$, $\uv{y} = \nabla y / \left| \nabla y \right|$,
\begin{eqnarray}
	\left| \nabla x \right| &=&   a \frac{\left| q_0 \right|}{\rho_0}\left| \nabla\psi_N \right|,\\
	\left| \nabla y \right| &=&  a \left| \frac{\partial \psi_N}{\partial \rho} \right|_{\rho_0} \left| \nabla \alpha \right|, 
\end{eqnarray} and
\begin{equation}
	k_p(\theta) \equiv \uv{b}\cdot\nabla\theta,
	\label{eqn:jacobdef}
\end{equation}
is \emph{not} a wavenumber but is the parallel gradient (called $\mathtt{gradpar}$ within \texttt{GS2}) and is, in general, a function of $\theta$.\footnote{
Except if the option $\mathtt{equal\_arc}$ is specified, in which case gridpoints are evenly spaced along the field line and $k_p$ is thus a constant.}
\subsection[The Equilibrium Flow Operator]{The Equilibrium Flow Operator $\vct{u}\cdot\nabla $}
\label{sec:equilibriumflowoperator}

In the case of an axisymmetric field, this operator, corresponding to convection by an equilibrium flow $\vct{u} = \omega R^2 \nabla\phi $, can be written very simply.\footnote{
It is worth noting that it is only possible to have a strong (that is, $\left| \vct{u}  \right|\sim v_{th} $) equilibrium flow where there is a high degree of axisymmetry \cite{vries2008effect}, and thus that the need for axisymmetry is not a restriction. }
Using our general expression for the gradient operator \ref{eqn:gradoperator}, we can write 
\begin{equation}
 \vct{u} \cdot \nabla = R^2 \omega \nabla \phi \cdot	\left( \uv{x}\left| \nabla x \right|\frac{\partial }{\partial x} + 
\uv{y}\left| \nabla y \right|\frac{\partial }{\partial y} +
\uv{b}k_p\frac{\partial }{\partial \theta}  \right)
\end{equation} The third term in brackets is clearly 0, and the first term is 0 as a result of axisymmetry. Using \eqref{eqn:alphadef} and axisymmetry we can write 
\begin{eqnarray}
	\nabla\phi\cdot\uv{y} &=&  \nabla \phi \cdot \frac{\nabla\alpha}{\left| \nabla\alpha \right|}
	%\nonumber\\
	%&=&
=	\frac{1}{\left| \nabla\alpha \right|}\nabla\phi \cdot\left( \nabla \phi + q \nabla\alpha + \theta \nabla q + \nabla\nu \right)\nonumber\\
	&=& \frac{1}{\left| \nabla\alpha \right| R^2 }, 
\end{eqnarray} so that
\begin{eqnarray}
	\vct{u}\cdot \nabla = \omega a \left| \frac{\partial \psi_N}{\partial \rho} \right|_{\rho_0} \frac{\partial }{\partial y}.
\end{eqnarray}
%Expanding $\omega $ around the surface $\psi=\psi_0 $:
%\begin{equation}
	%\omega = \omega_0 + \left. \frac{d \omega}{d \psi_N} \right|_{\psi_0}	\left( \psi_N - \psi_{N0}\right),
%\end{equation}
Using the local approximation \eqref{localomegagrad},
then transforming to a frame rotating with velocity $\omega_0$
%and using \eqref{eqn:xdef}
we can write 
\begin{equation}
	%\vct{u}\cdot \nabla = x \sigma_{\psi}\frac{\rho_0}{q_0}\left. \frac{d \omega}{d \rho} \right|_{\rho_0} \frac{\partial }{\partial y} = x \gamma_E \frac{\partial }{\partial y}.
	\vct{u}\cdot \nabla  = x \gamma_E \frac{\partial }{\partial y}.
\end{equation}
%where
%\begin{equation}
	%\sigma_{\psi\rho} = \mathrm{sign}\left[ \frac{\partial \psi}{\partial \rho} \right],
	%\label{eqn:signpsirhodef}
%\end{equation}
%and 

\subsection[The Parallel Streaming Operator]{The Parallel Streaming Operator $\uv{b}\cdot\nabla$}
%Trivially from \eqref{eqn:gradoperator} we have $\uv{b}\cdot\nabla = (  \uv{b}\cdot\nabla\theta)\, \partial / \partial \theta$.
In the \texttt{GS2} coordinate system $\uv{b}\cdot\nabla$ can be  written simply:

\begin{equation}
	\uv{b}\cdot\nabla = k_p \frac{\partial }{\partial \theta}.
	%\label{}
\end{equation}
The factor $k_p$ is calculated directly by the geometry module.

\subsection[The Nonlinear Term]{The Nonlinear Term $\left( c/B \right)\left( \vct{b}\times\nabla \gyroR{\field} \right) \cdot \nabla h_s$}

Using the expression for the gradient operator \eqref{eqn:gradoperator}, we may write 
%\begin{equation}
	%\uv{b}\times\nabla\gyroR{\field} = - \uv{y}\left| \nabla x  \right| \sin \upsilon \frac{\partial \gyroR{\field}}{\partial x} + \uv{x}\left| \nabla y \right|\sin \upsilon\frac{\partial \gyroR{\field}}{\partial y},
%\end{equation}
%which allows us to write 
\begin{eqnarray}
	\frac{c}{B}	\left( \vct{b}\times\nabla \gyroR{\field} \right) \cdot \nabla h_s &=& 
	\frac{c}{B}\frac{\left| \nabla x  \right| \left| \nabla y  \right|}{ \sin \upsilon}
	\left( \frac{\partial \gyroR{\field}}{\partial y} \frac{\partial h_s}{\partial x} -
	\frac{\partial \gyroR{\field}}{\partial x}\frac{\partial h_s}{\partial y}\right)
	\nonumber\\
	&=& \frac{c}{B}\frac{\left| q_0 \right|}{\rho_0}\frac{B}{B_a}\left| \frac{\partial \psi_N}{\partial \rho} \right|_{\rho_0}	\left( \frac{\partial \gyroR{\field}}{\partial y} \frac{\partial h_s}{\partial x} -
	\frac{\partial \gyroR{\field}}{\partial x}\frac{\partial h_s}{\partial y}\right)\qquad
	\label{eqn:nonlineartermsings2coordinates} \\
	&\equiv&   \mathcal{N}_s
\end{eqnarray}
where $\sin \upsilon \equiv (\uv{x} \times \uv{y}) \cdot \uv{b}$ (and thus $\left| \nabla\psi \right| \left| \nabla\alpha \right| = B \sin \upsilon$), and where we have defined the nonlinear term, calculated in Section \ref{sec:nonlinearterms}, as
\begin{equation}
	\mathcal{N}_s \equiv \kappa_x \frac{c}{B_a}\left(  \pd{\gyroR{\field}}{x}\pd{h}{y} - \pd{\gyroR{\field}}{y}\pd{h}{x}\right) 
	\label{poissonbracket},
\end{equation}
where 
\begin{equation}
	\kappa_x \equiv \frac{\left| q_0 \right|}{\rho_0}\left| \frac{\partial \psi_N}{\partial \rho} \right|_{\rho_0},
	\label{eqn:kxfacdef}
\end{equation}
and is called \texttt{kxfac} within \texttt{GS2}.

\subsection[The Radial Perturbed Flow]{The Radial Perturbed Flow $\left( c/B \right)\left(\vct{b}\times\nabla \gyroR{\field}\right) \cdot \nabla\psi$}

The radial component of the perturbed flow $\nabla\psi\cdot(\uv{b}\times\nabla\gyroR{\field}) $ can now be written very simply. Using the general expression for the $\nabla $ operator, \eqref{eqn:gradoperator}, we can write 
\begin{eqnarray}
	\frac{c}{B}(\uv{b}\times\nabla\gyroR{\field}) \cdot \nabla\psi
	&=&  \frac{c}{B}\left( \uv{b}\times\uv{y} \left| \nabla y \right|\frac{\partial \gyroR{\field} }{\partial y} \right)\cdot\nabla\psi\nonumber\\
	&=& \frac{c}{B}\uv{x}\cdot\nabla\psi\frac{\left| \nabla y \right|}{\sin \upsilon}\frac{\partial \gyroR{\field} }{\partial y}\nonumber\\
	&=& \frac{c}{B}\left| \nabla\psi \right| a \left| \frac{\partial \psi_N}{\partial \rho} \right|_{\rho_0} \frac{\left| \nabla\alpha \right|}{\sin \upsilon} \frac{\partial \gyroR{\field} }{\partial y}\nonumber\\
	&=& \frac{c}{aB_a}  \left| \frac{\partial \psi}{\partial \rho} \right|_{\rho_0} \frac{\partial \gyroR{\field} }{\partial y}.
\end{eqnarray}

\subsection[The Background Gradients]{The Background Gradients $\partial F_s / \partial \psi$}
Using the expression for the equilibrium distribution function \eqref{eqn:Fsdef}, we may write:
\begin{eqnarray}
	\frac{\partial F_s}{\partial \psi} &=&  
	\frac{\partial }{\partial \psi}\left[ n_s (\psi) \left( \frac{m_s}{2\pi T_s (\psi)} \right)^{3/2} e^{-\varepsilon_s/T_s} \right]\nonumber\\
	&=& F_s \left[ \frac{1}{n_s}\frac{\partial n_s}{\partial \psi} + \left( \frac{\varepsilon_s}{T_s} - \frac{3}{2} \right)\frac{1}{T_s}\frac{\partial T_s}{\partial \psi} \right]\nonumber\\
	&=& F_s \left[ \frac{1}{n_s}\frac{\partial n_s}{\partial \rho} + \left( \frac{\varepsilon_s}{T_s} - \frac{3}{2} \right)\frac{1}{T_s}\frac{\partial T_s}{\partial \rho} \right]\frac{\partial \rho}{\partial \psi}.
	%\label{eqn:gradientscalelengthdefs}
\end{eqnarray}
Using the local approximation, \eqref{eqn:Lnsdef} and \eqref{eqn:LTsdef}, we may write
\begin{equation}
	\frac{\partial F_s}{\partial \psi} =   - F_s \left[ \frac{a}{L_{ns}}+ \left( \frac{\varepsilon_s}{T_s} - \frac{3}{2} \right) \frac{a}{L_{Ts}}\right]\left| \frac{\partial \rho}{\partial \psi} \right|_{\psi_0} \sigma_\psi.
	\label{eqn:backgroundgradients}
\end{equation}

\section{The Gyrokinetic Equation in the \texttt{GS2} Coordinate System}

Having written the geometric operators in the \texttt{GS2} coordinate system we may now substitute them into the gyrokinetic equation \eqref{gyrokineticeq1}, giving 
%\begin{multline}
	%\left( \frac{\partial }{\partial t} + x \gamma_E \frac{\partial }{\partial y} \right)h_s
	%+ v_{\parallel} k_p \frac{\partial h_s}{\partial \theta}
	%+ \vct{V}_{Ds}\cdot \nabla h_s +  \mathcal{N}_s - \gyroR{C\left[ h_s \right]} = 	\\
	%\frac{Z_s e F_s}{T_s}\left( \frac{\partial }{\partial t} + x \gamma_E \frac{\partial }{\partial y} \right)\gyroR{\field} \\
	%+ \left\{ \left[ \frac{a}{L_{ns}}+\left( \frac{\varepsilon_s}{T_s}- \frac{3}{2} \right)\frac{a}{L_{Ts}} \right]\left. \frac{\partial\rho}{\partial\psi} \right|_{\psi_0}  
	%-\frac{ m_s I(\psi)}{T_s B} v_{\parallel} \frac{q_0}{\rho_0}\left| \frac{\partial \rho}{\partial \psi} \right|_{\psi_0}\gamma_E\right\}F_s \frac{c}{a B_a}  \left| \frac{\partial \psi}{\partial \rho} \right|_{\rho_0}\frac{\partial \gyroR{\field} }{\partial y}
%\end{multline}
%which reduces to 
\begin{multline}
	\left( \frac{\partial }{\partial t} + x \gamma_E \frac{\partial }{\partial y} \right)h_s
	+ v_{\parallel} k_p \frac{\partial h_s}{\partial \theta}
	+ \vct{V}_{Ds}\cdot \nabla h_s + \mathcal{N}_s - \gyroR{C\left[ h_s \right]} = 	\\
	\frac{Z_s e F_s}{T_s}\left( \frac{\partial }{\partial t} + x \gamma_E \frac{\partial }{\partial y} \right)\gyroR{\field} \\
	+ \left\{ \left[ \frac{a}{L_{ns}}+\left( \frac{\varepsilon_s}{T_s} -\frac{3}{2} \right)\frac{a}{L_{Ts}} \right]\sigma_\psi  
	- \frac{ m_s I(\psi)}{T_s B} v_{\parallel} \frac{q_0}{\rho_0}\gamma_E\right\}\frac{F_s c}{B_a a} \frac{\partial \gyroR{\field} }{\partial y}
	\label{eqn:gyrokineticeqgs2coordinates}
\end{multline}
where \texttt{GS2} always makes the choice $\sigma_\psi=1 $.

This is the equation that \texttt{GS2} solves in an axisymmetric toroidal system. The limits of a cylinder and a slab can easily be derived from this equation.
\texttt{GS2} can also solve the gyrokinetic equation in a non-axisymmetric system, for example, a stellarator. This, however, is not covered in this work.

\section{Normalisations}

As is standard, normalisations within \texttt{GS2}  are chosen such that the normalised quantities are of order unity. First, a set of normalising quantities are chosen which are easily determined for a real-life system, and then all quantities within the gyrokinetic equation are normalised appropriately. Perturbed quantities are scaled up by a factor of $\epsilon_{\mathrm{gk}} $ to keep them, in the general case, of order unity.

\subsection{Normalising Quantities}
\label{sec:normalisingquantities}

%\begin{table}[h]
	%\centering
	%\begin{tabularx}{14.2cm}{r c X}
		%\hline
		%\hline
		%Used for  & Quantity & Definition \\
		%\hline
		%Macro Lengths & $a$ & half radius of LCFS\\
		%Temperature & $T_r$ & temperature of ref. species\\
		%Mass & $m_r$ & mass of ref. species\\
		%Velocity & $v_{thr}$ & $\sqrt{2T_r/m_r}$\\
		%Time & $t_r$ & $a/v_{thr}$\\
		%Mag. Field & $B_a$ & toroidal magnetic field at the centre of the flux surface at the height of the magnetic axis\\
		%Cyclotron Freq. & $\Omega_r$ & $eB_a / m_r c$\\
		%Micro Lengths & $\rho_r$ & $v_{thr}/\Omega_r$\\
		%%Perturbed Fields & $\field_r$ & $ ea/T_r\rho_r$\\
		%%Perturbed Dist. Fn. & $h_r$ & 
		%\hline
	%\end{tabularx}
	%\caption{Normalising Quantities: the definitions of the quantities used in the \texttt{GS2} normalisations.}
	%\label{tab:normalisingquantities}
%\end{table}

\begin{table}[h]
	\centering
	\begin{tabularx}{4.5in}{ c X}
		\hline
		\hline
		Quantity & Definition \\
		\hline
		 $a$ & half diameter of the LCFS at the height of the magnetic axis \\
		 $T_r$ & temperature of ref. species\\
		 $m_r$ & mass of ref. species\\
		 $n_r$ & density of ref. species\\
		 $Z_r\equiv1$ & charge of ref. species\\
		 $v_{thr}$ & $\sqrt{2T_r/m_r}$\\
		 %$t_r$ & $a/v_{thr}$\\
		 $B_a$ & toroidal magnetic field at the centre of the flux surface at the height of the magnetic axis\\
		 $\Omega_r$ & $Z_r eB_a / m_r c$\\
		 $\rho_r$ & $v_{thr}/\Omega_r$\\
		%Perturbed Fields & $\field_r$ & $ ea/T_r\rho_r$\\
		%Perturbed Dist. Fn. & $h_r$ & 
		\hline
	\end{tabularx}
	\caption{Normalising Quantities: the definitions of the quantities used in the \texttt{GS2} normalisations.}
	\label{tab:normalisingquantities}
\end{table}

There are several things to bear in mind concerning the normalising quantities used in \texttt{GS2}, which are listed in Table \ref{tab:normalisingquantities}.
The first is that the reference species is not one of the species within the problem, for example ions or electrons, but is a hypothetical species to whose properties all the properties of the actual species are normalised. If, for example, the density or temperature of one of the actual species is chosen to be exactly 1.0, then the density or temperature of that species will happen to be the density or temperature of the reference species, but otherwise the density or temperature of the reference species will not correspond to any real quantity.
However, the properties of the reference species can easily be calculated using the normalised properties of any actual species and the real values of those properties for that actual species in the system of interest.
Note also that the reference charge, $Z_r$, is always equal to 1.
Thus the normalised charge of electrons must always be set to -1, and that of hydrogenic ions to 1.

The second is that the quantity $a $ does not actually appear in the original gyrokinetic equation, or in its derivation (see Ref.~\cite{flowtome1}). Thus, although many quantities are normalised to it, and although it is given the definition ``the half diameter of the LCFS at the height of the magnetic axis'', in certain simplified geometries, most notably the slab, the cylinder, and shifted circle geometry, it has no physical significance. In more complicated geometries, its only physical significance is in the calculation of some of the geometric factors listed in Table \ref{tab:fieldoperators}.

The third is that the normalising magnetic field $B_a $ will in many simpler magnetic geometries be simply equal to the magnetic field strength at the magnetic axis. Its rather involved definition becomes important in more complicated geometries.

\subsection{Normalised Quantities}

The normalised quantities used within \texttt{GS2} are laid out in Table \ref{tab:normalisedquantities}. Table \ref{tab:normalisedquantities} covers only the quantities and variables contained within the closed system of equations consisting of the gyrokinetic equation, the quasineutrality condition and Amp\`ere's law. It does not cover the myriad diagnostic quantities which can be output from \texttt{GS2}, which are not covered in this work, since we are concerned with only two: $Q$ and $\Pi$, whose normalisations are given in Chapter \ref{sec:fluxnorms}.

\begin{table}
	\centering
	\begin{tabular}{r c c l}
		\hline
		\hline
		Name & Quantity & Definition & \texttt{GS2}  Name\\
		\hline
		Eqbm. Dist. Fn. & $F_{Ns} $ & $F_s/(n_s/v_{ths}^3)$ \\
		Perturbed Dist. Fn. & $h_{Ns}$ & \( h_s/ (\rho_r/a)F_s\)\\
		Comp. Pert. Dist. Fn. & $g_{Ns}$ & \( g_s/ (\rho_r/a)F_s\) & $\mathtt{g}$\\
		Pert. Electric Potential & $\field_N$ & $\field/(\rho_r /a)(T_r/e)$& $\mathtt{phi}$\\ 
		%Time & $t_N$ & $t/t_r$\\
		Time & $t_N$ & $t/(a/v_{thr})$ & $\mathtt{time}$\\
		Parallel Velocity & $v_{\parallel N}$ & $v_\parallel / v_{ths}$ & $\mathtt{vpa}$\\
		Perp. Velocity Squared & $v_{\perp N}^2$ & $v_\perp^2 / v_{ths}^2$ & $\mathtt{verp2}$\\
		Thermal Velocities & $v_{thNs}$ & $v_{ths} / v_{thr} = \sqrt{T_{Ns}/m_{Ns}}$ & $\mathtt{stm}$\\
		Gradient Operator & $\nabla_N$ & $a\nabla$  & \\
		Parallel Gradient & $k_{pN} $ & $ a k_p $  & $\mathtt{gradpar}$\\
		Radial Coordinate & $x_N$ & $x/\rho_r$  & \\
		Poloidal Coordinate & $y_N$ & $y/\rho_r$  & \\
		Radial Wavenumber & $k_{xN}$ & $k_x \rho_r$  & $\mathtt{akx}$ \\
		Poloidal Wavenumber & $k_{yN}$ & $k_y \rho_r$  & $\mathtt{aky}$\\
		Magnetic Field & $B_N$ & $B/B_a$  & $\mathtt{bmag}$\\
		Magnetic Flux & $\psi_N$ & $\psi/a^2 B_a$  & \\
		Current & $I_N$ & $I/aB_a$ & \\
		Nonlinear Term & $\mathcal{N}_{Ns}$ & $\mathcal{N}_s / (c/B_a) (T_r/e) (F_s/a^2)$\\
		Collision Operator & $C_N$ & $C/(v_{thr}/a)(\rho_r/a)F_s$\\
		Energy & $\varepsilon_{Ns}$ & $\varepsilon_s /T_s$  & $\mathtt{energy}$\\
		Magnetic Moment & $\mu_{Ns}$ & $\mu/(T_s/2B_a)$ \\
		Lambda	 & $\lambda_{s}$ & $\mu_{Ns}/\varepsilon_{Ns}$  & $\mathtt{l}$\\
		Flow Shear & $\gamma_{EN}$ & $\gamma_E/(v_{thr}/a)$ & $\mathtt{g\_exb}$\\
		Temperature Gradients & $\kappa_s$ & $a/L_{Ts}$ & $\mathtt{tprim}$\\
		Density Gradients & $\kappa_{ns}$ & $a/L_{ns}$ & $\mathtt{fprim}$\\
		Temperatures  & $T_{Ns}$ & $T_s/T_r$ & $\mathtt{temp}$\\
		Densities  & $n_{Ns}$ & $n_s/n_r$ & $\mathtt{dens}$\\
		Masses  & $m_{Ns}$ & $m_s/m_r$ & $\mathtt{mass}$\\
		Charges  & $Z_{Ns}$ & $Z_s/Z_r=Z_s$ & $\mathtt{z}$\\
		\hline 
	\end{tabular}
	\caption{Normalised Quantities: definitions of the normalised quantities used within \texttt{GS2},  along with the names used for those quantities within \texttt{GS2}, where appropriate.}
	\label{tab:normalisedquantities}
\end{table}

\subsection{Normalisation of the Nonlinear Term}

Defining 

\begin{equation}
	\mathcal{N}_{Ns} = \kappa_x \left( \frac{\partial \gyroR{\field_N}}{\partial y_N}\frac{\partial h_{Ns}}{\partial x_N} - \frac{\partial \gyroR{\field_N}}{\partial x_N}\frac{\partial h_{Nx}}{\partial y_N} \right),
	\label{eqn:nonlineartermsnormalised}
\end{equation}
we find that

\begin{equation}
	\mathcal{N}_{Ns} = \mathcal{N}_{N}\left( \frac{c}{B_a}\frac{T_r}{e}\frac{F_s}{a^2} \right)^{-1}.
\end{equation}

%\subsection{Normalisation of the Magnetic Drifts}

\subsection[The Normalised Gyrokinetic Equation]{The Normalised Gyrokinetic Equation in the \texttt{GS2} Coordinate System}
Having defined a set of normalised quantities, we now substitute them into the gyrokinetic equation \eqref{eqn:gyrokineticeqgs2coordinates} to give 
\nopagebreak
\begin{multline}
	\left( \frac{v_{thr}}{a}\frac{\rho_r}{a}F_s \right) \left( \frac{\partial }{\partial t_N}+x_N\gamma_{EN}\frac{\partial }{\partial y_N} \right) h_{Ns} +
	\left( \frac{v_{thr}}{a}\frac{\rho_r}{a}F_s \right) \frac{v_{ths}}{v_{thr}} v_{\parallel N}k_{pN}\frac{\partial h_s}{\partial \theta} \\+
	\left( \frac{v_{thr}}{a}\frac{\rho_r}{a}F_s \right) \vct{V}_{DNs}\cdot \nabla_N h_{Ns}+
	\frac{c}{B_a}\frac{T_r}{e}\frac{F_s}{a^2} \mathcal{N}_{Ns} - 
	\left( \frac{v_{thr}}{a}\frac{\rho_r}{a}F_s \right) \gyroR{C_N\left[ h_{Ns} \right]} =\\
\frac{Z_s}{T_{Ns}}\left( \frac{v_{thr}}{a}\frac{\rho_r}{a}F_s \right) \left( \frac{\partial }{\partial t_N}+x_N\gamma_{EN}\frac{\partial }{\partial y_N} \right)\gyroR{\field_N} \\+
	\left\{ \left[ \kappa_{ns} - \left( \varepsilon_{Ns} - \frac{3}{2} \right)\kappa_s \right]\sigma_\psi - \frac{m_s v_{ths}v_{thr}}{T_s}\frac{I_N}{B_N}\frac{q_0}{\rho_0}\gamma_{EN} \right\}	\frac{c}{B_a}\frac{T_r}{e}\frac{F_s}{a^2} \frac{\partial \gyroR{\field_N}}{\partial y}.
\end{multline}  Using the identity 

\begin{equation}
	\frac{c}{B_a}\frac{T_r}{e}\frac{F_s}{a^2} = \frac{1}{2} \frac{v_{thr}}{a}\frac{\rho_r}{a}F_s,
\end{equation}
and dividing by $(v_{thr}/a)(\rho_r/a)F_s$, we obtain the normalised axisymmetric gyrokinetic equation in \texttt{GS2} coordinates:

\begin{multline}
\left( \frac{\partial }{\partial t_N}+x_N\gamma_{EN}\frac{\partial }{\partial y_N} \right) h_{Ns} +
v_{thNs} v_{\parallel N}k_{pN}\frac{\partial h_{Ns}}{\partial \theta} +
	 \vct{V}_{DNs}\cdot \nabla_N h_{Ns}+
	 \frac{1}{2}\mathcal{N}_{Ns} - 
	\gyroR{C_N\left[ h_{Ns} \right]} =\\
	\frac{Z_s}{T_{Ns}} \left( \frac{\partial }{\partial t_N}+x_N\gamma_{EN}\frac{\partial }{\partial y_N} \right)\gyroR{\field_N} \\+
	\left\{ \left[ \kappa_{ns} - \left( \varepsilon_{Ns} - \frac{3}{2} \right)\kappa_s \right]\sigma_\psi - \frac{2}{v_{thNs}}\frac{I_N}{B_N}\frac{q_0}{\rho_0}\gamma_{EN} \right\} \frac{1}{2}\frac{\partial \gyroR{\field_N}}{\partial y}.
	\label{eqn:gyrokineticeqnormalised}
\end{multline}

\subsection{The Normalised Quasineutrality Condition}

In a similar way, we can write the normalised quasineutrality condition: 
\begin{equation}
	\sum_s Z_s^2 \frac{n_{Ns}}{T_{Ns}} \varphi_N = \sum_s Z_s \int \frac{d^3\vct{v}}{v_{ths}^3} n_{Ns} \gyror{F_{Ns}h_{Ns}}.
	\label{eqn:quasineutralitynormalised}
\end{equation}

\section{Non-Trivial Geometric Operators}

The geometric operators in Table \ref{tab:fieldoperators} which are not trivial to calculate in the \texttt{GS2} coordinate system are calculated by the \texttt{GS2} geometry module on a case-by-case basis, depending on the specific magnetic configuration chosen.
These operators are $k_p$, $\vct{b}\times\left( \vct{b}\cdot\nabla\vct{b} \right)$, $\vct{b}\times\left( \nabla\ln B \right)$ and $k_{\perp}$.
In certain simplified geometries, for example, a slab, a cylinder or shifted circle geometry, they can be written as simple functions of $\theta $, but in the general case, for example a numerical equilibrium taken from experiment, they must be calculated from the data for each grid point in the simulation.
In this section, we describe how these operators are related to the normalised quantities that are calculated by the geometry module.

\subsection{The Parallel Gradient}

As noted earlier, the parallel streaming  operator is written
\begin{equation}
	\vct{b}\cdot \nabla = \vct{b}\cdot\nabla\theta\frac{\partial }{\partial \theta} = k_p (\theta)\frac{\partial }{\partial \theta}.
\end{equation}
The parallel gradient $k_p $, called $\mathtt{gradpar} $ within \texttt{GS2}, is calculated by the geometry module.

\subsection{The Magnetic Drifts}

The magnetic drifts are among those terms which are not rendered simple to express by the coordinate system used in \texttt{GS2}; hence, we do not expand $\vct{V}_{Ds} $ in the gyrokinetic equation \eqref{eqn:gyrokineticeqnormalised}.

For convenience of calculation, \texttt{GS2} splits up the magnetic drifts into four components in the following way. First we use Equation (127) from Ref.~\cite{flowtome1}:
\begin{equation}
	\frac{1}{c}\vct{j}\times\vct{B} = \nabla p,
\end{equation} where $\vct{j} = (c/4\pi) \nabla\times\vct{B}$ and $p$ is the total pressure, which expresses the balance between the magnetic field and the equilibrium pressure gradient (neglecting the high Mach terms), to write 
\begin{equation}
	\vct{B}\times\left( \vct{B}\cdot\nabla\vct{B} \right) = 
	4\pi \vct{B}\times\nabla p + \vct{B}\times\left[ \left( \nabla \vct{B} \right)\cdot \vct{B} \right].
\end{equation} Using the identities $\vct{B}\times\left( \vct{B}\cdot\nabla\vct{B} \right)= B^2 \vct{B} \times\left( \uv{b}\cdot\nabla\uv{b} \right)$ and $\vct{B}\times\left[ \left( \nabla \vct{B} \right)\cdot \vct{B} \right] = B\left( \vct{B}\times\nabla B \right)$ we can then write 

\begin{eqnarray}
	\vct{V}_{Ds}&=&  \frac{\uv{b}}{\Omega_s}\times\left( v_{\parallel}^2 \uv{b}\cdot\nabla \uv{b} + \frac{v_{\perp}^2}{2B}\nabla B \right)\nonumber\\
	&=& \frac{1}{\Omega_s}\left[ \frac{1}{B^2}\left( v_{\parallel}^2 + \frac{v_{\perp}^2}{2}  \right)\vct{B}\times\nabla B 
	+ \frac{ 4 \pi}{B^3 }v_{\parallel}^2\vct{B}\times\nabla p \right].
	\label{eqn:magneticdriftswithpressuregradient}
\end{eqnarray}
Using \eqref{eqn:not0eikonal} and normalising the magnetic drifts we may now write 
\begin{multline}
	\vct{V}_{DNs}\cdot\nabla_N h_{Ns} =\\ \frac{i \hat{h}_{Ns}}{\Omega_s}\frac{v_{ths}^{2}}{a v_{thr}}\left[ \frac{1}{B_N^2}\left( v_{\parallel N}^2 + v_{\perp N}^2 \right) \vct{B}_N\times\nabla_N B_N + \frac{v_{\parallel N}^2}{B_N^3}\frac{4\pi}{B_a^2}\vct{B}_N\times\nabla p_s  \right]\cdot \nabla_N S.
\end{multline}
Writing 
\begin{equation}
	\nabla_N S = \frac{a}{\rho_r}k_{yN}\frac{\partial \psi_N}{\partial \rho}\left( \nabla_N \alpha + \frac{k_{xN}}{k_{yN}\hat{s}}\nabla_N q \right),
\end{equation} 
and using the identity
\begin{equation}
	\frac{v_{ths}^2 B_N }{2 v_{thr} \Omega_s \rho_r} = \frac{1}{2}\frac{T_s}{T_r}\frac{1}{Z_s},
\end{equation}
we may write 
\begin{multline}
	\vct{V}_{DNs}\cdot\nabla_N h_{Ns} \\
	=\frac{i \hat{h}_{Ns} k_{yN}}{2}\frac{T_s}{T_r}\frac{1}{Z_s} 
\left[ 
\left( v_{\parallel N}^2 + v_{\perp N}^2 \right) 
\frac{2}{B_N^2}\frac{\partial \psi_N}{\partial \rho}
\left( \uv{b} \times \nabla_N B_N \right)\cdot
\left( \nabla_N \alpha + \frac{k_{xN}}{k_{yN} \hat{s}}\nabla_N q \right)
\right. \qquad \\ \left.
+ v_{\parallel N}^2 \frac{1}{B_N^3} \frac{\partial \psi_N}{\partial \rho}
\left(\uv{b} \times \nabla_N \beta_a \right)\cdot
\left( \nabla_N \alpha + \frac{k_{xN}}{k_{yN} \hat{s}}\nabla_N q \right)
\right],
\end{multline}
where $\beta_a = 8\pi p/B_a^2$.
Remembering that $\uv{b}\times\nabla p \cdot \nabla q = 0 $  because the equilibrium pressure is a flux function, we may now write 
\begin{eqnarray}
	\vct{V}_{DNs}\cdot\nabla_N h_{Ns}= i \frac{T_{Ns}}{Z_s}\omega_{DN} \hat{h}_{Ns}
	\label{eqn:omegaDNdef}
\end{eqnarray}
where
\begin{eqnarray}
	\omega_{DN}=	\frac{ k_{yN}}{2} 
\left[ 
\frac{v_{\perp N}^2}{2}
\left( \omega_{\nabla B N}+ \frac{k_{xN}}{k_{yN} \hat{s}}\omega_{\nabla B N}^0 \right)
\right.
+ 
\left.
v_{\parallel N}^2 
\left( \omega_{\kappa N}+ \frac{k_{xN}}{k_{yN} \hat{s}}\omega_{\kappa N}^0 \right)
\right] 	\label{eqn:magneticdrifts}
\end{eqnarray}
and where we have defined the four normalised magnetic drifts\footnote{ Known as $\mathtt{gbdrift} $, $\mathtt{gbdrift0} $, $\mathtt{cvdrift} $ and $\mathtt{cvdrift0} $ respectively within \texttt{GS2}.}
\begin{eqnarray}
	\omega_{\nabla B N}(\theta) &=&  
\frac{2}{B_N^2}\frac{\partial \psi_N}{\partial \rho}
\left( \uv{b} \times \nabla_N B_N \right)\cdot
\nabla_N \alpha 
\label{eqn:gbdriftdef},\\
	\omega_{\nabla B N}^0(\theta) &=&  
\frac{2}{B_N^2}\frac{\partial \psi_N}{\partial \rho}
\left( \uv{b} \times \nabla_N B_N \right)\cdot
\nabla_N q 
\label{eqn:gbdrift0def},\\
\omega_{\kappa N}(\theta) &=& 
 v_{\parallel N}^2 \frac{1}{B_N^3} \frac{\partial \psi_N}{\partial \rho}
\left(\uv{b} \times \nabla_N \beta_a \right)\cdot
\nabla_N \alpha + \omega_{\nabla B N}
\label{eqn:cvdriftdef},\\
\omega_{\kappa N}^0(\theta) &=& \omega_{\nabla B N }^0 
\label{eqn:cvdrift0def},
\end{eqnarray}
which are output by the \texttt{GS2} geometry module.

\subsection{The Perpendicular Wavenumber}

The perpendicular wave number appears in the Bessel functions which arise as a result of gyroaveraging spectral quantities.
Although the perpendicular wave number can be written simply in the \texttt{GS2} coordinate system: 
\begin{eqnarray}
	k_{\perp}^2 
	= \left| \nabla S \right|^2 
	=
	k_y^2 \left| \nabla y \right|^2
	+
	k_x^2 \left| \nabla x \right|^2 
	+ 2 k_x k_y \left| \nabla x \right| \left| \nabla y \right| \cos \upsilon
\end{eqnarray}
the  factors $\left| \nabla x \right| $,  $\left| \nabla y \right|$ and $\cos \upsilon$ are in  general functions of $\theta$ and  nontrivial to calculate. \texttt{GS2} uses the definition \eqref{eqn:Seikdef} to write $k_{\perp}^2 $ in terms of three numbers: 
\begin{eqnarray}
	k_{\perp}^2 (\theta)
  = 
	k_y^2\left| g_1 + 2 \frac{k_x}{k_y \hat{s}}g_2 + \left(  \frac{k_x}{k_y \hat{s}}\right)^2 g_3  \right|
	\label{eqn:kperp2def}
\end{eqnarray} where 
\begin{eqnarray}
	g_1(\theta) &=&  \left| \nabla_N \alpha \right|^2, \\
	g_2(\theta) &=&  \nabla_N \alpha \cdot \nabla_N q, \\
	g_3(\theta) &=&  \left| \nabla_N q \right|^2.
\end{eqnarray}
These three numbers are calculated by the \texttt{GS2} geometry module for the particular geometry in use.
It should be noted that $g_2 $ and $g_3 $ are both equal to 0 wherever  $\hat{s}=0 $.

\section{Geometric Factors in Specific Geometries}

\subsection{A Note on the ``$s$---$\alpha$'' Formalism}

As described by Miller \cite{miller1998noncircular}, the ``$s$---$\alpha$''  formalism \cite{greene1981second} is a method for perturbing a general magnetic equilibrium by changing only two parameters, $\partial p / \partial \psi $, the gradient of the pressure, and $\partial q / \partial \psi $, the gradient of the magnetic safety factor. Changing the pressure gradient causes the magnetic flux surfaces to shift one way or the other according to the Grad-Shafranov equation; this is known as the Shafranov shift. Changing the safety factor gradient (i.e. the magnetic shear) changes the angle between the magnetic field lines in one flux surface and the next, which has a dramatic effect on the stability of certain modes (as will be noted later on in this work). The ``$s$---$\alpha$'' formalism does not specify the magnetic equilibrium to be perturbed; hence, referring to ``$s$---$\alpha$'' geometry is misleading.

Two very commonly used geometric approximations which employ the ``$s$---$\alpha$'' formalism are shifted circle geometry and Miller geometry \cite{miller1998noncircular}. Shifted circle geometry, which is used in this work, is described in Section \ref{sec:shiftedcircles}.

\subsection{Shifted Circles}
\label{sec:shiftedcircles}
The shifted circle approximation describes a magnetic equilibrium composed of nested circular flux surfaces, generally, as here, taken in the limit of high aspect ratio. These flux surfaces may be shifted radially in the mid-plane by the Shafranov shift $\alpha_s$, and the field lines may be sheared from one surface to the next according to the magnetic shear parameter $\hat{s} $.
 The geometric factors used by \texttt{GS2} can be written explicitly for shifted circle geometry and are given in Table \ref{tab:shiftedcirclefactors}.
 It should be noted that the exact geometric factors which are implied by a ``shifted circle equilibrium'', or an ``$s$---$\alpha$'' equilibrium are not universally defined, and the different conventions can have significantly different results \cite{lapillonne2009clarifications} (indeed, several different options are available within \texttt{GS2}). Here we merely quote the values used in this study.

\begin{table}[h]
	\centering
	\begin{tabular}{c c l}
		\hline
		\hline
		Quantity & Value & \texttt{GS2} Name \\
		\hline
		$B_N$ & $\left( 1.0 + (r/R) \cos{\theta}  \right)^{-1}$ & $\mathtt{bmag}$ \\
		$\partial \rho / \partial \psi_N$ & $a/r$ & $\mathtt{drhodpsi}$ \\
		$\kappa_x$ & $1.0$ & $\mathtt{kxfac}$ \\
		$\omega_{\kappa N}$ & $(2a/R) \left( \cos{\theta} + (\hat{s}\theta - \alpha_s \sin{\theta}) \sin{\theta} \right)$ & $\mathtt{cvdrift}$ \\
		$\omega_{\kappa N}^0$ & $(2a/R)\hat{s}\sin{\theta}$ & $\mathtt{cvdrift0}$ \\
		$\omega_{\nabla B N}$ & $(2a/R) \left( \cos{\theta} + (\hat{s}\theta - \alpha_s \sin{\theta}) \sin{\theta} \right)$ & $\mathtt{gbdrift}$ \\
		$\omega_{\nabla B N}^0$ & $(2a/R)\hat{s}\sin{\theta}$ & $\mathtt{gbdrift0}$ \\
		$g_1$ & $1 + (\hat{s}\sin{\theta} - \alpha_s \sin{\theta})^2$ & $\mathtt{gds2}$ \\
		$g_2$ & $-\hat{s}\left( \hat{s}\theta - \alpha_s \sin{\theta} \right)$ & $\mathtt{gds21}$\\
		$g_3$ & $\hat{s}^2$ & $\mathtt{gds22}$\\
		$k_p$ & $a/Rq$ & $\mathtt{gradpar}$\\
	\hline	
	\end{tabular}
	\caption{Geometric Factors in Shifted Circle Geometry}
	\label{tab:shiftedcirclefactors}
\end{table}

%\subsection{The Cylinder}

%\begin{table}[h]
	%\centering
	%\begin{tabular}{c c l}
		%\hline
		%\hline
		%Quantity & Value & \texttt{GS2} Name \\
		%\hline
		%$B_N$ & $\left( 1.0 + (r/R) \cos{\theta}  \right)^{-1}$ & $\mathtt{bmag}$ \\
		%$\partial \rho / \partial \psi_N$ & $a/r$ & $\mathtt{drhodpsi}$ \\
		%$\kappa_x$ & $1.0$ & $\mathtt{kxfac}$ \\
		%$\omega_{\kappa N}$ & $(2a/R) $ & $\mathtt{cvdrift}$ \\
		%$\omega_{\kappa N}^0$ & $0$ & $\mathtt{cvdrift0}$ \\
		%$\omega_{\nabla B N}$ & $(2a/R) (1 - s) $ & $\mathtt{gbdrift}$ \\
		%$\omega_{\nabla B N}^0$ & $0$ & $\mathtt{gbdrift0}$ \\
		%$g_1$ & $1 + (\hat{s}\sin{\theta} - s \sin{\theta})^2$ & $\mathtt{gds2}$ \\
		%$g_2$ & $-\hat{s}^2\theta$ & $\mathtt{gds21}$\\
		%$g_3$ & $\hat{s}^2$ & $\mathtt{gds22}$\\

	%\hline	
	%\end{tabular}
	%\caption{Geometric Factors in Cylindrical Geometry}
	%\label{tab:cylinderfactors}
%\end{table}

\subsection{The Slab}

 The slab is perhaps the most simple magnetic geometry possible, consisting of straight magnetic field lines with a constant shear. It can be derived as a reduction of the shifted circle equilibrium, by setting $\alpha_s = 0 $, and letting $r \rightarrow \infty $, $R \rightarrow \infty $ but $ r/R \rightarrow 0 $.
 The geometric factors used by \texttt{GS2} can be written very simply for slab geometry and are given in Table \ref{tab:slabfactors}.

\begin{table}[h]
	\centering
	\begin{tabular}{c c l}
		\hline
		\hline
		Quantity & Value & \texttt{GS2} Name \\
		\hline
		$B_N$ & $1.0$ & $\mathtt{bmag}$ \\
		$\partial \rho / \partial \psi_N$ & $1.0$ & $\mathtt{drhodpsi}$ \\
		$\kappa_x$ & $1.0$ & $\mathtt{kxfac}$ \\
		$\omega_{\kappa N}$ & $0 $ & $\mathtt{cvdrift}$ \\
		$\omega_{\kappa N}^0$ & $0$ & $\mathtt{cvdrift0}$ \\
		$\omega_{\nabla B N}$ & $0 $ & $\mathtt{gbdrift}$ \\
		$\omega_{\nabla B N}^0$ & $0$ & $\mathtt{gbdrift0}$ \\
		$g_1$ & $1 + (\hat{s}\sin{\theta} )^2$ & $\mathtt{gds2}$ \\
		$g_2$ & $-\hat{s}^2\theta$ & $\mathtt{gds21}$\\
		$g_3$ & $\hat{s}^2$ & $\mathtt{gds22}$\\
		$k_p$ & $\mathtt{kp}$ & $\mathtt{gradpar}$\\

	\hline	
	\end{tabular}
	\caption{Geometric Factors in Slab Geometry. The quantity $\mathtt{kp}$ is an input parameter to the simulation.}
	\label{tab:slabfactors}
\end{table}

\subsection{The Cyclone Base Case}
\label{sec:cyclonebasecase}

The Cyclone Base Case is a special case of shifted circle geometry which has been used for benchmarks between gyrokinetic codes (see e.g. \cite{dimits:969}). In effect, it is shifted circle geometry with  the following choice of parameters: 
\begin{eqnarray}
	\frac{r}{R} &=&  0.18, \\
	a &=&  R, \\
	q &=&  1.4, \\
	\kappa_i &=&  6.9, \\
	\kappa_{ni} &=&  2.2, \\
	\alpha_s &=&  0, \\
	\hat{s } &=&  0.8.
\end{eqnarray}
All simulations in this work use the Cyclone Base Case (with $\hat{s}=0$) as a starting point, with the exception of Chapter \ref{sec:subcriticalfluctuations}, which considers the case of the slab.

\webonly{
\chapter{The \texttt{GS2} Algorithm}
\label{sec:gs2algorithm}
\section{The Linear Implicit Algorithm}
\label{sec:linearimplicitalgorithm}
\subsection{Equations}

The linear solver advances the closed system consisting of
the linearised gyrokinetic equation \eqref{eqn:gyrokineticeqnormalised}
for each species
and the quasineutrality condition \eqref{eqn:quasineutralitynormalised}.\footnote{
As well as the parallel and perpendicular components of 
Amp\`ere's Law in the electromagnetic case.}
Starting from \eqref{eqn:gyrokineticeqnormalised} and 
\eqref{eqn:quasineutralitynormalised} and dropping the 
nonlinear and perpendicular flow shear terms, the equations to be solved are

\begin{multline}
\frac{\partial h_{Ns}}{\partial t_N}  +
v_{thNs} v_{\parallel N}k_{pN}\frac{\partial h_s}{\partial \theta} +
	 \vct{V}_{DNs}\cdot \nabla_N h_{Ns}
	%+ \frac{1}{2}\mathcal{N}_{Ns}  
-	\gyroR{C_N\left[ h_{Ns} \right]} =\\
\frac{Z_s}{T_{Ns}} \frac{\partial \gyroR{\field_N}}{\partial t_N} +
	\left\{ \left[ \kappa_{ns} - \left( \varepsilon_{Ns} - \frac{3}{2} \right)\kappa_s \right]\sigma_\psi - \frac{2}{v_{thNs}}\frac{I_N}{B_N}\frac{q_0}{\rho_0}\gamma_{EN} \right\} \frac{1}{2}\frac{\partial \gyroR{\field_N}}{\partial y},
	\label{gyrokineticeqlinearsolve}
\end{multline}
and

\begin{equation}
	\sum_s Z_s^2 \frac{n_{Ns}}{T_{Ns}} \varphi_N = \sum_s Z_s \int \frac{d^3\vct{v}}{v_{ths}^3} n_{Ns} \gyror{F_{Ns}h_{Ns}}.
	\label{quasinlinearsolve}
\end{equation}
%where $y$ is defined in Section \ref{geometryapp} and all other quantities are defined in Section \ref{modelsec}.

%In order to avoid numerical errors
%from certain large scale Alfvenic fluctuations \cite{numata2010astrogk},
We start by introducing the complimentary distribution function:\footnote{
This is done for historical reasons, although the complimentary distribution has the pleasing property that $g=0$ is the exact solution of the long-wavelength shear-Alfv\`en-wave equation \cite{Tome}.}

\begin{equation}
	g_{Ns} = h_{Ns} - \frac{Z_s}{T_{Ns}}\gyroR{\varphi_{Ns}},
	\label{complimentarydistfndef}
\end{equation}
for which the evolution equations are:\footnote{Note that the collision operator must still be calculated in terms of $h_s$.}
\begin{multline}
\frac{\partial g_{Ns}}{\partial t_N}  +
v_{thNs} v_{\parallel N}k_{pN}\frac{\partial g_{Ns}}{\partial \theta} +
	 \vct{V}_{DNs}\cdot \nabla_N g_{Ns}
	%+ \frac{1}{2}\mathcal{N}_{Ns}  
-	\gyroR{C_N\left[ h_{Ns} \right]} =\\
-\frac{Z_s}{T_{Ns}} \left[ v_{thNs} v_{\parallel N}k_{pN}\frac{\partial \gyroR{\varphi}}{\partial \theta} +
	 \vct{V}_{DNs}\cdot \nabla_N \gyroR{\varphi}
 \right]\\ +
	\left\{ \left[ \kappa_{ns} - \left( \varepsilon_{Ns} - \frac{3}{2} \right)\kappa_s \right]\sigma_\psi - \frac{2}{v_{thNs}}\frac{I_N}{B_N}\frac{q_0}{\rho_0}\gamma_{EN} \right\} \frac{1}{2}\frac{\partial \gyroR{\varphi_N}}{\partial y},
	\label{gyrokineticeqlinearsolve}
\end{multline}
and

\begin{equation}
	\sum_s Z_s^2 \frac{n_{Ns}}{T_{Ns}} \varphi_N = \sum_s Z_s \int \frac{d^3\vct{v}}{v_{ths}^3} n_{Ns} F_{Ns}\gyror{g_{Ns}-\frac{Z_s}{T_{Ns}}\gyroR{\varphi_N}}.
	\label{quasinlinearsolve}
\end{equation}

We now
%move to the field-line-following coordinate system $(x,y,\theta)$, 
%where $x$ is defined in Section \ref{geometryapp} and $\theta$ is the poloidal angle,
%Fourier expanding all perturbed quanties in the perpendicular plane:
Fourier expand all perturbed quanties in the perpendicular plane using \eqref{eqn:kxkyeikonal} to give:

%\begin{equation}
	%\varphi \rightarrow \sum_{\vct{k}} \hat{\varphi}_{\vct{k}}\lp \theta \rp \exp \lsq \I \lp k_x x + k_y y\rp \rsq
	%\label{fourierexpandperturbed}
%\end{equation}

%to give 

\begin{multline}
\frac{\partial \hat{g}_{Ns}}{\partial t_N}  +
v_{thNs} v_{\parallel N}k_{pN}\frac{\partial \hat{g}_{Ns}}{\partial \theta} +
	 i \frac{T_{Ns}}{Z_s}\omega_{DN} \hat{g}_{Ns}
	%+ \frac{1}{2}\mathcal{N}_{Ns}  
-	\gyroR{C_N\left[ h_{Ns} \right]} =\\
-\frac{Z_s}{T_{Ns}} \left[ v_{thNs} v_{\parallel N}k_{pN}
\frac{\partial J_0\left( a_s \right)\hat{\varphi}}{\partial \theta} +
	 i \frac{T_{Ns}}{Z_s}\omega_{DN} J_0\left( a_s \right)\hat{\varphi}
 \right]\\ +
 i \frac{k_{yN}}{2}	\left\{ \left[ \kappa_{ns} - \left( \varepsilon_{Ns} - \frac{3}{2} \right)\kappa_s \right]\sigma_\psi - \frac{2}{v_{thNs}}\frac{I_N}{B_N}\frac{q_0}{\rho_0}\gamma_{EN} \right\}   J_0\left( a_s \right) \hat{\varphi}_N,
	\label{gyrokineticeqlinearsolvefourierexpanded}
\end{multline}
and

\begin{equation}
	\sum_s Z_s^2 \frac{n_{Ns}}{T_{Ns}} \varphi_N \left( 1 - \Gamma_s \right) = \sum_s Z_s \int \frac{d^3\vct{v}}{v_{ths}^3} n_{Ns} F_{Ns}\gyror{g_{Ns}}.
	\label{quasinlinearsolve}
\end{equation}
where 
$J_0$ is a Bessel function of the first kind and results from gyroaveraging a Fourier-transformed quantity,
$a_s \equiv \sqrt{T_{Ns}m_{Ns}}v_{\perp N}k_{\perp N}/ Z_s B_N = v_{\perp}k_{\perp}/\Omega_s$,
and the integrated Bessel function

\begin{equation}
	\Gamma_s(\theta)
	= 	 \int \frac{d^3\vct{v}}{v_{ths}^3}  F_{Ns} J_0^2\left( a_s \right)
	= \exp \lsq \frac{-k_{\perp}^2 \rho_s^2}{2} \rsq I_0 \lp \frac{-k_{\perp}^2 \rho_s^2}{2} \rp,
	\label{gammasdef}
\end{equation}
where $I_0$ is a modified Bessel function of the first kind\footnote{
The quantity $n_{Ns}Z_s^2 \Gamma_s/T_{Ns}$ is called $\mathtt{gamtot}$ in the code.}.

We proceed by collecting derivatives of each fluctuating quantity in \eqref{gyrokineticeqlinearsolve}, dropping circumflexes and writing: 

\begin{equation}
	\pdt{g_{Ns}} + a_{\theta} \pd{g_{Ns}}{\theta} + a_0 g_{Ns} = \\
	b_t \pdt{J_0\lp a_s \rp \varphi_N} + b_{\theta} \pd{J_0\lp a_s \rp \varphi_N}{\theta} + b_0 J_0\lp a_s \rp \varphi_N
	\label{collectgyrokineticeq}
\end{equation}
where 
\begin{eqnarray}
	a_{\theta}& = &v_{thNs} v_{\parallel N}k_{pN},\\
	a_0& =& i \frac{T_{Ns}}{Z_{s}}\omega_{DN},\\
	b_t& =& 0 ,\\
	b_{\theta}& =& -\frac{Z_{Ns}}{T_{Ns}}v_{thNs} v_{\parallel N}k_{pN} ,\\
	b_0& =& -i \omega_{DN}+ i \frac{k_{yN}}{2}	\left\{ \left[ \kappa_{ns} - \left( \varepsilon_{Ns} - \frac{3}{2} \right)\kappa_s \right]\sigma_\psi - \frac{2}{v_{thNs}}\frac{I_N}{B_N}\frac{q_0}{\rho_0}\gamma_{EN} \right\}.\qquad
\end{eqnarray}
For the rest of Chapter \ref{sec:gs2algorithm} we drop the subscript $N$ and work only in normalised quantities.

\subsection{Discretisation and Parallelisation}

The distribution function $g_s(\theta, k_x, k_y, \varepsilon, \lambda, \sigma)$ is stored on a six-dimensional grid
%as a function of discrete values of $\theta, k_x, k_y, \varepsilon, \lambda$ and $\sigma$. 
of size $\mathtt{ntheta}\times\mathtt{nakx}\times\mathtt{naky}\times\mathtt{n}\varepsilon\times\mathtt{n}\lambda\times2$.
The perturbed fields are functions only of $\theta, k_x$ and $k_y$.
The linearised gyrokinetic equation \eqref{collectgyrokineticeq} is in fact a 1D partial differential equation in $\theta$, 
with the values of the other coordinates appearing only as parameters; this means that its solution may be calculated separately for each value of $k_x, k_y, \varepsilon$ and $\lambda$, enabling very efficient parallelisation.
The field equation \eqref{quasinlinearsolve} requires integrating over velocity and species, but can still be solved separately for each value of $k_x$ and $k_y$.
Thus, a typical strategy when solving the linear problem would be to divide it among $\mathtt{nakx} \times \mathtt{naky} \times \mathtt{n\varepsilon} * \mathtt{n\lambda} * 2$ processes: each would solve \eqref{collectgyrokineticeq} independently, with communication only required for the solution of \eqref{quasinlinearsolve}.

The gyrokinetic equation \eqref{collectgyrokineticeq} is discretized using a compact stencil to keep its inversion matrix bidiagonal.
The calculation of the differences is complicated by the introduction of space centering and temporal implicitness parameters, $r_{\theta}$ and $r_t$, known as \texttt{bakdif} and \texttt{fexp} in the code.
The space centering parameter $r_{\theta}$ ranges from 0 (centered) to 1 (a first order upwind scheme), and the temporal implicitness parameter $r_t$ varies from 0 (fully implicit) to 1 (fully explicit).
With these parameters, the perturbed potential and its differences may be written

\begin{eqnarray}
	\pdt{\varphi} &=&\frac{1}{2}\left[ 
	\left( 1 - r_{\theta} \right) \frac{\varphi^{n+1}_{i} - \varphi^{n}_{i}}{\Delta t}+ 
	\left( 1 + r_{\theta} \right)\frac{\varphi^{n+1}_{i+1}-\varphi^{n}_{i+1}}{\Delta t }\right],\\
	\pd{\varphi}{z} &=& r_t \frac{\varphi^{n}_{i+1} - \varphi^{n}_{i}}{\Delta\theta}+\left( 1 - r_t \right)\frac{\varphi^{n+1}_{i+1}-\varphi^{n+1}_{i}}{\Delta\theta},\\
	\varphi &=& \frac{1}{2}\left\{ 
	\left( 1 - r_{\theta} \right)\left[ r_t \varphi^{n}_{i}+\left( 1 - r_t\right)\varphi^{n+1}_{i} \right]\right.\\\nonumber
	&&\left.\qquad+\left( 1+r_{\theta} \right)\left[ r_t \varphi^{n}_{i+1}+\left( 1-r_t \right)\varphi^{n+1}_{i+1}  \right] \right\},
	\label{differencingdefs}
\end{eqnarray}
and similarly for the distribution function.
The special choice of $r_{\theta}=0$ and $r_t=1/2$ corresponds to a space-centered, time-centered (trapezoidal) scheme, first proposed by Beam and Warming \cite{beam1976implicit},\footnote{Not to be confused with the Beam-Warming scheme, a different scheme from the same authors.} which is second order accurate and unconditionally stable.

Substituting the differences into \eqref{collectgyrokineticeq}, we obtain the discretized gyrokinetic equation (suppressing the species index $s$ and defining $\tilde{\varphi} \equiv J_0 \varphi$):

\begin{multline}
	\frac{1}{2}\left[ 
	\left( 1 - r_{\theta} \right) \frac{g^{n+1}_{i} - g^{n}_{i}}{\Delta t}+ 
	\left( 1 + r_{\theta} \right)\frac{g^{n+1}_{i+1}-g^{n}_{i+1}}{\Delta t }\right] \\ 
	+ a_{\theta} \left[ r_t \frac{g^{n}_{i+1} - g^{n}_{i}}{\Delta\theta}+\left( 1 - r_t \right)\frac{g^{n+1}_{i+1}-g^{n+1}_{i}}{\Delta\theta} \right] \\
	+  \frac{a_0}{2}\left\{ 
	\left( 1 - r_{\theta} \right)\left[ r_t g^{n}_{i}+\left( 1 - r_t\right)g^{n+1}_{i} \right]\right.
	\left.+\left( 1+r_{\theta} \right)\left[ r_t g^{n}_{i+1}+\left( 1-r_t \right)g^{n+1}_{i+1}  \right] \right\} \\
	= b_{\theta} \left[ r_t \frac{\tilde{\varphi}^{n}_{i+1} - \tilde{\varphi}^{n}_{i}}{\Delta\theta}+\left( 1 - r_t \right)\frac{\tilde{\varphi}^{n+1}_{i+1}-\tilde{\varphi}^{n+1}_{i}}{\Delta\theta}  \right]\\
	+  \frac{b_0}{2}\left\{ 
	\left( 1 - r_{\theta} \right)\left[ r_t \tilde{\varphi}^{n}_{i}+\left( 1 - r_t\right)\tilde{\varphi}^{n+1}_{i} \right]\right.
	\left.+\left( 1+r_{\theta} \right)\left[ r_t \tilde{\varphi}^{n}_{i+1}+\left( 1-r_t \right)\tilde{\varphi}^{n+1}_{i+1}  \right] \right\},
	\label{gyrokineticeqdiscretized}
\end{multline}
which may be written as

\begin{multline}
	A_{1} g_{i}^{n} + A_{2} g^{n}_{i+1} + B_{1} g^{n+1}_{i} + B_{2} g^{n+1}_{i+1} 
	= D_{1} \tilde{\varphi}_{i}^{n} + D_{2} \tilde{\varphi}^{n}_{i+1} + E_{1} \tilde{\varphi}^{n+1}_{i} + E_{2} \tilde{\varphi}^{n+1}_{i+1}
	\label{gyrokineticeqdiscretizedsimple}
\end{multline}
where

\begin{eqnarray}
	A_{1}&=& -\frac{1-r_{\theta}}{2\Delta t}- a_{\theta} \frac{r_t}{\Delta \theta} + \frac{a_0}{2}\left( 1 - r_{\theta} \right) r_t, \label{A1def} \\
	A_{2} &=&  - \frac{1+r_{\theta}}{2 \Delta t} + a_{\theta}\frac{r_t}{\Delta\theta} + \frac{a_0}{2}\left( 1 + r_{\theta} \right) r_t, \label{A2def} \\
	B_{1} &=& \frac{1 - r_{\theta}}{2 \Delta t} - a_{\theta}\frac{1-r_t}{\Delta \theta}+\frac{a_0}{2}\left( 1 - r_{\theta} \right)\left( 1 - r_t \right), \label{B1def} \\ 
	B_{2} &=& \frac{1 + r_{\theta}}{2 \Delta t} + a_{\theta}\frac{1-r_t}{\Delta \theta}+\frac{a_0}{2}\left( 1 + r_{\theta} \right)\left( 1 - r_t \right). \label{B2def}
\end{eqnarray}
The quantities $B_2^{-1}$, $B_2^{-1} B_1$, $A_2$ and $A_1$ are called $\mathtt{ainv}$, $\mathtt{r}$, $\mathtt{a}$ and $\mathtt{b}$ in the $\mathtt{dist\_fn}$ module of \texttt{GS2}. The definitions of $D_1, D_2, E_1$ and $E_2$ are unimportant as the RHS of \eqref{gyrokineticeqdiscretizedsimple} is calculated in a different way within the code (see Section \ref{sourcetermsec}). 

\subsection{Advancing Implicitly using Green's Functions}

In each iteration the new distribution function
must be calculated using \eqref{gyrokineticeqdiscretizedsimple}
(reproduced here for convenience),
and the new potential via the
discretized quasineutrality condition \ref{quasindiscretized}.

\begin{multline}
	A_{1} g_{i}^{n} + A_{2} g^{n}_{i+1} + B_{1} g^{n+1}_{i} + B_{2} g^{n+1}_{i+1} 
	= D_{1} \tilde{\varphi}_{i}^{n} + D_{2} \tilde{\varphi}^{n}_{i+1} + E_{1} \tilde{\varphi}^{n+1}_{i} + E_{2} \tilde{\varphi}^{n+1}_{i+1},
	\tag{\ref{gyrokineticeqdiscretizedsimple}}
\end{multline}

\begin{eqnarray}
	\sum_s \frac{  Z_s^2  n_s\lp 1 - \Gamma_s\rp}{T_s} \varphi^{n+1}=  \sum_{s,\varepsilon, \lambda, \sigma} Z_s \mathcal{J}_{\varepsilon,\lambda,\sigma} n_{s} J_{0,\varepsilon,\lambda} g^{n+1}_{s, \varepsilon, \lambda, \sigma}\nonumber
	%\\
	%&\equiv&
	\equiv \mathcal{V} g^{n+1},
	\label{quasindiscretized}
\end{eqnarray}
where $\mathcal{J}_{\varepsilon,\lambda,\sigma}$ are the weights for the velocity integral and where we have defined the velocity and species integration operator $\mathcal{V}$. Since this is an implicit scheme, solution of \eqref{gyrokineticeqdiscretizedsimple} requires $\tilde{\varphi}^{n+1}$, and solution of \eqref{quasindiscretized} requires $g^{n+1}$. Generically, this can be cast as a single matrix equation, writing \eqref{gyrokineticeqdiscretizedsimple} as 

\begin{equation}
	\doubleunderline{A} \cdot g^{n+1} + \doubleunderline{B} \cdot g^{n}
	= \doubleunderline{D} \cdot \tilde{\varphi}^{n+1} + \doubleunderline{E} \cdot \tilde{\varphi}^{n},
	\label{gyrokineticeqfullmatrix}
\end{equation}
and \eqref{quasindiscretized} as
\begin{equation}
	\doubleunderline{F} \cdot \varphi^{n+1} = \doubleunderline{G} \cdot g^{n+1}
	\label{quasinfullmatrix}
\end{equation}
so that
\begin{equation}
	g^{n+1} 	= \doubleunderline{M}^{-1}  \left( \doubleunderline{E} \cdot \tilde{\varphi}^{n} - \doubleunderline{B} \cdot g^{n} \right) 
	\label{fullmatrixequation}
\end{equation}
where $\doubleunderline{A}$ --- $\doubleunderline{F}$ are square matrices of size $N_{tot} =\mathtt{ntheta} \times \mathtt{nspec}\times \mathtt{n\varepsilon} * \mathtt{n\lambda} * 2$ and 
\begin{equation}
	\doubleunderline{M}= \left( \doubleunderline{A} - \doubleunderline{D} J_0 \doubleunderline{F}^{-1} \doubleunderline{G} \right)
	\label{fullimplicitmatrix}
\end{equation}
is the implicit response matrix. However, initial inversion of the (dense) response matrix would require $O\left( N_{tot}^{3} \right)$ operations, and subsequent timesteps would require $O\left( N_{tot}^{2} \right)$ operations: for any reasonable problem size the computational cost quickly becomes prohibitive.

Kotschenreuther et. al. \cite{gs2ref} realised that this cost could be avoided as follows.
Since the distribution function responds linearly to changes in the perturbed potential, the response of the distribution function to any arbitrary change in the perturbed potential can be calcuated as a superposition of the responses to a complete set of delta function changes to the perturbed potential: a Green's function approach. 
This allows the new field to be calculated \emph{before} the new distribution function, after which the new distribution function can be found using \eqref{gyrokineticeqdiscretizedsimple}.

The procedure is this. First define $\varphi^{*}$ as the change in the potential from one timestep to the next:
\begin{equation}
	\varphi^{*} \equiv \varphi^{n+1} - \varphi^{n}.
	\label{varphistardef}
\end{equation}
Next split $g^{n+1}$ into two parts:
\begin{equation}
	g^{n+1} \equiv g^{p} + g^{c},
	\label{splitdistfn}
\end{equation}
and define $g^{c}$ as that part of the new distribution function which results from the change in the potential $\varphi^{*}$:
\begin{equation}
	B_{1} g^{c}_{i} + B_{2} g^{c}_{i+1} 
	\equiv E_{1} \tilde{\varphi}^{*}_{i} + E_{2} \tilde{\varphi}^{*}_{i+1}.
	\label{ghomodef}
\end{equation}
Substituting \eqref{varphistardef}, \eqref{splitdistfn} and \eqref{ghomodef} into \eqref{gyrokineticeqdiscretizedsimple} we find that $g^{p}$ is that part of the new distribution function which results from the old distribution function and the old potential:\footnote{
The superscripts $p$ and $c$, which stand for \emph{predictor} and \emph{corrector}, are used here because the algorithm resembles a predictor-corrector scheme. However, as show by Belli \cite{belli2006studies}, the linear collisionless algorithm presented here is in fact mathematically equivalent to the implicit algorithm presented in Ref.~\cite{gs2ref}}
%The superscripts $h$ and $inh$, which stand for \emph{homogeneous} and \emph{inhomogeneous}, are inherited from the version of the algorithm described in Ref.~\cite{gs2ref} which, as shown for example by Belli \cite{belli2006studies}, is mathematically equivalent to the algorithm described here, the one actually implemented in \texttt{GS2}.
%However, the use of the words homogeneous and inhomogeneous is confusing and is avoided here, because the objects called the homogeneous and inhomogeneous solutions \emph{within the code itself} are actually the solutions of \eqref{gyrokineticeqhomogeneous} and \eqref{gyrokineticeqinhomogeneous}.}
\begin{equation}
	  B_{1} g^{p}_{i} + B_{2} g^{p}_{i+1} 
	= - A_{1} g_{i}^{n} - A_{2} g^{n}_{i+1} + D_{1}\tilde{\varphi}_{i}^{n} + D_{2} \tilde{\varphi}^{n}_{i+1} + E_{1} \tilde{\varphi}^{n}_{i} + E_{2} \tilde{\varphi}^{n}_{i+1}.
	\label{ginhomodef}
\end{equation}
The motivation for the definitions \eqref{varphistardef}, \eqref{splitdistfn} and \eqref{ghomodef} lies in the quasineutrality condition. Firstly let us define the response matrix, the response of the new distribution function at $i$ to a change in the potential at $j$,
\begin{equation}
	\left( \frac{\delta g}{\delta \tilde{\varphi}} \right)_{ij} = \left( B_1 \delta_{i,k} + B_2 \delta_{i,k+1} \right)^{-1}\left( E_1 \delta_{k,j} + E_2 \delta_{k,j+1} \right),
	\label{responsematrixintro}
\end{equation}
by rewriting \eqref{ghomodef} as
%\footnote{
%A little rearrangment will show that $\left( \delta g / \delta \tilde{\varphi} \right)_{ij} = \left( B_1 \delta_{i,k} + B_2 \delta_{i,k+1}) \right)^{-1}\left( E_1 \delta_{k,j} + E_2 \delta_{k,j+1}) \right)$.}
\begin{equation}
	g^{c}_{i} \equiv \left( \frac{\delta g}{\delta \tilde{\varphi}} \right)_{ij}	\tilde{\varphi}_{j}^{*}.
	\label{responsematrixdef}
\end{equation}
Secondly let us substitute the definitions \eqref{varphistardef}, \eqref{splitdistfn} and \eqref{responsematrixdef} into the quasineutrality condition \eqref{quasindiscretized} to give

\begin{equation}
	\sum_s \frac{  Z_s^2  n_s\lp 1 - \Gamma_s\rp}{T_s} \left( \varphi^{*}_{i} + \varphi^{n}_{i} \right) = \mathcal{V}\left( g^{p}_{i} + \left( \frac{\delta g}{\delta \tilde{\varphi}} \right)_{ij}\varphi^{*}_{j}\right) ,
	\label{quasindiscretized2}
\end{equation}
which can be rearranged to give
\begin{eqnarray}
	\underbrace{\left[ \sum_s \frac{  Z_s^2  n_s\lp 1 - \Gamma_s\rp}{T_s}\delta_{ij} -  \mathcal{V}\left( \frac{\delta g}{\delta \tilde{\varphi}} \right)_{ij}\right]}_{\doubleunderline{P}}
	\varphi^{*}_{i}  =  \mathcal{V}g^{p}_{i}  -  \sum_s \frac{  Z_s^2  n_s\lp 1 - \Gamma_s\rp}{T_s}\varphi^{n}_{i} ,\nonumber \\
	\implies \varphi^{*}_{i}  = \doubleunderline{P}^{-1} \left[ \mathcal{V}g^{p}_{i}  -  \sum_s \frac{  Z_s^2  n_s\lp 1 - \Gamma_s\rp}{T_s}\varphi^{n}_{i}\right] ,
	\label{quasindiscretized3}
\end{eqnarray}
where the matrix $\doubleunderline{P}^{-1}$ is a square matrix of size $\mathtt{ntheta}$, which must be calculated once at the beginning of a simulation.
Calculating the matrix $\doubleunderline{P}^{-1}$ is really a matter of calculating the response matrix $\left( \delta g / \delta \tilde{\varphi} \right)_{ij}$.
This could be done by direct calculation from $B_1,B_2,E_1$ and $E_2$, but this effort can be spared by noting that
if $\varphi^{*}_{j}$ is set equal to $\delta_{jk}$
then $g^{c}_{i}$, the solution of \eqref{responsematrixdef},
i.e., the solution of \eqref{ghomodef},
is actually the $k$th column of the response matrix.

The algorithm for \texttt{GS2} may now be written very simply indeed:
\begin{enumerate}
	\item Calculate each column of the response matrix by setting $\varphi^{*}_{j}=\delta_{jk}$ and solving \eqref{ghomodef} for $k=1,2... \mathtt{ntheta}$.
	\item Calculate the matrix $\doubleunderline{P}^{-1}$.
	\item Specify the initial values of $g$ and $\varphi$.
	\item Solve the equation \eqref{ginhomodef} to find $g^{p}$.
	\item Solve the equation \eqref{quasindiscretized3} to find $\varphi^{*}$ and hence $\varphi^{n+1}$.
	\item Solve the full gyrokinetic equation \eqref{gyrokineticeqdiscretizedsimple} to get $g^{n+1}$.
	\item Repeat steps 2--4 for every timestep.
\end{enumerate}
The effort of implementing this algorithm can be further reduced by noting that solving \eqref{ghomodef} is the same as solving \eqref{gyrokineticeqdiscretizedsimple} with $g^{n}$ and $\varphi^{n}$ set to zero, and $\varphi^{n+1}$ set to $\varphi^{*}$, and solving \eqref{ginhomodef} is the same as solving \eqref{gyrokineticeqdiscretizedsimple} with $\varphi^{n+1}$ set equal to $\varphi^{n}$.
And so now the algorithm may be written in a form that directly corresponds to the actual statements and subroutine calls in the code:\footnote{
With apologies for the corruption of the mathematical symbols by the use of the imperative programming style.} 
%\pagebreak
%\begin{minipage}
\needspace{16\baselineskip}
\begin{enumerate}
	\nopagebreak \item Set $k=1$.
	\nopagebreak \item Set $\varphi^{*}_{j}=\delta_{jk}$, $\varphi^{n+1}=\varphi^{*}$, $g^{n}=0$ and $\varphi^{n}=0$. \label{startinitial}
	\nopagebreak \item Solve the equation \eqref{gyrokineticeqdiscretizedsimple} for $g^{n+1}$.
	\nopagebreak \item Set $\left( \delta g / \delta \tilde{\varphi} \right)_{ik} = g^{n+1}_{i}$ \label{endinitial}
	\nopagebreak \item Repeat steps \ref{startinitial}--\ref{endinitial} for $k=2... \mathtt{ntheta}$.
	\nopagebreak \item Calculate the matrix $\doubleunderline{P}^{-1}$.
	\nopagebreak \item Specify the initial values of $g^n$ and $\varphi^n$.
	\nopagebreak \item Set $\varphi^{n+1}=\varphi^{n}$. \label{startmain}
	\nopagebreak \item Solve the equation \eqref{gyrokineticeqdiscretizedsimple} for $g^{n+1}$.
	\nopagebreak \item Set $g^{p}=g^{n+1}$.
	\nopagebreak \item Solve the equation \eqref{quasindiscretized3} to find $\varphi^{*}$.
	\nopagebreak \item Set $\varphi^{n+1} = \varphi^{n} + \varphi^{*}$.
	\nopagebreak \item Solve the equation \eqref{gyrokineticeqdiscretizedsimple} to get $g^{n+1}$. \label{endmain}
	\nopagebreak \item Repeat steps \ref{startmain}--\ref{endmain} for every timestep.
\end{enumerate}
%\end{minipage}
Thus, there are only three major steps that need to be implemented: solving the gyrokinetic equation \eqref{gyrokineticeqdiscretizedsimple}, solving the equation \eqref{quasindiscretized3} and calculating $\doubleunderline{P}^{-1}$. 
The latter two operations require the application of the velocity and species integration operator $\mathcal{V}$ (see Chapter \ref{sec:velocityspace}) and some linear algebra.
Solving the gyrokinetic equation itself requires careful treatment of the boundary conditions, as described in the next section.

\subsection{Solving the Discretized Gyrokinetic Equation}
\label{sec:solvingthediscretizedgyrokineticequation}

%\subsubsection{Boundary Conditions}

Solving \eqref{gyrokineticeqdiscretizedsimple} requires the correct treatment of the parallel boundary conditions.
The cases of trapped and passing particles are treated separately.
The boundary conditions may be summarized as follows:
\begin{itemize}
	\item For passing particles when the magnetic shear is zero, the boundary conditions are periodic (what leaves at one end must come back in at the other).
	\item For passing particles when the magnetic shear is finite, the boundary conditions are zero incoming.
	\item For trapped particles, the distribution function for particles arriving at a bounce point must equal the distribution function for particles leaving it. 
\end{itemize}
%\subsubsection{Passing Particles}
%When the magnetic shear $\hat{s}$ is zero, the parallel boundary conditions are periodic: 
The case of passing particles is complicated by
%what is known as the \emph{extended theta grid}.
%This arises becaus
the fact that
in the case of finite magnetic shear $\hat{s}$, the total radial derivative of the eikonal $dS/dx$ becomes a function of $\theta$ \cite{beer1995field}:

\begin{equation}
	\frac{dS}{dx}=k_{xt} = k_{x} - k_y \hat{s}\theta 
	\label{kxvaries}
\end{equation}
where $k_{x}$ is the ``actual'' radial wavenumber of a mode
%(i.e. the radial wavenumber at $\theta=0$
%and the quantity used to index $\varphi$ and $g$),
as defined in Section \ref{sec:spectralcoordinates}.\footnote{
%and $k_x$ is what goes into the eikonal
It should be noted at this point that Ref \cite{beer1995field} 
defines $k_{x}$ as $k_y\hat{s}\theta_0$,
which means that confusingly in \texttt{GS2} 
the ``actual'' values of the radial wavenumber 
are sometimes referred to as values of $\theta_0$.}
%The use of ``actual'' within the code is also confusing,
%as one could argue that the ``actual'' radial wavenumber
%at any given point is $k_x$, with $k_{x0}$ being merely an index.}:
%\begin{equation}
	%\varphi = \sum_{k_{x0}, k_y} \hat{\varphi}_{k_{x0}, k_y}\lp \theta \rp \exp \lsq \I \lp k_x x + k_y y\rp \rsq.
	%\label{fourierexpandperturbedwithmagneticshear}
%\end{equation}
The practical upshot of this is that while when magnetic shear is equal to zero the boundary condition is a very simple periodical one:
\begin{equation}
	g\left(  k_{x}, k_y, \theta=\theta_{max} \right) = g\left( k_{x}, k_y, \theta=-\theta_{max} \right) ,
	\label{periodicbc}
\end{equation}
the boundary condition when the magnetic shear is finite is such that the end of one field line with a given $k_{x}$ connects to the beginning of another field line with  the correct value of $k_{x}$ such that \emph{the value of $k_{xt}$ matches at the join}:
\begin{equation}
	g\left(  k_{x}, k_y, \theta=\theta_{max} \right) = g\left( k_{x} - 2 k_y \hat{s} \theta_{max}, k_y, \theta=-\theta_{max} \right),
	\label{twistandshift}
\end{equation}
where $2 \theta_{max}$ is the length of the parallel domain (equal to $\pi \times \mathtt{nperiod}$ in the code). This is known as the \emph{twist and shift} or \emph{linked} boundary condition. 
Implementing it requires that the quantity  $2 k_y \hat{s} \theta_{max} / \Delta k_x$
(where $\Delta k_x $ is the spacing in $k_{x}$ on the grid) be an integer:
in the code that integer is called  $\mathtt{jtwist}$.

\myfig{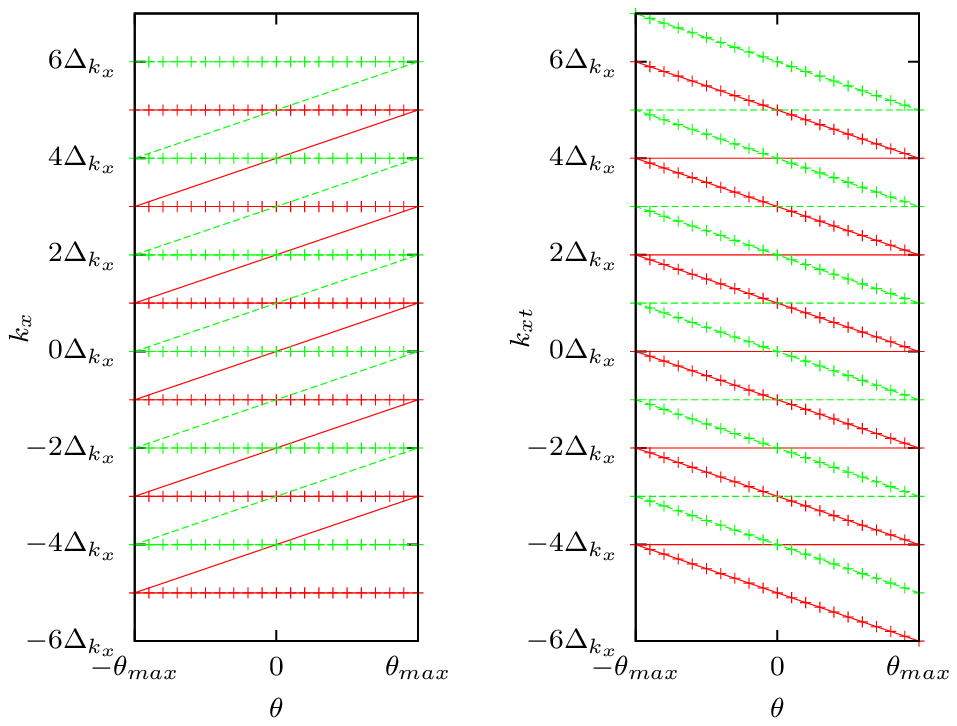}{The Extended Theta Grid}{An example of the extended theta grid where $\mathtt{jtwist}=2$.
The crosses show the gridpoints and the lines show the connections between them.
The \texttt{nakx} separate segments are connected into $\mathtt{jtwist}$ domains.
The left panel shows the domains vs. $\theta$ and $k_{x}$, which index the gridpoints.
A parallel segment with $k_{x} = n \Delta k_x$ connects at $\theta=\theta_{max}$ to a segment with $k_{x} = \left(n- \mathtt{jtwist}\right) \Delta k_x$ (see \eqref{twistandshift}).
The right panel shows the domains vs. $\theta$ and $k_{xt}$.
The quantity $k_{xt}$ varies with $\theta$ (see \eqref{kxvaries}).  At either end a parallel segment connects to the parallel segment that has the same $k_{xt}$ at the opposite end.}

Thus, the set of $\mathtt{nakx}$ radial modes are linked into a set of $\mathtt{nakx}/\mathtt{jtwist}$ parallel domains
%and the  extended theta grid runs from the beginning to the end of each of these interconnected parallel domains
(see \figref{extendedthetagrid}).
The gyrokinetic equation \eqref{gyrokineticeqdiscretizedsimple} is solved for each parallel domain separately.
Linearly, each connected parallel domain corresponds to a single ballooning mode;\footnote{ 
Note that if one wanted to simulate a single mode linearly, one could abandon this \emph{twist and shift} arrangement and extend the parallel domain sufficiently far for convergence by increasing \texttt{nperiod}.
This is the situation described in Ref.~\cite{gs2ref}.
However, the twist and shift system is necessary for nonlinear runs, as it is necessary to have a large number of $k_{x}$s to resolve the turbulent spectrum.}
the boundary conditions for the connected domain are thus the same as those for a single ballooning mode: the distribution function must vanish at either end \cite{gs2ref}.
Since \eqref{gyrokineticeqdiscretizedsimple} is only first order in $\theta$, this can only be enforced at one point in theta: physically this done by setting $g=0$ for incoming particles at either end of the domain: for particles with $\sigma>0$, $g=0$ for the lowest value of the extended theta grid, for particles with $\sigma<0$, $g=0$ for the highest value of the extended theta grid. 

For trapped particles, the situation is much simpler, and can be written as
\begin{eqnarray}
	g\left( \sigma = + 1,   \theta=\theta_{ub} \right) &=&  g\left(\sigma=-1, \theta=\theta_{ub} \right) , \nonumber\\
	g\left( \sigma = + 1,   \theta=\theta_{lb} \right) &=& g\left(\sigma=-1, \theta=\theta_{lb} \right),
	\label{trappedbc}
\end{eqnarray}
where $\theta_{ub},\theta_{lb}$ are the points where the trapped particles are reflected by the mirror effect.\footnote{Since the magnetic moment $\mu=mv_{\perp}^{2}/B$ of any particle is conserved, as a particle moves from the outboard side of a tokamak to the inboard, the increase in $B$ causes an increase in $v_{\perp}$. Since total energy $\varepsilon=m\left( v^{2}_{\pa} + v^{2}_{\perp} \right)$ of the particle is also conserved, its parallel velocity decreases until it reaches zero and changes sign --- the particle bounces back.}

To recapitulate, the boundary conditions must be solved differently for three different cases.
For the periodic case, the distribution function at one end must match the distribution at the other.
For the twist and shift case, the distribution function at one end of one parallel segment must match the other end of the connected parallel segment. 
At each end of the chain of connected parallel segments, zero incoming boundary conditions are applied.
For trapped particles, the distribution of particles arriving at each bounce point must match that of those leaving it. 
The exact routine for implementing these conditions for the twist and shift case is too tortuous to be covered here, but we will describe the general principle and the specific cases of trapped particles and passing particles in a periodic system.

The general principle follows the standard method for matching boundary conditions for a partial differential equation: first divide the equation \eqref{gyrokineticeqdiscretizedsimple} into a homogeneous and an inhomogeneous part, solve each part independently, and then combine them in a way that satisfies the boundary conditions.
Both parts of the equation are solved separately for each sign of velocity, so that there are four solutions for each parallel segment. 
For the periodic case and the case of trapped particles, these four solutions can then be combined to satisfy the boundary conditions of each segment separately. 
For the twist and shift case, satisfying the boundary conditions requires communication between each segment of the connected parallel domain, which leads to significant complications (as exemplified by the routine \texttt{init\_connected\_bc}).

The equation for the homogeneous part $g^{one}$ is 
\begin{equation}
	B_{1} g^{one}_{i} + B_{2} g^{one}_{i+1} = 0,
	\label{gyrokineticeqhomogeneous}
\end{equation}
and the equation for the inhomogeneous part $g^{new}$ is
\begin{multline}
	 B_{1} g^{new}_{i} + B_{2} g^{new}_{i+1} 
	= - A_{1} g_{i}^{n} - A_{2} g^{n}_{i+1} + D_{1} \tilde{\varphi}_{i}^{n} + D_{2} \tilde{\varphi}^{n}_{i+1} + E_{1} \tilde{\varphi}^{n+1}_{i} + E_{2} \tilde{\varphi}^{n+1}_{i+1}.
	\label{gyrokineticeqinhomogeneous}
\end{multline}
All terms on the RHS of \eqref{gyrokineticeqinhomogeneous} are grouped together and called the \emph{source}.
Both \eqref{gyrokineticeqhomogeneous} and \eqref{gyrokineticeqinhomogeneous} are solved first for $\sigma=-1$ ($v_{\parallel}<0$) and then for $\sigma=+1$ ($v_{\parallel}>0$). 

For passing particles, when $\sigma=-1$, the initial conditions are $g_{\mathtt{ntgrid}}^{one}=1$, $g^{new}_{\mathtt{ntgrid}}=0$, where the $\theta$ index runs from $-\mathtt{ntgrid}$ to $\mathtt{ntgrid}$ and 
\begin{equation}
	\mathtt{ntgrid}=(2\times\mathtt{nperiod} - 1)\times\mathtt{ntheta}.
	\label{<++>}
\end{equation}
Both equations are then solved starting at the highest theta and sweeping to the left. 
When $\sigma=+1$, the initial conditions are $g_{-\mathtt{ntgrid}}^{one}=1$, $g^{new}_{-\mathtt{ntgrid}}=0$. Both equations are then solved starting at the lowest theta and sweeping to the right.
This results in the generic solutions illustrated in \figref{genericsolutions}(a).

For trapped particles, when $\sigma=-1$, the initial conditions are $g_{\sigma=-1,\mathtt{ub}}^{one}=1$ and $g^{new}_{\sigma=-1,\mathtt{ub}}=0$, where $\mathtt{ub}$ is the index of the upper bounce point. Both equations are then solved starting at $\theta_{ub}$ and sweeping to the left. 
When $\sigma=+1$, the initial conditions are $g_{\sigma=+1,\mathtt{lb}}^{one}=g_{\sigma=-1,\mathtt{lb}}^{one}$ and $g_{\sigma=+1,\mathtt{lb}}^{new}=g_{\sigma=-1,\mathtt{lb}}^{new}$, where $\mathtt{lb}$ is the index of the lower bounce point. This ensures that the \eqref{trappedbc} is satisfied at the lower bounce point for any linear combination of $g^{one}$ and $g^{new}$. Both equations are then solved starting at $\theta_{lb}$ and sweeping to the right. 
This results in the generic solutions illustrated in \figref{genericsolutions}(b).

\myfig{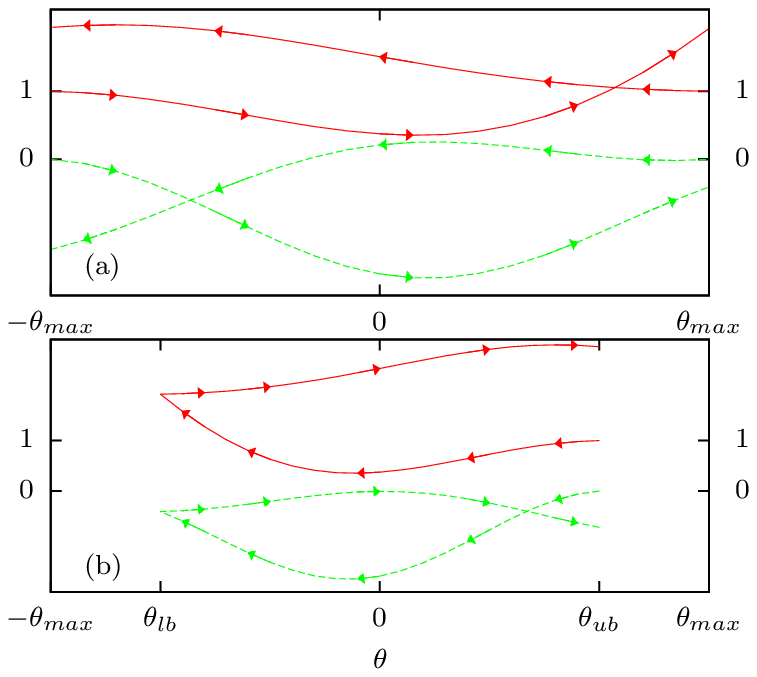}{Generic Solutions of the Gyrokinetic Equation}{Generic solutions of \eqref{gyrokineticeqhomogeneous} (red, solid) and \eqref{gyrokineticeqinhomogeneous} (green, dashed) for (a) passing particles and (b) trapped particles. The arrows indicate the sign of $v_{\pa}$.}

For trapped particles and for passing particles in a periodic system, the four solutions may now be combined to satisfy the boundary conditions \eqref{trappedbc} and \eqref{periodicbc} respectively.
Defining the total solution as

\begin{eqnarray}
	g^{n+1}_{\sigma=+1} =  g^{new}_{\sigma=+1} + \beta_+ g^{one}_{\sigma=+1} ,\nonumber\\
	g^{n+1}_{\sigma=-1} =  g^{new}_{\sigma=-1} + \beta_- g^{one}_{\sigma=-1},
	\label{totalsolution}
\end{eqnarray}
then for trapped particles the boundary condition at the lower bounce point requires $\beta_+ = \beta_-$, and the boundary condition at the upper bounce point requires that:
%and the boundary condition at the upper bounce point requires
\begin{eqnarray}
	  g^{new}_{\sigma=-1,ub} +\beta_- g^{one}_{\sigma=-1,ub}  = g^{new}_{\sigma=+1,ub}+ \beta_+ g^{one}_{\sigma=+1,ub},
	\label{upperbouncepointsolution}
\end{eqnarray}
which means that 
\begin{equation}
	\beta_+ = \beta_- = \frac{g^{new}_{\sigma=+1,ub}}{1 - g^{one}_{\sigma=+1,ub}},
	\label{beta1soln}
\end{equation}
remembering that $g_{\sigma=-1,\mathtt{ub}}^{one}=1$ and $g^{new}_{\sigma=-1,\mathtt{ub}}=0$.
For passing particles in a periodic system, we require:
\begin{eqnarray}
	g^{new}_{\sigma=\pm1,\mathtt{ntgrid}} +\beta_\pm g^{one}_{\sigma=\pm1,\mathtt{ntgrid}}  = g^{new}_{\sigma=\pm1,\mathtt{-ntgrid}}+ \beta_\pm g^{one}_{\sigma=\pm1,\mathtt{-ntgrid}},\nonumber\\
	\implies \beta_\pm = \frac{g^{new}_{\sigma=\pm 1,\mathtt{ntgrid}} - g^{new}_{\sigma=\pm 1,\mathtt{-ntgrid}}}{g^{one}_{\sigma=\pm1,\mathtt{-ntgrid}} - g^{one}_{\sigma=\pm1,\mathtt{ntgrid}}}.
	\label{periodicsolution}
\end{eqnarray}
In a similar manner, in the twist and shift case, a set of linear simultaneous equations can be constructed for the values of $\beta_+$ and $\beta_-$ from every segment of a connected domain to satisfy \eqref{twistandshift} at each connection and the zero incoming boundary at either end.

\subsection{Calculating the Source Term}
\label{sourcetermsec}

The only complication in calculating the source term (that is, the RHS of \eqref{gyrokineticeqinhomogeneous}) is the fact that the subcripts $i$ and $i+1$ switch with $\sigma$ --- depending on which way along the field line the equation is being solved. 
%hich is actually the RHS of \eqref{gyrokineticeqdiscretized}, is calculated
To reduce coding effort, the RHS of \eqref{gyrokineticeqinhomogeneous}
(which is actually the RHS of \eqref{gyrokineticeqdiscretized} plus the terms involving $g^n$), is tidied up in the following way. Defining

\begin{equation}
	\tilde{\varphi}^{av} = r_t  \tilde{\varphi}^{n} +   \left( 1 - r_t \right)\tilde{\varphi}^{n+1}  
	\label{phiavdef}
\end{equation}
and
\begin{equation}
	\tilde{\varphi}^{m} = \tilde{\varphi}^{av}_{i+1} - \tilde{\varphi}^{av}_{i}
	\label{phimdef}
\end{equation}
then the RHS of \eqref{gyrokineticeqinhomogeneous} is equal to
\begin{equation}
	\begin{cases}
		- A_{1} g_{i}^{n} - A_{2} g^{n}_{i+1}  + b_{\theta}\frac{\tilde{\varphi}^{m}}{\Delta\theta} + \frac{b_0}{2}\left[ \left( 1- r_{\theta} \right) \tilde{\varphi}^{av}_{i} + \left( 1 + r_{\theta} \right)\tilde{\varphi}^{av}_{i+1} \right]& \;\sigma=+1\\
		- A_{2} g_{i}^{n} - A_{1} g^{n}_{i+1}  + b_{\theta}\frac{\tilde{\varphi}^{m}}{\Delta\theta} + \frac{b_0}{2}\left[ \left( 1+ r_{\theta} \right) \tilde{\varphi}^{av}_{i} + \left( 1 - r_{\theta} \right)\tilde{\varphi}^{av}_{i+1} \right]& \;\sigma=-1.\\
	\end{cases}
	\label{source1}
\end{equation}
Defining
\begin{equation}
	\tilde{\varphi}^{p}= r_{\theta}^p \tilde{\varphi}^{av}_{i+1} + r_{\theta}^m \tilde{\varphi}^{av}_i,
	\label{phipdef}
\end{equation}
where the parameters $r_{\theta}^{p}$ and $r_{\theta}^{m}$ \footnote{Called \texttt{bdfac\_p} and \texttt{bdfac\_m} in the code.} are defined in the following way:
 
\begin{eqnarray}
	r_{\theta}^{p}=1 + \sigma r_{\theta}, \nonumber\\
	r_{\theta}^{m}=1 - \sigma r_{\theta}, 
	\label{rprmdef}
\end{eqnarray}
the expression \eqref{source1} can be written as 
\begin{equation}
	\begin{cases}
		- A_{1} g_{i}^{n} - A_{2} g^{n}_{i+1} + \frac{b_{\theta}\tilde{\varphi}^m}{\Delta\theta} + \frac{b_0 \tilde{\varphi}^p}{2}
& \sigma=+1\\
		- A_{2} g_{i}^{n} - A_{1} g^{n}_{i+1} + \frac{b_{\theta}\tilde{\varphi}^m}{\Delta\theta} + \frac{b_0 \tilde{\varphi}^p}{2}
& \sigma=-1.\\
\end{cases}
	\label{sourcefinal}
\end{equation}

\section{The Addition of the Nonlinear Terms}

\label{sec:nonlinearterms}
\subsection{Time Differencing}

In contrast to the linear parallel convection term $v\pa \uv{b} \cdot \nabla h_s$,
the nonlinear term $\mathcal{N} = \frac{c}{B}\uv{b}\times \nabla\gyroR{\varphi}  \cdot \nabla h_s $ is treated explicitly,
and a Courant-Friedrichs-Lewy (cfl) condition is used to keep the time step small enough for stability.
Accuracy is improved by the use of a linear multistep method: the third order Adams-Bashforth formula.
Thus, the nonlinear contribution to the discretized gyrokinetic equation \eqref{gyrokineticeqdiscretizedsimple}, which appears purely on the RHS and is added to the source calculated in Section \ref{sourcetermsec}, can be written \cite{numata2010astrogk} 
\begin{equation}
	\frac{23}{12}\mathcal{N}^n_{\vct{k}} - \frac{4}{3}\mathcal{N}^{n-1}_{\vct{k}} + \frac{5}{12}\mathcal{N}^{n-2}_{\vct{k}}. 
	\label{nonlinearcontribution}
\end{equation}
Where $\mathcal{N}_{\vct{k}}$ is the discrete Fourier transform of $\mathcal{N}$ in the perpendicular directions. It remains to calculate $\mathcal{N}$.

\subsection{Calculating the Nonlinear Term}

In the field line-following coordinates defined in Section \ref{sec:definingacoordinatesystem}, the nonlinear term may be written (see \eqref{eqn:nonlineartermsings2coordinates}):

\begin{equation}
	\mathcal{N} = \kappa_x \left(  \pd{\gyroR{\varphi}}{x}\pd{h}{y} - \pd{\gyroR{\varphi}}{y}\pd{h}{x}\right) 
	\label{poissonbracket}.
\end{equation}
The nonlinear term could be expressed and calculated as a complicated sum of the Fourier components of $\varphi$ and $g$. 
However, it
%is quicker and more natural
requires fewer operations \cite{orszag1970transform}
to
transform the fields and the distribution function
back into real space, calculate $\mathcal{N}$ using \eqref{poissonbracket} 
and then Fourier transform it to determine $\mathcal{N}_{\vct{k}}$.

Thus, the four partial derivatives in \eqref{poissonbracket} can be written:
%\footnote{The choice of names for 
%intermediate quantities follows that within the code.}:

\begin{eqnarray}
	 \pd{\gyroR{\varphi}}{x} = \mathcal{F}^{-1} \left[  k_x J_0 \hat{\varphi} \right],\\
	 \pd{h}{y} = \mathcal{F}^{-1} \left[ k_y \hat{g} + \frac{ Z_s k_y J_0}{T} \hat{\varphi} \right],\\
	 \pd{\gyroR{\varphi}}{y} = \mathcal{F}^{-1} \left[  k_y J_0 \hat{\varphi} \right],\\
	 \pd{h}{x} = \mathcal{F}^{-1} \left[ k_x \hat{g} + \frac{ Z_s k_x J_0}{T} \hat{\varphi} \right],\\
	\label{nonlinearterms}
\end{eqnarray}
temporarily restoring circumflexes to denote spectral quantities
and denoting the inverse Fourier transform defined in \eqref{eqn:kxkyeikonal} as $\mathcal{F}^{-1}$.
%As is noted in Ref.~\cite{dorlandimperialpres} t
This pseudo-spectral method of calculating the nonlinear term leads to spectral accuracy for the evaluation of the derivatives \cite{orszag1970transform}.

There is an important complication to this procedure which may be summarised as follows: energy from the truncated set of Fourier modes used in a given simulation would, had a larger number of modes been used, have been scattered via nonlinear interaction into modes with wavenumbers up to 1.5 times as large as the highest wavenumber in the truncated set.
If the above calculation were to be done using only the truncated set, that energy will be artificially added to the truncated set as a result of aliasing. 
This effect can be removed by performing both the inverse and forward Fourier transforms in the calculation using a larger grid whose highest wavenumber (in both the $k_x$ and $k_y$ directions) is greater than 1.5 times the highest wavenumber on the actual grid. 
Before the inverse transform to get the real space fields and distribution function the extra gridpoints are set to zero; 
after the forward transform of the calculated nonlinear term the energy in the extra wavenumbers is simply discarded.

\chapter{Velocity Space and The Collision Operator}
\begin{quote}
	\emph{This chapter is a summary of Chapter 4 of Ref.~\cite{barnes2009trinity} and of Refs.~\cite{abel2008linearized,numerics}.}
\end{quote}
\label{sec:velocityspace}
\section{Introduction}

The distribution function is a field in a five-dimensional parameter space, and the gyrokinetic equation must be solved in five dimensions.
It is of paramount importance to use the minimum possible resolution in every dimension (while still correctly resolving the physics) in order to keep the computational and memory requirements
of the simulation
within the limits of what is attainable.
The minimum resolution in real space is set by several factors, including the need to resolve both the driving scale and the dissipation scale.
The minimum resolution in velocity space is set by the need to resolve the small-scale structures that develop in velocity space  as a result of processes such as Landau damping and phase mixing \cite{schekcrete, Tome}.

This minimum can be reduced in two ways. Firstly, we repeat the observation that the gyrokinetic equation is solved separately for each velocity, and that the details of velocity space are important only for the velocity integration operator $\mathcal{V}$, and the collision operator $C$.
The resolution requirements for the integration operator can therefore be reduced by using spectrally accurate integration methods (although this complicates the implementation of the collision operator slightly).
Secondly, we note that most systems have a finite collisionality, which will lead to dissipation. 
 The introduction of a collision operator which models the operation of these collisions will cause a smoothing in velocity space, and a reduction of the minimum velocity resolution required.
 Since the rate of dissipation of turbulent energy is controlled only by the rate of its injection \cite{schekcrete}, and not by the collisionality, it is possible to have a collisionality slightly larger than that observed physically, leading to reduced velocity space structure and a further reduced minimum resolution, without affecting results such as the calculated heat flux.

 As well as smoothing velocity space structure, collisions are also responsible for the irreversible heating of the background equilibrium that results from the turbulence \cite{schekcrete, abel2008linearized}.
Since this heating must be included accurately in any attempt to attain a statistically steady state, it is desirable for the collisions to conserve energy, particles and momentum and to satisfy Boltzmann's H-Theorum \cite{abel2008linearized}. \texttt{GS2} now contains a collision operator which satisfies all these criteria and provides  diffusion in both parallel and perpendicular velocity dimensions \cite{abel2008linearized, numerics}.

This chapter describes the layout of the velocity space grid and the implementation of the collision operator as they currently stand, without justifying the choices that were made therein.

\section{The Velocity Space Grid}

Since the perturbed distribution function is independent of the  gyrophase, only two velocity space coordinates are required. 
In \texttt{GS2} we choose the kinetic energy\footnote{It should always be remembered that the choice of energy as a velocity coordinate 
necessitates the addition of another index to the distribution function: the sign of the parallel velocity, $\sigma$.
%This does not make the distribution function a six-dimensional function.
}
$\varepsilon$  and $\lambda = \mu/\varepsilon$ (dropping the species subscript $s$), a choice which eliminates all velocity derivatives from the collisionless gyrokinetic equation \cite{barnes2009trinity}.
The layout of the numerical grids of these two coordinates is chosen to result in a spectrally accurate velocity integration operator.

%\subsection{Energy Grid}

The energy integration operator can be written
\begin{equation}
	v_{th}^2 \int_{0}^{\infty} dx \; x^2,
	\label{energyintegrationop}
\end{equation}
where $x = v/v_{th}$.
 The energy integral is split into two sections, from 0 to $x_0$ and from $x_0$ to infinity, where $x_0$ is a free parameter, usually chosen to be $\gtrsim 2.5$. In the first section, the grid locations are chosen using Gauss-Legendre quadrature rules.
 The integral in the second section is performed by a change of variable to $y=x^2 - x_0^2$, so that the integral becomes
 \begin{equation}
	 \frac{1}{2}\int_{0}^{\infty} dy \; e^{-y} \left[ e^y \sqrt{y + x_0^2} G\left( x \right) \right],
	 \label{upperenergy}
 \end{equation}
 where $G\left( x \right)$ is the function to be integrated.
 In this interval, the grid locations are chosen using Gauss-Laguerre quadrature rules.

%\subsection{$\lambda$ Grid}

The choice of the grid layout for $\lambda$ is complicated by the existence of particles trapped within magnetic wells in certain curved magnetic geometries. We require there to be a value of $\lambda$ for each parallel grid point, corresponding to  the particle which bounces at that particular grid point. We also require there to be a concentration of grid points around the trapped/passing boundary in $\lambda$ --- $\varepsilon$ space. 

Accordingly, the $\lambda$ grid is divided into two regions: one for passing particles with $ 0 \leq \lambda < 1/B_{max}$ and one for trapped particles with $1/B_{max} < \lambda < 1/B_{min}$.
In the first region, the integration variable $\tilde{\xi} = \sqrt{1 - \lambda B_{max}}$ is chosen, and Gauss-Legendre quadrature rules are used to obtain the locations of the grid points. 
In the second region, values of $\lambda$ are chosen to correspond to the bounce points: for every value of $\theta$, a value of $\lambda$ is chosen such that $v_{\parallel}\left( \theta,\lambda \right)=0$.\footnote{
Note that in a stellarator with multiple magnetic wells, a single value of $\lambda$ may in fact correspond to several bounce points.}

An important observation must be made here, namely, that since there must be a value of $\lambda$ for every bounce point, 
the user has only limited freedom in choosing the total number of values of $\lambda$.
Specifically, the user may only specify the number of untrapped values of $\lambda$; the input parameter $\mathtt{ngauss}$ is equal to half this number. 
For example, with $\mathtt{ntheta} = 14$ and concentric circular flux surfaces, there are 6 bounce points corresponding to different values of $\lambda$, so $\mathtt{n}\lambda=6 + 2 \times \mathtt{ngauss}$.  The number of bounce points depends on geometry, but is usually around half $\mathtt{ntheta}$.

The velocity grid for the case of $\mathtt{ntheta} = 14$, $\mathtt{ngauss} = 4$, $\mathtt{negrid} \equiv \mathtt{n}\varepsilon = 8$ and Cyclone Base Case geometry (concentric circular flux surfaces with $r/R=0.18$ and $q=1.4$, see Section \ref{sec:cyclonebasecase}) is shown in \figref{vspacegrid}.

%\myfig{energylambdapoints}{The Layout of the Velocity Space Grid in $\varepsilon$ and $\lambda$}{The layout of the velocity space grid in $\varepsilon$ and $\lambda$ space}.
\myfig[4.5in]{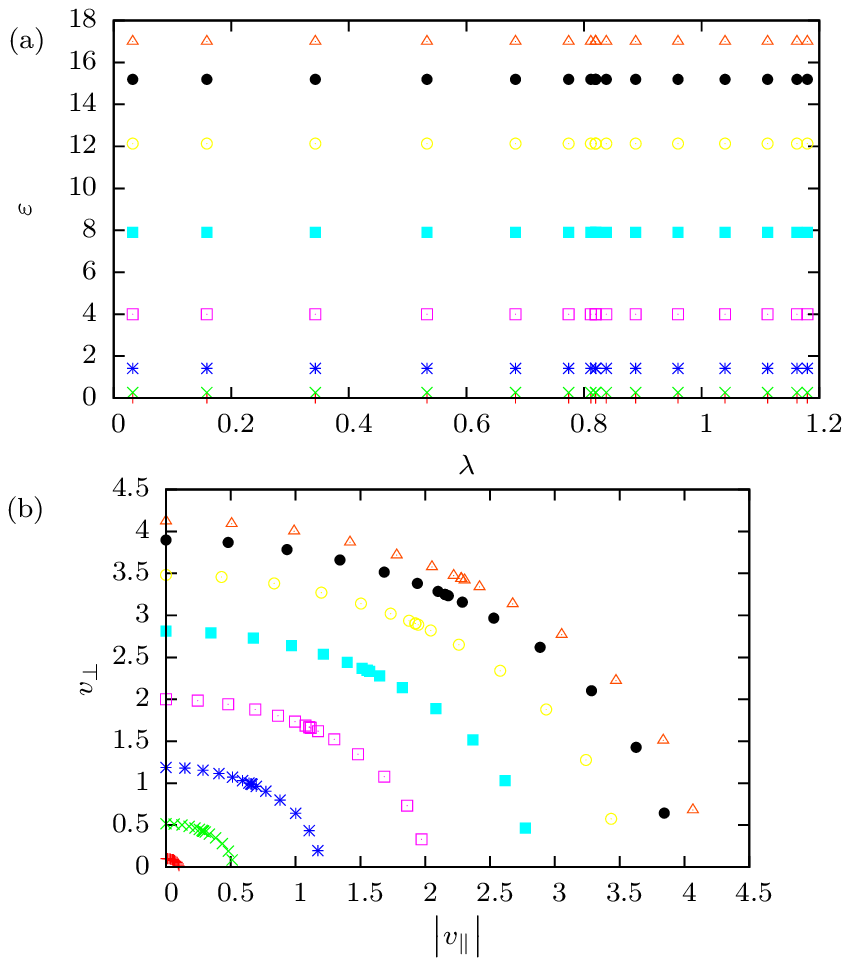}{The Layout of the Velocity Space Grid}{The layout of the velocity space grid corresponding to $\mathtt{ntheta}=14$, $\mathtt{ngauss}=4$ (corresponding to 14 values of $\lambda$), and $\mathtt{negrid}=8$ (i.e. 8 values of $\varepsilon$) in (a) $\varepsilon$ and $\lambda$ space and (b) $v_{\perp}$ and $\left|v_{\parallel} \right|$ space at the outboard midplane ($B=B_{min})$. It should be remembered that there are twice as many velocity grid points as displayed owing to the two signs of $v_{\parallel}$, indexed by $\sigma$.}

\section{The Collision Operator}
\label{sec:collisionoperator}
\subsection{Theory}

The collision operator defined in Ref.~\cite{abel2008linearized} causes diffusion in both the energy and $\lambda$ directions, while locally  conserving particles, energy and momentum. In its gyroaveraged spectral form, used in \texttt{GS2} \cite{numerics}, it can be written as the sum of five parts:
\begin{equation}
C_{GK}[\hat{h}] \equiv L[\hat{h}] + D[\hat{h}]
+ U_{L}[\hat{h}] + U_{D}[\hat{h}] + E[\hat{h}],
\label{eqn:cop}
\end{equation}
where

\begin{equation}
L[\hat{h}] \equiv \frac{\nu_{D}}{2}\pd{}{\xi}\left(1-\xi^{2}\right)\pd{\hat{h}}{\xi} - \frac{k^{2}v^{2}}{4\Omega_{0}^{2}}\nu_{D}\left(1+\xi^{2}\right)\hat{h}
\end{equation}
is the gyroaveraged Lorentz diffusion operator (i.e. it causes diffusion in $\lambda$ space) and

\begin{equation}
D[\hat{h}] \equiv \frac{1}{2v^{2}}\pd{}{v}\left(\nu_{\parallel}v^{4}F_{0}\pd{}{v}\frac{\hat{h}}{F_{0}}\right) - \frac{k^{2}v^{2}}{4\Omega_{0}^{2}}\nu_{\parallel}\left(1-\xi^{2}\right)\hat{h}
\end{equation}
is the gyroaveraged energy diffusion operator,

\begin{equation}
\begin{split}
U_{L}[\hat{h}] \equiv \nu_{D}F_{0}\Big(J_{0}(a)v_{\parallel}
\frac{\int d^{3}v \ \nu_{D}v_{\parallel}J_{0}(a)\hat{h}}{\int d^{3}v \ \nu_{D}v_{\parallel}^{2}F_{0}} 
+ J_{1}(a)v_{\perp}\frac{\int d^{3}v \ \nu_{D}v_{\perp}J_{1}(a)\hat{h}}{\int d^{3}v \ \nu_{D}v_{\parallel}^{2}F_{0}}\Big)
\label{eqn:scriptuL}
\end{split}
\end{equation}
and

\begin{equation}
\begin{split}
U_{D}[\hat{h}] \equiv -\Delta \nu F_{0}\Big(J_{0}(a)v_{\parallel}
\frac{\int d^{3}v \ \Delta \nu v_{\parallel}J_{0}(a)\hat{h}}{\int d^{3}v \ \Delta \nu v_{\parallel}^{2}F_{0}} 
+ J_{1}(a)v_{\perp}\frac{\int d^{3}v \ \Delta \nu v_{\perp}J_{1}(a)\hat{h}}{\int d^{3}v \ \Delta \nu v_{\parallel}^{2}F_{0}}\Big)
\label{eqn:scriptuD}
\end{split}
\end{equation}
 are the gyroaveraged momentum conserving corrections to the Lorentz and energy diffusion operators, and

\begin{equation}
E[\hat{h}] \equiv \nu_{E} v^{2} J_{0}(a) F_{0} \frac{\int d^{3}v \ \nu_{E} v^{2} J_{0}(a) \hat{h}}{\int d^{3}v \ \nu_{E} v^{4} F_{0}}
\label{eqn:scripte}
\end{equation}
 is the  gyroaveraged energy conserving correction. In addition, the electron collision operator has a piece corresponding to electron-ion collisions:

\begin{equation}
\begin{split}
	C_{GK}^{ei}[\hat{h}_{e}] = \nu_{D}^{ei} \Big(\frac{1}{2}\pd{}{\xi}\left(1-\xi^{2}\right)\pd{\hat{h}_{e}}{\xi}-\frac{k^{2}v^{2}}{4\Omega_{0_{e}}^{2}}\left(1+\xi^{2}\right)\hat{h}_{e}
	+\frac{2v_{\parallel}u_{\parallel}[\hat{h}_{i}]}{v_{th_{e}}^{2}}J_{0}(a_{e})F_{0e}\Big).
\label{eqn:Cei}
\end{split}
\end{equation}
In these equations $\xi\equiv v_{\parallel} / v$ is the pitch angle,  $J_{0}$ and $J_{1}$
are Bessel functions of the first kind, and $u_{\parallel}[\hat{h}_{i}]$ is the perturbed parallel ion flow 
velocity.  Definitions of the velocity-dependent collision frequencies $\nu_{D}$, $\Delta \nu$, 
$\nu_{\parallel}$, and $\nu_{E}$ are given in Ref.~\cite{abel2008linearized}.

The  local conservation properties of the collision operator  before being gyroaveraged do not survive the gyroaverage, which introduces nonlocal finite Larmor radius effects. However, if the Bessel functions are expanded in perpendicular wavenumber,
%\begin{equation}
	%C_{GK}[\hat{h}]  = C_{GK}^0[\hat{h}] + \mathcal{O}\left( k_{\perp} \rho \right),
	%\label{expandcollisionop}
%\end{equation}
the conservation properties are assured provided the first three moments of $C_{GK}^0[\hat{h}]$, that is, $C_{GK}[\hat{h}]$ with $k_{\perp}\rho$ set to 0 in the Bessel functions, vanish:
\begin{equation}
\int d^{3}v
\begin{pmatrix}
1 \\
v_{\parallel}\\
v^{2}
\end{pmatrix}
C_{GK}^{0}[\hat{h}]=0.
\label{eqn:momcons}
\end{equation}  

\subsection{Numerical Implementation}

The collision operator is implemented implicitly via two applications of Godunov splitting: starting with the result $\hat{h}^{n+1}$ of advancing the collisionless gyrokinetic equation (that is, the $\hat{h}^{n+1}$ of Sections \ref{sec:linearimplicitalgorithm} and \ref{sec:nonlinearterms}), which we redefine as $h^{\star}$, first the pitch angle scattering terms are applied:

\begin{equation}
	\frac{\hat{h}^{**}-\hat{h}^{*}}{\Delta t} 
	=  L[\hat{h}^{**}] + U_{L}[\hat{h}^{**}],\label{eqn:hstarstar}
\end{equation}
and then the energy diffusion terms are applied:		
\begin{gather}
	\frac{\hat{h}^{n+1}-\hat{h}^{**}}{\Delta t}
	=D[\hat{h}^{n+1}]
	+  U_{D}[\hat{h}^{n+1}] + E[\hat{h}^{n+1}].
\label{eqn:hnp1}
\end{gather}
This splitting technique of treating the collision operator separately to the collisionless gyrokinetic equation is accurate to first order in $\Delta t$. 
The combined operation can be expressed as the inversion and multiplication of two matrices:

\begin{gather}
	\hat{h}^{**} = \left[1-\Delta t \left( 
	L + U_L\right)\right]^{-1} \hat{h}^{*},
\label{eqn:Lsplit}
\\
\hat{h}^{n+1}=\left[1 - \Delta t \left(D
+ U_D+ E\right)\right]^{-1}\hat{h}^{**}.
\label{eqn:DMEsplit}
\end{gather} 
 The matrices $L$ and $D$ are kept tridiagonal through the use of a compact stencil for the velocity derivatives, but the matrices corresponding to the conservation operators are dense. 
 Application of the combined matrices $\left[1-\Delta t \left( L + U_L\right)\right]^{-1}$ and $\left[1 - \Delta t \left(D + U_D+ E\right)\right]^{-1}$ can, however, be speeded up using the Sherman-Morrison formula, which gives $\vct{x}$ in the matrix equation $M\vct{x} = \vct{b}$,
 provided $M$ can be written in the form $M = A + \mathbf{u}\otimes\mathbf{v}$, where where $\otimes$  is the tensor product. We identify $A$ with $L$ and $D$,
 and all the conservation operators can be written in the form $\vct{u}\otimes \vct{v}$.
 And so finally, $\hat{h}^{\star\star}$ and $\hat{h}^{n+1}$ can be calculated through two applications of the following formula:

\begin{equation}
	\vct{x} = \vct{y}_{2} - \left[\frac{\vct{v}_{2}\cdot \vct{y}_{2}}{1+\vct{v}_{2}\cdot\vct{z}_{2}}\right]\vct{z}_{2},
\label{eqn:hx}
\end{equation}
where

\begin{gather}
\vct{y}_{2} = \vct{y}_{0} - \left[\frac{\vct{v}_{0}\cdot\vct{y}_{0}}{1+\vct{v}_{0}\cdot\vct{s}_{0}}\right]\vct{s}_{0} - \left[\frac{\vct{v}_{1}\cdot\vct{y}_{0}}{1+\vct{v}_{1}\cdot\vct{w}_{0}}\right]\vct{w}_{0}, \label{eqn:sm1}\\
\vct{z}_{2} = \vct{z}_{0} - \left[\frac{\vct{v}_{0}\cdot\vct{z}_{0}}{1+\vct{v}_{0}\cdot\vct{s}_{0}}\right]\vct{s}_{0}- \left[\frac{\vct{v}_{1}\cdot\vct{z}_{0}}{1+\vct{v}_{1}\cdot\vct{w}_{0}}\right]\vct{w}_{0}. \label{eqn:sm2}
\end{gather}
and all quantities in these formulae are defined for each application in Table \ref{tab:sm}.
Before the application of the Lorentz operator,
the distribution function (at a given point in ($k_x$, $k_y$, $\theta$) space)
is rearranged in memory so that the $\lambda$ values for a given energy are
next to each other in memory.
Conversely, before the application of
the energy operator, the distribution function is rearranged in memory
so that the energy values for a given $\lambda$ are next to each other in memory.
This increases the speed of application of the formulas.
With the exception of $\vct{y}_m$, all quantities in these formulae are time independent so they only need to be calculated once at the beginning of each simulation.

\begin{table}
\begin{center}
%\begin{tabular} {| c || c | c |}
\begin{tabular} { c | c  c }
%\begin{tabular} { r | l  l }
\hline
\hline
Variable & Lorentz Op. & Energy Op.\\
\hline \hline
$A$ & $\ 1-\Delta t\left(L+U_L\right) \ $ & $\ 1-\Delta t\left(D+U_D+E\right) \ $\\
%\hline
$\vct{x}$ & $\hat{h}^{**}$ & $\hat{h}^{n+1}$ \\
%\hline
$\vct{b}$ & $\hat{h}^{*}$ & $\hat{h}^{**}$\\
%\hline
$A_{0}$ & $1-\Delta t L$ & $1-\Delta t D$\\
%\hline
$\vct{v}_{0}$ & $\nu_{D}v_{\perp}J_{1}$ & $-\Delta \nu v_{\perp}J_{1}$ \\
%\hline
$\vct{v}_{1}$ & $\nu_{D}v_{\parallel}J_{0}$ & $-\Delta \nu v_{\parallel}J_{0}$ \\
%\hline
$\vct{v}_{2}$ & $v_{\parallel}$ (electrons) & $\nu_{E}v^{2}J_{0}$ \\
& $0$ (ions) &\\
%\hline
$\vct{u}_{0}$ & $-\Delta t\vct{v}_0/d_{u}$ & $-\Delta t\vct{v}_0/d_{u}$ \\
%\hline
$\vct{u}_{1}$ & $-\Delta t\vct{v}_1/d_{u}$ & $-\Delta t\vct{v}_1/d_{u}$ \\
%\hline
$\vct{u}_{2}$ & $-\Delta t\nu_{D}^{ei}v_{\parallel}/d_{q}$ (electrons) & $-\Delta t\vct{v}_2/d_{q}$ \\
& $0$ (ions) &\\

%\hline
$d_{u}$ & $\int d^{3}v \ \nu_{D} v_{\parallel}^{2} F_{0}$ & $\int d^{3}v \ \Delta\nu v_{\parallel}^{2} F_{0}$ \\
%\hline
$d_{q}$ & $v_{th,e}^{2}/2$ & $\int d^{3}v \ \nu_{E} v^{4} F_{0}$ \\
\hline
\hline
& \multicolumn{2}{c}{Common}\\
\hline
\hline
 $A_1$ & \multicolumn{2}{c}{$A_0 + \vct{u}_0 \otimes \vct{v}_0$}\\
 $A_2$ & \multicolumn{2}{c}{$A_1 + \vct{u}_1 \otimes \vct{v}_1$}\\
 $\vct{y}_m$ & \multicolumn{2}{c}{$A_m^{-1} \vct{b}$}\\
 $\vct{z}_m$ & \multicolumn{2}{c}{$A_m^{-1} \vct{u}_2$}\\
 $\vct{w}_m$ & \multicolumn{2}{c}{$A_m^{-1} \vct{u}_1$}\\
 $\vct{s}_0$ & \multicolumn{2}{c}{$A_0^{-1} \vct{u}_0$}\\
\hline
\end{tabular}
\end{center}
\caption{Sherman-Morrison variable definitions for Lorentz and energy diffusion
operator equations.}
\label{tab:sm}
\end{table}
%The extended theta grid 

%which are easy to state but harder to implement.
%For passing particles the boundary condition is that there are no incoming particles.

Finally, the discretisation of the collision operator must be done in such a way as to preserve its conservation properties.
This can be achieved as long as certain conditions are satisfied, laid out in detail in Ref.~\cite{numerics},
most notably that the discrete differentiation operator must satisfy a discrete version of the Fundamental Theorem of Calculus, and a discrete version of integration by parts.
This, together with the need for a compact stencil and the fact that the velocity grids are unevenly spaced limits the choice of difference scheme, allowing only first order accuracy. The scheme used is:
\begin{equation}
\pd{}{x}G\pd{h}{x} \approx \frac{1}{w_{j}}\left(G_{j+1/2}\frac{h_{j+1}-h_{j}}{x_{j+1}-x_{j}}-G_{j-1/2}\frac{h_{j}-h_{j-1}}{x_{j}-x_{j-1}}\right),
\label{eqn:fd}
\end{equation}
where $w_{j}$ is the integration weight associated with $x_{j}$.  

\chapter{The Implementation of Flow Shear in \texttt{GS2}}
\label{flowshearapp}
\section{Introduction}
The implementation of equilibrium flow shear is achieved by an elegant trick which was announced at a conference \cite{gs2flowshear}, and which is described in an unpublished note \cite{gham_gs2vecshear_unpub}. Since the effect of this equilibrium flow shear is central to this work, the implementation merits a description in some detail.
%Here we describe the implementation right down to the function calls within the code to give the clearest possible description of how it works.

\section{Theory}
The system of equations \eqref{gyrokineticeq1} and \eqref{quasin} can be written as 
\begin{equation}
	\left( \frac{\partial }{\partial t} + x \gamma_E \frac{\partial }{\partial y} \right) g = L \left[ \frac{\partial }{\partial x},  \frac{\partial }{\partial y}, \frac{\partial }{\partial t} + x \gamma_E \frac{\partial }{\partial t} \right] g
	%\label{}
\end{equation}
where the  operator $L $ includes, among other things, the solution of the quasineutrality condition to obtain the potential from the distribution function.

We wish to solve this equation spectrally in both the $x $ and $y $ directions, but the explicit dependence on $x $ introduced by the flow shear prevents this. In order to eliminate the flow shear we write the eikonal in terms of a time varying wave number $k_x - \gamma_E k_y t $: 
\begin{equation}
	g = \sum_{k_x,k_y} \hat{g} e^{i\left[ \left( k_x - \gamma_E k_y t \right) x + k_y y \right]},
	\label{eqn:timevaryingkx}
\end{equation} which allows us to write 
\begin{equation}
	\frac{\partial \hat{g}}{\partial t} = L \left[ k_x - \gamma_E k_y t, k_y, \frac{\partial }{\partial t} \right]g.
	%\label{}
	\end{equation}

	We have eliminated the explicit dependence on $x $ at the cost of introducing an explicit dependence on $t $ in the operator $L $, and hence in the response matrices \eqref{responsematrixdef}. This would make it necessary to recalculate the response matrices at every time step, which is unacceptable.
	%This, however, can be dealt with as follows.
%Before proceeding, it is worth noting that the time derivative is a constant $k_x $: 
%\begin{equation}
	%\left. \frac{\partial \hat{g}}{\partial t}
	%\right|_{k_x}.
	%%\label{<++>}
%\end{equation}
We therefore define \(k_x^{\star}\) and \(\tilde{g}\) as follows
\cite{gham_gs2vecshear_unpub}: 

\begin{eqnarray}
k_x^{\star} \equiv k_x - \gamma_E k_y t,
\label{eqn:kxeuleriandef}
\end{eqnarray}
and

\begin{eqnarray}
\tilde{g}\lp k_x^{\star}, k_y, t \rp &\equiv& \hat{g}\lp k_x, k_y, t \rp \\
 &=&  \hat{g}\lp k_x^{\star} + \gamma_E k_y t, k_y, t \rp .
\end{eqnarray}
The equation is now 

\begin{eqnarray}
 \left.  \pdt{\tilde{g}}\right|_{k_x}  = L\lsq k_x^{\star}  ,k_y, \pd{}{t} \rsq \tilde{g}.
\end{eqnarray}
The explicit time dependence has been eliminated from the operator $L $, and the whole equation is now written in terms of $k_x^{\star} $, except for the fact that the derivative on the left-hand side is at constant $k_x $. The evaluation of this derivative is discussed in the next section.

\section{Implementation}

\subsection{Discretisation}

We now discretise in time so that

\begin{eqnarray}
	\left.  \pdt{\tilde{g}}\right|_{k_x} &=& 
\frac{\tilde{g}^{n+1}_{kx^{\star,n+1}} - \tilde{g}^{n}_{kx^{\star,n}}}{\Delta t} \nonumber\\  &=& 
L\lsq r_\theta k_x^{\star,n} + \left( 1 - r_\theta \right)k_x^{\star,n+1}  ,k_y, \pd{}{t} \rsq \left[ r_\theta \tilde{g}^{n}_{kx^{\star,n}} + \left( 1-r_\theta \right) \tilde{g}^{n+1}_{kx^{\star,n+1}}\right]. \;
\label{eqn:flowsheardiscretised}
\end{eqnarray}
where without loss of generality we now let $r_\theta=0$.
At this point we note that because in general \(\Delta k_x^{\star}\) is not equal to \( \gamma_E \times k_y \times \Delta t \), \(L^{n+1}\) will not be evaluated at \(k_x^{\star,n+1} \) but at the nearest discrete \( k_x^{\star}\) available.
The difference between this discrete value and the actual value of \( k_x^{\star}\) is stored in the variable $\mathtt{kxshift}$.
Rearranging \eqref{eqn:flowsheardiscretised} (and ignoring all other \(L\) dependences), we may write 

\begin{eqnarray}
\tilde{g}^{n}_{kx^{\star,n}} =
\lp 1 - \Delta t  L \lsq k_x^{\star,n+1}  \rsq \rp  \tilde{g}^{n+1}_{kx^{\star,n+1} }, 
\end{eqnarray}
and solving for \(\tilde{g}^{n+1}\), we obtain the equation 

\begin{eqnarray}
\label{discretised_v_shear}
\tilde{g}^{n+1}_{kx^{\star,n+1}} = 
\lp 1 - \Delta t  L \lsq k_x^{\star,n+1} \rsq \rp^{-1}  \tilde{g}^{n}_{kx^{\star,n}}. 
\end{eqnarray}

\subsection{Implementation}

Comparing \eqref{discretised_v_shear} to the equivalent equation with no velocity shear:

\begin{eqnarray}
\label{discretised_no_v_shear}
\hat{g}^{n+1}_{kx} = 
\lp 1 - \Delta t^{n+1/2}  L \lsq k_x \rsq \rp^{-1}  \hat{g}^{n}_{kx} ,
\end{eqnarray}
we note that
the only difference is that in equation \eqref{discretised_v_shear}, \(\tilde{g}^{n}\) is evaluated at a different radial wavenumber to \(L\) and \(\tilde{g}^{n+1}\). Thus, in order to get \texttt{GS2} to solve \eqref{discretised_v_shear} rather than \eqref{discretised_no_v_shear}, all that is necessary is an assignment step prior to the evaluation of \eqref{discretised_no_v_shear}:

\begin{eqnarray}
\label{g_assignment}
\tilde{g}^{n}_{kx^{\star}\lp t^{n} \rp} \rightarrow \tilde{g}^{n}_{kx^{\star}\lp t^{n + 1} \rp}
\end{eqnarray}
At the risk of anthropomorphising a computer code, \texttt{GS2} is 'tricked' into solving \eqref{discretised_v_shear} instead of \eqref{discretised_no_v_shear}, because where it is expecting \( \tilde{g}^{n}_{kx^{\star}\lp t^{n+1} \rp} \) on the right hand side, ie. \(\tilde{g}^{n}\) evaluated at the same radial wavenumber as \(L\) and \(\tilde{g}^{n+1}\) as in \eqref{discretised_no_v_shear}, it actually gets \( \tilde{g}^{n}_{kx^{\star}\lp t^{n} \rp} \) as in \eqref{discretised_v_shear}.

\subsection{Boundary Conditions}

The effect of the assignment step \eqref{g_assignment} is to shift all the grids of the distribution function and the fields by one grid point in the $k_x $ direction. At one end of the $k_x $ grid, the values of the grid are lost. At the other end, we choose a zero incoming boundary condition. This has important implications for both linear and nonlinear simulations. Linear simulations should only be run with one non-zero value of $k_x^{\star} $, and can only be run until this value falls off the grid\footnote{
Linear simulations with finite magnetic shear should, in fact, use the separate single mode implementation of flow shear which is not discussed here.}. 
Non-linear simulations should include high enough radial modes that the perturbations are damped by finite Larmour radius effects before they are swept off the grid.

} % end webonly

\mypart{A Bifurcation to a Reduced Transport State}
\label{sec:bifurcationtoareducedtransportstate}

\chapter{Introduction}
\label{sec:bifurcationintroduction}

\begin{quote}
	\center
	\emph{Chapters \ref{sec:bifurcationintroduction}--\ref{furtherobservationssec} are largely taken from Ref. \cite{highcock2011transport}.}
\end{quote}
\vspace{1cm}
\label{sec:bifurcationintro}
As described in Section \ref{sec:shearedflows}, a recent paper 
\cite{barnes2011turbulent} 
which investigated the effect of flow shears
several times larger than the linear growth rate 
of the ITG instability 
(one of the principal drivers of turbulence in fusion devices),
found that for relatively low temperature gradients 
there is a range of flow shear values where  
the turbulence is completely quenched.
The question remains how large shears in the equilibrium toroidal flow can be achieved, given finite  sources of angular momentum to drive the flow. 
In Ref.~\cite{barnes2011turbulent}, 
the turbulent flux of angular momentum  was calculated and found to be large at moderate flow shears; 
if it is too large, increasing the input of angular momentum will not generate strong enough shear in the flow. 

Ref.~\cite{barnes2011turbulent} considered the standard Cyclone Base Case \cite{dimits:969} (see Section \ref{sec:cyclonebasecase}), with the normalised inverse magnetic gradient scale length $\hat{s}=0.8$ and a range of values of the temperature gradient and the velocity shear. 
In Part \ref{sec:bifurcationtoareducedtransportstate}, which is published as Ref.~\cite{highcock2011transport}, 
we consider a similar regime to that described in Ref.~\cite{barnes2011turbulent},
with the difference that here the magnetic shear $\hat{s}=0$.  
%This choice is motivated by the experimental studies
%which have found that internal transport barriers 
%(local regions of very steep flow and temperature gradients)
%tend to form in regions of zero or low magnetic shear \cite{connor2004itbreview,vries2009internal}, 
%and by previous numerical results which show a lower ion thermal diffusivity
%at zero magnetic shear \cite{casson2009anomalous}.
The results reported in Part \ref{sec:bifurcationtoareducedtransportstate} show that
the range of flow velocity gradients and ion temperature gradients
where the turbulence is suppressed
is much larger than in the case of finite magnetic shear.
We demonstrate the existence of a transport bifurcation,
%from low to high values of the flow and temperature gradients
in which a positive feedback
between the increase in the flow gradient
and the suppression of the turbulence
provides the mechanism for a jump from low to high flow gradients,
with a simultaneous jump in the temperature gradient.\footnote{
Such a transition is in principle also possible with \(\hat{s}=0.8\),
but in a much smaller region of parameter space \cite{parra2011momentum}}  
We show that this transition results in a state
where the transport of heat is nearly neoclassical,
whilst the transport of momentum remains largely turbulent. 

In a fusion device, the quantities that can be controlled are the input of heat and toroidal angular momentum, which in steady state are proportional to their outgoing fluxes.
By varying these quantities, it is possible to vary the equilibrium gradients.
In contrast, in the local gyrokinetic simulations used in this study
the input parameters are the gradients, and the fluxes are calculated from the output. 
In order to demonstrate the existence of
a bifurcation in the gradients
at fixed values of the input fluxes,
we use local gyrokinetic simulations
to map out the dependence of the turbulent fluxes
on the temperature and flow gradients,
and then we add the neoclassical fluxes
and invert this numerically determined dependence
to find the gradients as functions of the total fluxes. 
We first do this by straightforward interpolation in the parameter space, then propose a simple parameterisation of the fluxes that fits the data and can be used to predict in which parameter regimes bifurcations can occur.

The rest of Part \ref{sec:bifurcationtoareducedtransportstate} is organised as follows.
In Chapter \ref{fluxessec}, we report a numerical scan in two parameters: the flow shear and the ion temperature gradient, calculating the turbulent heat and momentum fluxes over wide intervals of these parameters. 
%In Chapter \ref{subcriticalsec}, we describe the subcritical turbulence, driven by the ion temperature gradient (ITG) and the parallel velocity gradient (PVG), that is responsible for the turbulent fluxes.
In Chapter \ref{transitionsec}, we interpolate our results and determine how the temperature and flow gradients depend on the heat and momentum fluxes: a transport bifurcation is obtained as a result.  
In Chapter \ref{qbehaviour}, we construct a parametric model of the transport. 
This allows us, in Chapter \ref{furtherobservationssec}, to study the effect of the transport bifurcation  on the temperature gradient and investigate the range of heat and momentum flux values for which we expect transitions to occur.

\section{Model}

In Part \ref{sec:bifurcationtoareducedtransportstate}, we use the gyrokinetic equation \eqref{eqn:gyrokineticeqgs2coordinates}
to model the turbulence. The equation is solved by the simulation code \texttt{GS2}, as described in Part \ref{sec:modellingturbulence}. We 
vary
the perpendicular flow shear $\fsr$,
and $\kappa = R_0/L_T$.
Other parameters
are kept fixed:
\begin{equation}
	\frac{R_0}{L_n}=2.2,\; q_0 = 1.4,\; \frac{r_0}{R_0}=0.18.
	\label{params}
\end{equation}
The density gradient, the magnetic safety factor and the aspect ratio
are chosen to conform to the Cyclone Base Case \cite{dimits:969} (see Section \ref{sec:cyclonebasecase}),
while, as we explained above, the magnetic shear is set to 0.
Note that the value of $q_0$ effectively controls the strength of the PVG drive
(the term proportional to $q_0 \fsr$ on the right hand side of \eqref{eqn:gyrokineticeqgs2coordinates})
for a given velocity shear $\fsr$.
The relatively low value of $q_0$ in the Cyclone Base Case
reduces the destabilising effect of the PVG
--- compared, for example, to the Waltz Standard Case where $q_0=2.0$ \cite{kinsey2005flowshear}.
Collisions are included by means of the model collision operator
described in Section \ref{sec:collisionoperator}.
%which includes the scattering of particles
%in both pitch angle and energy \cite{numerics,abel2008linearized}. 
The numerical ion-ion collision frequency  $ \nu_{ii}^N = 0.01 v_{thi}/R_0$.
\label{sec:fluxnorms}
All fluxes will be reported in dimensionless units
by normalising them to the 
gyro-Bohm values 
\begin{equation}
	Q_{gB}=\frac{n_iT_i v_{thi} \rho_i^2}{2 \sqrt{2} R_0^2},\qquad  \Pi_{gB}=\frac{m_i n_i v_{thi}^2 \rho_i^2}{4 R_0}.
	\label{gpdefs}
\end{equation}
%The flow shear will be everywhere normalised to $ v_{thi} / R_0$;
%hence at this point we rescale $\fsr$ accordingly:

%\begin{equation}
	%\fsr \rightarrow \frac{v_{thi}}{R_0} \fsr.
	%\label{fsrnorm}
%\end{equation}

The resolution of the majority of simulations was \(64\times32\times14\times24\times8\times2\)
--- 
the number of gridpoints in the $k_{x0}$ and $k_y$ spectral directions,
in the $z$ spatial direction,
the average number of pitch angles (i.e. the number of values of  $\mu_i$ for a given $\varepsilon_i$, which varies with the parallel coordinate),
the number of gridpoints in energy space $\varepsilon_i$
and the two values of $\sigma$. 
This grid provided sufficient scale separation
in both spatial directions 
perpendicular to the magnetic field
for calculating the turbulent transport,
and sufficient resolution in velocity space
to calculate the velocity integrals in Maxwell's equations 
with the required accuracy. 
A discussion of the parallel resolution is given in Chapter \ref{sec:turbulencewithoutinstability}.

\chapter{Turbulent Fluxes}
\label{fluxessec}
\section{Heat Flux}
Using simulations in the manner described in Part \ref{sec:modellingturbulence}, 
the heat and momentum fluxes were calculated 
for $\tprim$ values in the range $4 \leq \tprim \leq 13$, 
vs flow shear values in the range $0\leq\fsr\leq2\gstwofsrnorm$.

\figref{gexbzero}(b) shows the heat flux vs the ITG in the absence of flow shear. 
When \(\fsr=0\), the critical temperature gradient
above which there is ITG-driven turbulence is \(\tprim=4.4\). 
When $\tprim$ is increased above this threshold,
the heat flux increases rapidly up to more than a hundred times the gyro-Bohm value.

\figref{fluxes} shows the turbulent heat and momentum fluxes vs the flow shear for different ITG values. 
As the flow shear is increased from 0, the heat flux initially responds weakly, 
either increasing or decreasing slightly.
As the flow shear $\fsr$ approaches 1, the heat flux dips sharply for all values of the ITG. 
For moderate temperature gradients (\(\tprim \lesssim 11\)),
the turbulence is suppressed altogether and the heat flux drops to zero. 
The suppression of turbulence also happens at finite magnetic shears \cite{barnes2011turbulent}, 
but for a significantly narrower range of $\tprim$.\footnote{
This is partly owing to the fact that at finite magnetic shear growing linear eigenmodes exist for non-zero flow shear, whereas at zero magnetic shear there are no such eigenmodes except when the flow shear is also zero; this is discussed further in Part \ref{sec:subcriticalturbulence}.
}

For larger $\fsr$, the heat flux starts to rise again. 
This increase is due to the PVG
--- we have verified that this effect disappears
if the term proportional to $q_0\fsr$
on the right hand side of eq \eqref{eqn:gyrokineticeqgs2coordinates}
is artificially set to 0.
This revival of the turbulent transport at large shears
was also observed in \cite{barnes2011turbulent}, 
but was absent from the quasilinear study conducted in \cite{waltz1997gyro}, 
which showed the turbulence being completely suppressed above a sufficiently large toroidal shear.

\myfig{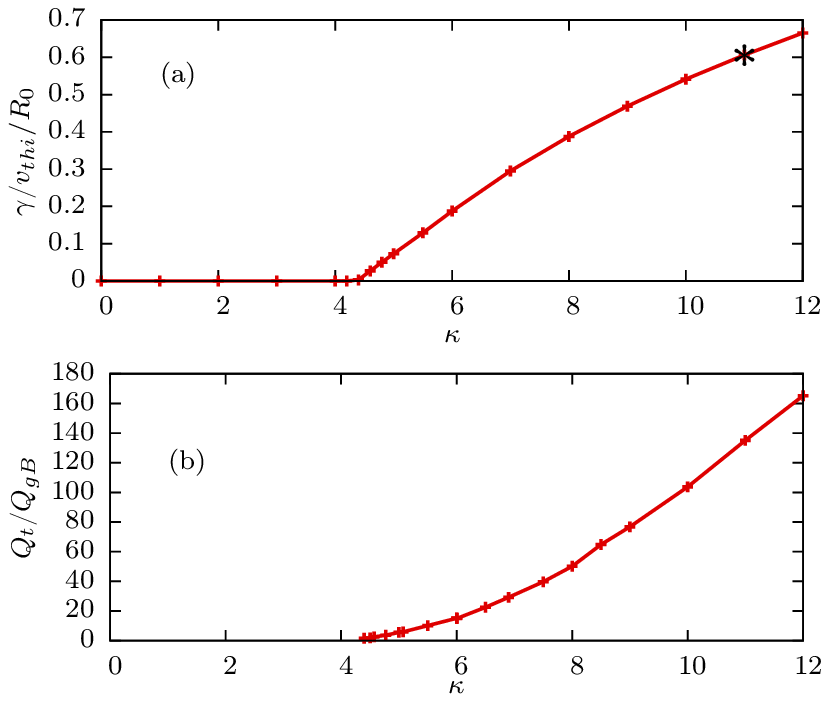}{The ITG Instability in the absence of Flow Shear}{ITG instability and turbulence at zero flow shear: (a) linear growth rate; (b) turbulent heat flux vs the normalised temperature gradient $\tprim=R_0/L_T$. The case of $\kappa=11$, further explored in \figref{transientgrowth}, is marked by *.}
\myfig{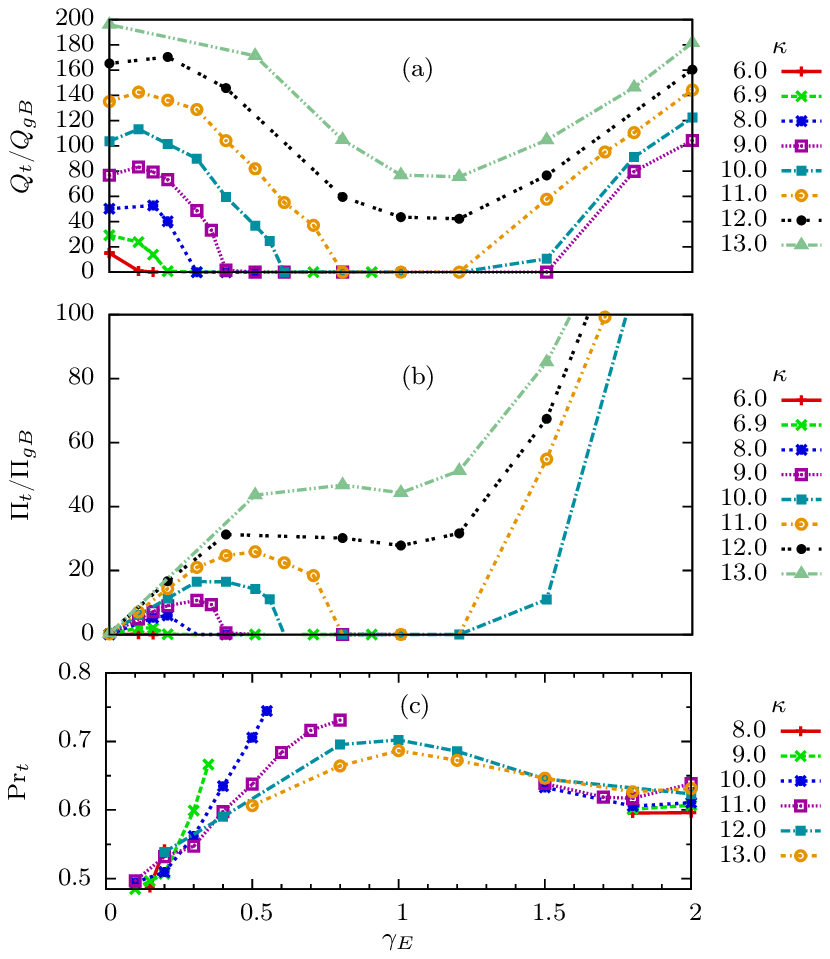}{Turbulent Heat and Momentum Fluxes}{Turbulent heat (a) and toroidal angular momentum (b) fluxes (normalised to gyro-Bohm values) as functions of flow shear for different values of the ion temperature gradient $\tprim$; (c) turbulent Prandtl number vs flow shear (for cases where the heat flux is non-zero).}

\section{Momentum Flux}

The momentum flux is zero when \(\fsr=0\) 
(as might be expected in an up-down symmetric case, 
\cite{parra2011up,camenen2009intrinsic}). 
As $\fsr$ increases, so does the momentum flux ---
a result of the turbulent viscosity.
However, when the flow shear reaches values at which it starts to suppress turbulence significantly, 
the trend is reversed and a remarkable situation arises 
in which increasing flow shear reduces the transport of momentum.
This behaviour persists until $\fsr$ reaches even larger values, 
and the turbulence is reignited by the PVG drive.
The positive correlation between the velocity shear and the momentum flux is then reestablished. 
It is the existence of the window of suppressed momentum transport 
at moderate $\fsr$ and $\tprim$ that will enable
the transport bifurcation analysed in Chapter \ref{transitionsec}.

\section{Turbulent Prandtl Number }
\label{prtsec}

There is a clear correlation between the heat and the momentum flux,
which is best quantified in terms of the turbulent Prandtl number

\begin{align}
	\mathrm{Pr}_t = \frac{\nu_t}{\chi_t} 
	= \frac{\Pi_t/\Pi_{gB}}{Q_t/Q_{gB}}\frac{\tprim}{\fsr} \frac{r_0}{\sqrt{2} R_0 q_0}
	\label{prtdef}
\end{align}
where the turbulent viscosity and the turbulent heat diffusivity are

\begin{align}
	&\nu_t = \frac{\Pi_t}{\Pi_{gB}} \frac{1}{\fsr} \frac{v_{thi} r_0  \rho_i^2}{4 q_0 R_0^2},\label{nutdef}\\
	&\chi_t = \frac{Q_t}{Q_{gB}}\oov{\tprim}\frac{v_{thi} \rho_i^2}{2\sqrt{2}R_0},\label{chitdef}
\end{align}
respectively.

The Prandtl number obtained from our simulations is plotted in Fig. \ref{fluxes}(c). 
There is a similar basic trend for all values of $\kappa$:
$\prt$ rises from approximately 0.5  when $\fsr\simeq0.1\gstwofsrnorm$,\footnote{
The low value of $\prt$ for small $\fsr$
has been observed before
and is sometimes referred to as a shear pinch 
\cite{dominguez1993anomalous,waltz2007gyrokinetic}.
It occurs because, at low $\fsr$,
the perpendicular flow shear can 
give rise to a contribution to the viscous stress 
that has the opposite sign to that of ITG-driven turbulence with zero flow shear,
reducing the overall diffusive transport.}
peaks at $\fsr\simeq1\gstwofsrnorm$ and then drops to approximately 0.6 when $\fsr\simeq2\gstwofsrnorm$. 
For those intermediate values of $\fsr$
where the turbulent fluxes are reduced or tend to 0,
$\prt$ rises sharply,
reaching just above 0.7  for low values of $\tprim$.
The location of this sharp rise varies with temperature gradient
similarly to the location where the turbulence is suppressed.

Although the Prandtl number does vary with both $\fsr$ and $\tprim$,
this dependence is relatively weak
compared to the individual dependence of $\nu_t$ and $\chi_t$ on these parameters.
Thus,  approximating $\prt\sim0.55$ everywhere 
keeps it within approximately 25\% of the true value for the entire range of $\fsr$.
This will prove convenient in constructing a model of turbulent transport presented in Chapter \ref{qbehaviour}.

\chapter{Transport Bifurcation}
\label{transitionsec}
\section{Possibility of Bifurcation}
In Chapter \ref{fluxessec},
we demonstrated the existence of a wide interval in $\fsr$
in which a sheared toroidal equilibrium flow completely suppresses turbulent transport. 
If such a suppression could be achieved experimentally in a tokamak,
confinement of energy would be dramatically improved. 
Unfortunately, while it is possible to specify the flow shear in numerical simulations,
in experiment it can only be varied indirectly by applying torque,
and the effect of that torque is strongly dependent
on how quickly the angular momentum escapes from the plasma.
If only a limited amount of torque can be injected and the momentum flux is too large,
it could be impossible to achieve flow shears that are large enough to suppress the turbulence.  
Fig \ref{fluxes}(b), however, suggests an intriguing possibility. 
For all temperature gradients, the momentum flux at first increases,
reaches a maximum and then decreases
before increasing again at higher flow shears.
There is, therefore, a parameter window in which
increasing the flow shear decreases the transport of momentum. 
If this were to happen, the momentum would build up,
increasing the flow shear, which would further decrease the transport of momentum. 
Such a positive feedback could lead to a bifurcation and so it might seem that high values of flow shear could be reached without excessive input of momentum,
a possibility which was discussed in \cite{waltz1995advances}.\footnote{
Note that although a similar mechanism was discussed in \cite{waltz1995advances}, 
that study considered flow shears
up to a maximum of approximately $0.08 c_s / a$, 
where $c_s=\sqrt{2T_e/m_i}$ is the sound speed 
and $a$ is the minor radius of the plasma (in this study $c_s=v_{thi}$ and $a=R_0$).
Thus they were considering a maximum in the momentum flux
which occurs at low flow shear and finite magnetic shear,
and which is discussed in more detail in \cite{barnes2011turbulent}.
As can be seen from Ref.~\cite{barnes2011turbulent}, 
the corresponding jump in flow shear is much smaller,
and the range of $\tprim$ values 
at which such a maximum in the momentum flux exists is much narrower,
than in the present case (\figref{fluxes}(b)).
} 
However, Fig \ref{fluxes}(b) shows the momentum flux at constant ion temperature gradient:
the temperature gradient 
%which
would not, in fact, stay constant during this process. 
Indeed, as soon as flow shear began to suppress the turbulent transport of momentum, it would also suppress the turbulent transport of heat, causing an increase in the temperature gradient as well as in the flow gradient. 
This increase in the temperature gradient would restore the turbulence to its former levels.
Nevertheless, we will show in this section that when neoclassical transport is also taken into account, a bifurcation is possible.

\section{Inverting the Problem}
\label{invertingsec}
What can actually be controlled in a steady-state experimental situation
is the flow of heat and momentum through a particular surface.  
This means that we must switch from using \(\fsr\) and \(\tprim\) as independent parameters 
to using the total fluxes of heat and momentum, $Q$ and $\Pi$ respectively. 

Let us consider an experimental set-up
where the heat and the momentum are injected by beams of neutral particles.
We assume that the energy and momentum from those neutral beams
are deposited uniformly across the torus.
We further assume that the beams are tangential to the magnetic axis and
we ignore corrections of order the aspect ratio of the device.
Then:

\begin{align}
	&\frac{Q}{Q_{gB}}\simeq \oov{Q_{gB}} \frac{V_f N_b  m_i v_b^2/2}{4\pi^2 r_0 R_0}
	= V_f \frac{2\sqrt{2} R_0 P_{\mathrm{NBI}}}{4\pi^2 r_0 n_i T_i v_{thi} \rho_i^2},\,
	\label{qexp}\\
	&\frac{\Pi}{\Pi_{gB}} \simeq \oov{\Pi_{gB}}  \frac{V_f N_b m_i v_b}{4\pi^2  r_0},\, \\
	&\frac{\Pi/\Pi_{gB}}{Q/Q_{gB}} \simeq \frac{v_{thi}}{\sqrt{2}v_b} = \lp\frac{T_i}{2 E_{\mathrm{NBI}}}\rp^{1/2}, \,
	\label{pioqexp}
\end{align}
where $N_b$ is the number of neutral beam particles injected per unit time,
$v_b$ is the beam particle velocity,
$P_{\mathrm{NBI}}=N_b m_i v_b^2 / 2$ is the beam power,
$E_{\mathrm{NBI}}=m_i v_b^2 / 2$ is the beam particle energy
and $V_f$ is the volume fraction of the plasma enclosed by the flux surface.
Thus, the total heat flux $Q$ is determined by the beam power $P_{\mathrm{NBI}}$ \eqref{qexp},
whereas the ratio of the total momentum angular flux to the total heat flux,
\(\Pi/Q\), is determined by the beam particle energy $E_{\mathrm{NBI}}$ \eqref{pioqexp}. 
We want to know whether, by varying $Q$ and $\Pi/Q$,  it is possible to reach,
or to trigger a transition to, a high-flow-shear regime where the turbulent transport is suppressed. 
Thus, it is necessary to
convert the dependence of $Q$ and $\Pi$ on $\tprim$ and $\fsr$,
determined from local simulations, 
to a dependence of $\tprim$ and $\fsr$ on $Q$ and $\Pi/Q$.

\section{Interpolation}
\label{interpolationsec}
The gyrokinetic code \texttt{GS2} gives the fluxes as a function of $\tprim$ and $\fsr$. 
Given the expense of the nonlinear simulations necessary to do this,
it is computationally challenging to use \texttt{GS2}
as a root finder to invert the problem and find the gradients as a function of the fluxes
\cite{barnes2009trinity,candy2009tgyro}.
Instead we interpolate within the set of data points
described in Chapter \ref{fluxessec}
(as well as additional simulations in parameter regions of particular interest)
to obtain the fluxes as continuous functions of the gradients;
these functions can then be inverted numerically to give
$\tprim$ and $\fsr$ as functions of the fluxes.

Interpolation in a multidimensional parameter space
is a nontrivial operation.
A standard technique is to use radial basis functions \cite{buhmann2001radial},
which weigh each data point
by its distance in parameter space from the point of interest
(after the parameter space has been normalised
to ensure that variation occurs on the same scale in each coordinate).
There are many choices of the function, or kernel,
which is used to calculate the relative importance of each data point.
Here we choose a linear kernel
(equivalent to linear interpolation in the case of only two points) \cite{buhmann2001radial}. 
Using this, the values of the fluxes at a point
\( \vct{p} = (\fsr, \tprim) \) can be calculated as follows:

\begin{align}
	Q_t\lp \vct{p} \rp = \sum_{\mathrm{j}} w_\mathrm{j}^Q \md{\vct{p} - \vct{p}_{\mathrm{j}}},\\
	\Pi_t\lp \vct{p} \rp = \sum_{\mathrm{j}} w_\mathrm{j}^\Pi \md{\vct{p} - \vct{p}_\mathrm{j}},
	\label{interpolation}
\end{align}
where \(\vct{p}_\mathrm{j}\) are the input gradients for the nonlinear simulation labelled $\mathrm{j}$,
and the weights \(w_\mathrm{j}^Q\) and \(w_\mathrm{j}^\Pi\) are calculated
so as to satisfy for all $\mathrm{j}$:
\begin{align}
Q_t(\vct{p}_\mathrm{j}) = Q_{t\mathrm{j}}\;, \Pi_t(\vct{p}_\mathrm{j}) = \Pi_{t\mathrm{j}}, 
	\label{weightcalc}
\end{align}
where $Q_{t\mathrm{j}}$ and $\Pi_{t\mathrm{j}}$ are the values of the fluxes obtained from simulation $\mathrm{j}$.

\section{Neoclassical Transport}
\label{neoclassicaltransportsec}

If the turbulent transport is successfully suppressed,
neoclassical (collisional) transport becomes important. 
The total fluxes are the sum of the neoclassical and turbulent contributions: 

\begin{align}
	 Q = Q_t+Q_n,\;\Pi = \Pi_t + \Pi_n,
	 \label{totalqp}
\end{align}
where the neoclassical fluxes are
\begin{align}
	&\frac{Q_n}{Q_{gB}}=\chi_n\tprim\frac{2 \sqrt{2}R_0  }{v_{thi} \rho_i^2 },
	\\
	& \frac{\Pi_n}{\Pi_{gB}} =\nu_n\fsr \frac{4  R_0^2 q_0  }{r_0 v_{thi} \rho_i^2},
\end{align}
and the neoclassical thermal diffusivity and viscosity are
\begin{align}
&	\chi_n\simeq 0.66 \lp\frac{ R}{r_0}\rp^{3/2}q_0^2\rho_i^2 \nu_{ii},
	\label{chindef}
	\\ 
&	\nu_n \simeq 0.1 q_0^2 \rho_i^2 \nu_{ii}. \label{nundef}
\end{align}
The formulae for
the neoclassical diffusivity
and viscosity
in  the large-aspect-ratio,
concentric-circular-flux-surface,
banana (\(\nu_{ii} \ll v_{thi}/Rq\)) regime,
were taken from \cite{hintonwong1985nit}. 
In \cite{hintonwong1985nit}, the ion-ion collision frequency is defined as
\begin{equation}
	\nu_{ii} = \frac{\sqrt{2 \pi}  Z_i^4 n_i e^4 \ln \Lambda}{T_i^{3/2} m_i^{1/2}}
	\label{nuiidef}
\end{equation}
(where $\ln \Lambda$ is the Coulomb logarithm). 
Thus, the neoclassical transport is a function
of the temperature and density,
and scales with these quantities 
in a different way to the turbulent transport.
To determine the ratio
between the neoclassical and turbulent transport
it is necessary to choose
specific values for the temperature and density.
Here, as in \cite{highcock2010transport}, 
we take
$\nu_{ii}=\nu_{ii}^N/2$.
Later, when we consider the case of specific 
tokamaks, we will use typical values of $n_i$ and $T_i$ 
to estimate $\nu_{ii}$.

Using the results \eqref{chindef} and \eqref{nundef}
the neoclassical Prandtl number
can be calculated from the parameters given in
\eqref{params}
and is found to be
\begin{equation}
	\mathrm{Pr}_n = \frac{\nu_n}{\chi_n} \simeq 0.01 \ll \prt.
	\label{prndef}
\end{equation}
Thus the neoclassical Prandtl number is much smaller than the turbulent Prandtl number (see Fig. \ref{fluxes}(c)).
In other words, the neoclassical transport of momentum is much smaller than the neoclassical transport of heat,
in contrast to turbulent transport,
for which the fluxes of momentum and heat are comparable.
While the formulae \eqref{nundef} and \eqref{chindef} are approximate,
we emphasise that the qualitative result
of the following section
is not affected by small changes in the values of \(\chi_n\) and \(\nu_n\),
provided that the property $\mathrm{Pr}_n \ll \mathrm{Pr}_t$ continues to hold.
In particular, this means that this qualitative result
is not affected by small changes in the value of $\nu_{ii}$.

\section{Bifurcation}
\label{transition}
\myfig{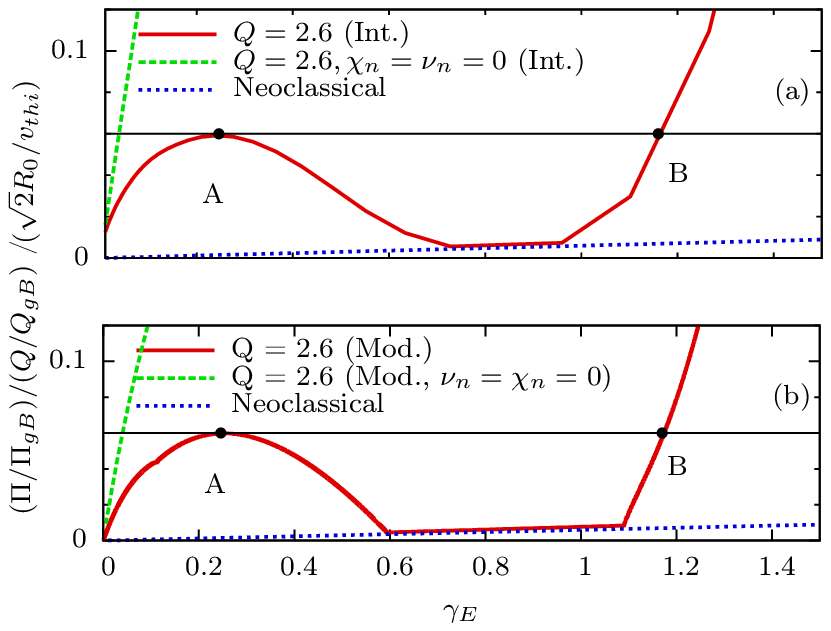}{Existence of a Bifurcation}{Momentum flux divided by the heat flux (each normalised by the corresponding gyro-Bohm estimate) vs the flow gradient at a constant value of the heat flux $Q=2.6Q_{gB}$ plotted using (a) interpolation from the data explained in Section \ref{interpolationsec} and (b) the parameterisation from Chapter \ref{qbehaviour}. Also shown are the neoclassical contribution to the momentum flux (dotted line) and the momentum flux at constant heat flux \emph{without} the neoclassical contribution (dashed line). }

In Fig. \ref{transitionpg}(a), the momentum flux is once more plotted against $\fsr$,
but this time keeping the heat flux constant at $Q/Q_{gB}=2.6$. 
The local maximum in the momentum flux still exists. 
Starting at this local maximum (point A), 
if the torque on the plasma was to be increased at constant $Q$, 
the flux of momentum could only increase 
if the flow shear were to jump to a much higher value (point B) 
where the PVG instability would drive turbulent momentum flux.
A bifurcation is manifest.

The inclusion of the neoclassical fluxes
is critical in obtaining this result. 
To illustrate this, \figref{transitionpg} also shows the momentum flux at constant heat flux
\emph{without} the neoclassical contribution to the fluxes, i.e., with $\chi_n=\nu_n=0$. 
If, in such a synthetic ``purely turbulent'' transport case,
the flow shear is increased at constant \(Q\),
the temperature gradient increases to maintain the turbulent heat flux,
and, as shown in \figref{transitionpg}, $\Pi/Q$ increases monotonically with \(\gamma_E\). 
The key property of neoclassical transport
that helps change this behaviour and produce a bifurcation
is that the neoclassical Prandtl number is much smaller than the turbulent Prandtl number. 
This means that as turbulent transport is suppressed,
the system goes through a regime
where the neoclassical contribution to the heat flux is significant,
while the neoclassical contribution to the momentum flux is not.
So as \(\Pi/Q\) is increased at constant \(Q\), the turbulence is suppressed, and the temperature gradient starts to rise, but as this happens more of the heat flux is transported neoclassically, and so the feedback loop breaks down: it is no longer necessary to increase the levels of turbulence to maintain the same \(Q\). 
The same is not true of the momentum: the neoclassical viscosity cannot make up for the lack of turbulence, and so the transport of momentum peaks and then falls.

At high flow shears the turbulent momentum flux increases rapidly once again; 
this turbulent flux is driven by the PVG,
as discussed in Chapter \ref{fluxessec}.
As a result, 
we observe that the bifurcation results in a turbulent state at B,
with a significant turbulent momentum flux. 
This is a substantial difference from the situation envisioned in \cite{waltz1997gyro},
where a transport bifurcation was predicted using a reduced quasilinear model.
Their model did not 
(and, being a quasilinear model, could not)
predict the existence
of the PVG-driven subcritical turbulence at high flow shears;
instead it predicted a bifurcation resulting in a non-turbulent state
where all transport was neoclassical. 
Thus a full nonlinear analysis is necessary
to describe accurately the reduced transport state produced by the bifurcation
that we find in a turbulent plasma.

Restoring dimensions,
we find that the bifurcation results
in a jump in the flow shear
from $0.25\,v_{thi}/R_0$
to $1.17\,v_{thi}/R_0$.
It is instructive to consider whether such values of shear 
appear in current devices.
Taking the measured values of the perpendicular flow shear, 
$T_i$ and $R_0$, in high-performance discharges in the JET and MAST tokamaks \cite{vries2009internal,field2004core},
we estimate that flow shears of up to approximately $v_{thi}/{R_0}$ have been recorded in both devices, and thus that values of shear of the order examined in this chapter have been observed.

\chapter{Parameterised Model}
\label{qbehaviour}
Through simple interpolation, without making any assumptions about the way the fluxes depend on the gradients, we have shown the existence of a bifurcation in the gradients at constant fluxes. 
However, even interpolating one line of constant $Q$ requires a very large number of data points in the region of the line. 
To produce the required data set for \figref{transitionpg}(a),
we had to perform about 350 nonlinear gyrokinetic simulations, each run to saturation.
In order to extend our understanding of the bifurcation,
and of the range of flux values for which similar bifurcations can occur,
we will first consider the behaviour of the turbulent fluxes in further detail
and then use this analysis to construct
a simple parameterised model of the turbulent fluxes $Q_t$ and $\Pi_t$
as functions of $\fsr$ and $\tprim$. 

\section{Modelling $Q_t$}
\label{modellingqtsec}

\subsubsection{Dependence of $Q_t$ on $\tprim$}

\myfig{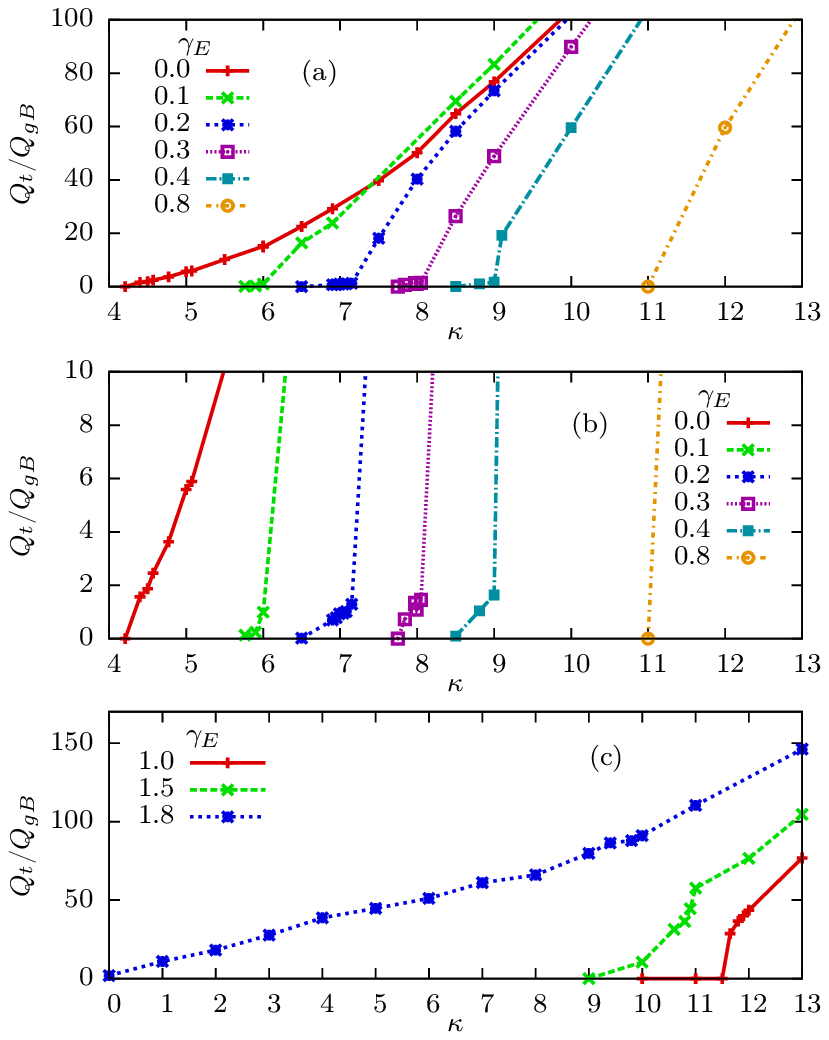}{Dependence of the Heat Flux on the Temperature Gradient}{ Turbulent heat flux vs the temperature gradient for different values of the flow shear, showing (a) low-shear values $\fsr<1\gstwofsrnorm$, (b) a close up of the low $Q_t$ region in (a), and (c) high-shear values $\fsr\geq1\gstwofsrnorm$}

We wish to construct a parameterised model of $Q_t$ as a function of $\fsr$ and $\tprim$.
In Chapter \ref{fluxessec} we described the dependence of $Q_t$ on $\fsr$;
we now consider the dependence of $Q_t$ on $\tprim$. 
Both experimentally \cite{manticastiffness} and numerically \cite{dimits:969},
it is usually found that $Q_t$ 
increases very sharply with $\tprim$ ---
a property known as stiff transport.
A recent experimental study \cite{manticastiffness} has
indicated that increasing the flow shear at low magnetic shear
might have the effect of reducing
the stiffness.
Fig. \ref{qvstprimbehaviour} shows that the effects of flow shear
on $\partial Q_t/\partial \tprim$ are in fact quite complex.
Let us consider the cases of low flow shear, $\fsr<1\gstwofsrnorm$, 
and high flow shear, $\fsr\gtrsim1\gstwofsrnorm$, separately.

For the case of $\fsr<1\gstwofsrnorm$, shown in \figref{qvstprimbehaviour}(a), the threshold in $\tprim$ above which turbulence can be sustained nonlinearly increases rapidly with $\fsr$, as the perpendicular shear suppresses the ITG instability. 
Above this threshold there are broadly three regions: low, intermediate and high $Q_t$.

At high $Q_t$, far above the threshold, 
for any value of flow shear
the heat flux eventually asymptotes to
the same dependence as it has at $\fsr=0$.
In other words, as the ITG drive becomes very strong,
the effect of flow shear becomes negligible.

At intermediate $Q_t$,
the heat flux rises rapidly from low values
to join the universal, high-$Q_t$ asymptotic. 
As $\fsr$ increases, because the threshold rises,
the heat flux rises more rapidly with $\tprim$ above the threshold;
thus at intermediate values of $Q_t$,
the stiffness $dQ_t/d\tprim$ increases with $\fsr$;
this is shown in \figref{stiffnessmodel}. 

At low $Q_t$, just above the threshold,
the heat flux rises very slowly,
that is, the stiffness $dQ_t/d\tprim$ is very low (see \figref{stiffnessmodel}).
This low $Q_t$ region is shown in detail in \figref{qvstprimbehaviour}(b).
Such is the sharpness of the transition
from the low-stiffness low-$Q_t$ region
to the high-stiffness intermediate-$Q_t$ region
that there appear to be two distinct thresholds.
The first threshold is the transition from no turbulence 
to non-zero turbulent transport.
Above the first threshold turbulence is present
but $Q_t$ rises slowly with $\tprim$ (the low-$Q_t$ region).
Above the second threshold $Q_t$ rises rapidly (the intermediate-$Q_t$ region).
These thresholds are plotted in \figref{modelq0}.
The low-stiffness region only exists for $0<\fsr<0.8\gstwofsrnorm$:
between $0.4<\fsr<0.8\gstwofsrnorm$ the first threshold joins the second
and the low-stiffness region disappears. 

At high flow shear, $\fsr\geq1$,
there are two principal differences to the case with low flow shear. 
Firstly, the low-stiffness region
is not present, and there is only one threshold.
Secondly,
the critical temperature gradient for turbulent heat transport starts to decrease
as the PVG drive re-enforces the ITG drive. 
When $\fsr=1.8$, the PVG drive is strong enough to drive turbulence at very low $\tprim$:
the threshold drops to zero.
As the threshold drops,
the transport stiffness at intermediate values of $Q_t$ decreases,
in a mirror image of the case at low flow shear.
At high $Q_t$, the heat flux still asymptotes to the universal $\fsr=0$ curve.

Thus, increasing $\fsr$ can both increase and decrease the nonlinear thresholds, 
and both increase and decrease the stiffness.
The rise and fall both in the thresholds
and the stiffness
can also be seen in
the finite-magnetic-shear simulations of 
\cite{barnes2011turbulent}. 
The principal differences between the zero-magnetic-shear case considered here 
and that case 
are that we have a higher value of the critical gradients for all $\fsr$,
and that we find a low-stiffness region at low values of $Q_t$ ---
a feature that seems to be absent at finite magnetic shear.

\myfig[4.5in]{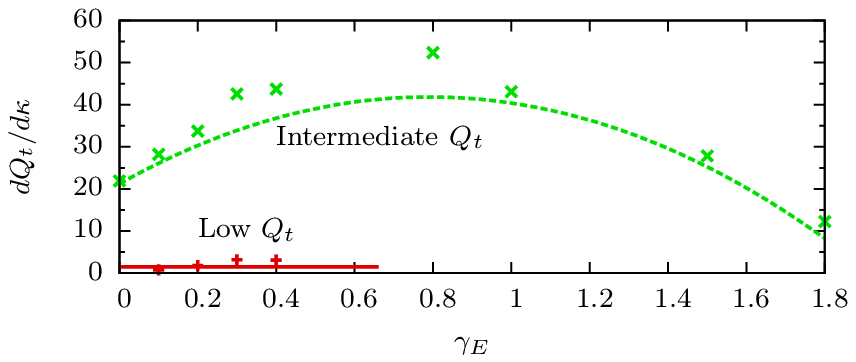}{The Effect of Flow Shear on Transport Stiffness}{The dependence of the heat transport stiffness $dQ/d\kappa$ on the flow shear in the low-$Q_t$ and intermediate-$Q_t$ regions, measured using simulations close to both thresholds (points, see Fig. \ref{qvstprimbehaviour}) and using the parameterised model of Section \ref{qtparameterisationsec} (lines).}
\myfig[4.5in]{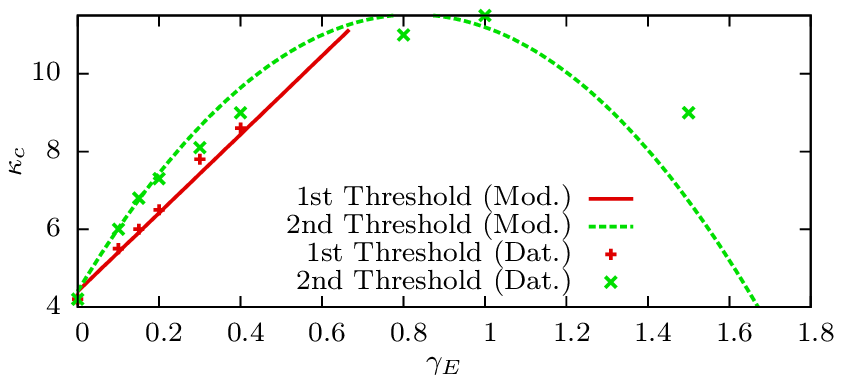}{The Effect of Flow Shear on the Critical Temperature Gradient}{The first and second critical thresholds, $\tprim_{c1}$ and $\tprim_{c2}$, measured using simulations close to both thresholds (points, see Fig. \ref{qvstprimbehaviour}), and using the parameterised model of Section \ref{qtparameterisationsec} (lines).}

\subsubsection{Parameterisation of $Q_t$}
\label{qtparameterisationsec}
In Ref.~\cite{parra2011momentum} a simple model
was used to characterise the behaviour of the turbulent heat flux,
and describes the qualitative behaviour of the heat flux reasonably well. 
Here, we describe a more complex model 
that reproduces most of the features described above
and is \emph{quantitatively} close to the interpolated fluxes.
We consider only
the low and intermediate $Q_t$ regions,
as the bifurcation occurs
near the boundary between these regions.
In order to describe $Q_t$ in these regions,
we have to parameterise the two thresholds and the transport stiffness $dQ_t/d\kappa$.

We parameterise the first and second critical thresholds
as linear and quadratic functions of $\fsr$ respectively:

\begin{align}
&	\tprim_{c1} = \tprim_0 + \alpha_1 \fsr,\\ 
&	\tprim_{c2} = \tprim_0 + \alpha_2 \fsr + \alpha_3 \fsr^2, 
\end{align}
where $\tprim_0=4.4$ is the nonlinear threshold for turbulence at $\fsr=0$ and the parameters $\alpha_1=10.1$, $\alpha_2=17.4$ and $\alpha_3=-17.0$ are chosen to fit the data.
The modelled thresholds are plotted next to the measured thresholds in Fig. \ref{modelq0}.
As with the observed thresholds,
the first threshold joins the second between \(0.4<\fsr<0.8\gstwofsrnorm\). 

Next we parameterise the transport stiffness $dQ_t/d\kappa$.
The measured values of $dQ_t/d\kappa$ in the first and second regions are shown in \figref{stiffnessmodel}.
Between the first and second thresholds (the low $Q_t$ region),
$d(Q_t/Q_{gB})/d\kappa$ is modelled as a constant, $\alpha_4$.
In the intermediate $Q_t$ region, remembering the observation that
$dQ_t/d\kappa$ broadly rises and falls with the nonlinear thresholds, we 
allow the stiffness to depend on $\tprim_{c2}$.
Thus,
\begin{align}
	&\oov{Q_{gB}}\dbd{Q_t}{\tprim} = \begin{cases}
		\alpha_4 & \tprim_{c1} < \tprim < \tprim_{c2},\\
	 \alpha_5 + \alpha_6 \kappa_{c2} & \tprim > \tprim_{c2}.
 \end{cases}
	\label{secondregionstiffness}
\end{align} 
The parameters are $\alpha_4=1.5$, $\alpha_5=3.0$ and $\alpha_6=8.0$.
The model of $dQ_t/d\tprim$ is shown in \figref{stiffnessmodel}.

Thus $Q_t$ is parameterised as a piecewise linear function of $\tprim$.
It is zero below the first threshold,
has gradient $\alpha_4$ above the first threshold
and gradient $\alpha_5 + \alpha_6 \kappa_{c2}$ above the second:
\begin{align}
	\frac{Q_t}{Q_{gB}} =
	\begin{cases}
	0 & \tprim < \tprim_{c1}, \\
	 \alpha_4 \lp \tprim - \tprim_{c1}\rp & \tprim_{c1} < \tprim < \tprim_{c2},\\
	 \alpha_4 \lp \tprim_{c2} - \tprim_{c1} \rp  +\lp\alpha_5 + \alpha_6 \kappa_{c2}\rp  \lp \tprim - \tprim_{c2}\rp 	& \tprim > \tprim_{c2} .
 \end{cases}
 \label{qtmodel}
\end{align}
The model is compared with the data points in Fig. \ref{qvstprimmod}. 
To reproduce the bifurcation of Section \ref{transition} in a quantitatively correct manner,
it is in fact sufficient to represent accurately the two thresholds and the low stiffness region.
However, by also parameterising the variation of $dQ/d\kappa$,
we have provided a model which uses six parameters to describe the (complicated) behaviour of the heat flux over a wider parameter regime with reasonable accuracy.

\myfig[4.5in]{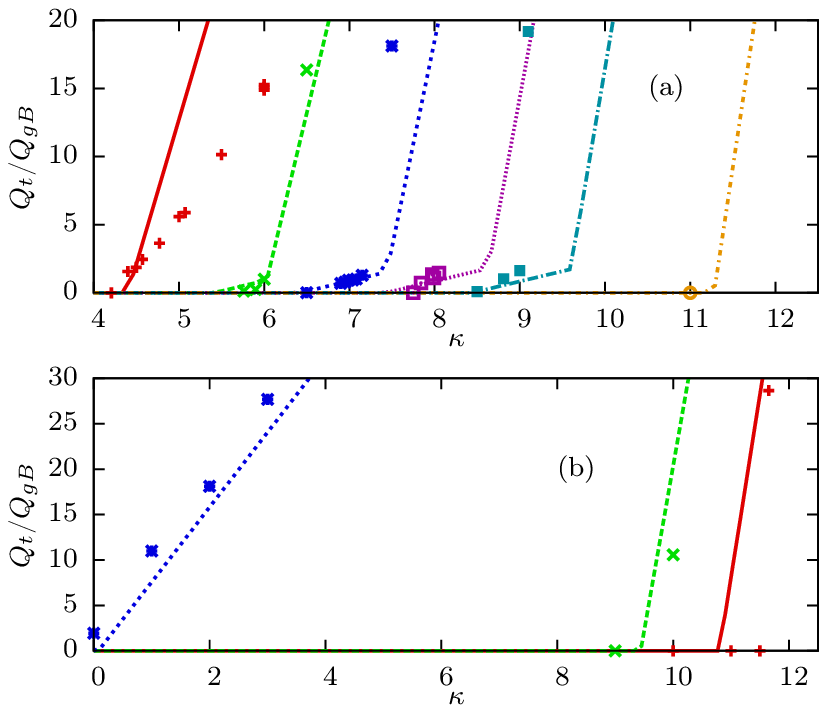}{The Modelled Heat Flux}{The modelled heat flux (lines) shown along with the simulated data points for (a) low flow shear and (b) high flow shear. Legends as in Figures \ref{qvstprimbehaviour}(a) and \ref{qvstprimbehaviour}(c).}

\section{Modelling $\Pi_t$}

While the turbulent Prandtl number does vary with $\fsr$ and $\tprim$,
as discussed in Section \ref{prtsec},
this variation is relatively weak.
Thus we model $\Pi_t$ by assuming a constant turbulent Prandtl number:

\begin{align}
	\label{pitmodel}
	\frac{\Pi_t}{\Pi_{gB}} = \frac{Q_t}{Q_{gB}} \frac{\fsr}{\tprim}\frac{\sqrt{2} R_0 q_0}{r_0}\prt,\qquad
	\prt = 0.55.
\end{align}
The same choice was made in Ref.~\cite{parra2011momentum}.

\section{The Modelled Bifurcation}

In Fig. \ref{transitionpg}(b) the model is used to replot 
the angular momentum flux at constant $Q/Q_{gB}=2.6$.
It has the same shape as the interpolated curve in \figref{transitionpg}(a),
and is quantitatively very close to it
at low flow shear ($\fsr < 1.0\gstwofsrnorm$). 
The agreement at higher flow shears is less good,
which reflects the fact that the second critical gradient is not really a quadratic.
Nonetheless, we consider this degree of precision adequate.
The model is used in Chapter \ref{furtherobservationssec} to explore further properties of the transition,
without the need for extremely large numbers of additional simulations.

\chapter{The Reduced Transport State}
\label{furtherobservationssec}

In Chapter \ref{transitionsec}, a transition was described leading to a reduced transport state. 
Here we use the parametric model that was designed in Chapter \ref{qbehaviour} to describe the properties of this state.

\section{Heat Flux at Constant $\Pi/Q$}
\myfig[3.5in]{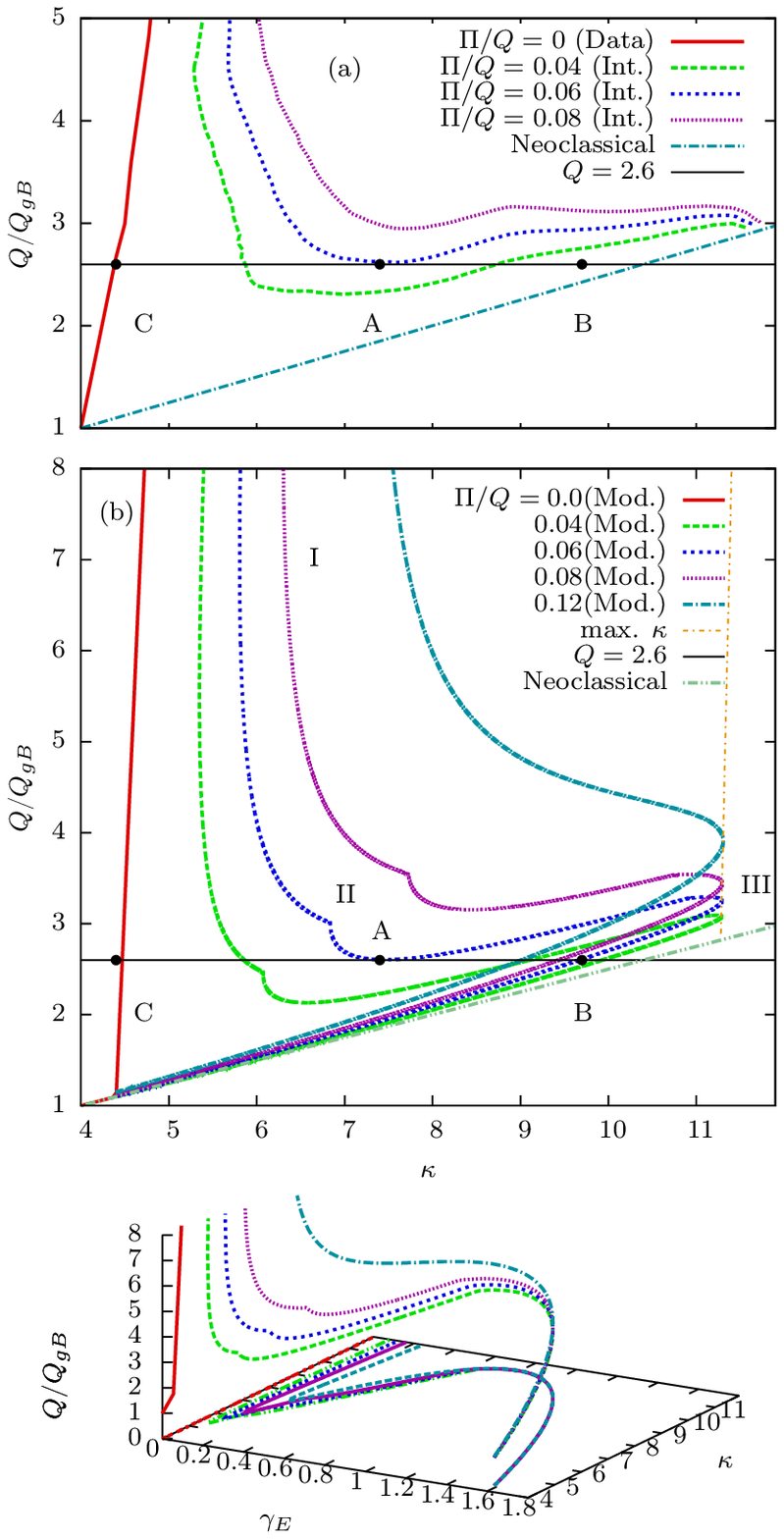}{The Temperature Gradient across the Transition}{Heat flux $Q$ vs the temperature gradient $\tprim$ at a constant ratio of the momentum flux to the heat flux $\Pi/Q$ (in units of $\sqrt{2} R_0 / v_{thi}$) plotted using (a) interpolation from the data (Sec \ref{interpolationsec}) and (b) the parameterised model (Sec \ref{qbehaviour}). Also shown is the neoclassical contribution to the heat flux. Interpolation in (a) is impossible near this neoclassical line, where the contours are closely spaced and $\Pi/Q$ is multivalued. (b) also shows the maximum possible temperature gradient at low $Q$ for a given $\Pi/Q$ (labelled ``max $\tprim$''). Fig. (c) plots the same curves as (b), showing both $Q$ against $\fsr$ and $\tprim$, and the curves projected on the $\fsr$-$\tprim$ plane, illustrating the increase in the flow shear along each curve of constant $\Pi/Q$. Points A and B in (a) correspond to points A and B in Fig. \ref{transitionpg}(a).}

\figref{transitionpg} showed the effect of varying $\Pi/Q$ whilst keeping $Q$ constant. 
In contrast, \figref{transitionqt} demonstrates the effect of varying $Q$ whilst keeping $\Pi/Q$ constant.
At high $Q$ and low temperature gradient (marked (I) in Fig. \ref{transitionqt}(b)),
increasing the heat flux has the effect of increasing the temperature gradient as might be expected:
this is the ordinary turbulent regime.
However, as $ Q/Q_{gB}$ is lowered below $\sim5$
there arises a counterintuitive situation where
decreasing $Q$ has the effect of raising the temperature gradient.
This is of course the combined effect
of the neoclassical Prandtl number
and the flow shear as described in Chapter \ref{transitionsec},
and also discussed in detail in \cite{parra2011momentum}.
In this region (marked (II) in \figref{transitionqt}(b)),
\(Q_t\) becomes comparable to \(Q_n\),
but as the heat flux is lowered,
the system cannot drop to a neoclassical state
because of the smallness of the neoclassical transport of momentum. 
Thus the constancy of the ratio $\Pi/Q$
(i.e., the large finite flux of momentum)
keeps the system in a turbulent state;
the temperature gradient and the flow shear rise,
and the curve representing $Q$ vs $\tprim$
becomes flatter and curls up
in order to maintain its distance from the neoclassical line. 
At large temperature gradients and flow shears
(region (III) in \figref{transitionqt}(b)),
the momentum flux increases rapidly due to the PVG drive
and the high flow gradient (see Fig \ref{fluxes}(b))
and so the curve rolls over and resumes its original downward trend. 
Eventually the heat flux asymptotes to the neoclassical value.

The last part of this trend, while physically obvious,
can only be seen using the modelled heat flux (Fig \ref{transitionqt}(b)),
as it is not feasible to interpolate
in the region near the neoclassical line
where the contours of constant $\Pi/Q$
are very closely spaced and \(\Pi/Q\) becomes multivalued. 

\section{Temperature Gradient Jump}

Interpolation cannot yield the temperature gradient after the transition directly:
as explained above, the contours of constant $\Pi/Q$ become too closely spaced
as they approach the neoclassical line in \figref{transitionqt}(a). 
However, the temperature gradient of the final state,
labelled B on both \figref{transitionpg} and \figref{transitionqt}(a,b),
can be calculated indirectly by using the value of $\fsr$ from \figref{transitionpg}(a)
and rearranging \eqref{prtdef} and \eqref{totalqp}
to give $\tprim$ as a function of $\fsr$, $\prt$, $\Pi/\Pi_{gB}$ and $Q/Q_{gB}$:

\begin{equation}
	\tprim = \frac{\prt \lp \sqrt{2} q_0 R_0 / r_0 \rp \fsr Q/Q_{gB}}
	{\Pi/\Pi_{gB} + \lp 4 R_0^2 q_0/v_{thi} r_0  \rho_i^2 \rp \fsr \lp \chi_n - \nu_n \rp }.
	\label{kappaform}
\end{equation}
Taking $\fsr=1.17$, $\prt=0.55$ yields a temperature gradient at point B of $\tprim=9.7$.
The temperature gradient at point A (the other side of the jump) is 7.4.
Alternatively, using our model,
the values of the temperature gradient can be read from \figref{transitionqt}(b) 
which results in an identical jump of $\tprim=7.4$ to $\tprim= 9.7$.
If we compare with the case of zero momentum input
and zero flow shear (point C on \figref{transitionqt}(a)),
we find that flow shear has enabled a total increase
in the temperature gradient of a factor \(9.7/4.5 = 2.2\).
If this were reproduced across the whole device, it could increase the core temperature by a factor of around 10, but this would require a more detailed 1D model (see \cite{barnes2010shear}), rather than the 0D model presented here.  
Note that while the jump in temperature gradient
caused by the bifurcation
(the transition from A to B)
is a striking feature,
a comparable contribution
to the increase of the temperature gradient 
arises from the incremental suppression of the turbulent transport
by the velocity shear (the difference between A and C).

\section{Neoclassical Heat Flux, Turbulent Momentum Flux}

It was noted earlier that the reduced transport state produced by the bifurcation is still turbulent,
with a momentum flux far greater than its neoclassical value (\figref{transitionpg}).
In \figref{transitionqt}(b), it can also be seen
that the reduced transport state does not lie exactly on the neoclassical line;
it does however lie very close to it,
which implies that the turbulent heat flux is much smaller than neoclassical.
Thus, the bifurcation takes the system into a state where
\emph{the heat is mostly transported neoclassically,
but the angular momentum is mostly transported by turbulence}.
The small ratio of $Q_t$ to $Q_n$
reflects the fact that the turbulence levels
in the reduced transport state are small.
The dominance of $\Pi_t$ over $\Pi_n$
given the low levels of turbulence
reflects the fact 
that the flow gradient is large
and the neoclassical momentum transport is very low
compared to the neoclassical heat transport
($\mathrm{Pr}_n \ll 1$).

\section{Bifurcation by lowering $Q/Q_{gB}$}

Starting from point A, if \(Q\) were to be reduced at constant \(\Pi/Q\),
the system would again be forced to jump to point B. 
In effect, what would happen is that the heat flux would become principally neoclassical;
the momentum flux would drop because $\mathrm{Pr}_n \ll \prt$, 
and a bifurcation would ensue in the manner described in Section \ref{transition}.
Thus, a decrease in the input heat could lead to a higher temperature gradient. 
We note, however, that \(Q\) is normalised to the gyro-Bohm value: \(Q_{gB} = \gstwoQnormflatbd\). 
Thus, decreasing \(Q/Q_{gB}\) could correspond to decreasing the deposition of heat,
but it could also correspond to an increase in temperature or density \cite{barnes2010shear}.

\section{Optimum $Q$}
\label{optimumqsec}
\figref{transitionqt}(b) shows that for every value of \(\Pi/Q\),
there is an optimum \(Q\) that gives a maximum temperature gradient
(the dotted line at $\tprim\sim11$ in Fig. \ref{transitionqt}(b)). 
The maximum temperature gradient increases with \(\Pi/Q\), but only slowly.  
The optimum value is the value of $Q$ such that:

\begin{equation}
	\left. \dbd{\kappa}{Q} \right|_{\Pi/Q} = 0
	\label{optqeq}
\end{equation}
and is plotted as a function of $\Pi/Q$ in \figref{transitionregion2}.
The maximum temperature gradient is effectively determined by the
maximum nonlinear critical temperature gradient $\kappa_c$ (see \figref{modelq0}). In Chapter 
\ref{sec:zeroturbulencemanifold} we will show how  this maximum of $\kappa_c$ 
can be dramatically increased.

\section{Transition Region}

Finally, we will show that there is in fact only a limited range  of both \(Q\) and \(\Pi/Q\)
where bifurcations can happen in the way described above.
To illustrate,  we observe that no transition can occur along the contour
\(\Pi/Q=0.12 \sqrt{2} R_0 / v_{thi}\) in \figref{transitionqt}(b):
if $\Pi/Q$ is kept constant at this value,
decreasing \(Q/Q_{gB}\) from greater than its optimum value of $4.0$
increases the temperature gradient smoothly up to its  maximum value. 
The existence of a bounded region where such transitions can happen
is studied in detail by the authors of Ref.~\cite{parra2011momentum},
who calculate, using a simple transport model,
the region in which transitions occur,
and derive a criterion necessary for their existence. 
We apply the analysis of Ref.~\cite{parra2011momentum}
using the model for the turbulent fluxes given in \eqref{qtmodel} and \eqref{pitmodel},
to determine the range of values of $Q$ and $\Pi/Q$
for which bifurcations of the form we have described can occur.
In essence, bifurcations can only occur
when there exist multiple values of $\fsr$ and $\tprim$
that give rise to the same values of $Q$ and $\Pi/Q$. 
From \figref{transitionqt}(b),
it is clear that this is only possible for values of $\Pi/Q$
where there is a minimum in the curve of constant $\Pi/Q$,
and for values of $Q$ which lie between this minimum and 
either a maximum in the curve or 
the point where the curve intercepts the neoclassical line.
Thus the region in which transitions are possible is bounded by the curve

\begin{equation}
	\left. \pd{Q}{\tprim} \right|_{\Pi/Q} = 0,
	\label{transitioncondition}
\end{equation}
and by the neoclassical transport curve.
This region is plotted in \figref{transitionregion2}. 

\begin{table}
\caption{Typical plasma properties from the ITER Profile Database \cite{roach2008profiledb}. The symbol $a$ denotes the minor radius of the device. The temperature was calculated as the mean of the core (TI0) and edge (TI95) temperatures.} 
\begin{center}
	\begin{tabular}{ l  r   r  r  r  r  r  r   }
		    \hline
		    \hline
				\textbf		& $T_i$ ($eV$) & $n_i$ ($m^{-3}$) & $R$ ($m$) & $B_T$ ($T$) & $a$ ($m$)   \\
			 \hline
DIII-D & 2.7e+03 & 9.0e+19& 1.7e+00& 1.7e+00& 6.1e-01 \\
JET & 2.6e+03& 6.3e+19& 2.9e+00& 2.4e+00& 9.3e-01\\
MAST & 0.6e+03& 3.9e+19& 8.0e-01& 5.3e-01& 5.6e-01\\
			 \hline
			 \hline
	\end{tabular}
\end{center}
\label{typicalvals}
\end{table}
In order to give a clearer meaning to this diagram,
we use equations (\ref{qexp} - \ref{pioqexp})  
and typical  properties of plasma devices
taken from the ITER Profile Database \cite{roach2008profiledb}
(and listed in Table \ref{typicalvals})
to replot the region in which transitions can happen
in terms of the input beam power and
beam particle energy,
$P_{\mathrm{NBI}}$ and $E_{\mathrm{NBI}}$.
To generate these plots,
we also calculate the collision frequency $\nu_{ii}$
self-consistently using \eqref{nuiidef}, 
to take account of the varying strength
of the neoclassical transport in each device.
The transition regions for each device
are displayed in \figref{pnbienbiregion},
along with a point indicating the 
typical values of
$P_{\mathrm{NBI}}$ and $E_{\mathrm{NBI}}$
in each device.
\figref{pnbienbiregion} shows that
in which transitions can happen 
lie within an order of magnitude 
of the typical values.
It should be stressed
that with the simplified magnetic geometry (Chapter \ref{sec:bifurcationintro}) and
model of neoclassical transport (Section \ref{neoclassicaltransportsec}) used in this study,
and with the many assumptions about the 
way the energy and momentum are deposited
in the plasma (Section \ref{invertingsec}),
closer agreement could not be expected.
In particular, the assumptions of Section \ref{invertingsec}
are likely to overestimate the applied torque
and hence overestimate the beam energy 
needed for a transition.

\myfig[4.8in]{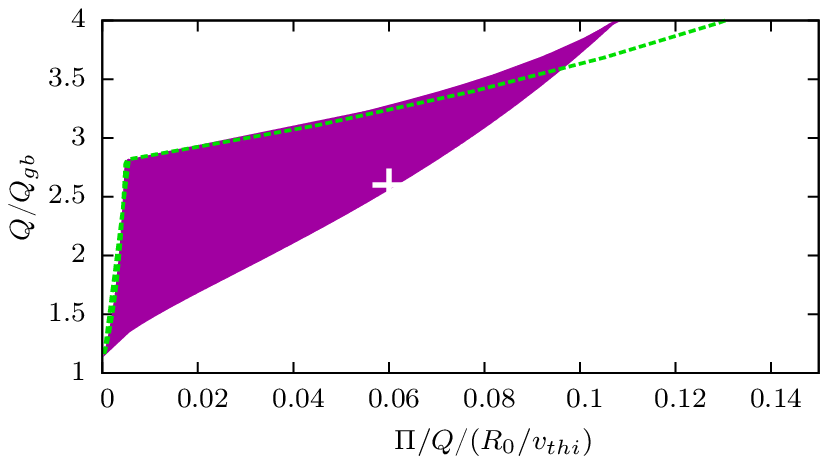}{The Transition Region and Optimum Normalized Heat Flux}{The region in which bifurcations can occur, calculated using the analysis of Ref.~\cite{parra2011momentum} and the parameterisation of the fluxes described in Chapter \ref{qbehaviour}. The cross indicates the location of the bifurcation described in Section \ref{transition}. The dashed line represents the optimum value of $Q$ for a given value of $\Pi/Q$, Eq. \eqref{optqeq}.}
\myfig[4.8in]{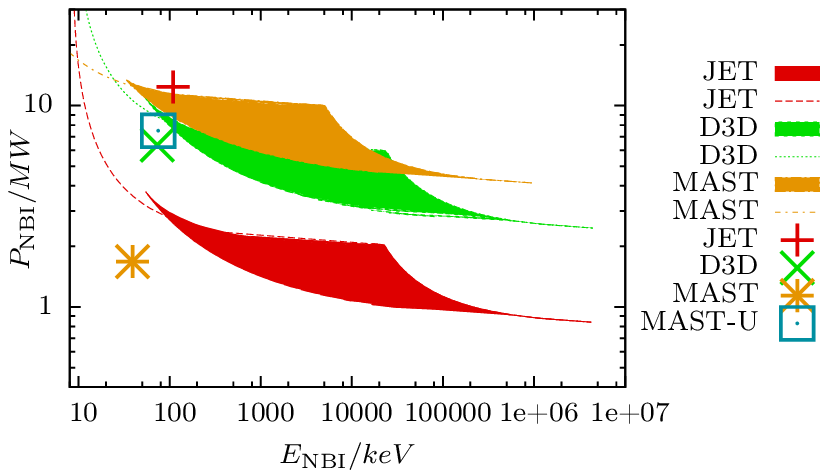}{The Transition Region and Optimum Beam Power for Real Devices}{The regions (shaded)
in which transitions can happen,
as a function of the beam power and the beam particle energy,
for three different devices.
The dashed lines indicate,
for each device,
the value of 
$P_{\mathrm{NBI}}$ for a given $E_{\mathrm{NBI}}$
which would lead to the optimum temperature gradient,
as described in Section \ref{optimumqsec}.
The points indicate typical values of 
$P_{\mathrm{NBI}}$ and $E_{\mathrm{NBI}}$
for each device.
The projected $P_{\mathrm{NBI}}$ and $E_{\mathrm{NBI}}$
for MAST Upgrade were taken from Ref.~\cite{ccfewebsite}.
}

\section{Summary}

 We have demonstrated the existence of a transport bifurcation to a high temperature gradient, high flow gradient state, and estimated the values of heat and momentum input at which we might expect such a transition to occur.
 In the reduced transport state the heat flux is nearly neoclassical but the momentum flux is large owing to the existence of turbulence driven by the PVG.
 It is this turbulence was effectively limits the maximum temperature gradient that can be reached in the bifurcation: owing to the stiffness of turbulent transport, this maximum gradient ($\kappa=9.7 $) is, in fact, close to the maximum nonlinear critical temperature gradient for turbulence ($\kappa_c\sim 11 $, see \figref{modelq0}).\footnote{The difference between the two is owing to the neoclassical transport.}
 In Part \ref{sec:subcriticalturbulence} we consider this PVG-driven turbulence in detail, and show that the maximum temperature gradient that can be reached without destabilising it can be greatly increased.

%\mypart{Subcritical Turbulence}
%\mypart{Turbulence and Flow Shear}
\mypart{Subcritical Turbulence}
\label{sec:subcriticalturbulence}

\chapter{Turbulence without Instability}
\label{sec:turbulencewithoutinstability}

%Before studying the implications
%of the results presented above, 
%we first discuss the nature of the fluctuations
%that give rise to the behaviour of
%the turbulent fluxes discussed in Section \ref{fluxessec}.

%\section{Transient Growth}

At zero flow shear, turbulence driven by the ion temperature gradient
results from the presence of unstable eigenmodes in the system.
The instability that is present in the system means that
arbitrarily small perturbations to the electric field
and the particle distribution function can grow to the
size needed for them to interact and form turbulence.
As was discussed in Section \ref{sec:shearedflows}, a recent paper \cite{barnes2011turbulent} which investigated
the effects of flow shear on turbulence driven by the ion temperature gradient
--- at a finite value of magnetic shear ($\hat{s}=0.8$) --- 
showed that at low non-zero values of the flow shear, unstable eigenmodes were present in the system considered
(see also the earlier results in slab geometry of
Artun. et. al. \cite{artun1993integral}),
but that for values of $\gamma_E$ higher than approximately 0.8, no such eigenmodes existed.
%This phenomenon was explained by Newton et. al. \cite{newton2010understanding} as follows.

In contrast to the results of Ref.~\cite{barnes2011turbulent}, we find in the case of zero magnetic shear that the growth rate is zero for \emph{all} non-zero values of $\gamma_E$.
With \(\fsr=0\), the ITG instability exhibits robust linear growth, 
with a threshold of \(\kappa=4.4\) (Fig. \ref{gexbzero}(a)).
However, for any finite value of the flow shear 
%there are no growing linear eigenmodes in the system:
perturbations grow only transiently before decaying (Fig \ref{transientgrowth}(a)).  While formally this means that $\fsr\rightarrow0$ is a singular limit,
there is in fact no physical discontinuity in behavior:
as $\fsr \rightarrow0$, the duration of the transient growth,
$\tau_{\gamma}$, tends to infinity (roughly as $1/\fsr$; see Fig. \ref{transientgrowth}(b)). 
Thus, there is a continuous transition at $\fsr=0$
from a transient mode that grows for 
an infinitely large time, to a growing eigenmode.
%The lack of growing eigenmodes when $\fsr>0$ stems from the wavenumber drift \eqref{radialshear}: 
%as time tends to infinity at finite flow shear and zero magnetic shear, 
%the radial wavenumbers increase inexorably through the shearing of the modes, 
%until they are extinguished by finite-Larmor-radius effects. 

The differences between the two cases may be
explained as follows. 
At finite magnetic shear,
growing eigenmodes are possible 
because while the perpendicular flow shear leads to an effective dependence of the radial wavenumbers on time (see equation \eqref{eqn:timevaryingkx}), 
the magnetic shear makes them dependent also on the position along the field line.
Therefore, for modes moving with a speed proportional to $\fsr/\hat{s}$, 
the magnetic shear cancels the effect of the velocity shear on $k_x\lp t\rp$ 
(as was also noted in Refs.~\cite{waltz1994toroidal,newton2010understanding}). 
When the magnetic shear becomes very small, or the flow shear very large,
the required mode velocity becomes much greater than the thermal speed of the ions 
and the cancellation is no longer possible 
because the mode cannot travel so fast \cite{newton2010understanding}.
In this situation, the radial wavenumber increases inexorably until the perturbations are damped, either by collisional dissipation in the fluid-like model of Ref.~\cite{newton2010understanding} or by finite-Larmor-radius effects in the gyrokinetic model (also used in this work) of Ref.~\cite{barnes2011turbulent}.
As a consequence, growing linear eigenmodes only exist for non-zero magnetic shear
and only up to a finite value of the flow shear \cite{barnes2011turbulent}.

\myfig[4.5in]{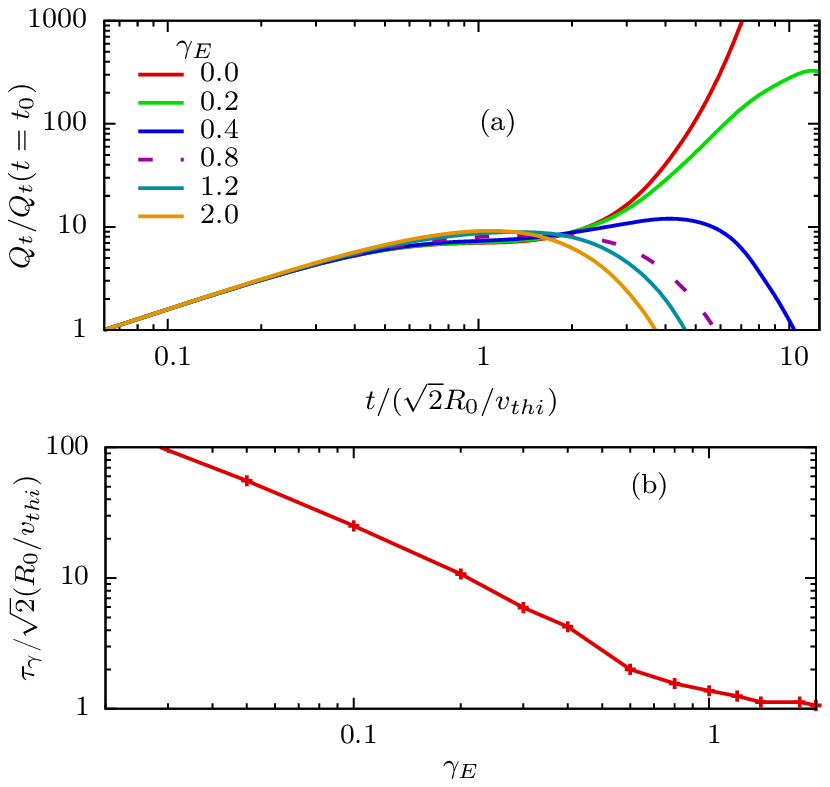}{Transient Growth of Linear Modes}{Transiently growing linear modes at $\tprim=11$.  (a) The heat flux as a function of time, normalised to its value at $t_0$, where $t_0=0.06\sqrt{2}R_0/v_{thi}$, so chosen to skip the short initial transient associated with a particular choice of initial condition.  All modes grow transiently for $\fsr>0$. (b) Duration of transient growth $\tau_{\gamma}$ from $t=t_0$ to the peak value of $Q_t$ vs flow shear.}
%\myfig{transientstrength}{Measures of the Strength of Transient Linear Growth}{Two measures of the strength of the transiently growing linear modes at $\kappa=11$: (a) the effective growth rate of the mode and (b) the number of exponentiations of the heat flux during the growing phase, both vs. the flow shear. 
%The apparent discontinuity at $\fsr=0$ in (a) is a result of taking the average growth rate of the heat flux 
%over $0 < t < \tau_{\gamma}$ rather than the initial or peak growth rate; 
%for all $\fsr>0$ the average will include a period where the growth rate tends to zero.
%}

%This situation differs somewhat from the case with finite magnetic shear $\hat{s}$ 
%\cite{barnes2011turbulent,artun1993integral}, 
%where there are linearly growing modes for a finite range of \(\fsr\). 

%\section{Subcritical Turbulence}
%
%\subsection{General Observations}

In a standard picture of ITG turbulence with $\fsr=0\gstwofsrnorm$ \cite{dimits:969},
the turbulence is driven by unstable linear modes.
With the exception of a narrow interval of temperature gradients
where self-generated zonal flows suppress the turbulence (the Dimits shift),
the presence and the amplitude of the turbulence
are largely determined by the presence and the growth rate of those unstable modes.
In the present case, by contrast, we have strong levels of turbulence sustained nonlinearly 
in a parameter regime where there are no linearly unstable modes, 
a phenomenon first discovered in tokamak turbulence 
in Refs.~\cite{barnes2011turbulent} and \cite{highcock2010transport}, 
but well known in various hydrodynamical contexts \cite{trefethen1993hydrodynamic,kerswell2005recent}. 
This phenomenon is known as \emph{subcritical turbulence}.

Subcritical turbulence differs from standard instability-driven turbulence in several important ways. 
Firstly, because there is no linear instability, 
turbulence will not grow from initial perturbations of arbitrarily small amplitude 
\cite{barnes2011turbulent}. 
Fluctuations must be initialised with sufficient amplitude (generally of the order of their amplitude in the saturated state)
in order for turbulence to be sustained; thus, the absence of sustained turbulence in a numerical experiment may merely indicate that the initial amplitude is insufficient, 
not that the plasma is quiescent. 

The second difference concerns the scales at which
the turbulent energy resides.
In ITG-driven supercritical turbulence ($\fsr=0$),
these scales are those where the linear drive injects energy ---
this tends to correspond to $k_{\pa} q_0 R_0 \sim 1$
and $k_y \rho_i$ relatively narrowly concentrated
around values of a fraction of unity
(at least for low values of $q_0$ and moderate temperature gradients;
see Ref.~\cite{barnes2011critically} and \figref{subcriticalspectrum}(a)).
In subcritical turbulence,
the preferred wavelengths appear to be
those at which the amplification of the transient modes is strongest.
In the case of turbulence where the PVG drive is dominant, 
we will show in Chapter \ref{sec:subcriticalfluctuations}  that the amplification 
is maximised along a line in Fourier space where 

\begin{equation}
	k_y \rho_i \simeq  \lp \frac{r_0}{R_0q_0} \rp^{1/3}\frac{k\pa R_0}{\sqrt{2}\fsr} \gstwofsrnorm.
	\label{maxamp}
\end{equation}
Fig. \ref{subcriticalspectrum}(b) shows the spectrum 
of strongly PVG-driven turbulence at high flow shear ($\fsr=2.2\gstwofsrnorm$).
We see that although the result \eqref{maxamp} 
is based on linear theory and
is derived for slab geometry,
it appears to describe the peak of the spectrum reasonably well. 
The spectrum extends to much higher 
parallel wavenumbers than in the standard ITG case; 
consequently, higher parallel resolution is necessary to resolve it.\footnote{Studies of the effect of increasing the parallel resolution showed that at values of flow shear $\fsr\gtrsim 1.2$, the parallel resolution used for the parameter scan described in
Chapter \ref{fluxessec} (14 gridpoints)
was insufficient and led to errors in the critical temperature gradient (above which subcritical turbulence could be sustained) of order 5-10\%. 
We do not consider this error
significant enough to merit a repeat of the parameter scan
with substantially higher parallel resolution
(although in principle such an exercise would be useful).}

\myfig{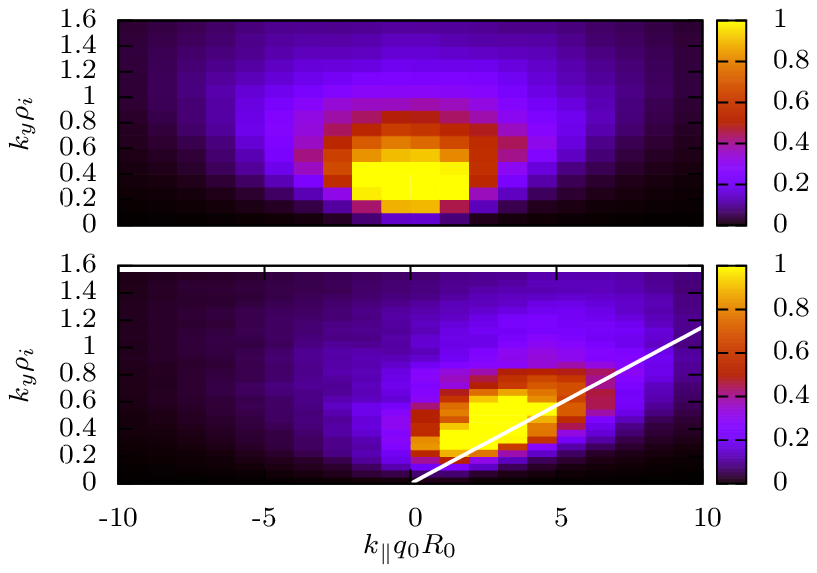}{The Spectrum of Subcritical Turbulence}{The spectrum of turbulent fluctuations for normal and subcritical turbulence: $k_y^2 \sum_{k_{x0}} \md{\hat{\varphi}_{k_{x0},k_y}}^2 $ (arbitrary units) vs $k_y$ and $k\pa$ for (a) $\fsr=0.0,\,\tprim=12$ (ITG turbulence with no flow shear) and (b) $\fsr=2.2,\,\tprim=12$ (subcritical, strong PVG-driven turbulence)
\footnote{It should be noted that as the sign of $k_y$ is opposite in {\texttt GS2} to the present work, these graphs were plotted via a parity transformation.}. Also shown in (b) is the line of maximum transient amplification \eqref{maxamp}.} 

Finally, faced with subcritical turbulence,
we are left without an intuitive way of estimating its saturation level
and, consequently, the turbulent fluxes. 
It is the presence of linear eigenmodes with a defined growth rate (positive or negative)
which has enabled the quasilinear modelling of the heat flux 
in many situations
without resorting to full nonlinear simulations
(see Ref. \cite{krommes2002fundamental} for an overview and Refs. \cite{bourdelle2007new,waltz2009gyrokinetic} for recent work).
When the growth of all modes is transient, 
such models will not work in their current form.
The question arises as to which characteristics
of the linear transient growth
are relevant in the resulting nonlinear state.

%\subsection{A Detailed Study of Transient Growth}
In order to answer this question
and, more importantly, the question of whether it is possible, without using nonlinear simulations,
to predict whether subcritical turbulence will be present for a given set of parameters,
we present in Chapter \ref{sec:subcriticalfluctuations}  a detailed
study --- published as Ref.~\cite{schekochihin2010subcritical} ---
of the transient growth of the fluctuations.
To make analytic solutions obtainable we work in a simplified slab geometry with straight magnetic field lines and periodic parallel boundary conditions which nevertheless contains the key physical effects of the ITG, the perpendicular flow shear and the PVG.
In such a system we may Fourier transform the fluctuations in the parallel direction so that each mode has a parallel wavenumber $k_{\parallel}$. 
We show that, defining the origin of time such that $k_x^{'}=0$  at $t=0$,\footnote{
The time varying radial wavenumber $k_x^{'}$ is defined in equation \eqref{eqn:kxeuleriandef}.
} the (now time-dependent) growth rate of the fluctuations is equal at $t=0$ to the growth rate of an unsheared mode with the same wavenumber.
%with $k_x=0$ and identical values of $k_y$ and the parallel wavenumber $k_{\parallel}$. 
As time increases, the growth rate of the fluctuations eventually reaches zero, at a time $t_0$ which can we calculate in certain limits. 
For times greater than $t_0$ the mode decays.

 To characterise the vigour of this transient growth we define the amplification exponent 
 \begin{equation}
	 N= \int_{0}^{t_0}\gamma(t) dt
	 \label{eqn:Ndef}
 \end{equation} 
 such that $e^N$ is the ratio between the size of the perturbed electric field at the time when it starts to decay and its size at time zero.
 The quantity $\gamma(t) $ is the time-dependent growth rate.
 We then define $N_{\mathrm{max}} $, the maximal amplification exponent, to be $N $ maximised across all $k_y $ and $k_{\parallel} $.
 We find that in general $N_{\mathrm{max}} $ depends on three parameters, the ratio between the parallel and perpendicular flow shears, $q/\epsilon $, the ratio between the ion temperature gradient and the perpendicular flow shear, $\eta_s $, and the ratio between the ion temperature gradient and the ion density gradient $\eta_i $ (although we do not investigate in great detail the effects of changing $\eta_i $).
 We find that with $\eta_s=0 $, $N_{\mathrm{max}}\sim 1 $ when $q/\epsilon\sim 7 $ and that $N_{\mathrm{max}} $ increases monotonically with $q/\epsilon $.
 We find that as $\eta_s\rightarrow 0 $, $N_{\mathrm{max}} $ tends to a constant determined by $q/\epsilon $, and that as $\eta_s $ becomes large, $N_{\mathrm{max}} $ increases monotonically with it.

 %Regarding subcritical turbulence, two clear predictions can be made from the results of Chapter \ref{sec:subcriticalfluctuations}.
 Regarding subcritical turbulence, a prediction can be made from the results of Chapter \ref{sec:subcriticalfluctuations}.
 If we assume that in order for subcritical turbulence to be sustained nonlinearly, fluctuations must grow somewhat before they decay in order to provide energy to be scattered into still-growing fluctuations, and if we assume that therefore $N_{\mathrm{max}}\gtrsim N_c $ for subcritical turbulence to be sustained, where $N_c$, the critical amplification exponent, is a number of order unity, the results predict that the PVG instability alone cannot drive turbulence for $q/\epsilon $ below a certain value determined by $N_c$ (for example, if $N_c\sim1$, the PVG drive would be ineffective for $q/\epsilon\lesssim 7$).
 %the predictions are as follows:
 %firstly, the results predict that the PVG cannot drive turbulence alone for $q/\epsilon\lesssim 7 $;
 %secondly, the results predict that when $q/\epsilon\lesssim 7 $, for any given temperature gradient $\kappa$ (provided the flow shear $\gamma_E $ is increased to a value such that $\eta_s $ is small) we may always obtain $N_{\mathrm{max}} \lesssim 1 $.
 %Thus, if we cast the problem of improving confinement as that of maximising the ion temperature gradient without igniting turbulence (i.e., maximising the critical temperature gradient), the optimal strategy is to minimise $q/\epsilon $ (subject to the constraints of the device) and then increase the perpendicular flow shear (i.e., increase the toroidal flow shear) to the maximum value possible.
%
%\begin{quote}
%!!!!!! The next paragraph is subject to change. !!!!!!
%\end{quote}

 %In fact, as we show in Chapter \ref{sec:zeroturbulencemanifold}, the optimal strategy is slightly different.
 In Chapter \ref{sec:zeroturbulencemanifold}, inspired by this prediction, we use full nonlinear simulations to map out, as a function of the three key parameters, $\gamma_E $, $q/\epsilon $ and $\kappa $, the surface within that parameter space which divides the regions where subcritical turbulence can and cannot be sustained: the zero-turbulence manifold.
 As in Part \ref{sec:bifurcationtoareducedtransportstate},
we take the Cyclone Base Case parameter regime \cite{dimits:969},
i.e., concentric circular flux surfaces with $\epsilon = 0.18 $,
inverse ion density scale length
 $R / L_{n}=2.2 $
 and ion to electron temperature ratio $T_i/T_e=1$ 
and magnetic shear $\hat{s}=0$.\footnote{
 A temperature ratio of 1 is appropriate for both lower power and future reactor-like conditions, but not for high-performance shots in current devices \cite{mantica2011ion,team1999alpha,petty1999dependence}.
 } 
The ratio $q/\epsilon $ is varied by varying $q $ alone. The resolution of all simulations in Chapter \ref{sec:zeroturbulencemanifold} was \(128\times128\times40\times28\times8\) (poloidal, radial, parallel, pitch angle, energy). 
Note that relatively high parallel resolution was used to resolve the PVG modes, for the reasons given above.

We cannot conclusively determine that there is a value of $q/\epsilon$ below which the PVG is unable to drive subcritical turbulence, but we are able to conclude that for $q/\epsilon\lesssim 7$, it is stable for all experimentally relevant values of the toroidal flow shear. 
 We find that, as a consequence of this,
 %, as predicted, the optimal strategy is to minimise $q/\epsilon $, but that contrary to prediction, for every value of $q/\epsilon $ and there is an optimum value of $\gamma_E $ which gives a maximum critical temperature gradient:
 %it is not possible to achieve an arbitrary temperature gradient by sufficiently increasing $\gamma_E $.
 %Nevertheless,
 at low $q/\epsilon $ the maximum critical temperature can be very high: exceeding 20 for values of $q/\epsilon \lesssim 6 $, 
that is, comparable to or higher than those experimentally observed for internal transport barriers \cite{vries2009internal,field2011plasma} in advanced high power operating regimes.

We also discover that the zero-turbulence manifold can be parameterised as $N_{\mathrm{max}}(\gamma_E,q/\epsilon,\kappa)=N_c$, 
where $N_{\mathrm{max}}$ is calculated from linear theory, but that $N_c$ is a function of $\gamma_E$. We are unable to explain this dependence.
Nonetheless, the parameterisation would allow the full zero-turbulence manifold to be calculated using linear simulations once $N_c(\gamma_E)$ had been determined using a set of nonlinear simulations at a single value of $q/\epsilon$: reducing the number of parameters that would be needed in the nonlinear calculation by one. 
 %If such a temperature gradient could be sustained across an entire device, a device only $<++> $ across would be sufficient for fusion.

\newcommand{\apprefeq}[1]{(equation (#1) of Ref.~\cite{schekochihin2010subcritical})}
\newcommand{\apprefeqnb}[1]{equation (#1) of Ref.~\cite{schekochihin2010subcritical}}
\newcommand{\appreffig}[1]{Fig. #1 of Ref.~\cite{schekochihin2010subcritical}}
\chapter{Subcritical Fluctuations}
\label{sec:subcriticalfluctuations}
\begin{quote}
	\emph{
	This chapter is largely taken from Ref.~\cite{schekochihin2010subcritical}, to which the author of this work was a major contributor.}
\end{quote}

\section{Introduction}

Where the magnetic shear is zero and the flow shear is finite, a fluctuation may only grow transiently before decaying. Nevertheless, it is possible, across a wide range of parameters, for this transient growth to be vigourous enough to allow subcritical turbulence to be sustained via nonlinear interactions. It is desirable, therefore, to be able to characterise the strength and duration of this transient growth, and to see whether it is possible to predict from that strength and duration whether or not turbulence could be sustained given a particular set of plasma parameters.

\section{The Flying Slab}
We will examine this transient growth using simplified slab geometry.
This allows progress to be made analytically in certain limits while still including the main physical features of the ITG drive, the PVG drive and the perpendicular shear suppression.
We will check and supplement these analytical results using the code  \texttt{AstroGK}  , which is a faster, simplified version of \texttt{GS2} designed for solving problems in slab geometry. The gyrokinetic equation in slab geometry can be simply obtained from the full axisymmetric equation by taking circular geometry and letting $R\rightarrow\infty $, $r\rightarrow\infty $, $r/R \rightarrow 0 $ (which eliminates particle trapping) and $R\omega_0\rightarrow u $, a constant. We further set $\alpha_s=0 $.

We are, in effect, looking at a finite piece of an infinite torus; within that finite piece, all field lines are locally flat. The only geometric parameter which remains is $\hat{s} $, which we set to 0. We note that the gyrokinetic equation \eqref{eqn:gyrokineticeqgs2coordinates} is in the frame rotating with velocity $\omega_0 $; our slab is therefore travelling at a finite speed $u $ around the infinite torus, and so we refer to our model as the flying slab approximation.

\subsection{ The Gyrokinetic Equation }

Starting from equation \eqref{eqn:gyrokineticeqgs2coordinates}, taking the limits given above, and defining 
\begin{equation}
	\frac{\partial }{\partial z} \equiv k_p \frac{\partial }{\partial \theta},
	\label{eqn:zdef}
\end{equation}
\begin{equation}
	x_{\mathrm{slab}}\equiv\frac{B}{\btor}x,
	\label{eqn:xslabdef}
\end{equation}
and
\begin{equation}
	S \equiv \frac{\bpol}{B}\frac{\partial R\omega}{\partial r} = \frac{\btor}{B}\gamma_E,
	\label{eqn:Sdef}
\end{equation}
we obtain the flying slab gyrokinetic equation (which we will solve for ions only):

\begin{multline}
	\left( \frac{\partial }{\partial t} + x_{\mathrm{slab}} S \frac{\partial }{\partial y} \right)h_i
	+ v_{\parallel}  \frac{\partial h_i}{\partial z}
	 + \mathcal{N}_i - \gyroR{C\left[ h_i \right]} = 	\\
	 \left( \frac{\partial }{\partial t} + x_{\mathrm{slab}} S \frac{\partial }{\partial y} \right)\gyroR{\tphi} \\
	+ \left\{ \left[ \frac{1}{L_{ni}}+\left( \frac{\varepsilon_i}{T_i} -\frac{3}{2} \right)\frac{1}{L_{Ti}} \right]  
	- \ \frac{2 v_{\parallel}}{v_{thi}^2} \frac{q}{\epsilon}S\right\}F_i \frac{\rho_i v_{thi}}{2} \frac{\partial \gyroR{\tphi} }{\partial y},
	\label{eq:gk_sh}
\end{multline}
where $\tphi\equiv Z_i e \varphi/T_i $ is the normalised potential.\footnote{
 It should be noted that in this chapter we are dealing with a much simpler system than elsewhere in this work and hence choose a different system of normalisations.
 } Once again we take the electrostatic limit with a  Boltzmann electron response, so that the system is closed by the quasi-neutrality condition: 
 \begin{equation}
	 \left( 1 + \frac{\tau}{Z_i} \right)\tphi = \frac{1}{n_i}\int d^2 \vct{v} \gyror{h},
	 \label{eq:quasineut_sh}
 \end{equation}
where $\tau \equiv T_i/T_e$.
For the remainder of this chapter we suppress the species indices and the $\mathrm{slab} $ label.

\subsection{Case of non-zero magnetic shear}

As a digression, we note that including a (locally) constant linear magnetic shear into the problem amounts to 
replacing in \eqref{eq:gk_sh}
\beq
Sx\frac{\dd h}{\dd y}\to
\lt(S + \frac{\wpar}{\LB}\rt)x\frac{\dd h}{\dd y},
\eeq
where $\LB$ is the scale length associated with the magnetic shear. 
This appears to introduce complications as we now have an ``effective shear'' 
that depends on the particle velocity $\wpar$. However, since the 
size of $\wpar$ is constrained by the Maxwellian equilibrium distribution, 
this term can be neglected provided $M_s=S\LB/\vth\gg1$. Under this assumption, 
the theory developed below applies without modification. We note
that the ``shear Mach number'' $M_s$ is precisely the parameter 
that is known to control linear stability and transient amplification 
of in the ITG-PVG-driven plasmas in the fluid (collisional) limit 
\cite{newton2010understanding}. 

\subsection{Shearing  frame}

The next step --- standard in treatments of systems with linear shear --- is to make 
a variable transformation $(t,\vr)\to(t',\vr')$ that removes the shear terms ($Sx\dd/\dd y$):
\beq
t'=t,\quad
x'=x,\quad
y'=y-Sxt,\quad
z'=z,
\eeq
and similarly for $(t,\vR)\to(t',\vR')$.
The Fourier transform can then be performed in the primed variables, so
\beq
\tphi = \sum_{\vk'} \tphi_{\vk'}(t') e^{i\vk'\cdot\vr'}
= \sum_{\vk'} \tphi_{\vk'}(t') e^{i\vk(\vk',t')\cdot\vr},
\eeq
where $k_x=k_x'-Sk_y't'$, $k_y=k_y'$, and $\kpar = \kpar'$ (we denote $\kpar\equiv k_z$). 
As usual in the gyrokinetic theory, working in Fourier space allows us to 
compute the gyroaverages in terms of Bessel functions:
\beq
\fl
\avgR{\tphi} = \sum_{\vk'} J_0(a(t'))\tphi_{\vk'}(t') e^{i\vk'\cdot\vR'},
\quad
\avgr{h} = \sum_{\vk'} J_0(a(t')) h_{\vk'}(t') e^{i\vk'\cdot\vr'},
\label{eq:avgs}
\eeq
where $a(t') = \kperp\wperp/\Omega_i = (\wperp/\Omega_i)\sqrt{(k_x'-Sk_y't')^2 + k_y^{\prime 2}}
= (k_y'\wperp/\Omega_i)\sqrt{1 + S^2t^{\prime\prime 2}}$, and we have shifted the origin 
of time: $t''=t'-S^{-1}k_x'/k_y'$. 

Finally, we rewrite the gyrokinetic system \eqref{eq:gk_sh}--\eqref{eq:quasineut_sh} 
in the new variables $(t'',\vk')$. Since we are interested only in the linear problem 
here, we will drop the nonlinearity. We also suppress all primes in the variables.
%and non-dimensionalise the potential $\tphi = Ze\dphi/T$. 
The result~is
%\bea
\begin{multline}
%\fl
\label{eq:gk_lin}
\frac{\dd h_\vk}{\dd t} + i\kpar\wpar h_\vk =
\\
\lt\{\frac{\dd}{\dd t}
-i\lt[\omn + \lt(\frac{w^2}{\vth^2}
%\right.\right.\right. \nonumber\\\left.\left.\left.
-\frac{3}{2}\rt)\omn\eta_i 
- \frac{\wpar}{\vth}\frac{q}{\ee}\,Sk_y\rhoi\rt]\rt\}F_0J_0(a(t))\tphi_\vk,
\end{multline}
\begin{equation}
\lt(1+\frac{\tau}{Z}\rt)\tphi_\vk = \frac{1}{n}\int d^3\vw\,J_0(a(t))h_\vk,
\label{eq:quasineut_lin}
%\eea
\end{equation}
where $\omn = k_y\rhoi\vth/2\Ln$ is the drift frequency and $\eta_i=\Ln/\LT$; 
%and $\omT = k_y\rhoi\vth/2\LT$ and $\omu = S k_y\rhoi$
%are the drift frequencies associated with density, temperature and velocity gradients. 
%Note that in this notation, 
the argument of the Bessel function is 
$a(t)=(\wperp/\vth)k_y\rhoi\sqrt{1 + S^2t^2}$. 
%$a(t)=(\wperp/\vth)\sqrt{k_y^2\rhoi^2 + \omu^2t^2}$. 

\subsection{Integral equation for the linearised problem}
\label{sec:lin_sh}

We integrate \eqref{eq:gk_lin} with respect to time, 
%\bea
%\fl
%\nonumber
%h_\vk(t) &=& \int_0^t dt' e^{-i\kpar\wpar(t-t')}
%\lt\{\frac{\dd}{\dd t'}
%-i\Biggl[\dots\Biggr]\rt\}F_0J_0(a(t'))\tphi_\vk(t')
%+ h_{\vk}(0) e^{-i\kpar\wpar t}\\
%\fl
%\label{eq:h_int}
%&=& F_0J_0(a(t))\tphi_\vk(t) - 
%i\int_0^t dt' e^{-i\kpar\wpar(t-t')}
%\lt\{\kpar\wpar + \Biggl[\dots\Biggr]\rt\}F_0J_0(a(t'))\tphi_\vk(t'),
%%\nonumber
%%\fl
%%&& + \lt[h_{\vk}(0) - F_0J_0(a(0))\tphi_\vk(0)\rt] e^{-i\kpar\wpar t}
%\eea
%where in the second line we have integrated by parts and 
assume the initial fluctuation amplitude small compared to values to which 
it will grow during the subsequent time evolution, 
rescale time  $|\kpar|\vth t \to t$, denote $\tbar=t-t'$, 
%insert \eqref{eq:h_int} into \eqref{eq:quasineut_lin} and  
use \eqref{eq:quasineut_lin}, in which the velocity integrals  
involving the Maxwellian $F_0=n\,e^{-v^2/\vth^2}/(\pi\vth^2)^{3/2}$ 
are done in the usual way, 
%\bea
%\int\frac{d\twpar}{\sqrt{\pi}}\,e^{-\twpar^2 - i\twpar\tbar} 
%= e^{-\tbar^2/4},\\
%\int\frac{d\twpar}{\sqrt{\pi}}\,\twpar\,e^{-\twpar^2 - i\twpar\tbar} 
%= -\frac{i\tbar}{2}\,\mathrm{sgn}(\kpar)\,e^{-\tbar^2/4},\\
%\int\frac{d\twpar}{\sqrt{\pi}}\,\twpar^2 e^{-\twpar^2 - i\twpar\tbar} 
%= \lt(\frac{1}{2} - \frac{\tbar^2}{4}\rt)e^{-\tbar^2/4},\\
%\int \frac{d^2\tvwperp}{\pi}\,J_0(\twperp\sqrt{2\lambda})J_0(\twperp\sqrt{2\lambda'})\,
%e^{-\twperp^2} = \Gamma_0(\lambda,\lambda'),\\ 
%\int \frac{d^2\tvwperp}{\pi}\,\twperp^2 J_0(\twperp\sqrt{2\lambda})J_0(\twperp\sqrt{2\lambda'})\,
%e^{-\twperp^2} = \Lambda(\lambda,\lambda')\Gamma_0(\lambda,\lambda'),
%\eea
%where $\twpar=\wpar/\vth$, $\twperp=\wperp/\vth$, $\lambda(t) = (k_y^2\rhoi^2 + \zS^2t^2)/2$,  
%$\lambda'=\lambda(t') = (k_y^2\rhoi^2 + \zS^2t^{\prime 2})/2$, $\zS = S k_y\rhoi/\kpar\vth$ 
and obtain finally the following integral equation for $\tphi(t)$:
\bea
\nonumber
\fl
\lt(1+\frac{\tau}{Z} - \Gamma_0(\lambda,\lambda)\rt)\tphi(t) 
= \int_0^t d\tbar\,e^{-\tbar^2/4}
\lt\{\lt(\frac{q}{\ee}\,\zS-1\rt)\frac{\tbar}{2}\rt.\\ 
-\lt. i\zn\lt[1 + \eta_i\lt(\Lambda(\lambda,\lambda') - 1 - \frac{\tbar^2}{4}\rt)\rt]\rt\}
\Gamma_0(\lambda,\lambda')\tphi(t-\tbar),
\label{eq:inteq}
\eea
where $\zn=\omn/|\kpar|\vth = k_y\rhoi/2|\kpar|\Ln$ is the normalised 
drift frequency, $\zS = S k_y\rhoi/\kpar\vth$ the normalised shear parameter, and 
%(it is {\it not} a typo that $\zn$ contains $|\kpar|$ while $\zS$ 
%can depend both on the magnitude and the sign of $\kpar$). 
%and $\eta_i=\omT/\omn=\Ln/\LT$. 
\beq
\fl
\Gamma_0(\lambda,\lambda') = e^{-(\lambda+\lambda')/2}I_0(\sqrt{\lambda\lambda'}),\quad
\Lambda(\lambda,\lambda') = 1-\frac{\lambda+\lambda'}{2} + \sqrt{\lambda\lambda'}\,
\frac{I_1(\sqrt{\lambda\lambda'})}{I_0(\sqrt{\lambda\lambda'})}, 
\eeq 
where 
$\lambda(t) = (k_y^2\rhoi^2 + \zS^2t^2)/2$,  
$\lambda'=\lambda(t') = (k_y^2\rhoi^2 + \zS^2t^{\prime 2})/2$, 
and $I_0$ and $I_1$ are modified Bessel functions of the first kind.  

\Eqref{eq:inteq} is the master equation for the linear time evolution of the plasma fluctuations 
driven by the ITG (the $\zn\eta_i$ term) and the PVG (the $(q/\ee)\zS$ term).

\section{Solution for the case of strong shear}
\label{sec:pure_pvg}

We will first consider the maximally simplified case of pure PVG drive (strong shear). 
This is a good quantitative approximation to the general case 
if $\zn, \eta_i\zn \ll (q/\ee)\zS$, which in terms of the basic dimensional 
parameters of the problem translates into 
\beq
\frac{qS}{\ee} \gg \frac{\vth}{\Ln}, \frac{\vth}{\LT}. 
\label{eq:pure_pvg}
\eeq
This is the regime into which the plasma is pushed as the flow shear is increased
--- under certain conditions, the transition can occur abruptly, via 
a transport bifurcation (see Part \ref{sec:bifurcationtoareducedtransportstate}). Besides being, 
therefore, physically the most interesting, this limit 
also has the advantage of particular analytical transparency 
(the more general case including ITG will be considered in \secref{sec:incl_itg}). 

Thus, neglecting all terms that contain $\zn$ and $\eta_i$, \eqref{eq:inteq} becomes
\beq
\fl
\lt(1+\frac{\tau}{Z} - \Gamma_0(\lambda,\lambda)\rt)\tphi(t) 
= \frac{1}{2}\int_0^t d\tbar\,\tbar\,e^{-\tbar^2/4}
\lt(\frac{q}{\ee}\,\zS-1\rt)
\Gamma_0(\lambda,\lambda')\tphi(t-\tbar).
\label{eq:inteq_pvg}
\eeq

\subsection{Short-time limit: the PVG instability}
\label{sec:pvg}

Let us first consider the case in which the velocity shear is unimportant 
except for the PVG drive, i.e., we can approximate 
$\lambda\approx\lambda'\approx k_y^2\rhoi^2/2$ and so there is no time 
dependence in the Bessel functions in \eqref{eq:inteq_pvg}.
We would also like to be able to assume $t\gg1$ 
so that the time integration in \eqref{eq:inteq_pvg} can be extended to $\infty$. 
Formally this limiting case is achieved by ordering $k_y\rhoi\sim1$ and 
$1\ll t \ll 1/\zS$ 
(in dimensional terms, this is equivalent to $St\ll1$, $\kpar\vth t\gg1$
and $\kpar\vth/S\gg1$). 
Physically, this regime will be realised in the initial stage of evolution 
of the fluctuations and, as we are about to see, will require a large enough $q/\ee$.  
 
Under this ordering, we can seek solutions to \eqref{eq:inteq_pvg} 
in the form $\tphi(t) = \tphi_0\exp(-i\zz t)$, where 
$\zz = \omega/|\kpar|\vth$ is the nondimensionalised complex frequency 
and $\omega$ its dimensional counterpart. The time integral in \eqref{eq:inteq_pvg} 
can be expressed in terms of the plasma dispersion function, which 
satisfies \cite{fried1961pdf}
\beq
\label{eq:Z}
\Zfn(\zz) %= \frac{1}{\sqrt{\pi}}\int_{-\infty}^{+\infty}dz\,\frac{e^{-z^2}}{z-\zz} 
= i\int_0^\infty d\tbar\,e^{i\zz\tbar - \tbar^2/4},\qquad
\Zfn'(\zz) = -2[1+\zz\Zfn(\zz)].
\eeq
%(not to be confused with $Z$ in the ion charge $Ze$). 
With the aid of these formulae, \eqref{eq:inteq} is readily converted 
into a transcendental equation for $\zz$: 
\beq
\label{eq:pvg}
\frac{1+\tau/Z}{\Gamma_0(\lambda)} - 1 = 
\lt(\frac{q}{\ee}\,\zS - 1\rt) [1+\zz\Zfn(\zz)], 
\eeq
where $\Gamma_0(\lambda) = e^{-\lambda} I_0(\lambda)$, 
$\lambda=k_y^2\rhoi^2/2$ and $\zS=S k_y\rhoi/\kpar\vth$.

%\subsubsection{Destabilisation of the sound wave.}

\Eqref{eq:pvg} is simply the dispersion relation for the ion acoustic wave modified 
by the PVG drive term. This point is probably best 
illustrated by considering the cold-ion/long-wavelength limit 
$\tau\ll1$, $\zz=\omega/|\kpar|\vth\gg1$, $\lambda=k_y^2\rho^2/2\ll1$. 
Then $\Gamma_0(\lambda)\approx1$, 
$1+\zz\Zfn(\zz)\approx -1/2\zz^2 + i\zz\sqrt{\pi}\,e^{-\zz^2}$, 
and so, restoring dimensions in \eqref{eq:pvg}, we get
\beq
\zz^2 \approx \frac{Z}{2\tau}\lt(1 - \frac{q}{\ee}\,\zS\rt)
\quad\Rightarrow\quad
\omega \approx \pm\kpar c_s\lt(1 - \frac{qS}{\ee}\frac{k_y\rhoi}{\kpar\vth}\rt)^{1/2},
\label{eq:sound_wave}
\eeq
where in the last expression, we have restored dimensions and denoted 
$c_s=(Z/2\tau)^{1/2}\vth=(ZT_e/m_i)^{1/2}$, the sound speed.  
When $(q/\ee)\zS$ is sufficiently large, the sound wave is destabilised 
and turns into the PVG instability. Note that it loses its real frequency 
in this transition.

\begin{figure}[t]
	\dpanel{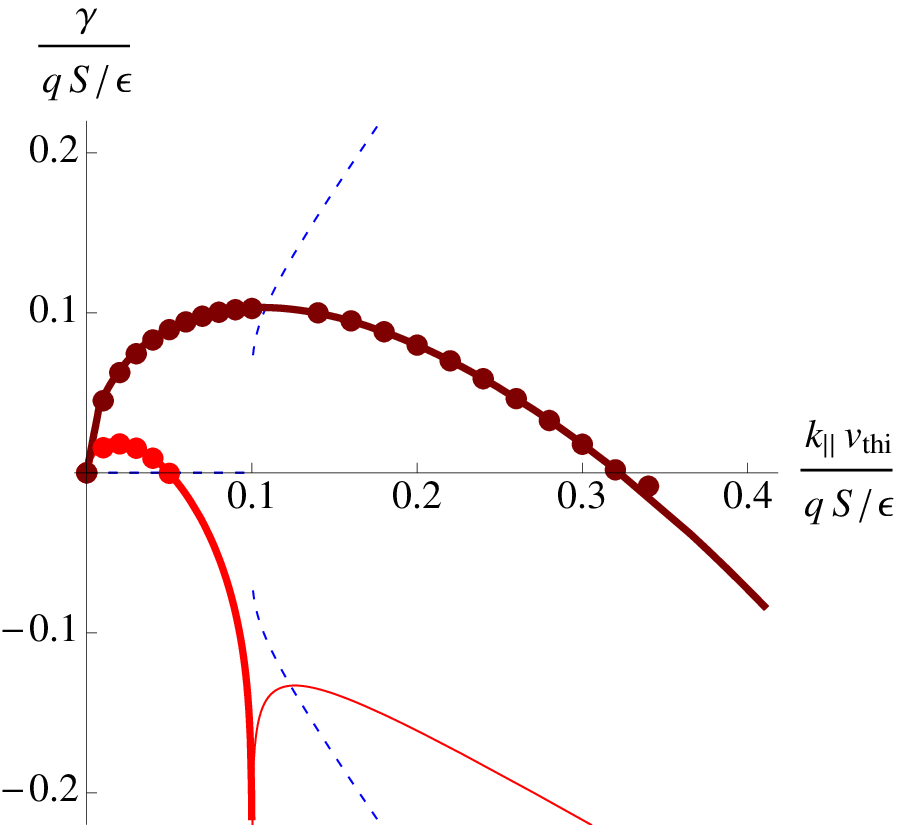}{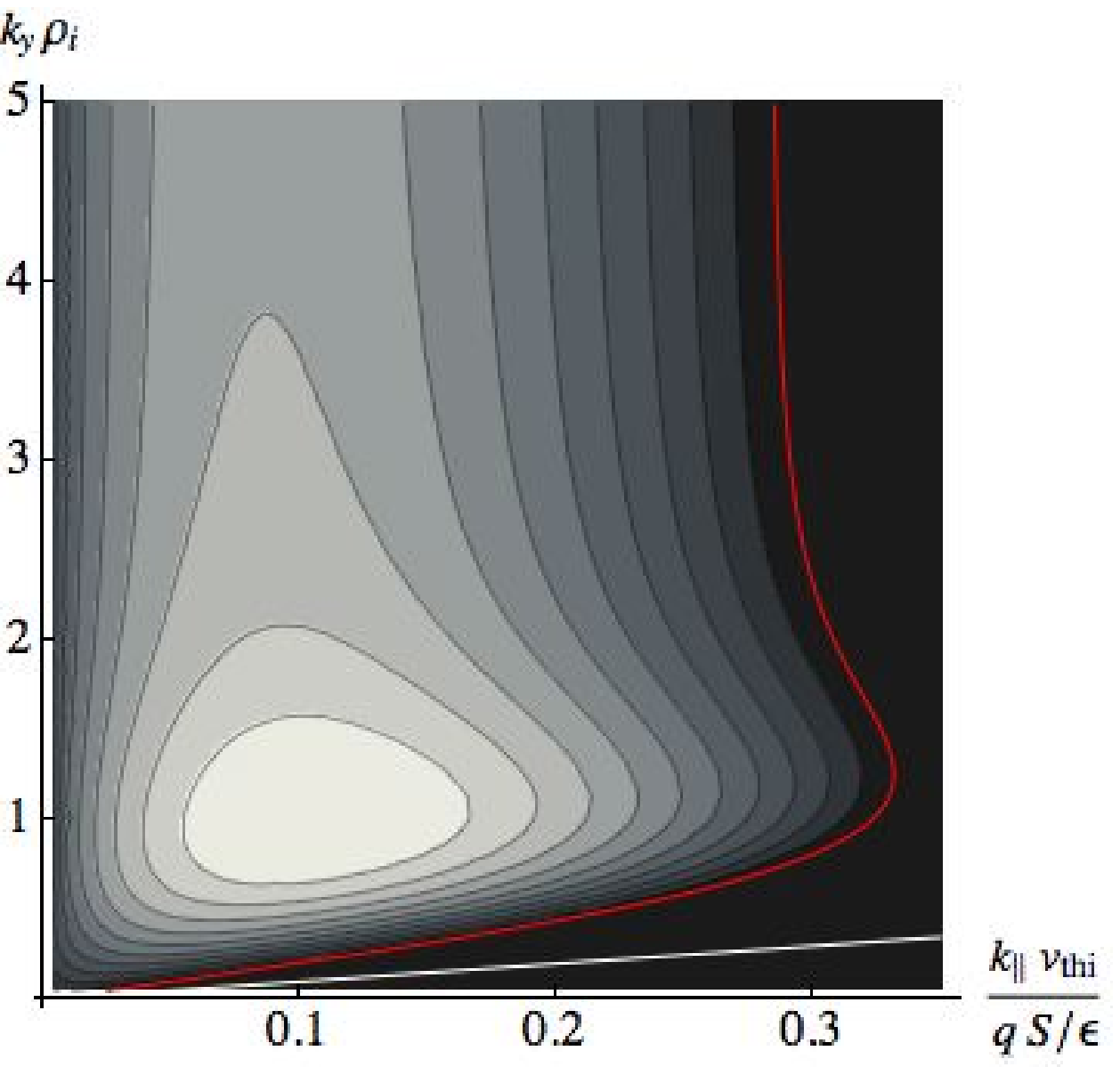}
%\centerline{\includegraphics{pvg_num}
%\hskip0.1in
%\includegraphics{pvgcont}}
\caption{The PVG Instability}{\label{fig:pvg} {\em Left panel:} 
Growth rate $\gamma = \Im\,\omega$ (normalised by $qS/\ee$) 
vs.\ $\kpar\vth/(qS/\ee)$. Here $\tau/Z=1$, and the two curves 
are for $k_y\rhoi=0.1$ (red) and $k_y\rhoi=1$ (brown).
The growth rate becomes independent of $k_y$ for $k_y\rhoi\gg1$ 
(see end of Appendix B.1 of Ref.~\cite{schekochihin2010subcritical}) and all curves for large $k_y\rhoi$ are 
very close to each other and similar in shape to $k_y\rhoi=1$ (see right panel). 
The mode has no real frequency for $\kpar\vth/(qS/\ee)<k_y\rhoi$;
for $\kpar\vth/(qS/\ee)>k_y\rhoi$, it turns into a damped sound wave: 
the corresponding frequencies and the damping rate are 
shown only for $k_y\rhoi=0.1$, as thin blue (dashed) and red (solid) lines, 
respectively. The discrete points show growth rates calculated by direct linear numerical 
simulation using the gyrokinetic code \texttt{AstroGK} \cite{numata2010astrogk}. 
{\em Right panel:} Contour plot of $\gamma/(qS/\ee)$ vs.\ $\kpar\vth/(qS/\ee)$ 
and $k_y\rhoi$. Only positive values are plotted, black means $\gamma<0$.
The red curve shows the stability boundary \apprefeq{B.1}. 
The white line shows the boundary \apprefeq{B.2} 
between PVG modes (above it) and sound waves (below it).
}
\end{figure}

Some further (elementary) analytical considerations of the PVG 
instability are given in Appendix B.1 of Ref.~\cite{schekochihin2010subcritical}. Here it will suffice 
to notice that if the (dimensional) growth rate 
$\gamma=\Im\,\omega$ is scaled by $qS/\ee$ and $\kpar$ 
by $(qS/\ee)/\vth$, their mutual dependence is universal for all values of the 
velocity shear or $q/\ee$, namely, 
\beq
\gamma = \frac{q S}{\ee}\,f\lt(k_y\rhoi,\frac{\kpar\vth}{qS/\ee}\rt).
\label{eq:gamma_gen_pvg}
\eeq 
Once this rescaling is done, the dispersion relation \eqref{eq:pvg} no longer 
contains any parameters (except $\tau/Z$, which we can safely take to be order unity). 
The growth rate, obtained via numerical solution 
of \eqref{eq:pvg} with $\tau/Z=1$, is plotted in \figref{fig:pvg}. 
The maximum growth rate is $\gamma_\mathrm{max}\approx 0.10(qS/\ee)$. 
This peak value is reached when $k_y\rhoi\approx1.0$ and $\kpar\approx 0.10(qS/\ee)/\vth$, 
i.e., at $(q/\ee)\zS\approx 10$. 

The conclusion is that, at least in the initial stage 
of their evolution, plasma fluctuations in a significant part of the 
wavenumber space ($k_y$ and $\kpar$ are constrained by \apprefeq{B.1}) 
are amplified by the PVG. 
This amplification does not, however, go on for a long time as the approximation 
we adopted to derive the dispersion relation \eqref{eq:pvg} breaks down when 
$\zS t\sim k_y\rhoi$ (or $St\sim 1$ if the dimensional units of time are restored)
--- this gives $\gamma t\sim 0.1 (q/\ee)$, so, 
realistically, after barely one exponentiation. 
The key question is what happens after that. 
We will see shortly that all modes will eventually decay and 
that the fastest initially growing modes are in fact not quite 
the ones that will grow the longest or get maximally amplified. 

\subsection{Long-time limit: transient growth}
\label{sec:longt}

Let us now investigate the long-time limit, $\zS t\gg 1$, $\zS t\gg k_y\rhoi$ 
(or, in dimensional form, $St\gg1$, $k_y\rhoi St\gg1$).  
In this limit, the kernel involving the Bessel function in \eqref{eq:inteq_pvg} 
simplifies considerably: we have $\lambda\approx\zS^2t^2/2\gg1$, 
$\lambda'\approx\zS^2(t-\tbar)^2/2\gg1$ and so 
\beq
\label{eq:Gamma0_longt}
\Gamma_0(\lambda,\lambda')\approx 
\frac{e^{-(\sqrt{\lambda}-\sqrt{\lambda'})^2/2}}{\sqrt{2\pi\sqrt{\lambda\lambda'}}}
\approx \frac{e^{-\zS^2\tbar^2/4}}{\sqrt{\pi}|\zS|t}.
\eeq
Working to the lowest nontrivial order in $1/t$, we can now rewrite 
\eqref{eq:inteq_pvg} as follows
\beq
\fl
\lt(1+\frac{\tau}{Z}\rt)|\zS|t\tphi(t) 
= \frac{1}{2\sqrt{\pi}}\int_0^\infty d\tbar\,\tbar\,e^{-(\zS^2+1)\tbar^2/4}
\lt(\frac{q}{\ee}\,\zS-1\rt)\tphi(t-\tbar).
\label{eq:longt}
\eeq
We will seek a solution to this equation in the form 
\beq
\tphi(t) = \tphi_0 \exp\lt[\int_0^t dt'\gbar(t')\rt],
\label{eq:phit}
\eeq
where $\gbar(t)=\gamma(t)/|\kpar|\vth$ is the effective time-dependent 
growth rate (nondimensionalised) and $\gamma(t)$ the dimensional version 
of it (remember that time is scaled by $|\kpar|\vth$). 
Because of the exponential in the kernel, the memory of the time-history 
integral in the right-hand side of \eqref{eq:longt} is limited, so 
$\tbar\ll t$ and we will be able to make progress by expanding
\beq
\tphi(t-\tbar) = \tphi_0 \exp\lt[\int_0^t dt'\gbar(t') - \tbar\gbar(t) 
+ \frac{\tbar^2}{2}\,\gbar'(t) + \dots\rt].
\label{eq:phi_exp}
\eeq
We will assume that this expansion can be truncated; the resulting 
solution will indeed turn out to satisfy 
$\gbar'(t)\ll1$, with all higher-order terms even smaller. 
Substituting \eqref{eq:phi_exp} into \eqref{eq:longt}, we obtain an implicit 
transcendental equation for $\gbar(t)$: 
\beq
\lt(1+\frac{\tau}{Z}\rt)|\zS|t 
= \frac{1}{2\sqrt{\pi}}\int_0^\infty d\tbar\,\tbar\,e^{-\tbar\gbar(t)-(\zS^2+1)\tbar^2/4}
\lt(\frac{q}{\ee}\,\zS-1\rt).
\label{eq:zeta}
\eeq
This equation can be written in a compact form by invoking once again  
the plasma dispersion function \eqref{eq:Z}: denoting 
$\tg(t)=\gbar(t)/\sqrt{\zS^2+1}$, we get
\beq
\lt(1+\frac{\tau}{Z}\rt)\sqrt{\pi}(\zS^2+1)|\zS|t =
\lt(\frac{q}{\ee}\,\zS-1\rt) [1+i\tg(t)\Zfn(i\tg(t))] 
\label{eq:tz}
\eeq
--- effectively, a time-dependent dispersion relation, reminiscent 
of the PVG dispersion relation \eqref{eq:pvg}. 
%This is, of course, quite easy solve numerically, but we will make a few 
%important inferences analytically before doing that. 

\subsubsection{Transient growth.}
\label{sec:trans}

First of all, since the effective time-dependent growth rate 
$\gbar(t)$ appears with a negative sign in the exponential in 
\eqref{eq:zeta}, it is clear that as time increases, 
(the real part of) $\gbar(t)$  
must decrease and indeed go negative because 
the right-hand side has to keep up with the increasing left-hand side.
Therefore, fluctuations will eventually decay. However, 
if $\Re\,\gbar(t)$ is positive for some significant initial period of time, 
there can be a substantial transient amplification. 

We can determine 
the time $t_0$ when the transient growth ends by setting 
$\Re\,\tg(t_0) = 0$ in \eqref{eq:tz}. We immediately find that 
$\Im\,\tg(t_0) = 0$ as well and that 
\beq
t_0 =\frac{(q/\ee)\zS-1}{(1+\tau/Z)\sqrt{\pi}(\zS^2+1)|\zS|}.
\label{eq:t0}
\eeq
%where we have restored dimensions to $t_0$ to make explicit its dependence on $\kpar$. 
%If we hold $\kpar$ fixed, we can find 
%$k_y$ for which the transient growth last longest by maximising 
%\eqref{eq:t0} with respect to $\zS$: 
%assuming $q/\ee\gg1$, letting $\dd t_0/\dd\zS=0$ 
%gives $\zS\approx(\ee/2q)^{1/3}$ and so, for these modes,  
%$t_0 \approx (q/\ee)/[|\kpar|\vth (1+\tau/Z)\sqrt{\pi}]$. 
%This suggests that the modes with small $\kpar$ and also 
%small $k_y\rhoi\approx (\ee/2q)^{1/3}\kpar\vth/S$ will 
%grow for the longest time. 
The dependence of $t_0$ on $k_y$ and $\kpar$ 
--- via $\zS=Sk_y\rhoi/\kpar\vth$ 
and via the time normalisation factor of $|\kpar\vth|$ 
--- tells us which modes grow longest. The interesting question, however, 
is rather which modes get maximally amplified 
during this transient growth. 

\myfiga{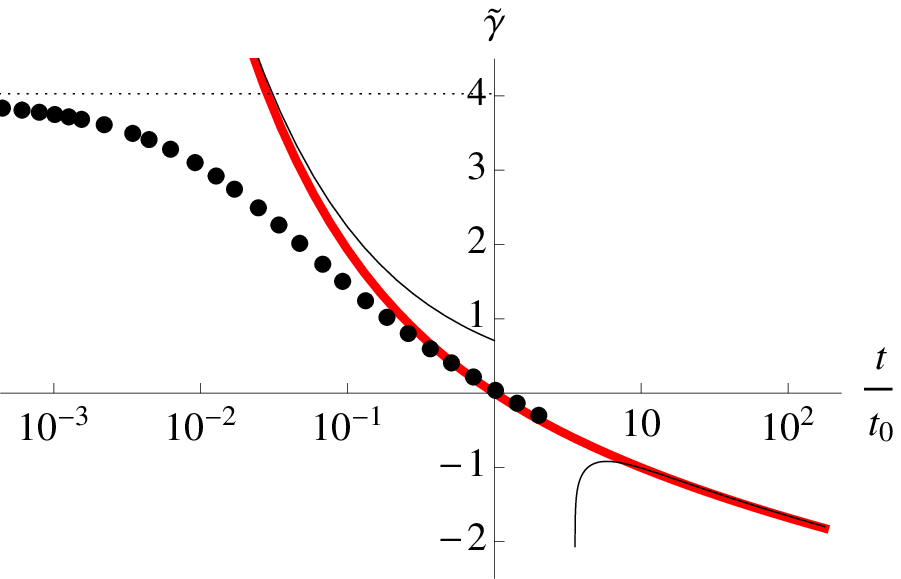}{fig:gamma}{Time Evolution of the Transient Growth Rate}{
Time evolution of the effective growth rate: the red (bold) line is   
$\tg(t) = \gamma(t)/|\kpar|\vth\sqrt{\zS^2+1}$ 
obtained as a numerical solution of \eqref{eq:pvgt} and 
plotted vs.\ $t/t_0$ (the time axis is logarithmic in base 10). 
The black (thin) lines are the asymptotics 
\eqref{eq:growth} and \eqref{eq:decay_precise}.
The discrete points show the time evolution of the effective growth rate 
obtained in a direct linear numerical simulation using the gyrokinetic 
code \texttt{AstroGK} \cite{numata2010astrogk} with
$1/\Ln=1/\LT=0$, $q/\ee=50$, $\tau/Z=1$, $k_y\rhoi=1$, $\kpar\vth/S=0.5$. 
The short-time-limit PVG growth rate 
for this case (obtained by solving \eqref{eq:pvg}) is shown as a dotted 
horizontal line. The time is normalised using \eqref{eq:t0} for $t_0$.} 

\subsubsection{Maximal amplification.}
\label{sec:Nmax}

\begin{figure}[t]
\dpanel{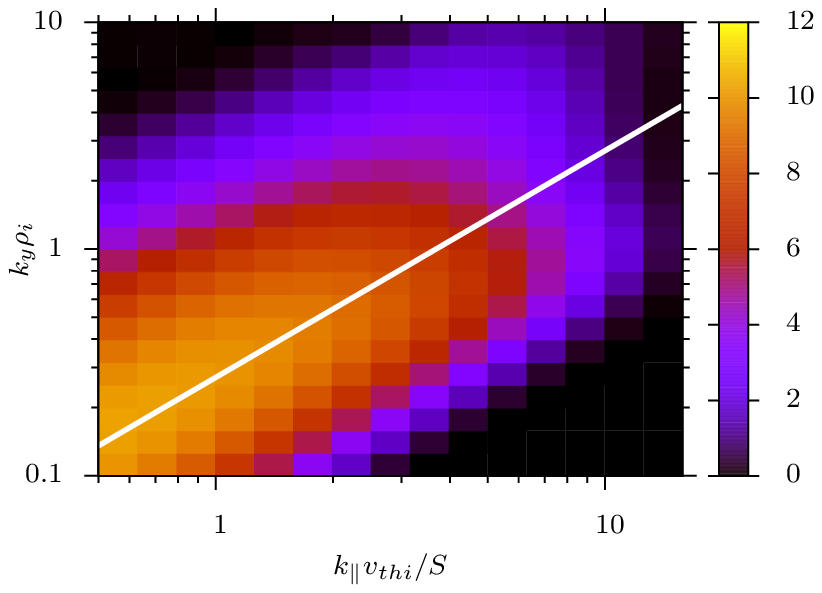}{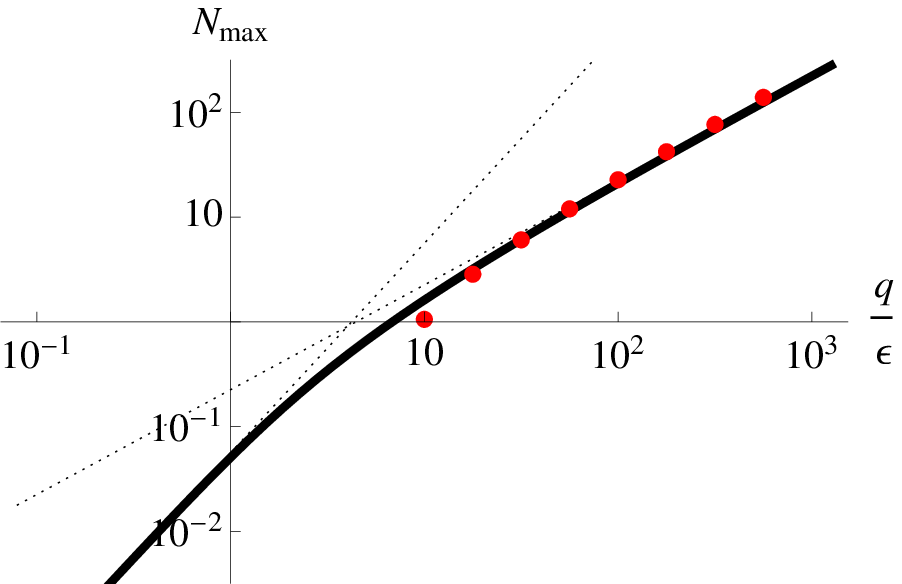}
%\centerline{\epsfig{file=max_N_graph_pvg.eps,width=2.75in}
%\hskip0.25in
%\epsfig{file=Nmaxvsqe.eps,width=3in}}
\caption{The PVG Amplification Exponent}{\label{fig:kspace} 
{\em Left panel:} The amplification exponent $N$ vs.\ $\kperp$ and $\kpar$ 
in the same \texttt{AstroGK} simulation as used to produce the 
discrete points in \figref{fig:gamma}. The straight line 
shows the relationship between the wavenumbers given by 
the second formula in \eqref{eq:Nmax}.
{\em Right panel:} The maximum amplification exponent $\Nmax$ 
[$N$ given by \eqref{eq:N}, maximised with respect to $\zS$] vs.\ $q/\ee$.
The dotted lines show the $q/\ee\gg1$ asymptotic [see \eqref{eq:Nmax}] 
and the $q/\ee\ll1$ asymptotic, the latter straightforwardly 
obtained from \eqref{eq:N}: $\zS\approx2(\ee/q)$, $\Nmax\approx0.11(q/\ee)^2/(1+\tau/Z)$ 
(but this is purely formal because the long-time conditions $St_0\gg1$ 
and $k_y\rhoi St_0\gg1 $ will be broken in this regime, except for extremely long wavelengths).
The discrete points show $\Nmax$ obtained via an \texttt{AstroGK} numerical parameter 
scan: $\kpar\vth/S=0.5$, varying $\kperp\rhoi$ and holding all
other parameters fixed as in \figref{fig:gamma} 
(note that the asymptotic results do not in fact depend on $\kpar$ or $S$, although 
the quality of the long-time asymptotic does).} 
\end{figure}

The total amplification factor 
is given by $e^N$, where $N = \int_0^{t_0}dt\gbar(t)$ is the number 
of exponentiations experienced by the mode during its growth period. 
In order to determine this, we need to know the time evolution 
of $\gbar(t)$ up to $t=t_0$. 
Using \eqref{eq:t0}, it is convenient to rewrite \eqref{eq:tz} as follows 
\beq
\frac{t}{t_0} = 1 + i\tg\,\Zfn(i\tg).
\label{eq:pvgt}
\eeq 
When $t\ll t_0$, the solution is found by expanding the 
plasma dispersion function in $\tg\gg1$: 
\beq
\frac{t}{t_0}\approx \frac{1}{2\tg^2}
\quad\Rightarrow\quad
\tg\approx \sqrt{\frac{t_0}{2t}}.
\label{eq:growth}
\eeq
This asymptotic is not valid when $t$ approaches $t_0$. 
More generally, \eqref{eq:pvgt} has a solution 
$\tg=\tg(t/t_0)$, whose functional form is independent of 
any parameters of the problem. It is plotted in \figref{fig:gamma} 
together with the asymptotic \eqref{eq:growth} and with a direct 
numerical solution showing how the transition between the 
short-time (\secref{sec:pvg}) and long-time limits occurs. 
The amplification exponent is easily found:
\beq
\fl
N = \int_0^{t_0}dt\gbar(t) 
= t_0\sqrt{\zS^2+1}\int_0^1 d\xi\,\tg(\xi)
\approx 0.45\,\frac{(q/\ee)\zS-1}{(1+\tau/Z)\sqrt{\zS^2+1}\,|\zS|},
\label{eq:N}
\eeq
where we have used \eqref{eq:t0} for $t_0$ and computed the 
integral numerically. It is clear from \eqref{eq:growth} that the 
integral converges on its lower limit and is not dominated by it, 
so it does not matter that we cannot technically use \eqref{eq:pvgt} 
for short times (as confirmed by comparison with a direct numerical 
calculation in the right panel of \figref{fig:kspace}). 

According to \eqref{eq:N}, $N$ depends on both wavenumbers via $\zS$ only 
(a plot of this dependence will be given in \figref{fig:N}). 
Assuming $q/\ee\gg1$, we find that the amplification exponent is maximised 
for $\zS \approx (\ee/q)^{1/3}$, giving\footnote{Note that this is well outside 
of the wavenumber domain populated by the damped sound waves, 
$\zS<\ee/q$ (see Appendix B.1 of Ref.~\cite{schekochihin2010subcritical}).} 
\beq
\Nmax \approx 
0.45\,\frac{q/\ee}{1+\tau/Z}
\quad \mathrm{for} \quad 
k_y\rhoi\approx \lt(\frac{\ee}{q}\rt)^{1/3}\frac{\kpar\vth}{S}.
\label{eq:Nmax}
\eeq
These results are illustrated in the left panel of \figref{fig:kspace}.
The amplification time for the maximally amplified modes 
identified in \eqref{eq:Nmax} is, from \eqref{eq:t0}, 
\beq
t_0 \approx \frac{q/\ee}{|\kpar|\vth(1+\tau/Z)\sqrt{\pi}}
\approx \frac{(q/\ee)^{2/3}}{|Sk_y\rhoi|(1+\tau/Z)\sqrt{\pi}},
\label{eq:t0_max} 
\eeq
where we have restored dimensions to make explicit the dependence 
of $t_0$ on $\kpar$.

Thus, we have learned that an entire family of modes, 
characterised by a particular (linear) relationship between $k_y$ and $\kpar$,
given in \eqref{eq:Nmax}, will eventually enjoy the same 
net amplification, even though, as follows from the results 
of \secref{sec:pvg}, they were not the fastest initially 
growing modes and some within this equally amplified family 
started off growing more slowly than others or even decaying. 
The more slowly growing modes are the longer-wavelength ones and,  
according to \eqref{eq:t0_max}, they compensate for their 
sluggishness with longer growth times. 

Note that \eqref{eq:t0_max} confirms that the long-time limit is 
analytically reasonable because for large~$q/\ee$, 
\eqref{eq:t0_max} formally satisfies $k_y\rhoi St_0\sim(q/\ee)^{2/3}\gg1$ 
and also $St_0\gg1$, provided $\kpar\vth/S\ll q/\ee$. 
The latter condition is marginally broken by the 
fastest initially growing modes: indeed,  
in \secref{sec:pvg}, we saw that they had $\kpar\vth/S\sim 0.1 q/\ee$, 
so for them, $St_0\sim 3$, not really a large number.
For all longer-wavelength modes, $t_0$ is safely 
within the domain of validity of the long-time limit.

Note also that, as is shown at the end of Appendix B.1 of Ref.~\cite{schekochihin2010subcritical}, the time-dependent 
dispersion relation \eqref{eq:pvgt} and its consequences derived above can be 
obtained directly from the PVG-instability dispersion relation \eqref{eq:pvg} 
simply by restoring its $k_x$ dependence, setting $k_x=Sk_yt$, taking the 
short-wavelength and long-time limit ($k_x\rhoi\gg1$, $St\gg1$), 
and assuming $\ee/q\ll\zS\ll1$. This calculation underscores the fundamental
simplicity of the physics of the transient amplification: perturbations 
initially destabilised by the PVG are eventually swept by the perpendicular 
velocity shear into a stable region of the wavenumber space. 

\subsubsection{Limits on short and long wavelengths.}
\label{sec:UV}

We have seen that modes with parallel wavenumbers up to 
$\kpar\vth/S\sim q/\ee$ can be transiently amplified. 
From \eqref{eq:Nmax}, we conclude that of these, 
the maximally amplified ones will have perpendicular wavenumbers 
up to $k_y\rhoi\lesssim (q/\ee)^{2/3}$, i.e., 
$k_y\rhoi$ can be relatively large --- unlike in the short-time limit 
treated in \secref{sec:pvg}, where the modes with 
%there was no constraint on the relationship between $k_y$ and $\kpar$ and 
$k_y\rhoi\sim1$ grew the fastest (although large $k_y\rhoi$ were also 
unstable).

It should be understood that, while there is no ultraviolet 
cutoff in our theory that would limit the wavenumbers of the growing modes 
(in either direction), 
such a cutoff does of course exist in any real system. 
In the parallel direction, those $\kpar$ that were strongly damped 
in the short-time limit (see \apprefeqnb{B.1}) are unlikely to recover 
in the long-time limit. In the perpendicular direction, the cutoff 
in $k_y\rhoi$ will come from the collisional damping, which, in 
gyrokinetics, contains a spatial diffusion (see, e.g., \cite{abel2008linearized}), 
and from the electron Landau damping, which we have lost by using 
the Boltzmann electron response (see \eqref{eq:quasineut_sh}) and 
which should wipe out large $k_y$ and $\kpar$. 

On the infrared (long-wavelength) side of the spectrum, we have no 
cutoffs either. In a slab, these would be provided by the dimensions 
of the periodic box. In a real plasma, the cutoffs are set by 
the scales at which the system can no longer be considered 
homogeneous (in a tokamak, these are the equilibrium-gradient scale lengths 
and the minor radius for the perpendicular scales and the connection 
length $qR$ for the parallel scales). 

\subsubsection{Significant amplification threshold.}
\label{sec:threshold}
If we maximise \eqref{eq:N} without assuming $q/\ee\gg1$ (with the caveat 
that the long-time limit asymptotics are at best marginally valid then), 
we obtain a more general curve than \eqref{eq:Nmax}, plotted in the right 
panel of \figref{fig:kspace}. We may define a critical threshold for 
significant amplification: $\Nmax=1$ when $q/\ee\approx7$. 
The role of this threshold will be discussed in Chapter \ref{sec:zeroturbulencemanifold}. 

\subsubsection{Long-time decay.}
\label{sec:decay}

Finally, we obtain the long-time asymptotic decay law. 
Let us seek a solution of \eqref{eq:pvgt} 
such that $t\gg t_0$ and $\tg\ll-1$. Then 
\beq
\frac{t}{t_0} \approx 2\sqrt{\pi}\,|\tg| e^{\tg^2}
\quad\Rightarrow\quad
\tg\approx -\sqrt{\ln\frac{t}{t_0}}.
\label{eq:decay}
\eeq
Since $\tg$ is only root-logarithmically large, the quality of this 
asymptotic is rather poor. If we insist on a more precise decay law, 
we can retain small corrections in \eqref{eq:decay} and 
get what turns out, upon a numerical test,
to be a reasonably good approximation (see \figref{fig:gamma}):
\beq
\tg \approx -\sqrt{\ln\frac{t/(2\sqrt{\pi}t_0)}{\sqrt{\ln(t/2\sqrt{\pi}t_0)}}}.
\label{eq:decay_precise}
\eeq
For the maximally amplified modes (see \eqref{eq:Nmax}), 
the dimensional damping rate is $\gamma(t)\approx|\kpar|\vth\tg(t)$ 
with $t_0$ given by \eqref{eq:t0_max}.  
This tells us is that the decay 
is just slightly faster than exponential at the rate of order~$S$. 
The longest-wavelength modes decay the slowest, after having 
being amplified the longest.  

\section{Solution including ITG}
\label{sec:incl_itg}

Let us now generalise the results obtained in \secref{sec:pure_pvg} to 
include non-zero (i.e., non-negligible) density and temperature gradients. 
This means that we restore the terms involving $\zn$ and $\eta_i$ in 
the general integral equation \eqref{eq:inteq}. 

\subsection{Short-time limit: the ITG-PVG dispersion relation}
\label{sec:itg}

The short-time limit, introduced at the beginning of \secref{sec:pvg}
for the case of pure PVG, is treated in an analogous fashion for the 
general ITG-PVG case.
An analysis of the solutions of the resulting dispersion relation is 
useful in that its results assist 
physical intuition in ways relevant for some of the forthcoming 
discussion, but it is not strictly necessary for us to have them 
in order to work out how transient growth happens in the 
presence of the ITG drive. For this analysis therefore we refer the interested reader to Appendix B.2 of Ref.~\cite{schekochihin2010subcritical}. 

\subsection{Long-time limit}

We now continue in the same vein as in \secref{sec:longt} and 
consider the long-time limit ($St\gg1$, $k_y\rhoi St\gg1$), in which 
we can simplify the kernels involving the Bessel functions in \eqref{eq:inteq} 
by using \eqref{eq:Gamma0_longt} 
%with $\lambda\approx\zS^2t^2/2\gg1$, $\lambda'\approx\zS^2(t-\tbar)^2/2\gg1$ 
and also 
\beq
\Lambda(\lambda,\lambda')\approx \frac{1}{2} - \frac{(\sqrt{\lambda}-\sqrt{\lambda'})^2}{2}
\approx \frac{1}{2} - \frac{\zS^2\tbar^2}{4}. 
\eeq
This allows us to rewrite \eqref{eq:inteq} in the form that 
generalises \eqref{eq:longt}: 
%\bea
\begin{multline}
\fl
\lt(1+\frac{\tau}{Z}\rt)|\zS|t\tphi(t) 
= \frac{1}{\sqrt{\pi}}\int_0^\infty d\tbar\,e^{-(\zS^2+1)\tbar^2/4}
\lt\{\lt(\frac{q}{\ee}\,\zS-1\rt)\frac{\tbar}{2}\rt.\\ 
\qquad\qquad-\lt. i\zn\lt[1-\eta_i\lt(\frac{1}{2} + \frac{(\zS^2+1)\tbar^2}{4}\rt)\rt]\rt\}
\tphi(t-\tbar).
\label{eq:longt_itg}
\end{multline}
%\eea
As in \secref{sec:longt}, we 
seek a solutions to this equation in the form \eqref{eq:phit}, 
taking $\tbar\ll t$ and expanding the delayed potential under 
the integral according to \eqref{eq:phi_exp}. 
The result is the generalised form of \eqref{eq:zeta}:
\bea
\nonumber
\fl
\lt(1+\frac{\tau}{Z}\rt)|\zS|t 
&=& \frac{1}{\sqrt{\pi}}\int_0^\infty d\tbar\,e^{-\tbar\gbar(t)-(\zS^2+1)\tbar^2/4}
\lt\{\lt(\frac{q}{\ee}\,\zS-1\rt)\frac{\tbar}{2}\rt.\\ 
&&\qquad\qquad-\lt. i\zn\lt[1-\eta_i\lt(\frac{1}{2} + \frac{(\zS^2+1)\tbar^2}{4}\rt)\rt]\rt\}.
\label{eq:zeta_itg}
\eea
Using again the plasma dispersion function \eqref{eq:Z} to 
express the time integrals and introducing the complex scaled frequency 
$\tz(t)=i\gbar(t)/\sqrt{\zS^2+1}$, we get
\bea
\fl
\nonumber
\lt(1+\frac{\tau}{Z}\rt)\sqrt{\pi(\zS^2+1)}|\zS|t &=&  
\lt[\frac{(q/\ee)\zS-1}{\sqrt{\zS^2+1}} - \eta_i\zn\tz(t)\rt]
[1+\tz(t) \Zfn(\tz(t))]\\ 
&&+\,\, (\eta_i-1)\,\zn \Zfn(\tz(t)),
\label{eq:tz_itg}
\eea
a time-dependent dispersion relation, which is the generalisation of \eqref{eq:tz}. 

\subsubsection{Transient growth.} The general argument that
the real part of $\gbar(t)$ (i.e., the effective time-dependent growth rate) 
must eventually decrease and so fluctuations will, in the end, decay, applies 
to \eqref{eq:zeta_itg} similarly to the way it did to \eqref{eq:zeta} 
(see \secref{sec:trans}), although this decay need not (and, as we will see, 
will not) be monotonic. 
The time $t_0$ when the transient growth ends is now determined as follows.  
Let $\Im\,\tz(t_0) = 0$ and $\tz(t_0)=\tz_0$ (real!). Then from  
\eqref{eq:tz_itg} taken at $t=t_0$, we find the real frequency $\tz_0$ by demanding 
that the imaginary part of the right-hand side vanish --- this means that 
the coefficient in front of $\Zfn(\tz_0)$ must be zero, because, for real $\tz_0$, 
$\Im\,\Zfn(\tz_0) = \sqrt{\pi}\, e^{-\tz_0^2}$. This condition gives
\beq
\eta_i\zn\tz_0^2 - \frac{(q/\ee)\zS-1}{\sqrt{\zS^2+1}}\,\tz_0 - (\eta_i-1)\zn = 0,
\eeq
whence
\beq
%\fl
%\tz_0 = \frac{1}{\sqrt{\zS^2+1}}\lt\{\frac{(q/\ee)\zS-1}{2\zn\eta_i} \pm 
%\sqrt{\lt[\frac{(q/\ee)\zS-1}{2\zn\eta_i}\rt]^2 + \lt(1-\frac{1}{\eta_i}\rt)(\zS^2+1)}\rt\}.
\tz_0 = \frac{(q/\ee)\zS-1 \pm \sqrt{[(q/\ee)\zS-1]^2 + 4\eta_i(\eta_i-1)\,\zn^2\,(\zS^2+1)}}{2\eta_i\zn\sqrt{\zS^2+1}}.
\label{eq:freq}
\eeq
Substituting this solution into the real part of \eqref{eq:tz_itg} and taking 
advantage of the already enforced vanishing of the coefficient in front 
of $\Zfn(\tz_0)$, we get %, again restoring dimensions to $t_0$,  
\beq
t_0 = \frac{(q/\ee)\zS - 1 + \sqrt{[(q/\ee)\zS - 1]^2 
+ \eS^2\,(1-1/\eta_i)\,\zS^2\,(\zS^2+1)}}{2(1+\tau/Z)\sqrt{\pi}(\zS^2+1)|\zS|},
\label{eq:t0_itg}
\eeq
where we have replaced $\eta_i^2\zn^2 = \eS^2\zS^2/4$ 
with $\eS = \vth/\LT S$ a new parameter that measures 
the strength of the ITG drive relative to the velocity shear.
Note that we picked the ``$-$'' mode in \eqref{eq:freq} because the ``+'' mode 
is not amplified ($t_0<0$, assuming $\eta_i>1$). 
\Eqref{eq:t0_itg} is the generalisation of \eqref{eq:t0}, 
to which it manifestly reduces when $\eS=0$ and with which it shares the property 
that the transient growth time depends on $k_y$ and $\kpar$ 
only via $\zS$ and the time normalisation factor $|\kpar|\vth$. 
%While \eqref{eq:t0_itg} has a somewhat more complicated parameter 
%dependence than \eqref{eq:t0}, it is obvious from it that larger temperature 
%or density gradients (larger $\eta_i$ or $\zn$) will prolong transient 
%growth --- the PVG and ITG drives amplify each other's effect.

\subsubsection{General dispersion relation.} We can now recast 
the general ITG-PVG case in a form that shows explicitly how it reduces 
to the case of pure PVG drive studied in \secref{sec:longt}. First we note 
that the transient growth termination time \eqref{eq:t0_itg} can be 
rewritten as
\beq
\fl
t_0 = \frac{1}{2}\lt[1 + \sigma
\sqrt{1+ \lt(1-\frac{1}{\eta_i}\rt)\chi^2}\rt]\tpvg,
\qquad
\chi = \frac{\eS\,\zS \sqrt{\zS^2+1}}{(q/\ee)\zS - 1},
\label{eq:chi}
\eeq
where $\tpvg$ is given by \eqref{eq:t0}, $\eS\zS=2\eta_i\zn$, $\eS=\vth/\LT S$
and $\sigma = \mathrm{sgn}[(q/\ee)\zS - 1]$.\footnote{If $(q/\ee)\zS < 1$, 
$\tpvg<0$ and $\chi<0$, but the ITG mode can still have transient growth, $t_0>0$.}  
Then the time-dependent dispersion relation \eqref{eq:tz_itg} can be manipulated into 
the following form:
\beq
\fl
\lt[1 + \sigma\sqrt{1+ \lt(1-\frac{1}{\eta_i}\rt)\chi^2}\rt]
\frac{t}{t_0} =
(2 - \chi\,\tz) [1+\tz\,\Zfn(\tz)] 
+ \lt(1-\frac{1}{\eta_i}\rt)\chi\,\Zfn(\tz).
\label{eq:itgt}
\eeq
The analogous equation for the case of pure PVG drive, 
\eqref{eq:pvgt}, is recovered when $\chi\ll1$, which 
means $\eS\ll q/\ee$ (cf.\ \eqref{eq:pure_pvg}) and $\eS\zS\ll1$. 
In this limit, the behaviour of fluctuations in the presence 
of both PVG and ITG drives is well described by 
the results of \secref{sec:longt}. Before discussing the general case, 
it is useful to consider the opposite extreme of weak velocity shear.

\subsection{Case of weak shear} 
\label{sec:weakS}

Let $\eS\gg q/\ee$ and $\eS\zS\gg1$, 
so $\chi\gg1$ (note that the same limit is also achieved for $\zS\to\infty$). 
%i.e., for very long parallel wavelengths: $\kpar\vth/S\ll k_y\rhoi$). 
Then the $\chi$ dependence falls out of \eqref{eq:itgt}: 
\beq
\frac{t}{t_0} = - \tz [1+\tz\,\Zfn(\tz)] + \Zfn(\tz),
\label{eq:weakS}
\eeq
where we have discarded the $1/\eta_i$ terms by assuming $\eta_i\gg1$. 
The transient growth termination time in this limit is, from \eqref{eq:chi}, 
\beq
t_0 = \frac{1}{2}\,\chi\,\tpvg = \frac{\eS}{2(1+\tau/Z)\sqrt{\pi(\zS^2+1)}}.
\label{eq:t0_weakS}
\eeq
When $t\ll t_0$, we find the solution of \eqref{eq:weakS} by expanding in 
$\tz\gg1$. It turns out that it consists of a large real frequency and 
an exponentially small growth rate: \eqref{eq:weakS} becomes
\beq
\fl
\frac{t}{t_0} \approx - \frac{1}{2\tz} - i\sqrt{\pi}\,\tz^2\,e^{-\tz^2}
\quad\Rightarrow\quad
\tz \approx -\frac{t_0}{2t} + 
i\,2\sqrt{\pi}\lt(\frac{t_0}{2t}\rt)^4 e^{-(t_0/2t)^2}.
\label{eq:growth_weakS}
\eeq
More generally, for finite values of $t/t_0$, 
the solution of \eqref{eq:weakS} is $\tz=\tz(t/t_0)$, with a functional 
form independent of the parameters of the problem. This solution it 
plotted in \figref{fig:weakS}. It turns out that at 
$t\approx 0.15\, t_0$, the growth rate increases sharply, 
reaches a finite maximum and then decreases towards zero, 
which it reaches at $t=t_0$, whereupon growth turns to decay.\footnote{This 
implies that perturbations first grow due to the ITG-PVG 
instability (at $St\ll1$), then slow down to exponentially small 
growth rates, then (at $St\gg1$) grow vigorously again before finally starting 
to decay at $t=t_0$. The intermediate period of virtually zero growth, 
which is a feature both of the weak-shear regime and of the general 
case (see \secref{sec:finiteS}) and 
may appear strange at first glance, can be traced to the 
$k_x$ dependence of ITG-PVG growth rates at long parallel wavelengths
--- this is explained in Appendix B.2 of Ref.~\cite{schekochihin2010subcritical}.}

%\begin{figure}[t]
%\centerline{\epsfig{file=gamma_weakS.eps,width=3.5in}}
%\caption{\label{fig:weakS} 
\myfiga{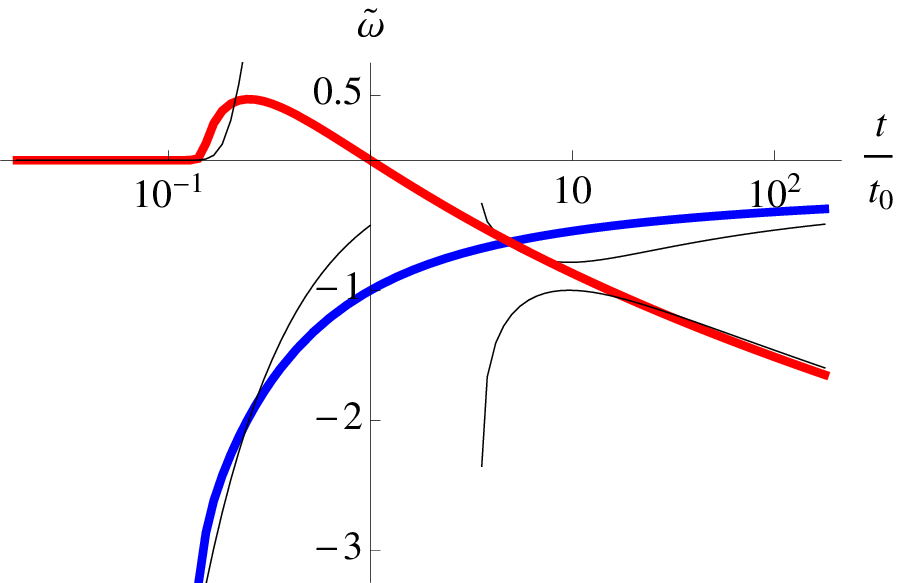}{fig:weakS}{Transient Growth with Weak Shear}{
Time evolution of the effective growth rate and frequency: 
the red (upper bold) line 
is  $\tg(t) = \Im\,\omega(t)/|\kpar|\vth\sqrt{\zS^2+1}$ 
and the blue (lower bold) line is 
$\Re\,\omega(t)/|\kpar|\vth\sqrt{\zS^2+1}$, both   
obtained as a numerical solution of \eqref{eq:weakS} and 
plotted vs.\ $t/t_0$ (the time axis is logarithmic$_{10}$). 
The black (thin) lines show the growth rate and frequency 
given by the asymptotics \eqref{eq:growth_weakS} 
and \apprefeqnb{C.9} (the latter taken in the limit 
$\chi\to\infty$ and $\eta_i\to\infty$).} 
%\end{figure}

Thus, there is a period of strong transient amplification, which lasts 
for a finite fraction of time $t_0$. The amplification exponent is 
\beq
N = \int_0^{t_0}dt\gbar(t) 
= t_0\sqrt{\zS^2+1}\int_0^1 d\xi\,\Im\,\tz(\xi)
\approx 0.057\,\frac{\eS}{1+\tau/Z},
\label{eq:N_weakS}
\eeq
where we have used \eqref{eq:t0_weakS} and calculated the value of the 
integral under the curve in \figref{fig:weakS} numerically
(note that since the growth rate is exponentially small at $t\ll t_0$, 
the precise lower integration limit is irrelevant). 
Remarkably, unlike in the case of the PVG drive (see \eqref{eq:Nmax}), 
the amplification exponent has no wavenumber dependence 
at all. Also unlike in the PVG case, it does depend on the shear and on the 
temperature gradient: $N\propto\eS=\vth/\LT S$. 

To recapitulate, 
we have found that, at low velocity shear, all modes are amplified by a large (and the same) 
factor before decaying eventually. Their transient amplification time
is given by \eqref{eq:t0_weakS}. Restoring dimensions, \eqref{eq:t0_weakS} 
and \eqref{eq:N_weakS} are
\beq
\fl
t_0 \approx \frac{\vth/(\LT S)}{2(1+\tau/Z)\sqrt{\pi(S^2k_y^2\rhoi^2 + \kpar^2\vth^2)}},
\quad
N \approx 0.057\,\frac{\vthi/(\LT S)}{1+\tau/Z}.
\label{eq:t0_weakS_dim}
\eeq
The transient growth lasts for a very long time at low $S$ and the 
longest-growing modes are the long-wavelength ones. 
The limit $S\to 0$ is singular in the sense that for arbitrarily small but 
non-zero $S$ all modes eventually decay, while for $S=0$, the indefinitely 
growing linear ITG instability is recovered ($t_0=\infty$, $N=\infty$). 

We have already made the point (in \secref{sec:UV}) that while our theory 
does not limit the transiently growing wavenumbers from above, 
a fuller description of the plasma will. 
%The same considerations apply 
%to the case of transient ITG growth. 

%\begin{figure}[t]
%\centerline{\epsfig{file=gamma_itg.eps,width=3.5in}}
\myfiga{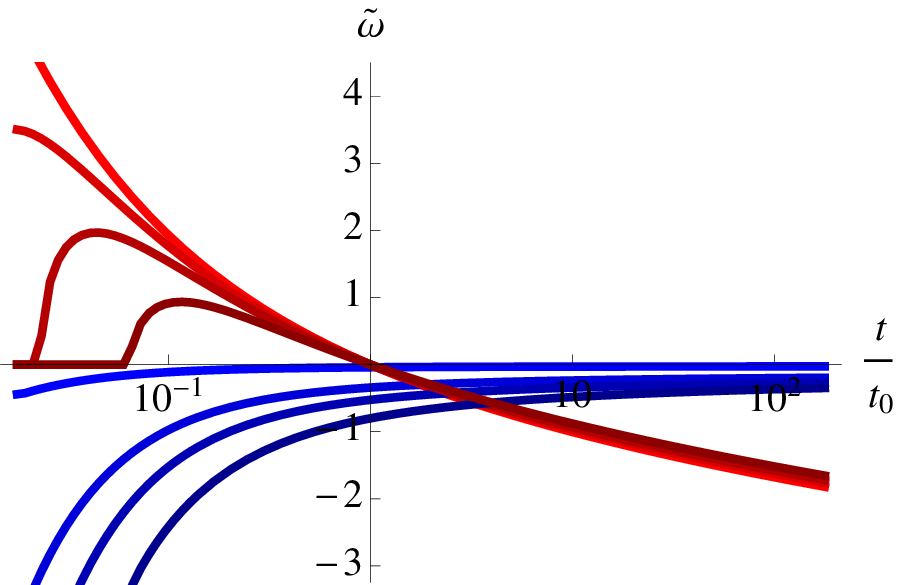}{fig:gamma_itg}{Transient Growth for Combined ITG-PVG Drive}{
Effective normalised growth rates $\Im\,\tz(t/t_0)$ (red, top) 
and frequencies $\Re\,\tz(t/t_0)$ (blue, bottom) 
obtained via numerical solution of \eqref{eq:itgt} with 
$\eta_i=5$ and $\chi=0.1,1,2,10$ (from top/lighter to bottom/darker curves).
See \appreffig{C1} for a more detailed depiction of the 
$\chi=1$ case.} 
%\end{figure}

\begin{figure}[t]
\dpanel{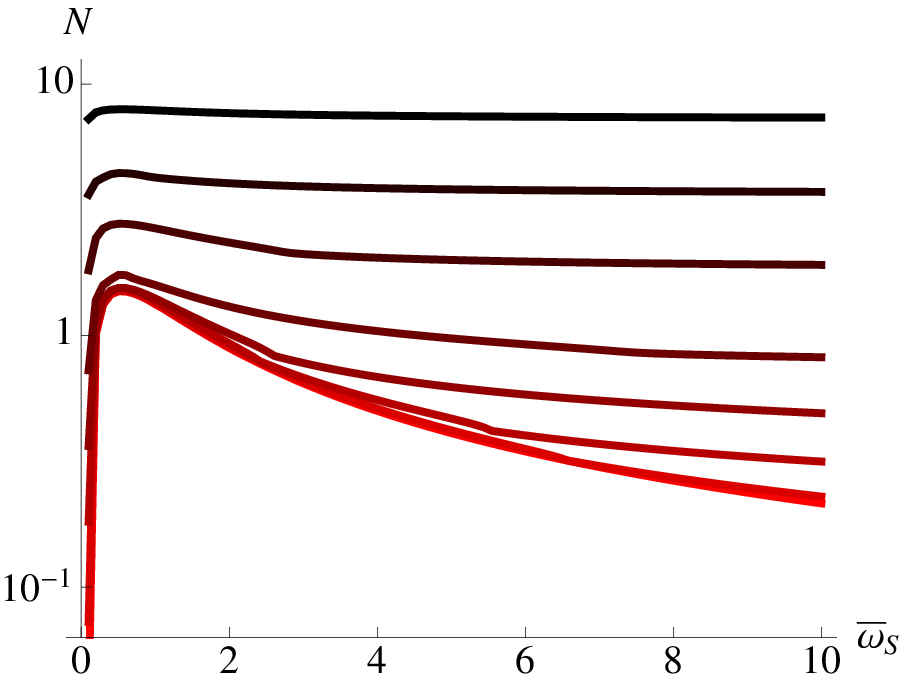}{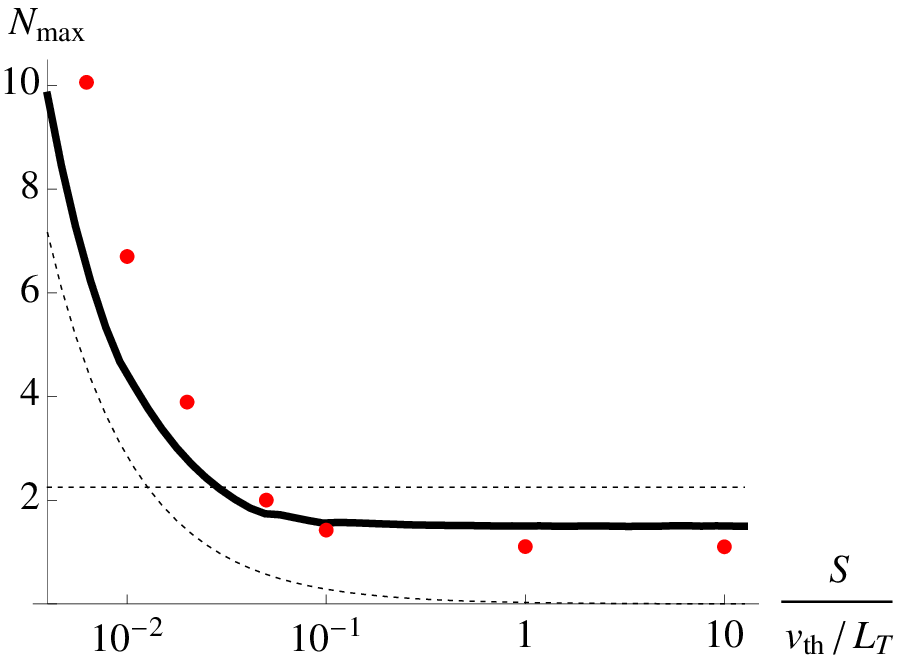}
%\centerline{\parbox{3in}{\epsfig{file=Nlines.eps,height=2.2in}}
%\parbox{3in}{\epsfig{file=Nmax.eps,height=2.2in}}}
\caption{The PVG-ITG Amplification Exponent}{\label{fig:N} The amplification exponent $N$, given by \eqref{eq:N_itg} with 
$\tz$ the solution of \eqref{eq:itgt} for $\eta_i=5$, $q/\ee=10$ and $\tau/Z=1$.  
{\em Left panel:} $N$ vs.\ $\zS$ for 
$\eS=1,2,5,10,20,50,100,200$ (from bottom to top curves, in darkening shades of red). 
The numerical results for the same cases are shown in \figref{fig:N2D}. 
The $\eS=0$ case (pure PVG drive) was virtually indistinguishable 
from $\eS=1$ when plotted (not shown here). 
The lowest-$\eS$ behaviour is well described by \eqref{eq:N}, 
the highest-$\eS$ by \eqref{eq:N_weakS} (in the latter case, except 
for corrections associated with finite $\eta_i$, which are easy 
to compute if they are required).
{\em Right panel:} The maximal amplification exponent $N$ vs.\ 
the normalised shear ($\eS^{-1}=S\LT/\vth$). 
The maximum $N(\zS)$ is reached at $\zS\approx 0.54$, independently of $\eS$. 
The dotted lines are the asymptotics 
\eqref{eq:Nmax} and \eqref{eq:N_weakS}. The finite offsets between the 
asymptotics and the exact curve are due to the fact that the asymptotics 
were calculated in the limits $\eta_i\gg1$ and $q/\ee\gg1$, while for the 
exact solution we used relatively moderate values of these parameters.
The discrete points show $\Nmax$ obtained via an \texttt{AstroGK} 
numerical parameter scan varying $\kperp\rhoi$ and $S$ (i.e., $\eS$) while 
holding $\kpar\LT=0.02$ and the rest of the parameters fixed at the same values 
as quoted above.} 
\end{figure}

%\begin{figure}[p!]
%\centerline{\epsfig{file=itg_to_pvg.eps,height=8in}}
\myfiga[6in]{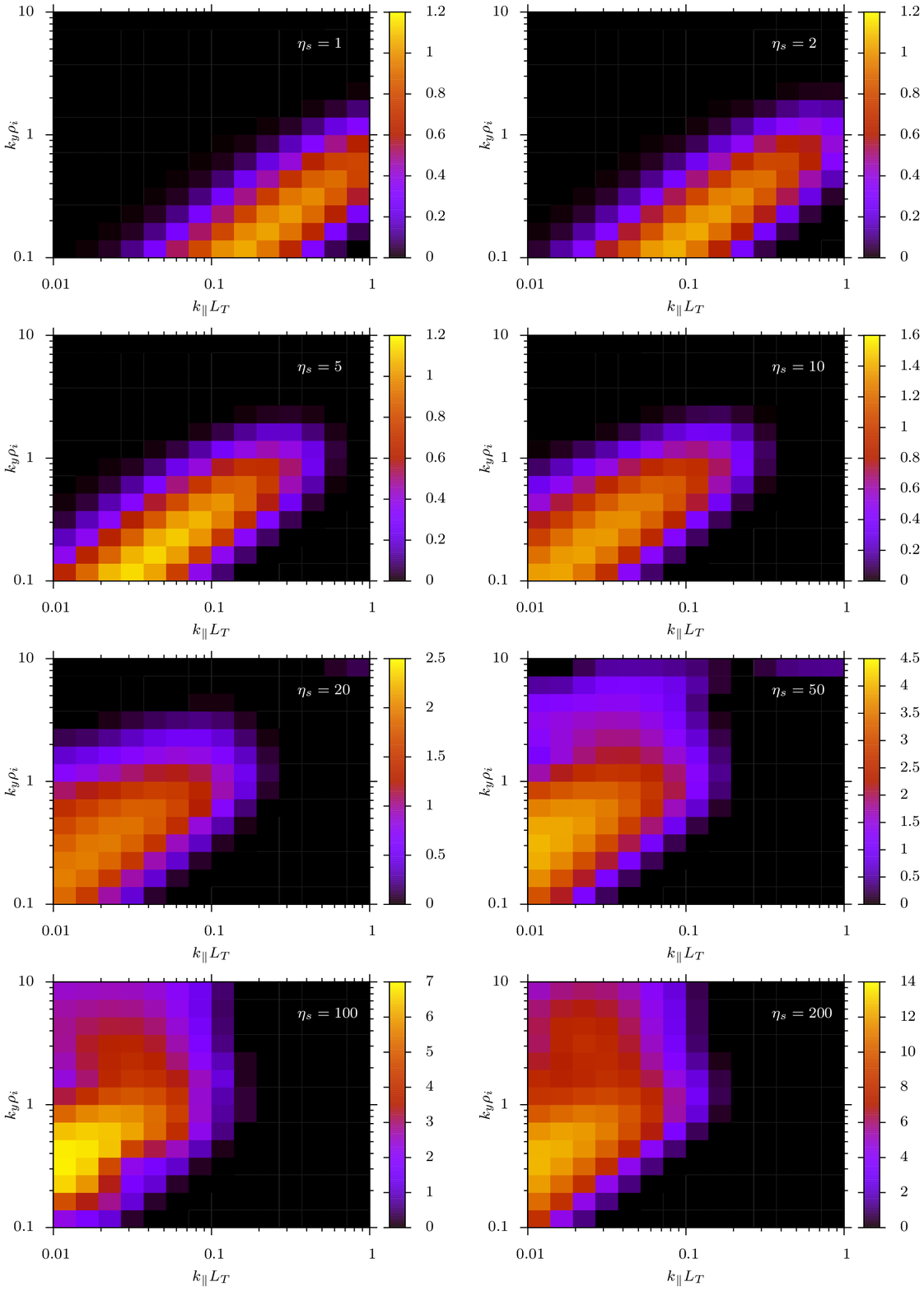}{fig:N2D}{Parameter dependence of the Transient Amplification}{
The amplification exponent $N$ vs.\ $\kpar\LT$ 
and $k_y\rhoi$, obtained numerically using \texttt{AstroGK} for the same  
parameters as the curves in the left panel of \figref{fig:N}.} 
%\end{figure}

\subsection{Case of finite shear}
\label{sec:finiteS}

In the intermediate regimes between large and small $\chi$ 
(i.e., weak and strong shear), the solutions of \eqref{eq:itgt}
transit from the weak-shear form described in \secref{sec:weakS} 
to the pure-PVG case treated in \secref{sec:longt}. 
\Figref{fig:gamma_itg} shows the time-dependent growth rates 
and frequencies for several values of $\chi$. As $\chi$ decreases 
(i.e., $S$ increases), the peak of the growth rate moves 
further into the past and the growth rate asymptotes 
to the pure-PVG case (\figref{fig:gamma}). 
It is not hard to convince oneself analytically that this is indeed 
what ought to happen. Further asymptotic considerations on this subject are given in Appendix C of Ref.~\cite{schekochihin2010subcritical}. 
The long-time decay asymptotic is also derived there (it is exactly 
analogous to that found in \secref{sec:decay}).

Similarly to our previous calculations, the amplification exponent is 
\beq
N(\zS) %= \int_0^{t_0}dt\gbar(t) 
= t_0(\zS)\sqrt{\zS^2+1}\int_0^1 d\xi\,\Im\,\tz(\xi,\chi(\zS)),
\label{eq:N_itg}
\eeq
where $\tz$ is the solution of \eqref{eq:itgt}. 
The wavenumber dependence enters via the $\zS$ dependence 
of $t_0$ and of $\chi$ (see \eqref{eq:chi}).
The numerically computed amplification exponent as a function 
of $\zS$ and of $\eS$ is plotted in the left panel of \figref{fig:N} 
for $\eta_i=5$ and $q/\ee=10$ 
(these are representative of the values encountered 
in the more realistic numerical studies of tokamak 
transport \cite{barnes2011turbulent,highcock2010transport,highcock2011transport}). 
The pure-PVG case treated in \secref{sec:longt} remains 
a good approximation up to values of $\eS$ of order $10$. 
After that, there is a transition towards the weak-shear limit 
(\secref{sec:weakS}), accompanied by the 
loss of wavenumber dependence as $\eS$ 
is increased to values of order $100$. 
The constant of proportionality between $k_y$ and $\kpar$ 
for the maximally amplified modes (i.e., $\zS$ at which $N$ is 
maximised) does not appear to depend on $\eS$, although 
at large $\eS$, the maximum is increasingly weak.  
The maximal amplification exponent is plotted in the 
right panel of \figref{fig:N}. It is perhaps worth pointing 
out the qualitative similarity between this plot and figure~1 of 
Ref.~\cite{highcock2010transport}, obtained from gyrokinetic simulations 
in full tokamak geometry.  

Finally, the amplification exponent as a function of $k_y$ and $\kpar$, 
obtained in direct (linear) numerical simulations, is shown in \figref{fig:N2D}
(the parameters are the same as in the ``theoretical'' \figref{fig:N}, left panel).  
The transition from the PVG curve \eqref{eq:Nmax} to the flat wavenumber 
dependence \eqref{eq:N_weakS} is manifest, as are the limits of 
applicability of our approximations in the wavenumber space. Note 
the different normalisation of the parallel wavenumber here ($\kpar\LT$, characteristic 
of ITG) compared to the left panel of \figref{fig:kspace} ($\kpar\vthi/S$, characteristic 
of PVG). Hence the drift towards higher $\kpar\LT$ as $\eS=\vthi/S\LT$ decreases 
towards the PVG-dominated regime, where the parallel scale of maximally amplified 
modes is set by the shear rather than the temperature gradient.

%\section{Discussion}
%\label{sec:conc}

%Our main findings are already itemised in a succinct way in the Abstract. 
%Here we give a qualitative summary and discuss possible implications 
%for turbulence and transport. 

\section{Qualitative summary of the linear results}
\label{sec:qualit} 

In a gyrokinetic plasma with radial gradients of temperature and parallel velocity, 
both gradients are sources of free energy and so will drive the growth of fluctuations 
(ITG and PVG instabilities). The typical growth rate is of order $\gamma\sim \vth/\LT$ for ITG  
and $\gamma\sim qS/\ee$ for PVG (see \eqref{eq:gamma_gen_pvg}), 
or the mean square of the two if they are comparable (see \apprefeqnb{B.10}). 
Because the mean plasma velocity is toroidal, it always has both a parallel and 
a perpendicular component (the latter a factor of $q/\ee$ smaller than the former). 
The shear in the perpendicular ($\vE\times\vB$) velocity is stabilising and causes 
all modes to decay eventually, so the fluctuation growth is transient --- it is always transient 
in the limit, considered here, of zero magnetic shear and it is transient for large enough 
velocity shear $S$ when the magnetic shear is finite \cite{newton2010understanding,barnes2011turbulent}. 
If the linear physics provides sufficiently vigorous and lasting amplification of finite initial 
perturbations, 
the system is able to sustain nonlinearly a saturated (subcritical) turbulent state
(see Chapter \ref{sec:turbulencewithoutinstability}). Therefore, the interesting question is how much transient amplification 
should be expected to occur and on what time scale. 

In the preceding sections, we have addressed this question mathematically, with the 
results summarised by \figsref{fig:kspace}, \ref{fig:N} and \ref{fig:N2D} 
(see also \eqref{eq:Nmax}, \eqref{eq:t0_max} and \eqref{eq:t0_weakS_dim}). 
Very roughly, these results can be explained as follows. The effect of the perpendicular shear 
is to produce a secular increase with time of the radial wavenumber, 
$k_x(t)\sim Sk_y t$. When this becomes large enough, the instability is killed by 
Landau damping (see discussion at the end of Appendix B.1 of Ref.~\cite{schekochihin2010subcritical}). 
If we estimate that this happens after $t_0\sim S^{-1}$ (i.e., 
for $k_x(t_0)\rhoi\sim1$, assuming $k_y\rhoi\sim1$), 
we may conclude that initial perturbations will be amplified by a factor 
of $e^N$, where the amplification exponent is 
\beq
N\sim\gamma t_0 \sim \frac{\vthi}{\LT S}~~\mathrm{for~ITG\quad and}\quad  
N\sim \frac{q}{\ee}~~\mathrm{ for~PVG}.
\eeq
Thus, the shear quenches the ITG amplification 
--- but $N$ cannot fall below the shear-independent level associated with the PVG (\figref{fig:N}, right panel). 
This is indeed the case (see \eqref{eq:Nmax} and \eqref{eq:t0_weakS_dim}), 
although, strictly speaking, one has to take into account the dependence of the quenching effect 
on the perpendicular and parallel wavenumbers --- long-wavelength modes grow more slowly, 
but for a longer time; in the case of PVG, there is also a preferred relationship 
$k_y\rhoi\sim (\ee/q)^{1/3}\kpar\vthi/S$ for the most strongly amplified modes 
(see \secref{sec:longt} and \figref{fig:N2D}). While these wavenumber dependences 
are likely to be important in the analysis of the resulting turbulent state and the associated 
transport, they effectively cancel out in the expression for the amplification exponent
(because $\gamma\propto k_y\rhoi$, $t_0\propto 1/k_y\rhoi$ at long wavelengths) 
and the results of the qualitative argument that we have given hold true.  

It is instructive to compare these results with the conclusions of a long-wavelength 
fluid ITG-PVG theory presented in \cite{newton2010understanding} (for the case of finite magnetic shear). 
In that regime, perpendicular shear, by effectively increasing $k_x(t)$, also caused eventual damping of 
the fluctuations, but this time via collisional viscosity. Therefore, to estimate the 
transient growth time $t_0$, one must set $\gamma \sim \nuii k_x^2(t_0)\rhoi^2 
\sim \nuii S^2k_y^2\rhoi^2t_0^2$, where $\nuii$ is the ion collision rate. 
Then, ignoring wavenumber dependences again,
$t_0\propto\gamma^{1/2}S^{-1}$, so the amplification exponent is 
\beq
N\sim\gamma t_0\propto\frac{\gamma^{3/2}}{S}
\propto\frac{1}{S}~~\mathrm{ for~ITG\quad and}\quad  
N \propto \sqrt{S}~~\mathrm{ for~PVG}.
\eeq
Thus, the $\vE\times\vB$ velocity shear again quenches the ITG instability, 
but once $S$ is large enough for the PVG drive to take over, the amplification exponent
actually grows as $\sqrt{S}$, the result obtained rigorously by \cite{newton2010understanding} --- in contrast 
with the shear-independent $N\sim q/\ee$ that we have found in the kinetic regime. 
The practical conclusion from this is that it should be easier 
to obtain states of reduced transport \cite{barnes2011turbulent,highcock2010transport,parra2010plausible} 
in weakly collisional, kinetic plasmas.

\chapter{The Zero-Turbulence Manifold}
\label{sec:zeroturbulencemanifold}
\newcommand{\kapdefeq}{\kappa =}
\newcommand{\kapcom}{\kappa}
\newcommand{\kapcritcom}{\kappa_c}
\begin{quote}
	\centering
	\emph{
	This chapter is published as Ref.~\cite{highcock2012zero}.}
\end{quote}

 At zero magnetic shear, the turbulence is subcritical for all nonzero values of the flow shear: there are no linearly unstable eigenmodes,
 and sustained turbulence is the result of nonlinear interaction between linear modes which grow only transiently before decaying.
 In Chapter \ref{sec:subcriticalfluctuations},
which studied this transient growth in slab geometry,
it was demonstrated that at large velocity shears the maximal amplification 
exponent of a transiently growing perturbation before it decays is proportional
to the ratio of the PVG to the perpendicular flow shear.
In a torus, this quantity is equal to the ratio of the toroidal to poloidal magnetic field components, or $q/\epsilon $, where $q $ is the magnetic safety factor and $\epsilon $ is the inverse aspect ratio.
Therefore, if we conjecture that a certain minimum amplification exponent is required for sustained turbulence, the results of Chapter \ref{sec:subcriticalfluctuations} predict that there should be a value of $q/\epsilon $ below which the PVG drive is rendered harmless.
Below that value of $q/\epsilon $, it should be possible to maintain an arbitrarily high temperature gradient
without triggering turbulent transport provided a high enough perpendicular flow shear can be achieved.

In this chapter, motivated by the possibility of reduced transport at low values of $q/\epsilon$,
we use nonlinear simulations to map out \emph{the zero-turbulence manifold},
the surface in the parameter space that divides the regions where turbulent transport can and cannot be sustained.
The parameter space we consider is ($\gamma_E $, $q/\epsilon $, $\kapcom $). As ever, we set the magnetic shear to zero, the regime we expect to be most amenable to turbulence quenching by shear flow \cite{barnes2011turbulent, highcock2010transport, highcock2011transport, parra2010plausible, vries2009internal, sips1998operation, mantica2011ion}.

We discover that reducing $q/\epsilon $ is indeed uniformly beneficial to maintaining high temperature gradients in a turbulence-free regime,
and that values of $\kapcom$ can be achieved
that are comparable to those experimentally observed for internal transport barriers \cite{vries2009internal,field2011plasma}.

In the next sections, having presented our methodology, we will describe these results and discuss their physical underpinnings, as well as their implications for confinement in a toroidal plasma. We will show that linear theory of subcritical fluctuations \cite{schekochihin2010subcritical} can, with certain additional assumptions, provide good predictions of the nonlinear results.

\myfig{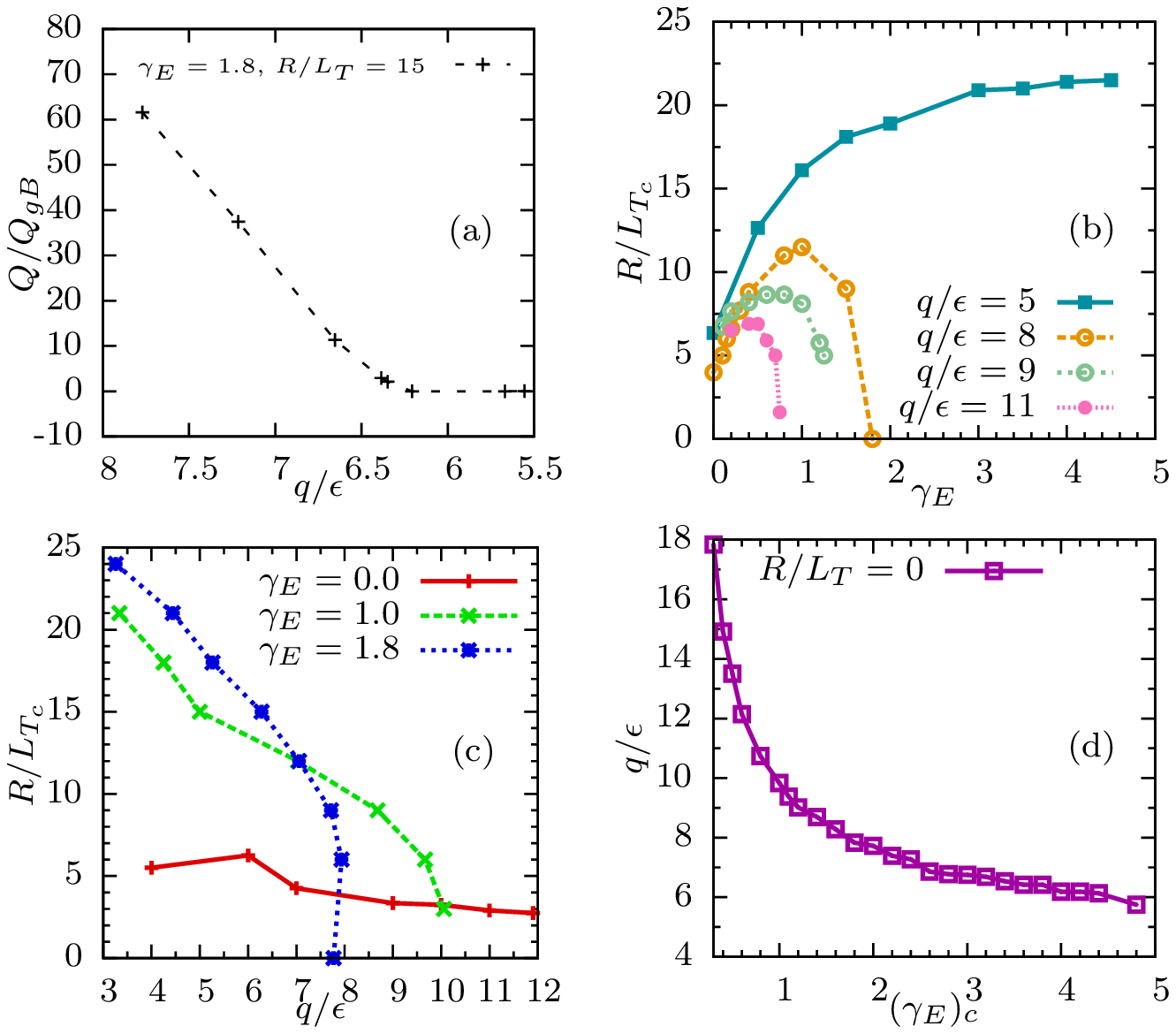}{Sections of the Zero Turbulence Manifold}{(a) The simulations used to find the point on the manifold $\gamma_E=1.8$, $\kapcom=15$, $q/\epsilon=6.3$, showing the heat flux vs. $q/\epsilon$ at ($\gamma_E=1.8$, $\kapcom=15$). The point on the manifold is the point where the heat flux drops to zero.  (b-d) Sections through the critical manifold with parameters as indicated.
Turbulence cannot be sustained for $\kapcom<\kapcritcom$ in (b,c), or for $\gamma_E<\gamma_{Ec}$ in (d).
The data points were found as illustrated in (a), and used to generate the manifold shown in \figref{criticalcurve3dall}.
}
\section{Finding the Boundary}
We wish to determine, in a three-dimensional parameter space ($\gamma_E$, $q/\epsilon$, $\kapcom$),  the boundary  between the regions where turbulence can and cannot be sustained nonlinearly.
We cover this space using four scans with constant $q/\epsilon $  (\figref{projections}(b)), three scans with constant $\gamma_E$ (\figref{projections}(c)) and one scan with  constant $\kapcom$ (\figref{projections}(d)).
For each of these cases, we consider multiple values of a second parameter and
find the value of the third parameter corresponding to the zero-turbulence boundary.
The boundary is defined as the point where both the turbulent heat flux and the turbulent momentum flux vanish.
Thus, the location of each single point on the boundary is determined using on the order of ten nonlinear simulations.
An example of this procedure is shown in \figref{projections}(a).
In total, we performed more than 1500 simulations to produce the results reported below.

Because the turbulence that we are considering is subcritical, there is always a danger that a simulation might fail to exhibit  a turbulent stationary state because of an insufficient initial amplitude
\cite{baggett1995mostly,kerswell2005recent}.
As we are not here concerned with the question of critical initial amplitudes
we will consider a given set of parameters to correspond to a turbulent state if such a state can be sustained starting with a large enough perturbation.
Therefore, all simulations are initialised with high-amplitude noise.
They are then run to saturation; close to the boundary a simulation may need to run for up to $t\sim 1000 R/v_{thi}$ to achieve this.

The critical curves obtained in this manner are plotted in \figrefs{projections}(b-d).
These curves, which effectively give the critical temperature gradient $\kapcritcom$ as a function of $\gamma_E$ and $q/\epsilon$,
are then used to interpolate a surface, the zero-turbulence manifold, plotted in \figref{criticalcurve3dall}.
The interpolation is carried out using radial basis functions with a linear kernel \cite{buhmann2001radial} (see Chapter \ref{transitionsec}).

\myfig{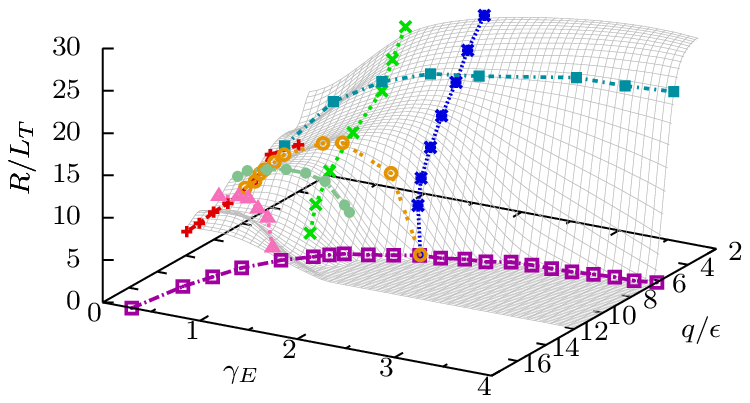}{The Zero Turbulence Manifold}{ The zero-turbulence manifold. Turbulence can be sustained at all points outside the manifold (that is, at all points with a higher temperature gradient and/or higher value of $q/\epsilon$ than the nearest point on the manifold).  This plot is made up from the sections shown in \figref{projections}(b-d) (heavy lines) and the manifold interpolated from them (thin grey mesh).}
\myfig{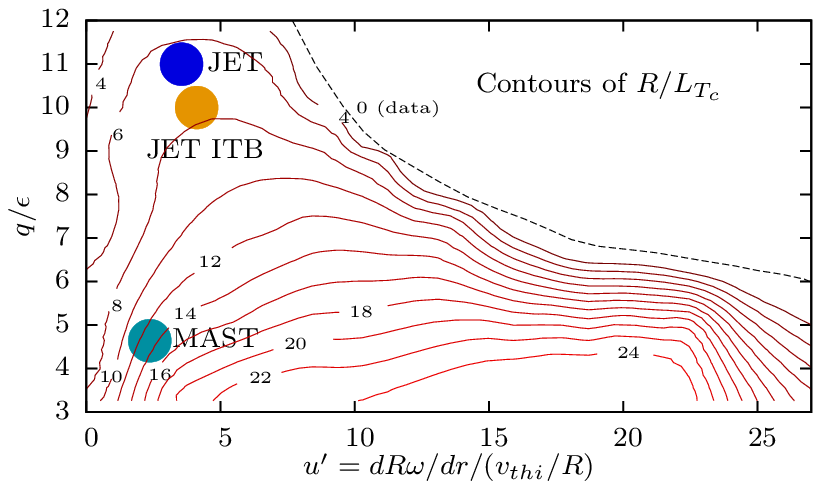}{Contours of the Zero Turbulence Manifold}{Contours of the zero-turbulence manifold plotted against the toroidal flow shear $u'=dR\omega/dr / (v_{thi}/R)=\gamma_E/(q/\epsilon)$. The contours indicate the value $\kapcom=\kapcritcom$ below which turbulence is quenched.
From top to bottom, the circles indicate approximate values of $u'$ and $q/\epsilon$, corresponding to Ref.~\cite{mantica2011ion} (JET; $R/L_{T}\sim 8$), Ref.~\cite{vries2009internal} (JET ITB; $R/L_{T}\sim 17$; note large discrepancy, see text) and Ref.~\cite{field2011plasma} (MAST ITB; $R/L_T\sim 10$).
}

\section{The Zero-Turbulence Manifold} The results of the scan described above are displayed in \figrefs{projections}(b-d). These three figures show, at fixed values of either $\gamma_E $, $\kapcom $ or $q/\epsilon $, the threshold in either $\kapcom $ or $q/\epsilon $ below which turbulence cannot be sustained; they are, in effect, sections through the zero-turbulence manifold.

Considering first \figref{projections}(b), we see that, at fixed $q/\epsilon $, the critical gradient  $\kapcritcom $ first rises with $\gamma_E $, as the perpendicular flow shear suppresses the ITG-driven turbulence, and then falls --- in most cases to 0 --- as the PVG starts to drive turbulence instead. This phenomenon was discussed at length in Chapters \ref{fluxessec} and \ref{qbehaviour}.
 Thus, for every $q/\epsilon $, there is an optimum value of the perpendicular flow shear $\gamma_E $
(and hence of the toroidal shear $u' $) for which the critical temperature gradient $\kapcritcom $ is maximised.
We see that reducing
$q/\epsilon$
increases
the maximum $\kapcritcom$ that can be achieved without igniting turbulence.
\figref{projections}(c) shows that this rule applies for all considered values of flow shear.\footnote{
The increase at $\gamma_E=0 $ cannot, of course, be due to reduction of the PVG;
we assume that this occurs because of the simultaneous reduction of the maximum parallel length scale $qR$ in the system, leading to weaker ITG turbulence; see \cite{barnes2011critically}.}
This is to be expected, because lower $q/\epsilon $ means weaker PVG relative to the perpendicular shear,
allowing higher values of the perpendicular flow shear to suppress the ITG before the PVG drive takes over.

Lastly, \figref{projections}(d) shows the threshold in $\gamma_E $ above which the PVG can drive turbulence alone, without the help of the ITG;
in other words, even configurations with a flat temperature profile would be unstable.
At very high $q/\epsilon $, already a very small flow shear will drive turbulence; as $q/\epsilon $ decreases, higher and higher values of $\gamma_E $ are required for the PVG turbulence to be sustained.
It cannot be conclusively determined from this graph whether, 
as suggested by linear theory,
there is 
a finite critical value of $q/\epsilon $ below which PVG turbulence cannot be sustained, i.e., a nonzero value of $q/\epsilon $ corresponding to $\gamma_{Ec}\rightarrow \infty $.
However, for $q/\epsilon\lesssim 7  $, the critical $\gamma_E $ is far above what might be expected in an experiment,\footnote{
By order of magnitude, $\gamma_E\sim M/q$, where $M$ is the Mach number of the toroidal flow. Thus, values of $\gamma_E$ much above unity are unlikely to be possible.
} and so the $\gamma_E\rightarrow\infty$ limit is somewhat academic.
A definite conclusion we may draw is that at experimentally relevant values of shear, pure PVG-driven turbulence cannot be sustained for $q/\epsilon\lesssim 7$.

The zero-turbulence manifold interpolated from the numerical data points is displayed in \figref{criticalcurve3dall}.
The manifold comprises three main features: a ``wall'' where the critical temperature gradient increases dramatically at low $q/\epsilon $; a ``spur'' at low $\gamma_E $, jutting out to high $q/\epsilon $
(where, as $\gamma_E$ increases, the ITG-driven turbulence is suppressed somewhat before the PVG drive becomes dominant),
and finally the curve where the manifold intercepts the plane $\kapcom=0 $, whose shape is described above.

\section{Practical Implications and Comparison with Experiment} 
In order to illustrate better the implications of our findings for confinement,
we plot, in \figref{manifoldcontoursqedwdr}, contours of
$\kapcritcom$ versus $q/\epsilon $
and the toroidal flow shear $u' = dR\omega/dr/(v_{thi}/R) $.
The basic message is clear:
the lower the value of $q /\epsilon $, the higher the temperature gradient that can be achieved without igniting turbulence.
Once we have obtained the lowest possible value of $q/\epsilon $,
there is an optimum value of $u' $ which will lead to that maximum $\kapcritcom $.
We note that the dependence of this optimum value of $u' $ on $q/\epsilon $
is not as strong as the dependence of the optimum value of $\gamma_E$ on $q/\epsilon$
(clearly this must be so because $u'=(q/\epsilon)\gamma_E $).
In a device with an optimised value of $q/\epsilon $, a near maximum critical temperature gradient would be achievable for $u'\gtrsim 5$, shears comparable to those observed in experiment \cite{mantica2011ion,field2011plasma,vries2009internal}.

%[``Mantica2011'', ``vth'', 380120.0735545084, ``uprim'', 3.5515093622289657, ``r_over_lT'', 8.749999999999996, ``q/eps'', 11.35135135135135, ``eps'', 0.12333333333333334]
%[``Vries2009'', ``vth'', 821153.2103221639, ``uprim'', 4.110073440102465, ``r_over_lT'', 17, ``q/eps'', 10.18867924528302]
%[``Field2011'', ``vth'', 294439.742, ``uprim'', 2.3720055301107217, ``r_over_lT'', 10, ``q/eps'', 4.642857142857142]

While simulation results obtained for Cyclone Base Case parameters are not suitable for detailed quantitative comparison with real tokamaks, it is appropriate to ask whether our results are at all compatible with experimental evidence.
For an internal transport barrier (ITB) in MAST, Ref.~\cite{field2011plasma} reports $R/L_T\sim 10 $ at $q/\epsilon\sim 4.6 $ 
and $u'\sim 2.4 $.\footnote{
The ratio of toroidal to poloidal field in MAST can be smaller on the outboard side, so the effective value of $q/\epsilon$ for locating this case on the zero-turbulence manifold might be smaller than quoted.}
This is comparable to the critical values shown in \figref{manifoldcontoursqedwdr}. In JET, Ref.~\cite{mantica2011ion} reports $R/L_T\sim8 $ at $q/\epsilon\sim 11 $ and $u'\sim 3.6 $, again reasonably close to what we would have predicted.
However, an ITB in JET studied by Ref.~\cite{vries2009internal} achieved $R/L_T\sim17 $ at $q/\epsilon\sim10 $ and  $u'\sim4.1 $ --- substantially higher than our $R/L_{Tc} $ at the same values of $q/\epsilon $ and $u' $. Note, however, that Ref.~\cite{vries2009internal}
reports that the shear was dominated by an enhanced poloidal flow, an effect which is not included in our numerical model.

\section{Relation to Linear Theory} Since the mapping of the zero-turbulence manifold using nonlinear simulations is computationally expensive, we may ask whether linear theory can predict marginal stability.
 The question is also interesting in terms of our theoretical understanding of subcritical plasma turbulence.
It is clear that in a situation where perturbations grow only transiently,
existing methods based on looking for marginal stability of the fastest growing eigenmode will not be applicable.
In Chapter \ref{sec:subcriticalfluctuations}, we considered these transiently growing modes in a sheared slab, and posited a new measure of the vigour of the transient growth: $N_{\mathrm{max}} $, the maximal amplification exponent, defined as the number of e-foldings of transient growth a perturbation experiences before starting to decay, maximised over all wavenumbers.
 It appears intuitively clear that in order for turbulence to be sustained,
 transient perturbations must interact nonlinearly before they start to decay.
 We may then assume that a saturated turbulent state will exist if $N_{\mathrm{max}}\gtrsim N_c $, where $N_c $ is some threshold value of order unity. The zero-turbulence manifold is then the surface $N_{\mathrm{max}}(\gamma_E,q/\epsilon,\kapcom)=N_c $.

 We now test this idea by calculating $N_{\mathrm{max}} $ for linear ITG-PVG-driven transient perturbations in a slab, using the code \texttt{AstroGK} \cite{numata2010astrogk} to solve the linearised gyrokinetic equation, as done in Chapter \ref{sec:subcriticalfluctuations}. \figref{fit}(a) shows that for each value of $\gamma_E $ and a range of $q/\epsilon $, it is possible to choose $N_c(\gamma_E) $ such that the equation $N_{\mathrm{max}}(\gamma_E,q/\epsilon,\kapcom)=N_c $ correctly reproduces the critical curve $\kapcritcom(q/\epsilon) $ obtained as a section of the zero-turbulence manifold at that value of $\gamma_E $.
However, $N_c $ does have a strong dependence on $\gamma_E $, shown in \figref{fit}(b), ranging from $N_c\lesssim 0.5 $ at $\gamma_E \gtrsim 2 $ to $N_c \rightarrow \infty $ as $\gamma_E\rightarrow 0 $ (the latter is an expected result: at $\gamma_E=0 $, there is a growing eigenmode, so either $N_{\mathrm{max}}=\infty $ or there is no growth at all).
It is not clear if $N_c $ tends to a finite limit as $\gamma_E\rightarrow \infty $, but, similarly to the existence of a critical value of $q/\epsilon $ as $\gamma_E\rightarrow\infty $, this is a somewhat academic question because such a limit would be achieved (or not) at $\gamma_E $ too large to be experimentally achievable.

 The practical conclusion of this exercise is that
all that appears to be required to determine the two-dimensional dependence of $\kapcritcom $ on $\gamma_E $ and $q/\epsilon $ is finding $N_c(\gamma_E) $ using a nonlinear scan at a single value of $q/\epsilon $; thus, the number of parameters in the nonlinear scan is reduced by one.

\myfig{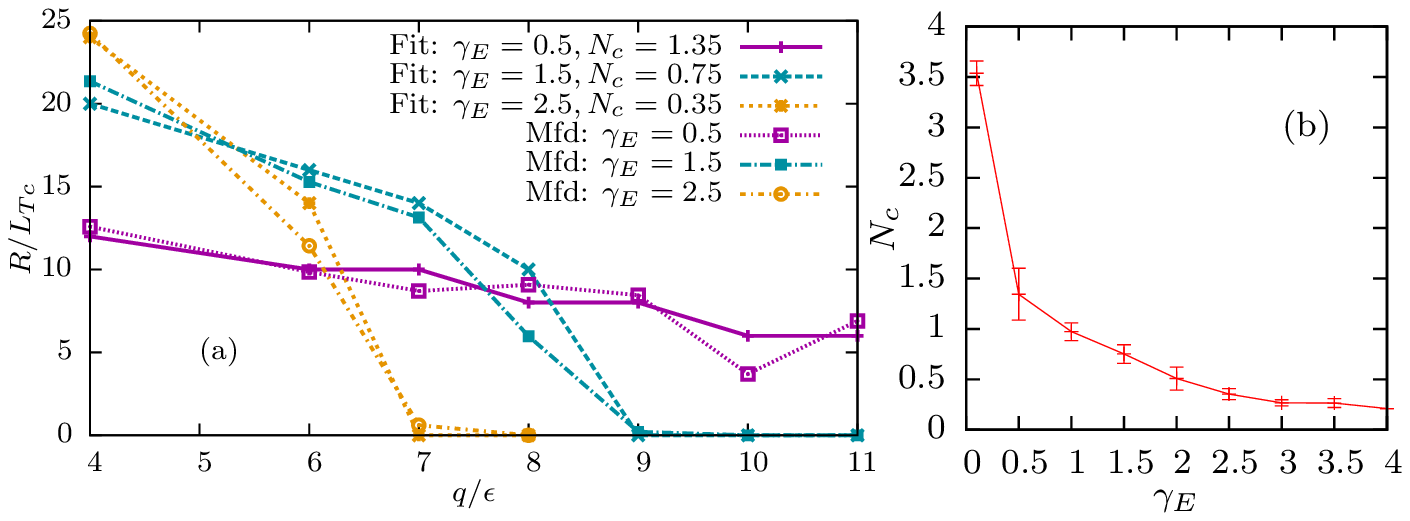}{Linear Calculation of the Zero Turbulence Manifold}{(a) The critical temperature gradient $\kapcritcom$ vs. $q/\epsilon$ for different values of $\gamma_E$, showing both $\kapcritcom$ obtained from the interpolated manifold, and $\kapcritcom$ such that $N_{\mathrm{max}} = N_c$, with $N_c$ suitably chosen for each $\gamma_E$, as shown in (b).} 
\section{Discussion}  In this chapter we have presented two key results.
Firstly, and principally, we have calculated
the shape of the zero-turbulence manifold, the surface that divides the regions in the parameter space ($\gamma_E $, $q/\epsilon $, $\kapcom $) where subcritical turbulence can and cannot be nonlinearly sustained.
We have described the shape of this manifold and its physical origins, and presented its two implications for confinement in toroidal plasmas: that reducing the ratio $q/\epsilon $,
i.e.,
increasing the ratio of the poloidal to the toroidal magnetic field,
improves confinement at every nonzero value of $\gamma_E $, and that at fixed $q/\epsilon $, there is an optimum value of $\gamma_E $ (that is, an optimum value of the toroidal flow shear $u'= dR\omega/dr/(v_{thi}/R) $) at which the critical temperature gradient is maximised,
 in some instances to values comparable 
to those observed in internal transport barriers \cite{vries2009internal,field2011plasma}.
How to calculate the heat and momentum fluxes that would need to be injected
in order for such optimal temperature gradients to be achieved
was discussed in Ref.~\cite{parra2010plausible}.

Secondly, we have shown that the zero-turbulence manifold can be parameterised as $N_{\mathrm{max}}(\gamma_E, q/\epsilon, \kapcom) = N_c(\gamma_E) $, where $N_{\mathrm{max}} $ is the maximal amplification exponent of linear transient perturbations (calculated from linear theory) and $N_c $ must be fit to the data.
Thus, using a single scan at constant $q/\epsilon $ to determine $N_c(\gamma_E) $ appears to be sufficient for calculating the full two-parameter dependence of the critical temperature gradient.
Obviously, the need to fit $N_c(\gamma_E) $ indicates a limitation of our current theoretical understanding of the criterion for sustaining subcritical turbulence in a sheared toroidal plasma.
The results reported here provide an empirical constraint on future theoretical investigations.

\part{Summary and Discussion}
\label{sec:summaryanddiscussion}
\chapter{Summary and Discussion}

 We have presented several key results in this thesis.
We have used nonlinear simulations to show that at zero magnetic shear, a sheared toroidal flow can completely suppress turbulence across a range of flow and temperature gradients,
and can lead to a transport bifurcation from low to high temperature gradients.
We have identified subcritical PVG-driven turbulence as being a factor limiting the temperature gradients that can be achieved in such a transition.
We have investigated the properties of the transient linear modes which underly this turbulence,
and we have shown that its virulence can be diminished by reducing the ratio of the destabilising parallel velocity gradient to the stabilising perpendicular gradient (which in a tokamak is equal to $q/\epsilon $).
We have used nonlinear simulations to calculate, as a function of the parameters ($\gamma_E $, $q/\epsilon $, $R/L_T $), the surface dividing the regions where subcritical turbulence can and cannot be sustained, and from this result we have shown that for low enough values of $q/\epsilon $, logarithmic temperature gradients $R/L_T > 20 $ can be maintained without generating turbulence.

From these results there naturally arise to further lines of enquiry: practical and theoretical. We begin with the practical. From the results given above we may infer that tokamak confinement could be greatly increased in the limit of low magnetic shear, low $q/\epsilon $ and high perpendicular flow shear.
However, since all three of these quantities must necessarily be a function of the minor radius, and since these parameters may only be controlled indirectly in a fusion device, it would be necessary to use a 1-dimensional turbulent transport solver such as \texttt{TRINITY} \cite{barnes2009trinity}, with proper models of momentum and heat input, and a more realistic magnetic geometry, to demonstrate more conclusively that such a performance gain could be realised.

Another avenue for future investigations is determining the dependence of the zero turbulence boundary on some of the parameters that were held fixed in this work:
$T_i/T_e$,
magnetic shear, and, more generally, the shape of the flux surfaces, density gradient, inverse aspect ratio $\epsilon $ (separately from $q $), etc. Mapping out the dependence just on $\gamma_E $, $q/\epsilon $ and $\kapcom $
took approximately 1500 nonlinear simulations at a total cost of around 4.5 million core hours. Adding even two or three more parameters to the search would take computing requirements beyond the limit of resources today, but not of the near future.

Theoretically, we have demonstrated the existence, at high flow shear, of subcritical turbulence: turbulence which exists in the absence of linear eigenmodes.
We have described the transiently growing modes which give rise to such turbulence, and developed a criterion based upon them which had been partially successful in determining under which circumstances they may give rise to turbulence nonlinearly. The theoretical challenge which naturally arises is that of developing a full picture of the mechanism of the nonlinear interaction between the modes and consequently a robust criterion for when turbulence can and cannot be sustained.
If such a criterion were to be developed, it would lend the ability to model high-flow-shear subcritical turbulence to the quasi linear mixing-length-based models which are the workhorse of transport modelling today, as well as being a significant advance in our theoretical understanding of subcritical turbulence.

It is clear that there are many questions arising from this work which need to be considered. However, these questions aside, we believe that we have provided strong evidence that sheared toroidal flows, with a judicious choice of magnetic geometry, can be used to achieve high confinement of energy, and hence high performance, in fusion devices.

%\include{thesis-ch1}
%\include{thesis-ch2}
%\include{thesis-ch3}

%\appendix

%\include{thesis-appendix}

\backmatter

\clearpage

{
\raggedright % justification likes to break URLs

% Hack: point the PDF link at the top of the page, rather than below
% the title as \bibliography does.  Yeah, this is ridiculously pedantic
% and will (hopefully) not affect the printed page at all... :-)
\phantomsection
\addcontentsline{toc}{chapter}{Bibliography}
\nobibintoc
\bibliographystyle{unsrt}
\bibliography{references}

\begin{thebibliography}{100}

\bibitem{rebhan2002thermonuclear}
E.~Rebhan and G.~Van~Oost.
\newblock Thermonuclear burn criteria.
\newblock {\em Transactions of Fusion Science and Technology}, 41(2T):16--26,
  2010.

\bibitem{maisonnier2007power}
D.~Maisonnier, D.~Campbell, I.~Cook, L.~Di~Pace, L.~Giancarli, J.~Hayward,
  A.~Li~Puma, M.~Medrano, P.~Norajitra, M.~Roccella, et~al.
\newblock Power plant conceptual studies in {Europe}.
\newblock {\em Nucl. Fusion}, 47:1524, 2007.

\bibitem{uknationalstatistics}
{Energy consumption in the United Kingdom, 2011}.
\newblock {\em UK National Statistics}, 2012.

\bibitem{internationalenergyagency}
Key world energy statistics 2010.
\newblock {\em The International Energy Agency}, 2010.

\bibitem{cook2001safety}
I.~Cook, G.~Marbach, L.~Di~Pace, C.~Girard, and N.~P. Taylor.
\newblock Safety and environmental impact of fusion.
\newblock {\em European Fusion Development Agreement (EFDA) Report EFDA-S-RE-1.
  EUR (01) CCE-FU$\backslash$ FTC}, 8(5), 2001.

\bibitem{ongena2010prospects}
J.~Ongena and G.~Van~Oost.
\newblock Prospects for fusion power.
\newblock {\em Transactions of Fusion Science and Technology}, 41(2T):3--15,
  2010.

\bibitem{alexandroff1935topologie}
P.~Alexandroff and H.~Hopf.
\newblock {\em Topologie: Vol.: 1}.
\newblock Jul. Springer, 1935.

\bibitem{todd1993build}
T.~N. Todd.
\newblock How to build a tokamak.
\newblock In {\em Plasma Physics: an Introductory Course}, volume~1, page 443,
  1993.

\bibitem{wesson2004tokamaks}
J.~Wesson.
\newblock {\em Tokamaks}.
\newblock Oxford University Press, 2004.

\bibitem{hirsch2008major}
M.~Hirsch, J.~Baldzuhn, C.~Beidler, R.~Brakel, R.~Burhenn, A.~Dinklage,
  H.~Ehmler, M.~Endler, V.~Erckmann, Y.~Feng, et~al.
\newblock Major results from the stellarator {Wendelstein 7-AS}.
\newblock {\em Plasma\ Phys.\ Control.\ Fusion}, 50:053001, 2008.

\bibitem{motojima2003recent}
O.~Motojima, N.~Ohyabu, A.~Komori, O.~Kaneko, H.~Yamada, K.~Kawahata,
  Y.~Nakamura, K.~Ida, T.~Akiyama, N.~Ashikawa, et~al.
\newblock Recent advances in the {LHD} experiment.
\newblock {\em Nucl. Fusion}, 43:1674, 2003.

\bibitem{hastie1993plasma}
R.~J. Hastie.
\newblock Plasma particle dynamics.
\newblock In {\em Plasma Physics: an Introductory Course}, volume~1, page~5,
  1993.

\bibitem{blank2010guiding}
H.~J. de~Blank.
\newblock Guiding centre motion.
\newblock {\em Transactions of Fusion Science and Technology}, 41(2T):61--68,
  2010.

\bibitem{helander2002ctm}
P.~Helander and D.~J. Sigmar.
\newblock {\em Collisional Transport in Magnetized Plasmas}.
\newblock Cambridge University Press, 2002.

\bibitem{hintonwong1985nit}
F.~L. Hinton and S.~K. Wong.
\newblock Neoclassical ion transport in rotating axisymmetric plasmas.
\newblock {\em Phys.\ Fluids}, 28:3082, 1985.

\bibitem{galambos1995commercial}
J.~D. Galambos, L.~J. Perkins, S.~W. Haney, and J.~Mandrekas.
\newblock Commercial tokamak reactor potential with advanced tokamak operation.
\newblock {\em Nucl. Fusion}, 35:551, 1995.

\bibitem{colas1998internal}
L.~Colas, X.~L. Zou, M.~Paume, J.~M. Chareau, L.~Guiziou, G.~T. Hoang,
  Y.~Michelot, and D.~Gresillon.
\newblock Internal magnetic fluctuations and electron heat transport in the
  {Tore Supra} tokamak: Observation by cross-polarization scattering.
\newblock {\em Nucl. Fusion}, 38:903, 1998.

\bibitem{horton1980fluid}
W.~Horton and R.~D. Estes.
\newblock Fluid simulation of ion pressure gradient driven drift modes.
\newblock {\em Plasma\ Phys.}, 22:663, 1980.

\bibitem{waltz1988three}
R.~E. Waltz.
\newblock Three-dimensional global numerical simulation of ion temperature
  gradient mode turbulence.
\newblock {\em Phys.\ Fluids}, 31:1962, 1988.

\bibitem{fonck1989ion}
R.~J. Fonck, R.~Howell, K.~Jaehnig, L.~Roquemore, G.~Schilling, S.~Scott, M.~C.
  Zarnstorff, C.~Bush, R.~Goldston, H.~Hsuan, et~al.
\newblock Ion thermal confinement in the enhanced-confinement regime of the
  {TFTR} tokamak.
\newblock {\em Phys.\ Rev.\ Lett.}, 63(5):520--523, 1989.

\bibitem{wootton1990fluctuations}
A.~J. Wootton, B.~A. Carreras, H.~Matsumoto, K.~McGuire, W.~A. Peebles, C.P.
  Ritz, P.~W. Terry, and S.~J. Zweben.
\newblock {Fluctuations and anomalous transport in tokamaks}.
\newblock {\em Physics of Fluids B: Plasma Physics}, 2:2879, 1990.

\bibitem{cowley1991considerations}
S.~C. Cowley, R.~M. Kulsrud, and R.~Sudan.
\newblock {Considerations of ion-temperature-gradient-driven turbulence}.
\newblock {\em Phys.\ Fluids B}, 3:2767, 1991.

\bibitem{kotschenreuther1995quantitative}
M.~Kotschenreuther, W.~Dorland, M.~A. Beer, and G.~W. Hammett.
\newblock Quantitative predictions of tokamak energy confinement from
  first-principles simulations with kinetic effects.
\newblock {\em Phys.\ Plasmas}, 2:2381, 1995.

\bibitem{carreras1997progress}
B.A. Carreras.
\newblock {Progress in anomalous transport research in toroidal
  magneticconfinement devices}.
\newblock {\em IEEE Trans. on Plasma Sci.}, 25(6):1281--1321, 1997.

\bibitem{dimits:969}
A.~M. Dimits, G.~Bateman, M.~A. Beer, B.~I. Cohen, W.~Dorland, G.~W. Hammett,
  C.~Kim, J.~E. Kinsey, M.~Kotschenreuther, A.~H. Kritz, L.~L. Lao,
  J.~Mandrekas, W.~M. Nevins, S.~E. Parker, A.~J. Redd, D.~E. Shumaker,
  R.~Sydora, and J.~Weiland.
\newblock Comparisons and physics basis of tokamak transport models and
  turbulence simulations.
\newblock {\em Phys.\ Plasmas}, 7:969, 2000.

\bibitem{dorlandETG}
W.~Dorland, F.~Jenko, M.~Kotschenreuther, and B.~N. Rogers.
\newblock Electron temperature gradient turbulence.
\newblock {\em Phys.\ Rev.\ Lett.}, 85:5579, 2000.

\bibitem{jenko:1904}
F.~Jenko, W.~Dorland, M.~Kotschenreuther, and B.~N. Rogers.
\newblock Electron temperature gradient driven turbulence.
\newblock {\em Phys.\ Plasmas}, 7:1904, 2000.

\bibitem{dannert2005gyrokinetic}
T.~Dannert and F.~Jenko.
\newblock Gyrokinetic simulation of collisionless trapped-electron mode
  turbulence.
\newblock {\em Phys.\ Plasmas}, 12:072309, 2005.

\bibitem{burrell1997effects}
K.~H. Burrell.
\newblock {Effects of E x B velocity shear and magnetic shear on turbulence and
  transport in magnetic confinement devices}.
\newblock {\em Phys.\ Plasmas}, 4(5):1499, 1997.

\bibitem{wagner1984development}
F.~Wagner, G.~Fussmann, T.~Grave, M.~Keilhacker, M.~Kornherr, K.~Lackner,
  K.~McCormick, E.~R. M{\"u}ller, A.~St{\"a}bler, G.~Becker, et~al.
\newblock Development of an edge transport barrier at the {H-mode} transition
  of {ASDEX}.
\newblock {\em Phys.\ Rev.\ Lett.}, 53(15):1453--1456, 1984.

\bibitem{putterich2009evidence}
T.~P{\"u}tterich, E.~Wolfrum, R.~Dux, and C.~F. Maggi.
\newblock Evidence for strong inversed shear of toroidal rotation at the
  edge-transport barrier in the {ASDEX Upgrade}.
\newblock {\em Phys.\ Rev.\ Lett.}, 102(2):25001, 2009.

\bibitem{synakowski1999comparative}
E.~J. Synakowski, M.~A. Beer, R.~E. Bell, K.~H. Burrell, B.~A. Carreras, P.~H.
  Diamond, E.~J. Doyle, D.~Ernst, R.~J. Fonck, P.~Gohil, et~al.
\newblock Comparative studies of core and edge transport barrier dynamics of
  {DIII-D and TFTR} tokamak plasmas.
\newblock {\em Nucl. Fusion}, 39:1733, 1999.

\bibitem{oost2003turbulent}
G.V. Oost, J.~Ad{\'a}mek, V.~Antoni, P.~Balan, J.~A. Boedo, P.~Devynck,
  I.~{\v{D}}uran, L.~Eliseev, J.~P. Gunn, M.~Hron, et~al.
\newblock Turbulent transport reduction by {E$\times$ B} velocity shear during
  edge plasma biasing: recent experimental results.
\newblock {\em Plasma\ Phys.\ Control.\ Fusion}, 45:621, 2003.

\bibitem{wagner1984regime}
F.~Wagner, G.~Becker, K.~Behringer, D.~Campbell, A.~Eberhagen, W.~Engelhardt,
  G.~Fussmann, O.~Gehre, J.~Gernhardt, G.~v. Gierke, et~al.
\newblock Regime of improved confinement and high beta in neutral-beam-heated
  divertor discharges of the asdex tokamak.
\newblock {\em Phys.\ Rev.\ Lett.}, 49:1408--1412, Nov 1982.

\bibitem{akers2003transport}
R.~J. Akers, J.~W. Ahn, G.~Y. Antar, L.~C. Appel, D.~Applegate, C.~Brickley,
  C.~Bunting, P.~G. Carolan, C.~D. Challis, N.~J. Conway, et~al.
\newblock Transport and confinement in the {Mega Amp{\`e}re Spherical Tokamak
  (MAST)} plasma.
\newblock {\em Plasma\ Phys.\ Control.\ Fusion}, 45:A175, 2003.

\bibitem{connor2004itbreview}
J.~W. Connor, T.~Fukuda, X.~Garbet, C.~Gormezano, V.~Mukhovatov, M.~Wakatani,
  et~al.
\newblock {A review of internal transport barrier physics for steady-state
  operation of tokamaks}.
\newblock {\em Nucl. Fusion}, 44:R1, 2004.

\bibitem{wolf2003internal}
R.~C. Wolf.
\newblock {Internal transport barriers in tokamak plasmas*}.
\newblock {\em Plasma\ Phys.\ Control.\ Fusion}, 45:R1, 2003.

\bibitem{synakowski1997local}
E.~J. Synakowski, S.~H. Batha, M.~A. Beer, M.~G. Bell, R.~E. Bell, R.~V. Budny,
  C.~E. Bush, P.~C. Efthimion, T.~S. Hahm, G.~W. Hammett, et~al.
\newblock Local transport barrier formation and relaxation in reverse-shear
  plasmas on the {Tokamak Fusion Test Reactor}.
\newblock {\em Phys.\ Plasmas}, 4:1736, 1997.

\bibitem{waltz1994toroidal}
R.~E. Waltz, G.~D. Kerbel, and J.~Milovich.
\newblock {Toroidal gyro-Landau fluid model turbulence simulations in a
  nonlinear ballooning mode representation with radial modes}.
\newblock {\em Phys.\ Plasmas}, 1:2229, 1994.

\bibitem{waltz1997gyro}
R.~E. Waltz, G.~M. Staebler, W.~Dorland, G.~W. Hammett, M.~Kotschenreuther, and
  J.~A. Konings.
\newblock {A gyro-Landau-fluid transport model}.
\newblock {\em Phys.\ Plasmas}, 4(7):2482--2496, 1997.

\bibitem{dimits2001parameter}
A.~M. Dimits, B.~I. Cohen, W.~M. Nevins, and D.~E. Shumaker.
\newblock {Parameter dependences of ion thermal transport due to toroidal ITG
  turbulence}.
\newblock {\em Nucl. Fusion}, 41:1725, 2001.

\bibitem{kinsey2005flowshear}
J.~E. Kinsey, R.~E. Waltz, and J.~Candy.
\newblock Nonlinear gyrokinetic turbulence simulations of e x b shear quenching
  of transport.
\newblock {\em Phys.\ Plasmas}, 12:062302, 2005.

\bibitem{camenen2009impact}
Y.~Camenen, A.~G. Peeters, C.~Angioni, F.~J. Casson, W.~A. Hornsby, A.~P.
  Snodin, and D.~Strintzi.
\newblock {Impact of the background toroidal rotation on particle and heat
  turbulent transport in tokamak plasmas}.
\newblock {\em Phys.\ Plasmas}, 16:012503, 2009.

\bibitem{roach2009gss}
C.~M. Roach, I.~G. Abel, R.~J. Akers, W.~Arter, M.~Barnes, Y.~Camenen, F.~J.
  Casson, G.~Colyer, J.~W. Connor, S.~C. Cowley, et~al.
\newblock {Gyrokinetic simulations of spherical tokamaks}.
\newblock {\em Plasma\ Phys.\ Control.\ Fusion}, 51:124020, 2009.

\bibitem{barnes2011turbulent}
M.~Barnes, F.~I. Parra, E.~G. Highcock, A.~A. Schekochihin, S.~C. Cowley, and
  C.~M. Roach.
\newblock Turbulent transport in tokamak plasmas with rotational shear.
\newblock {\em Phys.\ Rev.\ Lett.}, 106(17):175004, 2011.

\bibitem{highcock2010transport}
E.~G. Highcock, M.~Barnes, A.~A. Schekochihin, F.~I. Parra, C.~Roach, and S.~C.
  Cowley.
\newblock Transport bifurcation in a rotating tokamak plasma.
\newblock {\em Phys. Rev. Lett.}, 105:215003, Nov 2010.

\bibitem{catto1973parallel}
P.J. Catto, M.N. Rosenbluth, and C.S. Liu.
\newblock {Parallel velocity shear instabilities in an inhomogeneous plasma
  with a sheared magnetic field}.
\newblock {\em Phys.\ Fluids}, 16:1719, 1973.

\bibitem{artun1993integral}
M.~Artun, J.~V.~W. Reynders, and W.~M. Tang.
\newblock {Integral eigenmode analysis of shear flow effects on the ion
  temperature gradient mode}.
\newblock {\em Physics of Fluids B: Plasma Physics}, 5:4072, 1993.

\bibitem{newton2010understanding}
S.~L. Newton, S.~C. Cowley, and N.~F. Loureiro.
\newblock {Understanding the effect of sheared flow on microinstabilities}.
\newblock {\em Plasma\ Phys.\ Control.\ Fusion}, 52:125001, 2010.

\bibitem{baggett1995mostly}
J.S. Baggett, T.A. Driscoll, and L.N. Trefethen.
\newblock {A mostly linear model of transition to turbulence}.
\newblock {\em Phys.\ Fluids}, 7:833, 1995.

\bibitem{sips1998operation}
A.~C.~C. Sips, Y.~Baranov, C.~D. Challis, G.~A. Cottrell, L.~G. Eriksson,
  C.~Gormezano, C.~Gowers, C.~M. Greenfield, J.C.M. Haas, M.~Hellerman, et~al.
\newblock Operation at high performance in optimized shear plasmas in {JET}.
\newblock {\em Plasma\ Phys.\ Control.\ Fusion}, 40:1171, 1998.

\bibitem{vries2009internal}
P.~C. Vries, E.~Joffrin, M.~Brix, C.~D. Challis, K.~Cromb{\'e}, B.~Esposito,
  N.~C. Hawkes, C.~Giroud, J.~Hobirk, J.~L{\"o}nnroth, et~al.
\newblock {Internal transport barrier dynamics with plasma rotation in JET}.
\newblock {\em Nucl. Fusion}, 49:075007, 2009.

\bibitem{flowtome1}
I.~G. Abel et~al.
\newblock Multiscale gyrokinetics for rotating tokamak plasmas {I}:
  {Fluctuations} and transport.
\newblock {\em Plasma Phys. Control. Fusion, in preparation}, 2012.

\bibitem{frieman1982nge}
E.~A. Frieman and L.~Chen.
\newblock {Nonlinear gyrokinetic equations for low-frequency electromagnetic
  waves in general plasma equilibria}.
\newblock {\em Phys.\ Fluids}, 25:502, 1982.

\bibitem{sugama1998neg}
H.~Sugama and W.~Horton.
\newblock Nonlinear electromagnetic gyrokinetic equation for plasmas with large
  mean flows.
\newblock {\em Phys.\ Plasmas}, 5:2560, 1998.

\bibitem{catto1987ion}
P.J. Catto, I.B. Bernstein, and M.~Tessarotto.
\newblock {Ion transport in toroidally rotating tokamak plasmas}.
\newblock {\em Phys.\ Fluids}, 30:2784, 1987.

\bibitem{peeters2007toroidal}
A.~G. Peeters, C.~Angioni, and D.~Strintzi.
\newblock {Toroidal momentum pinch velocity due to the coriolis drift effect on
  small scale instabilities in a toroidal plasma}.
\newblock {\em Phys.\ Rev.\ Lett.}, 98:265003, 2007.

\bibitem{casson2010gyrokinetic}
F.J. Casson, A.~G. Peeters, C.~Angioni, Y.~Camenen, W.~A. Hornsby, A.~P.
  Snodin, and G.~Szepesi.
\newblock Gyrokinetic simulations including the centrifugal force in a rotating
  tokamak plasma.
\newblock {\em Phys.\ Plasmas}, 17:102305, 2010.

\bibitem{hammett1993developments}
G.~W. Hammett, M.~A. Beer, W.~Dorland, S.~C. Cowley, and S.~A. Smith.
\newblock Developments in the gyrofluid approach to tokamak turbulence
  simulations.
\newblock {\em Plasma\ Phys.\ Control.\ Fusion}, 35(8):973, 1993.

\bibitem{gyrokineticswebsite}
W.~Dorland, E.~G. Highcock, M.~Barnes, G.~W. Hammett, G.~Colyer, et~al.
\newblock {Gyrokinetic Simulations Project}.
\newblock {\em http://gyrokinetics.sourceforge.net/}, 2009.

\bibitem{jenko2001nonlinear}
F.~Jenko and W.~Dorland.
\newblock Nonlinear electromagnetic gyrokinetic simulations of tokamak plasmas.
\newblock {\em Plasma\ Phys.\ Control.\ Fusion}, 43:A141, 2001.

\bibitem{gorler2011global}
T.~G{\"o}rler, X.~Lapillonne, S.~Brunner, T.~Dannert, F.~Jenko, F.~Merz, and
  D.~Told.
\newblock The global version of the gyrokinetic turbulence code {GENE}.
\newblock {\em J.\ Comp.\ Phys.}, 2011.

\bibitem{candy2003eulerian}
J.~Candy and R.~E. Waltz.
\newblock An {Eulerian} gyrokinetic-{Maxwell} solver.
\newblock {\em J.\ Comp.\ Phys.}, 186(2):545--581, 2003.

\bibitem{fahey2004gyro}
M.R. Fahey and J.~Candy.
\newblock {GYRO}: A 5-d gyrokinetic-{Maxwell} solver.
\newblock In {\em Proceedings of the 2004 ACM/IEEE conference on
  Supercomputing}, page~26. IEEE Computer Society, 2004.

\bibitem{gkwRef}
A.~G. Peeters, Y.~Camenen, F.~J. Casson, W.~A. Hornsby, A.~P. Snodin,
  D.~Strintzi, and G.~Szepesi.
\newblock The nonlinear gyro-kinetic flux tube code {GKW}.
\newblock {\em Comp.\ Phys.\ Comm.}, 180:2650, 2009.

\bibitem{beer1995field}
M.~A. Beer, S.~C. Cowley, and G.~W. Hammett.
\newblock Field-aligned coordinates for nonlinear simulations of tokamak
  turbulence.
\newblock {\em Phys.\ Plasmas}, 2:2687, 1995.

\bibitem{gs2ref}
M.~Kotschenreuther, G.~W. Rewoldt, and W.~M. Tang.
\newblock Comparison of initial value and eigenvalue codes for kinetic toroidal
  plasma instabilities.
\newblock {\em Comp.\ Phys.\ Comm.}, 88:128, 1995.

\bibitem{dorland-microinstabilities}
W.~Dorland and M.~Kotschenreuther.
\newblock {Microinstabilities in Axisymmetric Configurations}.

\bibitem{belli2006studies}
E.A. Belli.
\newblock {\em Studies of numerical algorithms for gyrokinetics and the effects
  of shaping on plasma turbulence (Ph.D. Thesis)}.
\newblock Ann Arbor, 2006.

\bibitem{abel2008linearized}
I.~G. Abel, M.~Barnes, S.~C. Cowley, W.~Dorland, and A.~A. Schekochihin.
\newblock {Linearized model Fokker--Planck collision operators for gyrokinetic
  simulations. I. Theory}.
\newblock {\em Phys.\ Plasmas}, 15:122509, 2008.

\bibitem{numerics}
M.~Barnes, I.~G. Abel, T.~Tatsuno, A.~A. Schekochihin, S.~C. Cowley, and
  W.~Dorland.
\newblock Linearized model {Fokker-Planck} collision operators for gyrokinetic
  simulations, {II.} numerics.
\newblock {\em Phys.\ Plasmas}, 16:072107, 2008.

\bibitem{numata2010astrogk}
R.~Numata, G.G. Howes, T.~Tatsuno, M.~Barnes, and W.~Dorland.
\newblock Astrogk: Astrophysical gyrokinetics code.
\newblock {\em J.\ Comp.\ Phys.}, 229(24):9347--9372, 2010.

\bibitem{michaelthesis}
M.~Barnes.
\newblock {\em Trinity: A Unified Treatment of Turbulence, Transport, and
  Heating in Magnetized Plasmas ; Ph.D. Thesis (U. Maryland)}.
\newblock 2008.

\bibitem{gs2flowshear}
G.~W. Hammett, W.~Dorland, N.~F. Loureiro, and T.~Tatsuno.
\newblock {\em Bull. Am. Phys. Soc}, 2006.
\newblock Abstract VP1.136.

\bibitem{hazeltine2003plasma}
R.D. Hazeltine and J.D. Meiss.
\newblock {\em Plasma confinement}.
\newblock Dover Pubns, 2003.

\bibitem{kruskal1958equilibrium}
M.D. Kruskal and R.~M. Kulsrud.
\newblock Equilibrium of a magnetically confined plasma in a toroid.
\newblock {\em Phys.\ Fluids}, 1:265, 1958.

\bibitem{orszag1970transform}
S.A. Orszag.
\newblock Transform method for the calculation of vector-coupled sums:
  Application to the spectral form of the vorticity equation.
\newblock {\em J.\ Atmos.\ Sci.}, 27:890--895, 1970.

\bibitem{platzman1960spectral}
G.W. Platzman.
\newblock The spectral form of the vorticity equation.
\newblock {\em J.\ Atmos.\ Sci.}, 17:635--644, 1960.

\bibitem{vries2008effect}
P.~C. De~Vries, A.~Salmi, V.~Parail, C.~Giroud, Y.~Andrew, T.M. Biewer,
  K.~Cromb{\'e}, I.~Jenkins, T.~Johnson, V.~Kiptily, et~al.
\newblock Effect of toroidal field ripple on plasma rotation in {JET}.
\newblock {\em Nucl. Fusion}, 48:035007, 2008.

\bibitem{miller1998noncircular}
R.~L. Miller, M.~S. Chu, J.~M. Greene, Y.~R. Lin-Liu, and R.~E. Waltz.
\newblock Noncircular, finite aspect ratio, local equilibrium model.
\newblock {\em Phys.\ Plasmas}, 5:973, 1998.

\bibitem{greene1981second}
J.~M. Greene and M.~S. Chance.
\newblock The second region of stability against ballooning modes.
\newblock {\em Nucl. Fusion}, 21:453, 1981.

\bibitem{lapillonne2009clarifications}
X.~Lapillonne, S.~Brunner, T.~Dannert, S.~Jolliet, A.~Marinoni, L.~Villard,
  T.~G{\"o}rler, F.~Jenko, and F.~Merz.
\newblock Clarifications to the limitations of the s-$\alpha$ equilibrium model
  for gyrokinetic computations of turbulence.
\newblock {\em Phys.\ Plasmas}, 16:032308, 2009.

\bibitem{Tome}
A.~A. Schekochihin, S.~C. Cowley, W.~Dorland, G.~W. Hammett, G.~G. Howes,
  E.~Quataert, and T.~Tatsuno.
\newblock {Astrophysical gyrokinetics: kinetic and fluid turbulent cascades in
  weakly collisional astrophysical plasmas}.
\newblock {\em Astrophys.\ J.\ Suppl.}, 182:310, 2009.

\bibitem{beam1976implicit}
R.M. Beam and R.~F. Warming.
\newblock An implicit finite-difference algorithm for hyperbolic systems in
  conservation-law form.
\newblock {\em J.\ Comp.\ Phys.}, 22(1):87--110, 1976.

\bibitem{barnes2009trinity}
M.~A. Barnes, I.~G. Abel, W.~Dorland, T.~G{\"o}rler, G.~W. Hammett, and
  F.~Jenko.
\newblock Direct multiscale coupling of a transport code to gyrokinetic
  turbulence codes.
\newblock {\em Phys.\ Plasmas}, 17:056109, 2010.

\bibitem{schekcrete}
A.~A. Schekochihin, S.~C. Cowley, W.~Dorland, G.~W. Hammett, G.~G. Howes, G.~G.
  Plunk, E.~Quataert, and T.~Tatsuno.
\newblock Gyrokinetic turbulence: a nonlinear route to dissipation through
  phase space.
\newblock {\em Plasma\ Phys.\ Control.\ Fusion}, 50:124024, 2008.

\bibitem{gham_gs2vecshear_unpub}
G.~W. Hammett.
\newblock Notes on adding equilibrium scale {ExB} shear to {GS2}.
\newblock 2006.

\bibitem{highcock2011transport}
E.~G. Highcock, M.~Barnes, F.~I. Parra, A.~A. Schekochihin, C.~M. Roach, and
  S.~C. Cowley.
\newblock Transport bifurcation induced by sheared toroidal flow in tokamak
  plasmas.
\newblock {\em Phys.\ Plasmas}, 18:102304, 2011.

\bibitem{parra2011momentum}
F.~I. Parra, M.~Barnes, E.~G. Highcock, A.~A. Schekochihin, and S.~C. Cowley.
\newblock Momentum injection in tokamak plasmas and transitions to reduced
  transport.
\newblock {\em Phys.\ Rev.\ Lett.}, 106:115004, Mar 2011.

\bibitem{parra2011up}
F.I. Parra, M.~Barnes, and A.G. Peeters.
\newblock Up-down symmetry of the turbulent transport of toroidal angular
  momentum in tokamaks.
\newblock {\em Phys.\ Plasmas}, 18:062501, 2011.

\bibitem{camenen2009intrinsic}
Y.~Camenen, A.~G. Peeters, C.~Angioni, F.~J. Casson, W.~A. Hornsby, A.~P.
  Snodin, and D.~Strintzi.
\newblock {Intrinsic rotation driven by the electrostatic turbulence in up-down
  asymmetric toroidal plasmas}.
\newblock {\em Phys.\ Plasmas}, 16:062501, 2009.

\bibitem{dominguez1993anomalous}
R.~R. Dominguez and G.~M. Staebler.
\newblock {Anomalous momentum transport from drift wave turbulence}.
\newblock {\em Phys.\ Fluids B}, 5:3876, 1993.

\bibitem{waltz2007gyrokinetic}
R.~E. Waltz, G.~M. Staebler, J.~Candy, and F.~L. Hinton.
\newblock {Gyrokinetic theory and simulation of angular momentum transport}.
\newblock {\em Phys.\ Plasmas}, 14:122507, 2007.

\bibitem{waltz1995advances}
R.~E. Waltz, G.~D. Kerbel, J.~Milovich, and G.~W. Hammett.
\newblock {Advances in the simulation of toroidal gyro-Landau fluid model
  turbulence}.
\newblock {\em Phys.\ Plasmas}, 2:2408, 1995.

\bibitem{candy2009tgyro}
J.~Candy, C.~Holland, R.~E. Waltz, M.~R. Fahey, and E.~Belli.
\newblock Tokamak profile prediction using direct gyrokinetic and neoclassical
  simulation.
\newblock {\em Phys.\ Plasmas}, 16:060704, 2009.

\bibitem{buhmann2001radial}
M.D. Buhmann.
\newblock {Radial basis functions}.
\newblock {\em Acta Numerica}, 9:1--38, 2001.

\bibitem{field2004core}
A.~R. Field, R.~J. Akers, D.~J. Applegate, C.~Brickley, P.~G. Carolan,
  C.~Challis, N.~J. Conway, S.~C. Cowley, G.~Cunningham, N.~Joiner, et~al.
\newblock Core heat transport in the mast spherical tokamak.
\newblock In {\em Proc. 20th IAEA Fusion Energy Conf. on Fusion Energy 2004
  (Vilamoura, 2004)}. IAEA, Vienna, 2004.

\bibitem{manticastiffness}
P.~Mantica, D.~Strintzi, T.~Tala, C.~Giroud, T.~Johnson, H.~Leggate, E.~Lerche,
  T.~Loarer, A.~G. Peeters, A.~Salmi, S.~Sharapov, D.~Van~Eester, P.~C.
  de~Vries, L.~Zabeo, and K.-D. Zastrow.
\newblock Experimental study of the ion critical-gradient length and stiffness
  level and the impact of rotation in the {JET} tokamak.
\newblock {\em Phys. Rev. Lett.}, 102:175002, Apr 2009.

\bibitem{barnes2010shear}
M.~Barnes, F.~I. Parra, E.~Highcock, A.~A. Schekochihin, S.~C. Cowley, and
  C.~M. Roach.
\newblock Shear flow suppression of turbulent transport and self-consistent
  profile evolution within a multi-scale gyrokinetic framework.
\newblock {\em IAEA, Korea}, 2010.
\newblock Abstract THC/P4-01.

\bibitem{roach2008profiledb}
C.~M. Roach, M.~Walters, R.~V. Budny, F.~Imbeaux, T.~W. Fredian, M.~Greenwald,
  J.~A. Stillerman, D.~A. Alexander, J.~Carlsson, J.~R. Cary, et~al.
\newblock {The 2008 Public Release of the International Multi-tokamak
  Confinement Profile Database}.
\newblock {\em Nucl. Fusion}, 48:125001, 2008.

\bibitem{ccfewebsite}
{Culham Centre for Fusion Energy Website}.
\newblock {\em http://www.ccfe.ac.uk/}, 2011.

\bibitem{trefethen1993hydrodynamic}
L.~Trefethen, A.~Trefethen, S.~Reddy, and T.~Driscoll.
\newblock {Hydrodynamic stability without eigenvalues}.
\newblock {\em Science}, 261(5121):578--584, 1993.

\bibitem{kerswell2005recent}
R.~R. Kerswell.
\newblock Recent progress in understanding the transition to turbulence in a
  pipe.
\newblock {\em Nonlinearity}, 18:R17, 2005.

\bibitem{barnes2011critically}
M.~Barnes, F.~I. Parra, and A.~A. Schekochihin.
\newblock Critically balanced ion temperature gradient turbulence in fusion
  plasmas.
\newblock {\em Phys.\ Rev.\ Lett.}, 107(11):115003, 2011.

\bibitem{krommes2002fundamental}
J.~A. Krommes.
\newblock {Fundamental statistical descriptions of plasma turbulence in
  magnetic fields}.
\newblock {\em Phys.\ Rep.}, 360(1-4):1--352, 2002.

\bibitem{bourdelle2007new}
C.~Bourdelle, X.~Garbet, F.~Imbeaux, A.~Casati, N.~Dubuit, R.~Guirlet, and
  T.~Parisot.
\newblock {A new gyrokinetic quasilinear transport model applied to particle
  transport in tokamak plasmas}.
\newblock {\em Phys.\ Plasmas}, 14:112501, 2007.

\bibitem{waltz2009gyrokinetic}
R.~E. Waltz, A.~Casati, and G.~M. Staebler.
\newblock {Gyrokinetic simulation tests of quasilinear and tracer transport}.
\newblock {\em Phys.\ Plasmas}, 16:072303, 2009.

\bibitem{schekochihin2010subcritical}
A.~A. Schekochihin, E.~G. Highcock, and S.~C. Cowley.
\newblock Subcritical fluctuations in rotating gyrokinetic plasmas.
\newblock {\em Plasma Phys. Control. Fusion, in press}, 2012.
\newblock (e-print arXiv:1111.4929).

\bibitem{mantica2011ion}
P.~Mantica, C.~Angioni, B.~Baiocchi, M.~Baruzzo, M.~N.~A. Beurskens, J.~P.~S.
  Bizarro, R.~V. Budny, P.~Buratti, A.~Casati, C.~Challis, et~al.
\newblock Ion heat transport studies in {JET}.
\newblock {\em Plasma\ Phys.\ Control.\ Fusion}, 53:124033, 2011.

\bibitem{team1999alpha}
{The JET Team}.
\newblock Alpha particle studies during {JET DT} experiments.
\newblock {\em Nucl. Fusion}, 39(11):1619--1625, 1999.

\bibitem{petty1999dependence}
C.~C. Petty, M.~R. Wade, J.~E. Kinsey, R.~J. Groebner, T.~C. Luce, and G.~M.
  Staebler.
\newblock Dependence of heat and particle transport on the ratio of the ion and
  electron temperatures.
\newblock {\em Phys.\ Rev.\ Lett.}, 83:3661--3664, Nov 1999.

\bibitem{field2011plasma}
A.~R. Field, C.~Michael, R.~J. Akers, J.~Candy, G.~Colyer, W.~Guttenfelder,
  Y.~Ghim, C.~M. Roach, and S.~Saarelma.
\newblock Plasma rotation and transport in mast spherical tokamak.
\newblock {\em Nucl. Fusion}, 51:063006, 2011.

\bibitem{fried1961pdf}
B.~D. Fried and S.~D. Conte.
\newblock {\em {The plasma dispersion function: the Hilbert transform of the
  Gaussian}}.
\newblock Academic Press, 1961.

\bibitem{parra2010plausible}
F.~I. Parra, M.~Barnes, E.~G. Highcock, A.~A. Schekochihin, and S.~C. Cowley.
\newblock Momentum injection in tokamak plasmas and transitions to reduced
  transport.
\newblock {\em Phys.\ Rev.\ Lett.}, 106:115004, 2011.

\bibitem{highcock2012zero}
E.~G. {Highcock}, A.~A. {Schekochihin}, S.~C. {Cowley}, M.~{Barnes}, F.~I.
  {Parra}, C.~M. {Roach}, and W.~{Dorland}.
\newblock Zero-turbulence manifold in a toroidal plasma.
\newblock {\em Phys. Rev. Lett., submitted}, 2012.
\newblock arXiv-eprint 1203.6455.

\end{thebibliography}
}

\end{document}